\documentclass[12pt,preprint]{aastex}
\usepackage{natbib}
\usepackage{graphicx}

\newcommand{\sub}[1]{\ensuremath{_{\mbox{\scriptsize#1}}}}
\newcommand{\upp}[1]{\ensuremath{^{\mbox{\scriptsize#1}}}}

\newcommand{\noprint}[1]{}

\DeclareGraphicsExtensions{.ps}

\newlength{\stampwidth}
\begin{document}

\slugcomment{To appear in ApJ}
\title{\emph{Spitzer} IRS Spectra of Debris Disks in the Scorpius-Centaurus OB Association}
\author{Hannah Jang-Condell\altaffilmark{1}, 
  Christine H.~Chen\altaffilmark{2}, 
  Tushar Mittal\altaffilmark{2,3},
  P.~Manoj\altaffilmark{4},
  Dan Watson\altaffilmark{5},
  Carey M.~Lisse\altaffilmark{2,6},
  Erika Nesvold\altaffilmark{7,8},
  Marc Kuchner\altaffilmark{7}
}
\altaffiltext{1}{Department of Physics and Astronomy, University of Wyoming, 
Laramie, WY 82071}
\altaffiltext{2}{Space Telescope Science Institute, 3700 San Martin Dr., 
Baltimore, MD 21218}
\altaffiltext{3}{Department of Earth and Planetary Science, University of California, Berkeley, CA 94720}
\altaffiltext{4}{Department of Astronomy \& Astrophysics,  Tata Institute of Fundamental Research, Homi Bhabha Rd, Mumbai  400005, India}
\altaffiltext{5}{Department of Physics and Astronomy, University of Rochester, 
Rochester, NY 14627}
\altaffiltext{6}{Johns Hopkins University Applied Physics Laboratory,
 Laurel, MD 20723}
\altaffiltext{7}{NASA Goddard Space Flight Center, Greenbelt, MD}
\altaffiltext{8}{University of Maryland Baltimore County, Baltimore, MD 21250}

\begin{abstract}
We analyze Spitzer/IRS spectra of 110 B-, A-, F-, and G-type stars 
with optically thin infrared excess in the Scorpius-Centaurus
(ScoCen) OB association.  The age of these stars ranges from 
11-17 Myr.  We fit the infrared excesses observed 
in these sources by Spitzer/IRS and Spitzer/MIPS to simple dust 
models according to Mie theory.  We find that nearly all the objects 
in our study can be fit by one or two belts of dust.  Dust around 
lower mass stars appears to be closer in than around higher 
mass stars, particularly for the warm dust component in the two-belt 
systems, suggesting mass-dependent evolution of debris disks 
around young stars.
For those objects with stellar companions, all dust
distances are consistent with trunction of the debris disk by 
the binary companion.
The gaps between several of the two-belt 
systems can place limits on the planets that might lie between 
the belts, potentially constraining the mass and locations of planets 
that may be forming around these stars.  
\end{abstract}


\section{Introduction}

High contrast imaging surveys using adaptive optics (AO) enabled large 
telescopes are beginning to discover Jovian planets in nascent planetary 
systems. These surveys typically target nearby ($<$100 pc), young ($<$300 Myr) 
stars because atmospheric modeling of Jovian planets predicts that their 
self-emission is bright when they are young and fades with time,
because of their captured heat of formation and ongoing gravitational contraction
\citep{marley07}. Debris disks are dusty disks around main sequence stars that 
are typically discovered via excess thermal infrared emission above 
the stellar photosphere. \emph{Spitzer} MIPS surveys of young clusters and 
field stars indicate that young stars generally possess larger infrared excess 
than old stars \citep{2006Su_etal,carpenter09a}. In fact, the first images 
of Jovian exoplanets were made in the dusty debris disk systems, HR 8799 
\citep{Marois2010} and $\beta$ Pictoris \citep{Lagrange2010}. Mid- to far-infrared 
excess was originally discovered toward these targets more than two decades
ago using the \emph{IRAS} satellite
\citep{1984Aumann,backman93}.

Studies of the spatial distribution of dust in debris disks can provide 
constraints on the presence of planets. For example, SED modeling of the 
thermal emission from the dust around HR 8799 indicates the presence of two 
dust populations generated by two separate parent body belts, one interior to 
10 AU and another beyond 100 AU \citep{su09,chen09}. The system of four $\sim$10
$M_{Jup}$ planets discovered using the Keck Telescope lies between these two 
dust populations at distances of 15, 24, 38, and 68 AU \citep{Marois2010}. 
Similarly, SED modeling of \emph{IRAS} excesses toward $\beta$ Pictoris 
indicates that the dust is located at distances $>$20 AU \citep{backman93} and
the Jovian planet $\beta$ Pic b lies inside the central clearing at 10 AU 
\citep{Lagrange2010}. SMA observations of $\beta$ Pic have further revealed a 
ring of millimeter-sized grains at $\sim$94 AU that has been identified as the 
location of the main reservoir of dust-producing planetesimals 
\citep{wilner11}. The architecture of the $\beta$ Pic system is consistent with
creation of dust by collisions among parent bodies in the main belt, the larger
of which spiral inward under Poynting-Robertson drag until they encounter 
$\beta$ Pic b and are scattered out of the system.  
A study of strong silicate emission at 8-13
$\mu$m of $\beta$ Pic finds evidence for belts at 6, 16, and 30 AU
\citep{2004Okamoto+}.  These belts may also be sculpted by
$\beta$ Pic b, and possibly additional as-yet-undiscovered planets.  
The infrared spectrum of the dust in $\beta$ Pic support this picture
\citep{2007Chen+}.
Stellar activity and starspots in young stars make the detection of 
planets by radial velocities and transits infeasible,
especially in later type stars, meaning that 
direct imaging and gap characterization
may be some of the best methods of 
finding planets in young stars, particularly in pre-main sequence 
stars.   

Simple black body modeling of $\sim$500 debris disks observed with the 
\emph{Spitzer} IRS and MIPS at 70 $\mu$m indicates that the excess from 
one-third of the targets can be described using a single temperature black
body with a median grain temperature, $T_{gr}$ $\sim$ 180 K, and the excess
from two-thirds of the targets can be described using a two temperature
black body model with median grain temperatures, $T_{gr}$ $\sim$ 80 K and 
340 K \citep{2014Chen_etal}. In these systems, the presence of Jovian planets could 
naturally explain how planetesimals populations are (1) dynamically excited
leading to collisions between parent bodies and (2) sculpted into rings. 
However, coagulation N-body simulations of 'self-stirred' disks suggest that 
significantly smaller Pluto-sized objects may also induce collisions between
parent bodies \citep{2004KenyonBromley}. A detailed census of Jovian-mass planets in
debris disks is needed to determine the role that Jovian planets play in
exciting and sculpting parent body belts. Gemini South and VLT have recently
commissioned GPI and SPHERE, second-generation coronagraphs, that are
expected to take of census of planets with masses $>$1 $M_{Jup}$ around
nearby, young stars \citep{Macintosh2014}. 

Prime targets for these searches will be young stars in the Scorpius-Centaurus 
OB association (ScoCen). ScoCen is the closest OB association to the Sun with 
typical stellar distances of $\sim$100 - 200 pc and contains three subgroups: 
Upper Scorpius (US), Upper Centaurus Lupus (UCL), and Lower Centaurus Crux 
(LCC), with estimated ages of
$\sim$11 Myr, $\sim$15 Myr, and 
$\sim$17 Myr, \citep{2012PecautMamajekBubar,2002Mamajek_etal} respectively.
Several hundred candidate members
have been identified to date, although the association probably contains 
thousands of low-mass members. Member stars with spectral-type F and earlier 
have been identified using moving group analysis of \emph{Hipparcos} positions,
parallaxes, and proper motions \citep{deZeeuw1999}, while later-type members have
been identified using youth indicators (i.e., high coronal X-ray activity and 
large lithium abundance; \citep{Preibisch2008, Slesnick2006}). Jovian mass planets 
have already been discovered in two ScoCen debris disk systems thus far. VLT 
NaCO differential imaging at L'-band has revealed the presence of a 
5.2 $M_{Jup}$ planet at 56 AU (0.62$\arcsec$) from HD 95086, an A8V member of 
LCC \citep{Rameau2013}. Magellan AO + Clio2 differential imaging at J-, Ks- and 
L'-bands has revealed the presence of an 11 $M_{Jup}$ planet at 650 AU from 
HD 106906, a F5 member of LCC \citep{Bailey2014}. 

We report here
the results of a study modeling the \emph{Spitzer} IRS and MIPS 70
$\mu$m SEDs of all of the debris disks around B- through G-type ScoCen members
observed during the \emph{Spitzer} cryogenic
mission.  Our scientific goal is to better constrain 
the location of debris dust and infer the presence of planets and their
orbital properties where possible. We list the targets for the sample, along
with their spectral types, distances, and subgroup memberships in Table 1.
For debris disks with two belts, the width of a gap between the belts can 
provide important constraints on the
mass of a planet orbiting within the disk 
\citep{Quillen2006, Chiang2009}. \cite{Nesvold2014} used the 3D collisional debris 
disk model SMACK to derive a relationship between gap width, planet mass, 
stellar age, and disk optical depth. We use this relationship to determine 
which of the two-belt ScoCen targets are consistent with a planet on a circular
orbit, and which require multiple planets or eccentric planet orbits. For the 
systems consistent with a single non-eccentric planet, we place an upper limit 
on the mass of the putative perturbing planet.

\section{Observations}

The infrared properties of nearby, young stars in ScoCen were systematically 
explored during the \emph{Spitzer} cryogenic mission. Observers used the MIPS 
mid-IR photometric camera
to search for infrared excess at 24 and/or 70 $\mu$m around ScoCen members 
selected based on \emph{Hipparcos} astrometry \citep{deZeeuw1999}, 
color-magnitude diagrams \cite{preibisch99,Preibisch2002}, and x-ray surveys
\citep{walter94,martin98,preibisch98,kunkel99,kohler00}. They followed-up
excess targets using the IRS mid-IR spectrometer at 5-35 $\mu$m
to search for solid-state emission features 
and characterize the shapes of the SEDs. Taken together, the \emph{Spitzer} 
MIPS photometry indicated that approximately one-quarter of the $\sim$600 
ScoCen stars observed using MIPS possess infrared excess 
\citep{2006Su_etal,carpenter09b,2011Chen_etal,2012Chen_etal}. For stars with ages 10 - 20 Myr, 
stellar evolutionary models suggest that stars with spectral type earlier than 
F are main sequence while stars later than F are not. The infrared excess 
properties of late-type stars in Upper Sco are consistent with
gas-rich, optically thick T Tauri stars 
while those of early-type stars in Upper Sco and early- and solar-type stars in
UCL and LCC are consistent with
gas-depleted, optically thin debris disks.
Because our models are only valid for optically thin disks, we
focus only on the debris disks.

The \emph{Spitzer} IRS spectra for ScoCen members possess a diverse array
of properties. For example, the spectrum of the F3/F5 LCC member HIP 63975 
(HD 113766) shows prominent 10 and 20 $\mu$m silicate emission features
consistent with the presence of forsterite, enstatite, olivine, and
pyroxene-rich dust, generated by the destruction of a $\gtrsim300$
km radius asteroid.
Detailed modeling of the IRS spectrum suggests that the dust in this system is
located in two cold
belts located at 4 - 9 AU and 30 - 80 AU from the star,
plus a warm belt at 1.8 AU
\citep{2008Lisse+}. By contrast, the IRS spectra of debris disks around B- and
A-type stars in Upper Sco reveal rising continuua without strong solid-state emission
features. The dust in these systems has been modeled using a single temperature
black body \citep{Dahm2009}. We have collected all of the IRS spectra of ScoCen
US, UCL, and LCC debris disks and analyzed their spectra
self-consistently. In Figure 1, we plot the $K_{s}$-[24] color as a function of
$J-H$ color (as a proxy for spectral type) for all of the sources observed using
MIPS and overlay the targets analyzed here in red. We note that the IRS spectra
for 26 disk-bearing members of the $\sim$11 Myr old Upper Sco have been 
modeled in detail by \citet{Dahm2009}; for self-consistency, we independently 
model the spectra of all of the debris disks in their sample but do not reanalyze 
those of the primordial disks. We further note that some MIPS UCL and LCC excess 
sources were not observed using the IRS. 

\begin{figure}[bt]
\centerline{\includegraphics[width=4in]{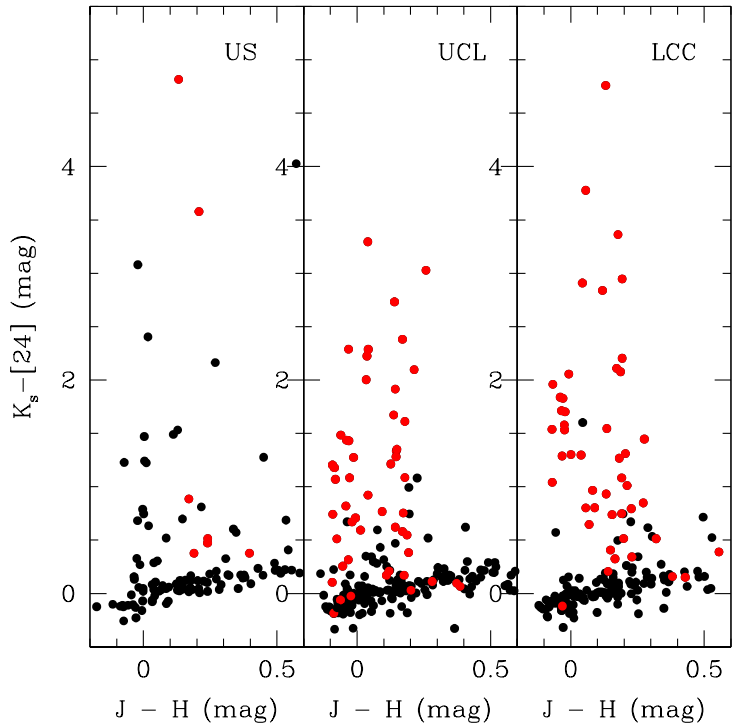}}
\caption{\label{fig:colorcolor}
Color-color plot of all ScoCen targets surveyed with Spitzer.  
We plot in black all of the objects surveyed with MIPS
\citep{2006Su_etal,2008Carpenter_etal,carpenter09b,2011Chen_etal,2012Chen_etal},
and in red the objects whose IRS spectra are studied here.}
\end{figure}

We drew the calibrated IRS spectra for our targets from the \emph{Spitzer} IRS 
Debris Disk Catalog \citep{2014Chen_etal}. Calibrated IRS 
low-resolution spectra typically
possess absolute calibration uncertainties of 5\% while calibrated MIPS 24
$\mu$m fluxes typically possess calibration uncertainties of 2\%
\citep{engelbracht07}. Therefore, the spectra in the Debris Disk catalog are
pinned to the MIPS 24 $\mu$m fluxes as reported in the \emph{Spitzer} Enhanced
Imaging Products (SEIP)
Catalog\footnote{http://irsa.ipac.caltech.edu/data/SPITZER/Enhanced/Imaging/overview.html} 
to improve the absolute calibration of the data. The IRS Debris Disk catalog 
contains not only the spectra for debris disks but also their repeatability 
uncertainties, $\sigma_{{\rm IRS},\lambda}$, estimated using the difference 
between spectra obtained at two separate nod positions. Since the repeatability
uncertainty can vary substantially from pixel-to-pixel, inconsistent with our
understanding of the instrument, we averaged the repeatability uncertainty in
quadrature over the nearest 5 points with boxcar weighting to smooth out
anomalously low or high values. To wit, if $\sigma_{0,i}$ is the IRS
repeatability error at $\lambda_i$, then 
\begin{equation}
\sigma_{{\rm IRS},i}^2 = 
\displaystyle\sum_{j=i-2}^{i+2} \frac{\sigma_{0,i}^2}{5}.
\end{equation}

One source, HIP 77911, was observed only at high resolution with IRS\@.
For this source, we used the CASSIS (Cornell AtlaS of Spitzer/IRS Sources)
optimal reduction of the data \citep{CASSIS}.  The spectrum
appears different from that of the other sources because the
spectral resolution is higher and the wavelength coverage does not
extend shortward of 10 microns.  

We photosphere-subtracted our spectra using our own stellar photosphere models. 
For stars whose MIPS data were analyzed by \cite{2011Chen_etal,2012Chen_etal},
we adopted stellar spectral types, effective temperatures ($T\sub{eff}$), visual
extinctions ($A_{V}$), and luminosities ($L_*$) published therein. 
Stellar properties for sources not analyzed in these papers were 
taken from the references indicated in Table \ref{tab:starprops}.
Then, we
selected Kurucz model atmospheres\footnote{http://www.stsci.edu/hst/observatory/cdbs/k93models.html}
\citep{Kurucz1979}
consistent with the listed effective temperatures, assuming solar abundances and
surface gravities, $\log g=4.0$. Next, we reddened the stellar model SEDs assuming the
Cardelli, Clayton, \& Mathis interstellar extinction law and 
$A_{V} = 3.1 E(B-V)$. Finally, we normalized these stellar atmospheres to the
MIPS 24 $\mu$m predictions given by \cite{2011Chen_etal,2012Chen_etal}, 
with the same 3\% photosphere calibration error used
in those works. The total uncertainty in the photosphere-subtracted 
spectrum is 
\begin{equation}\label{eq:errexcess}
\sigma_{{\rm excess},\lambda} = 
\sqrt{\sigma_{{\rm IRS},\lambda}^2 + \sigma_{{\rm phot},\lambda}^2}
\end{equation}
where $\sigma_{\rm phot}=3\% \times F_{\rm phot}$.  
In Figures \ref{specfig0}-\ref{specfig4},
we show the reduced 
and photosphere-subtracted spectra of our sources.  


\setlength{\stampwidth}{0.24\textwidth}
\newcounter{subfig}
\renewcommand{\thefigure}{\arabic{figure}\alph{subfig}}
\stepcounter{subfig}

\begin{figure}
\parbox{\stampwidth}{
\includegraphics[width=\stampwidth]{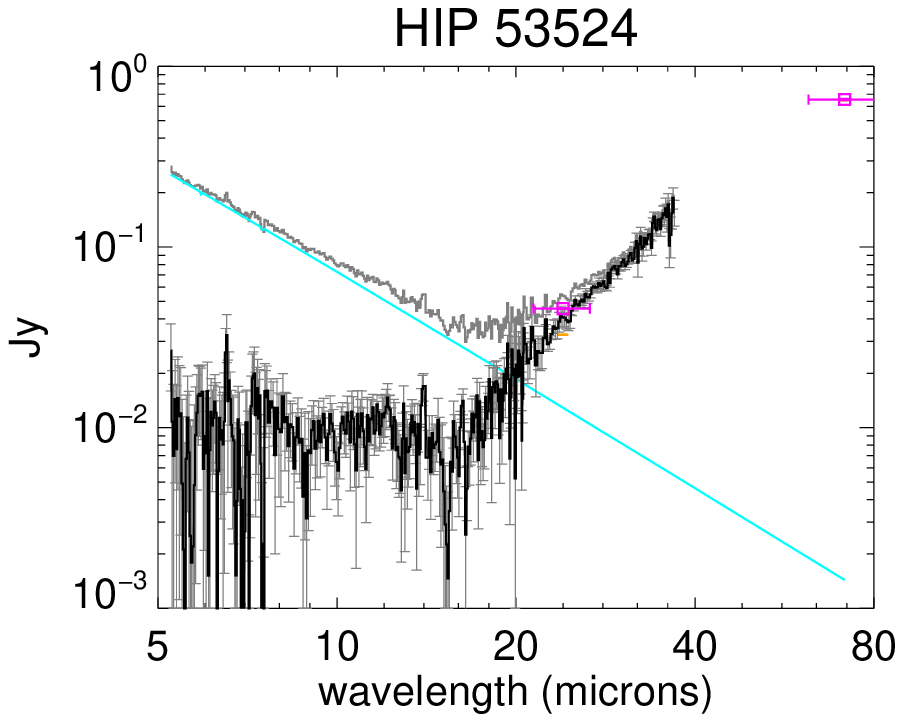} }
\parbox{\stampwidth}{
\includegraphics[width=\stampwidth]{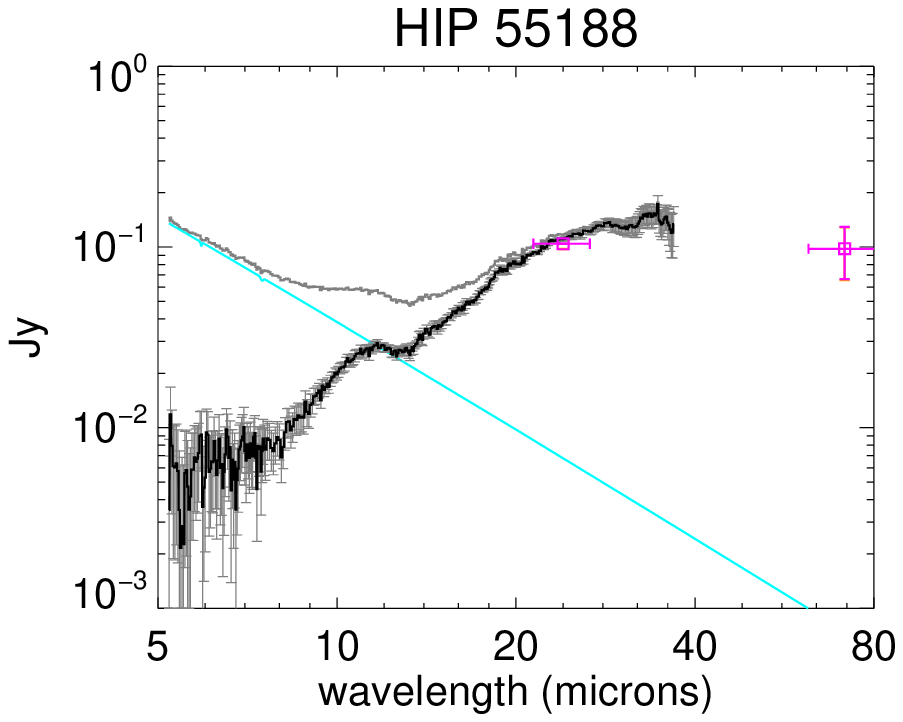} }
\parbox{\stampwidth}{
\includegraphics[width=\stampwidth]{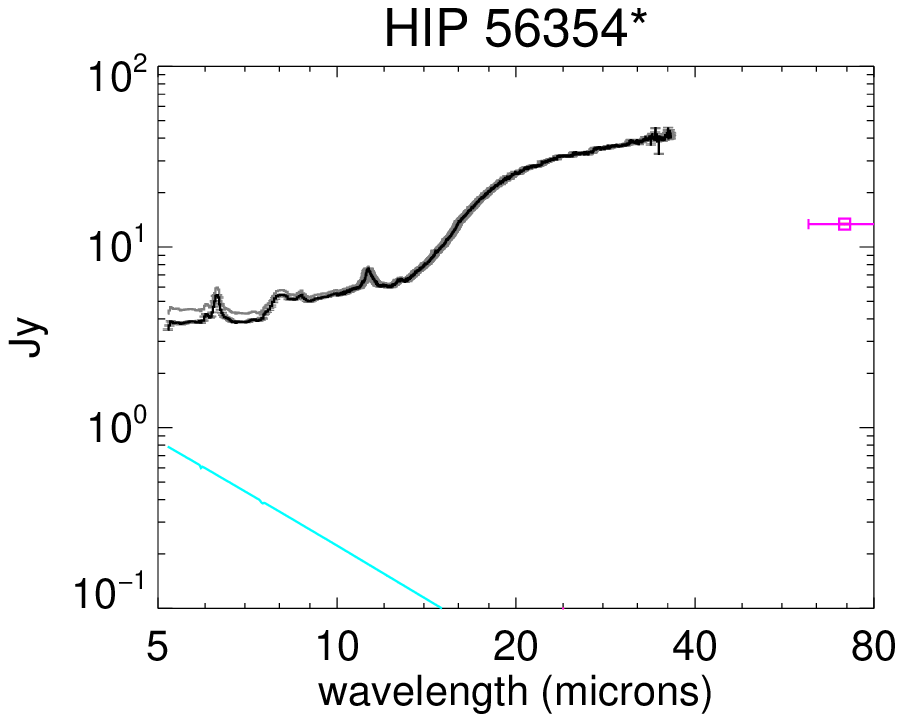} }
\parbox{\stampwidth}{
\includegraphics[width=\stampwidth]{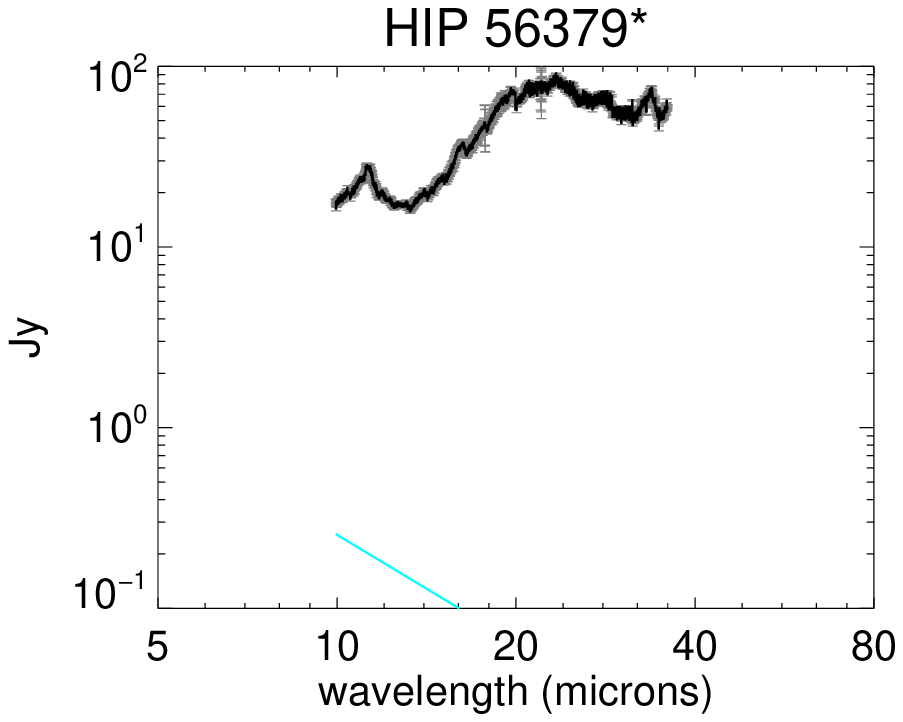} }
\\
\parbox{\stampwidth}{
\includegraphics[width=\stampwidth]{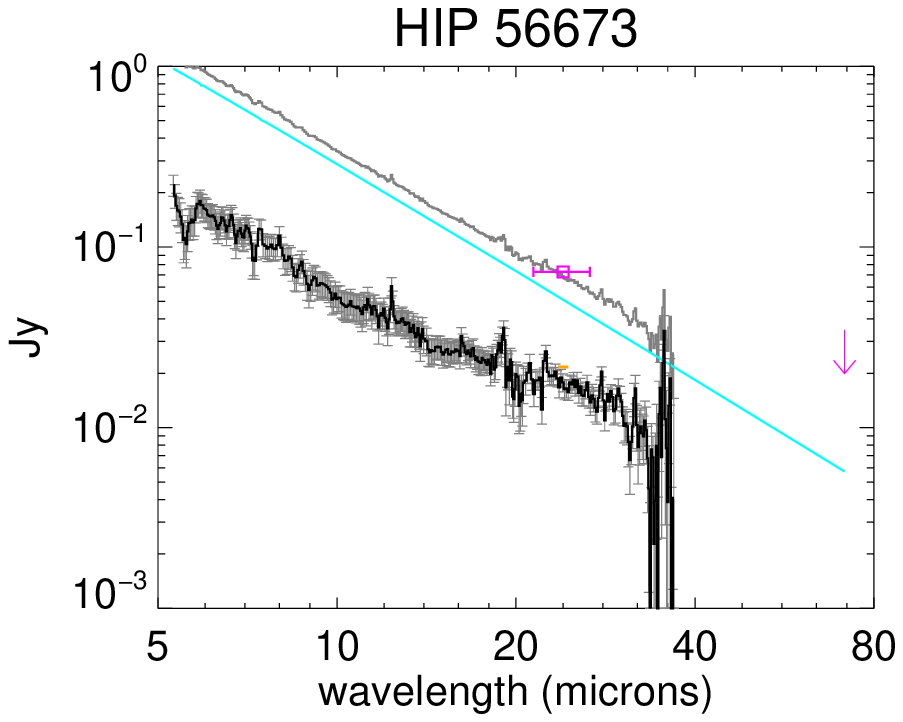} }
\parbox{\stampwidth}{
\includegraphics[width=\stampwidth]{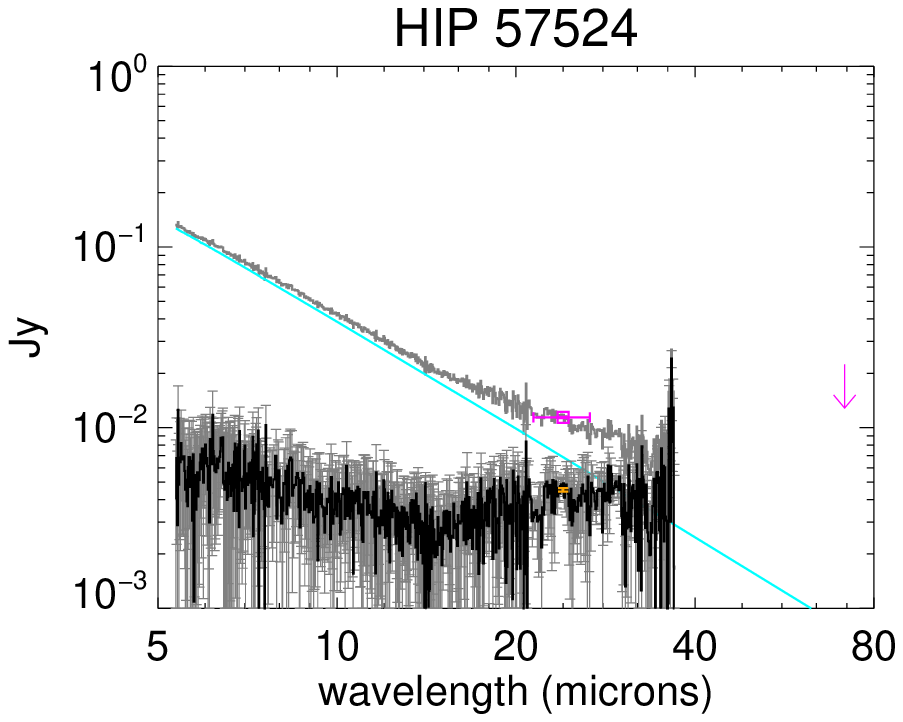} }
\parbox{\stampwidth}{
\includegraphics[width=\stampwidth]{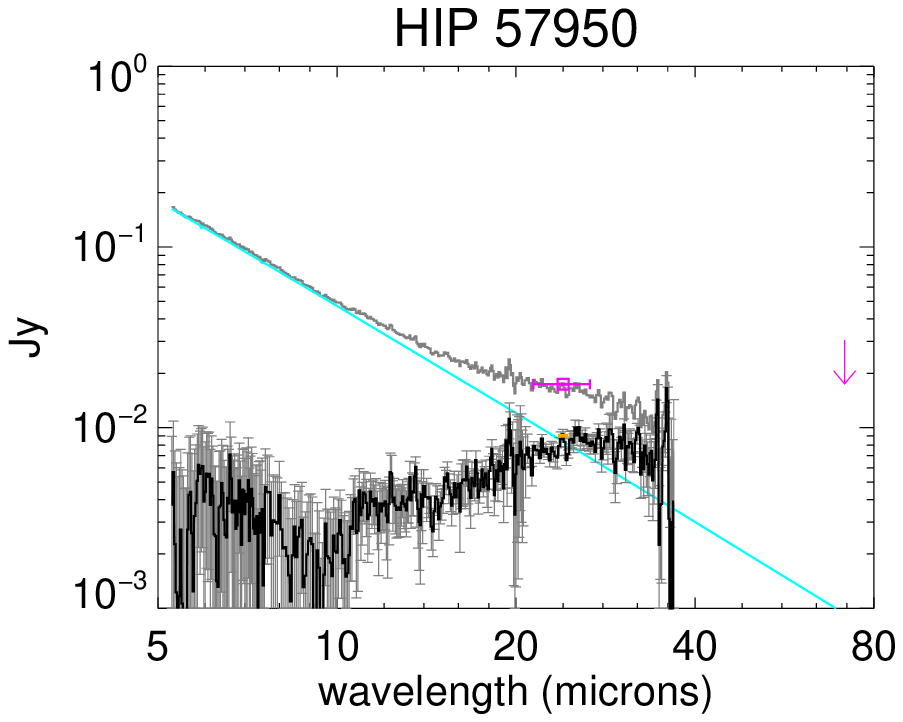} }
\parbox{\stampwidth}{
\includegraphics[width=\stampwidth]{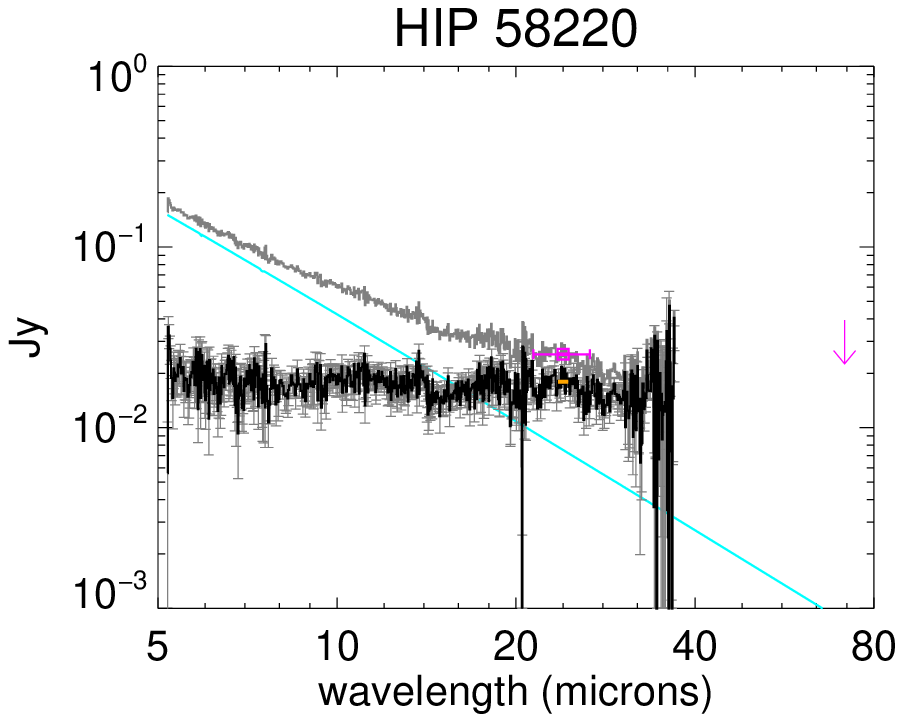} }
\\
\parbox{\stampwidth}{
\includegraphics[width=\stampwidth]{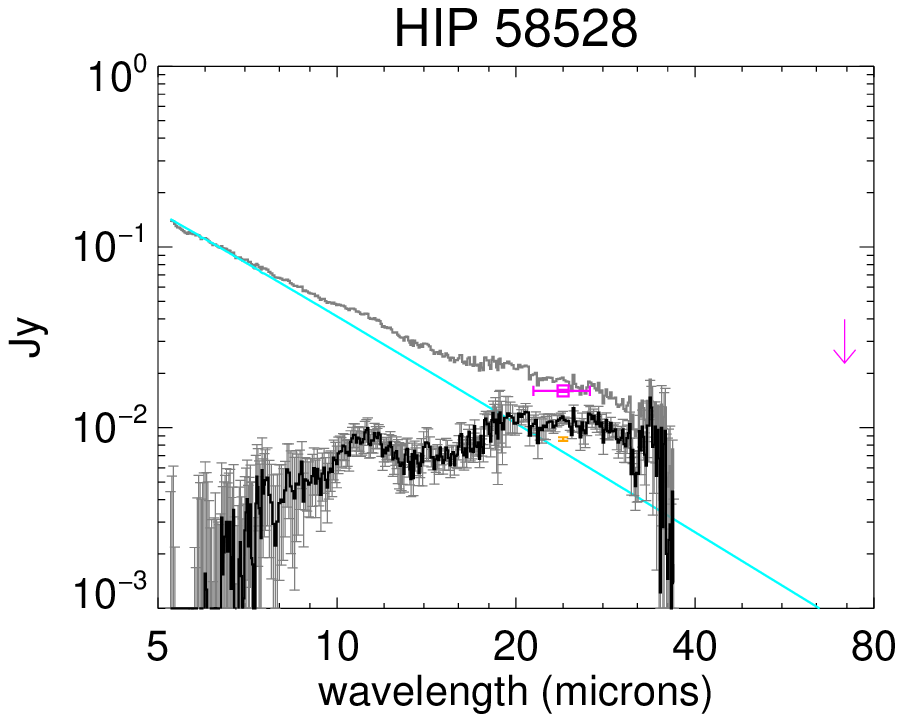} }
\parbox{\stampwidth}{
\includegraphics[width=\stampwidth]{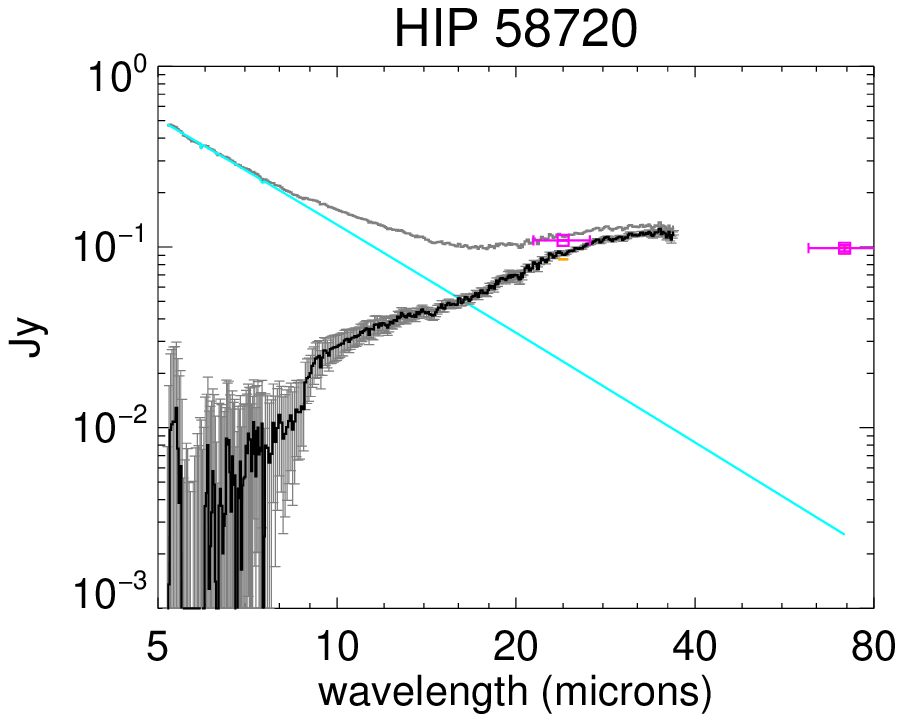} }
\parbox{\stampwidth}{
\includegraphics[width=\stampwidth]{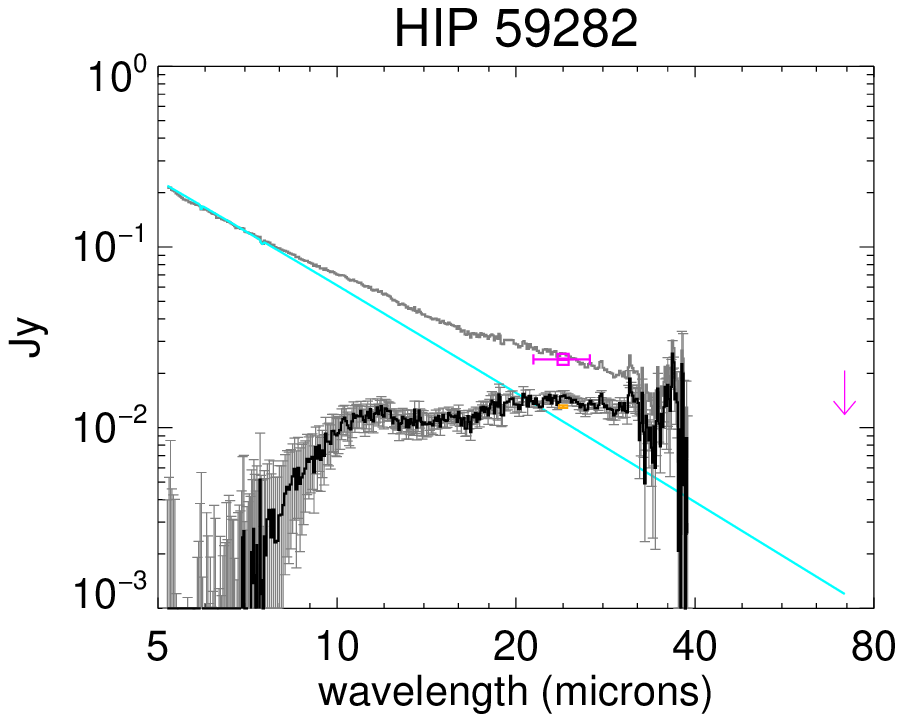} }
\parbox{\stampwidth}{
\includegraphics[width=\stampwidth]{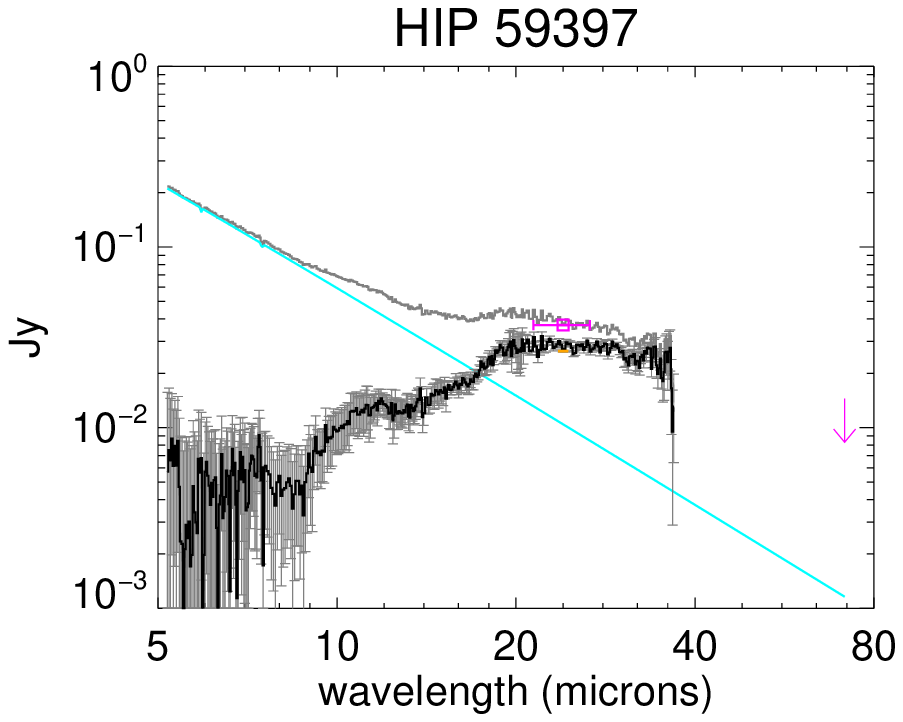} }
\\
\parbox{\stampwidth}{
\includegraphics[width=\stampwidth]{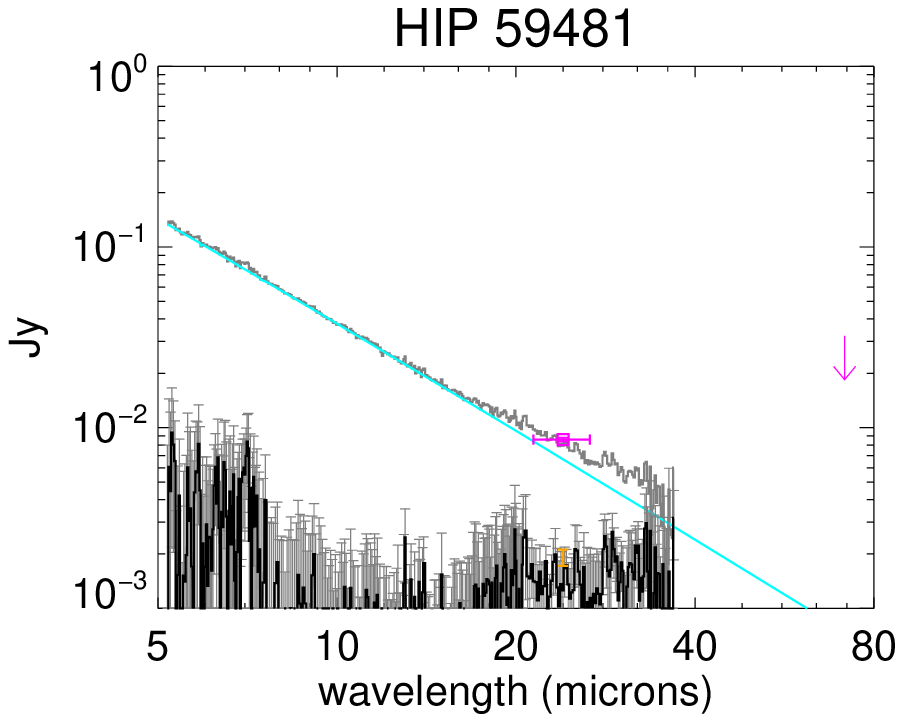} }
\parbox{\stampwidth}{
\includegraphics[width=\stampwidth]{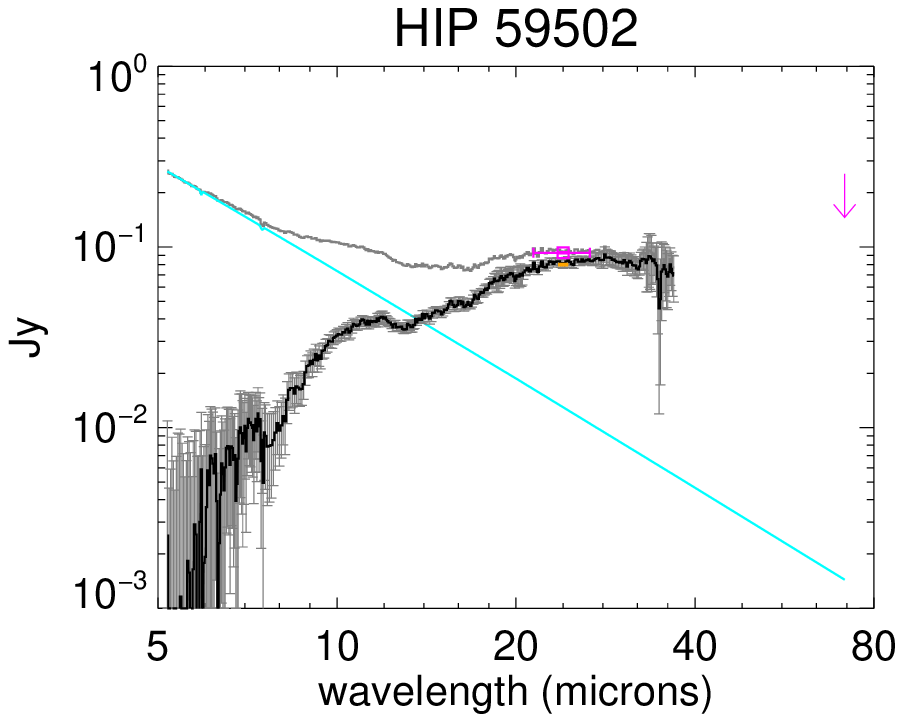} }
\parbox{\stampwidth}{
\includegraphics[width=\stampwidth]{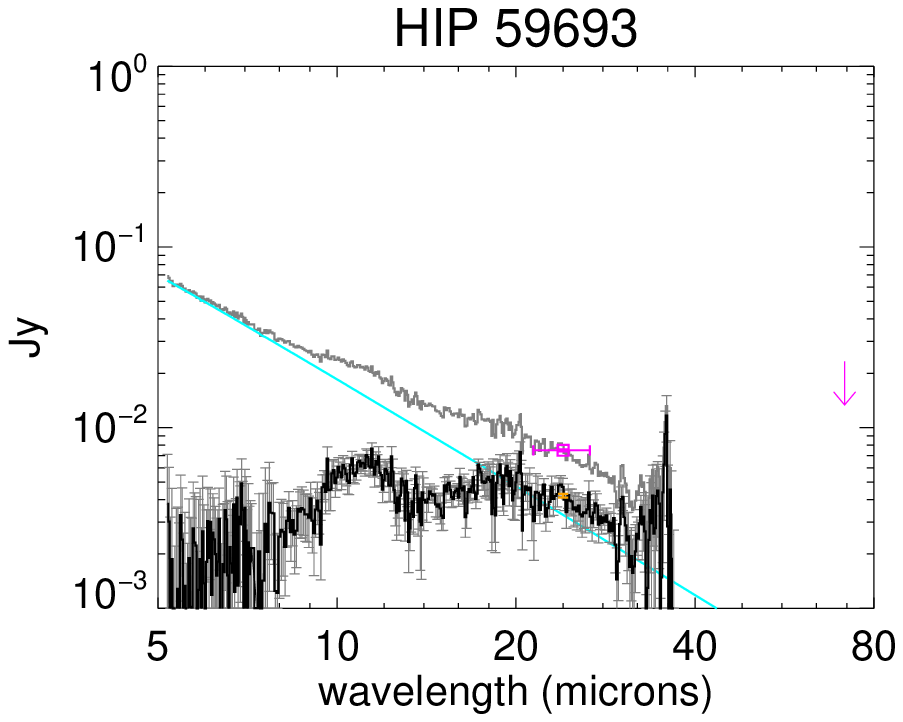} }
\parbox{\stampwidth}{
\includegraphics[width=\stampwidth]{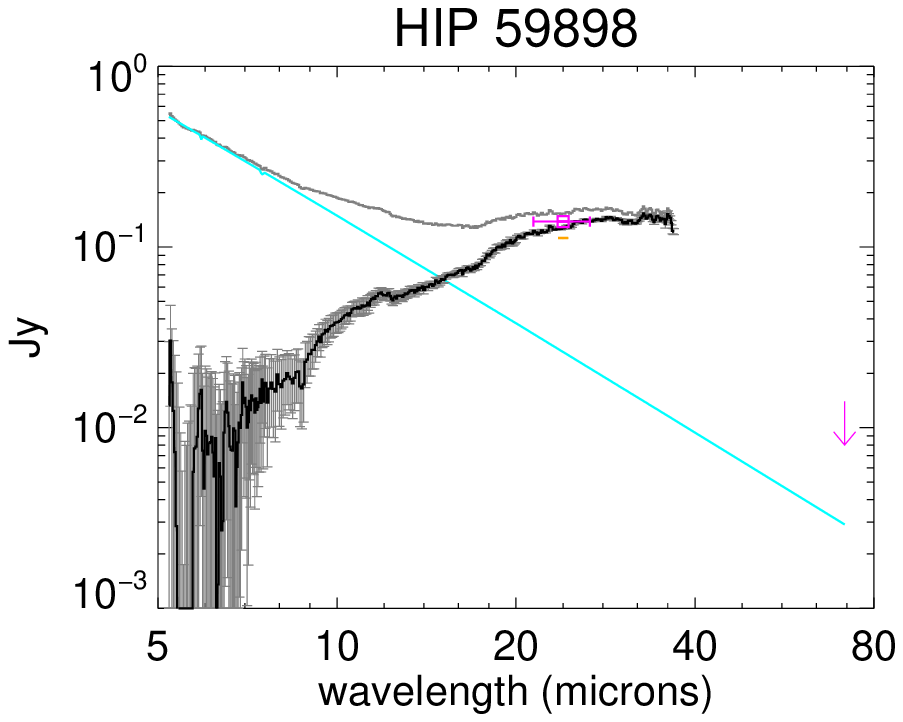} }
\\
\parbox{\stampwidth}{
\includegraphics[width=\stampwidth]{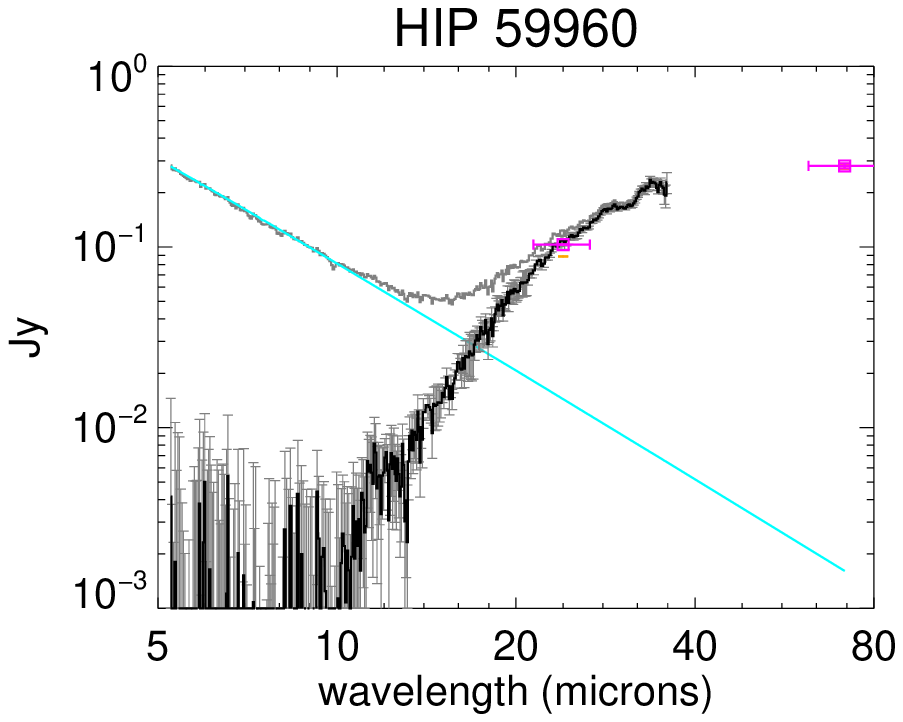} }
\parbox{\stampwidth}{
\includegraphics[width=\stampwidth]{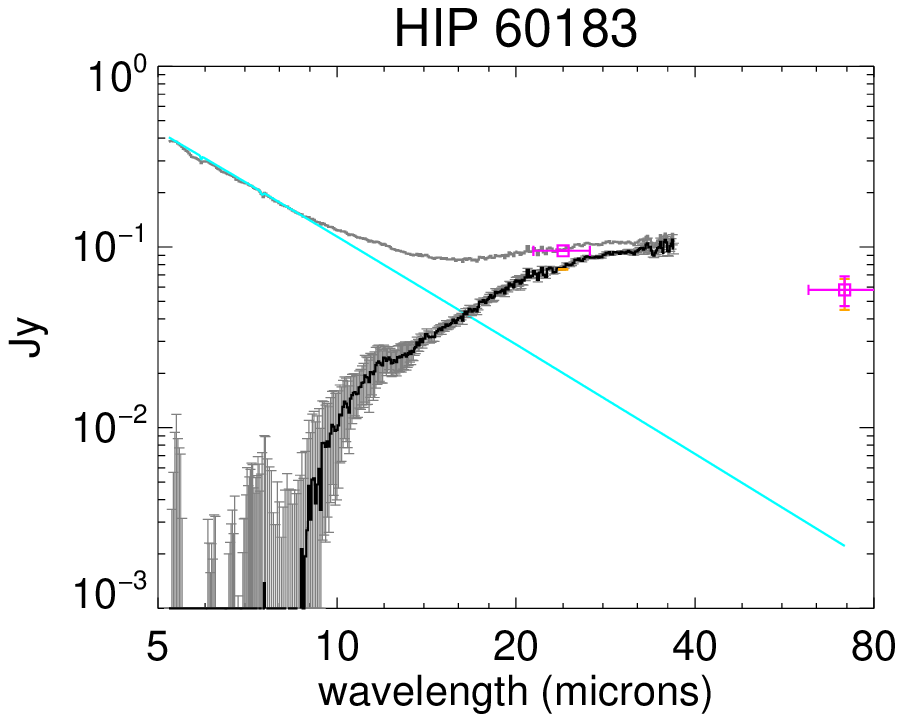} }
\parbox{\stampwidth}{
\includegraphics[width=\stampwidth]{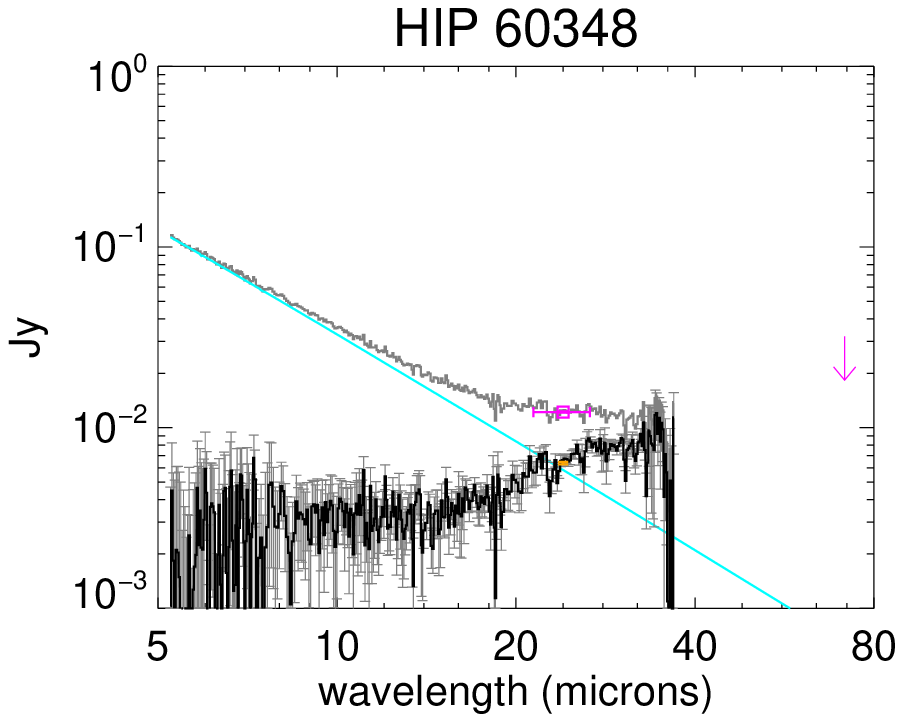} }
\parbox{\stampwidth}{
\includegraphics[width=\stampwidth]{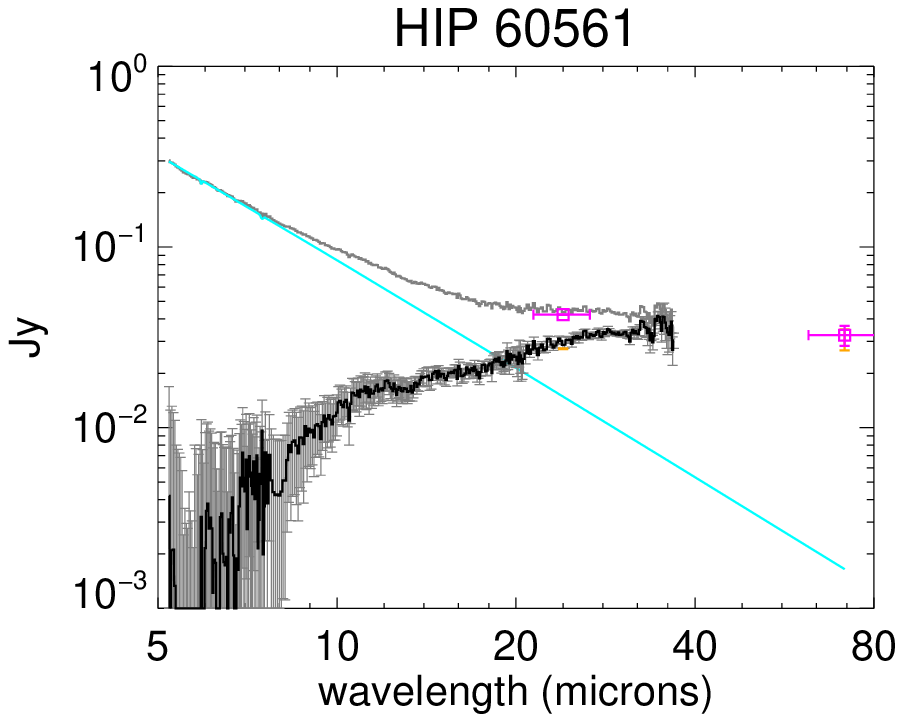} }
\\
\parbox{\stampwidth}{
\includegraphics[width=\stampwidth]{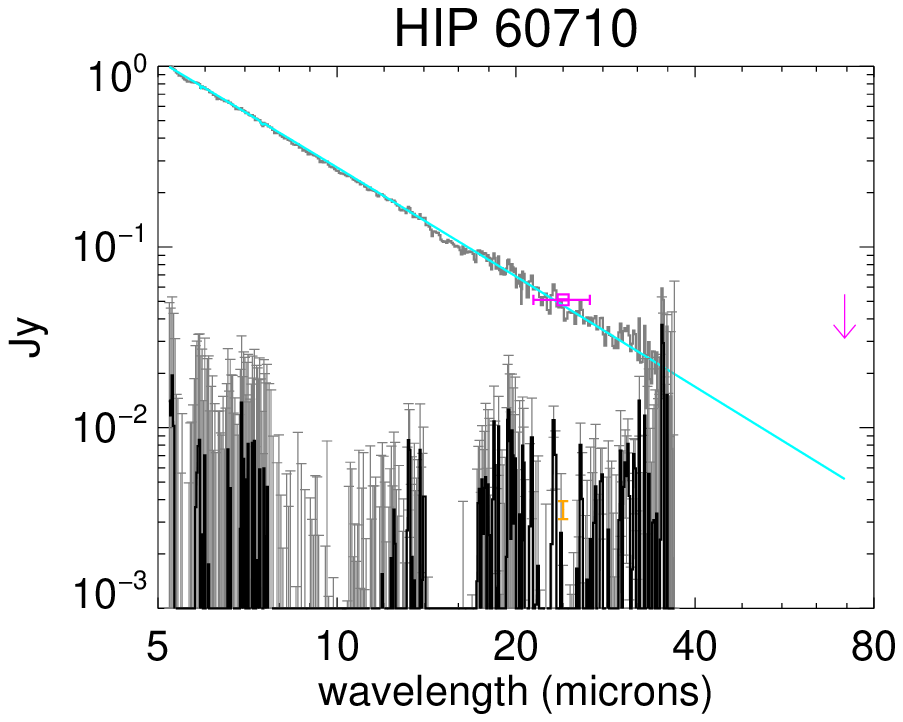} }
\parbox{\stampwidth}{
\includegraphics[width=\stampwidth]{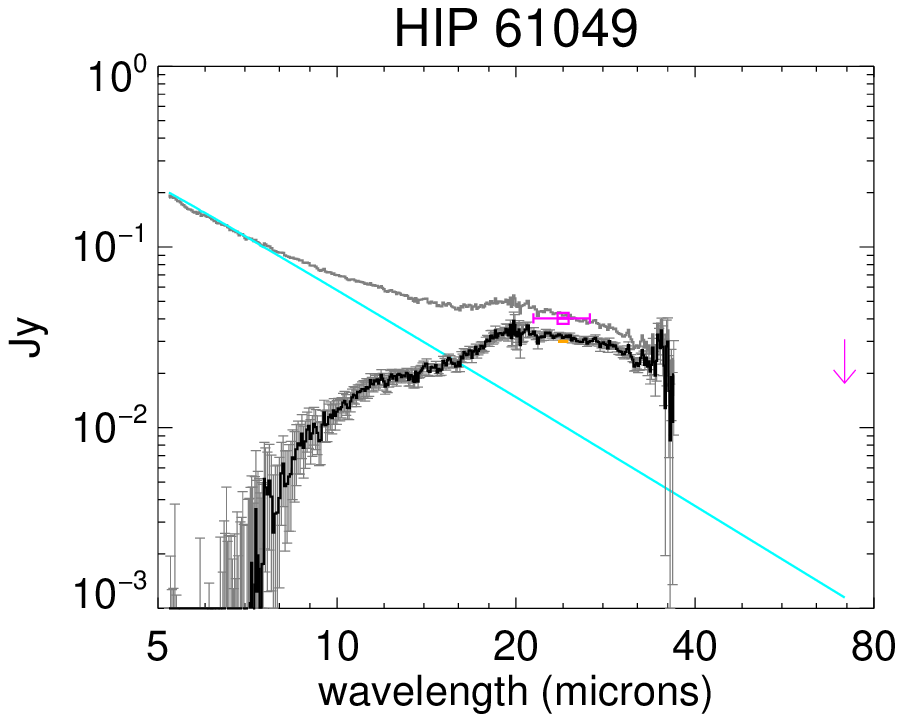} }
\parbox{\stampwidth}{
\includegraphics[width=\stampwidth]{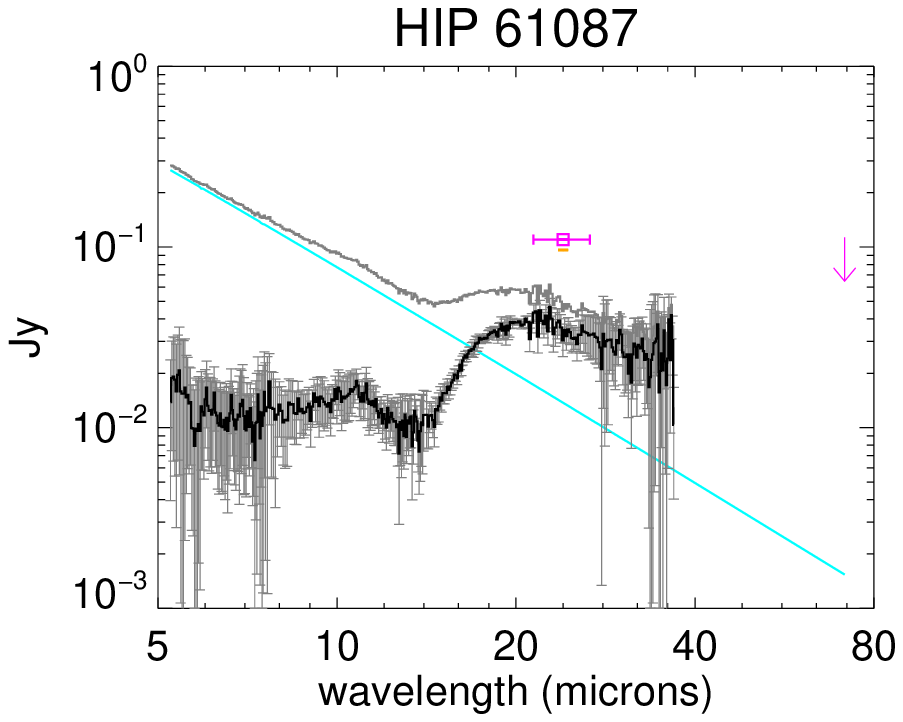} }
\parbox{\stampwidth}{
\includegraphics[width=\stampwidth]{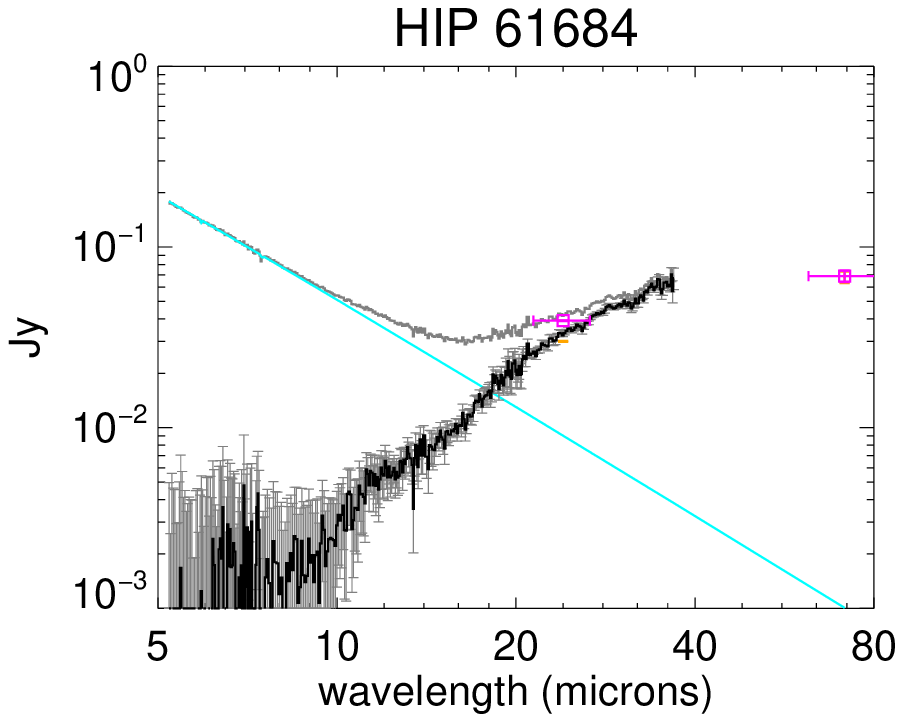} }
\\
\caption{ \label{specfig0}
\protect\input{speccaption}
}
\end{figure}
\addtocounter{figure}{-1}
\stepcounter{subfig}
\begin{figure}
\parbox{\stampwidth}{
\includegraphics[width=\stampwidth]{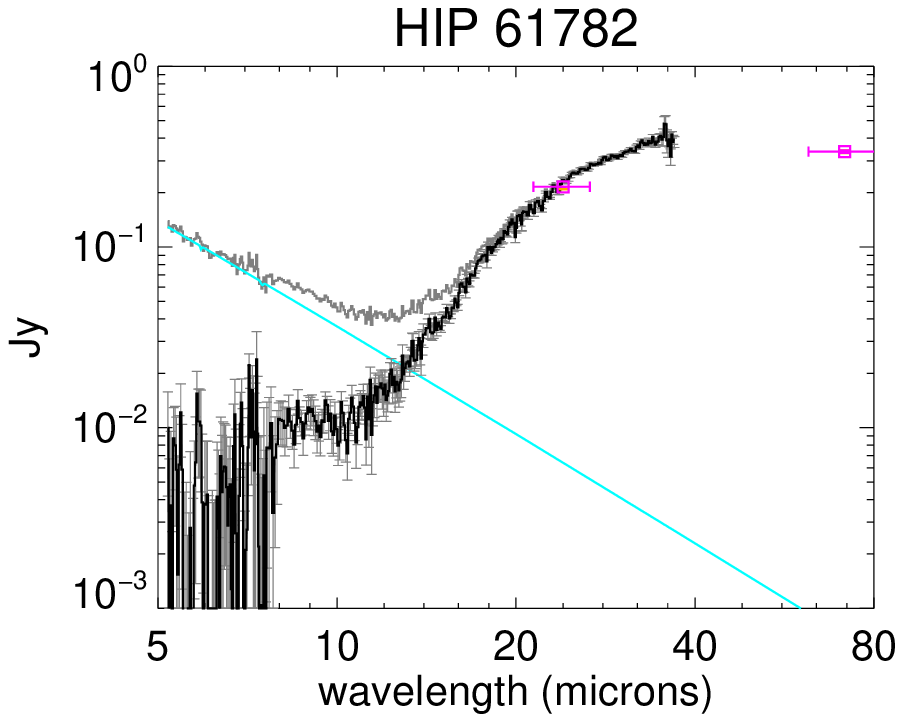} }
\parbox{\stampwidth}{
\includegraphics[width=\stampwidth]{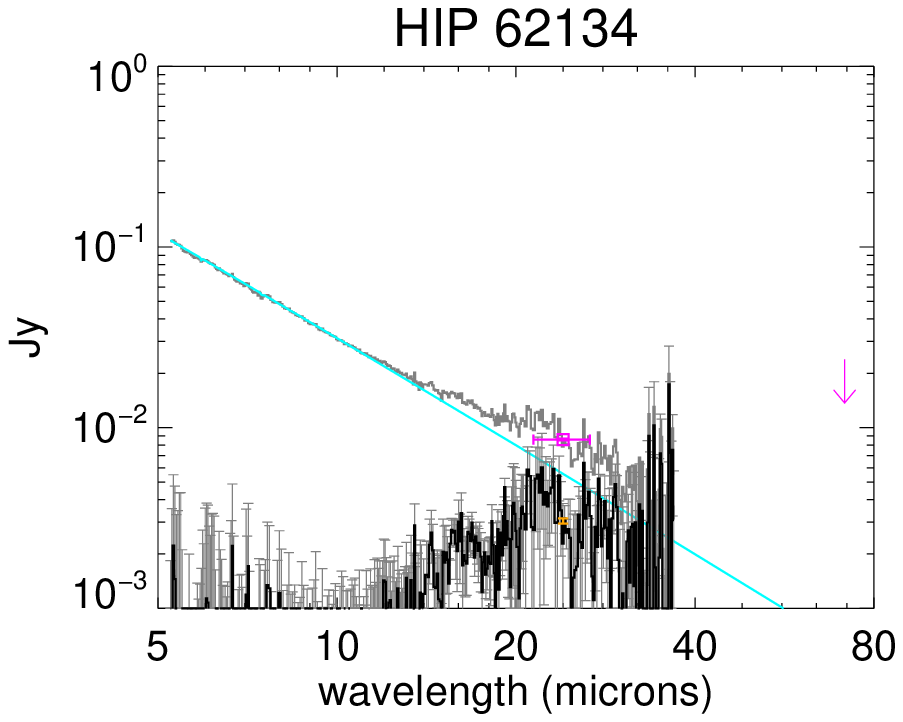} }
\parbox{\stampwidth}{
\includegraphics[width=\stampwidth]{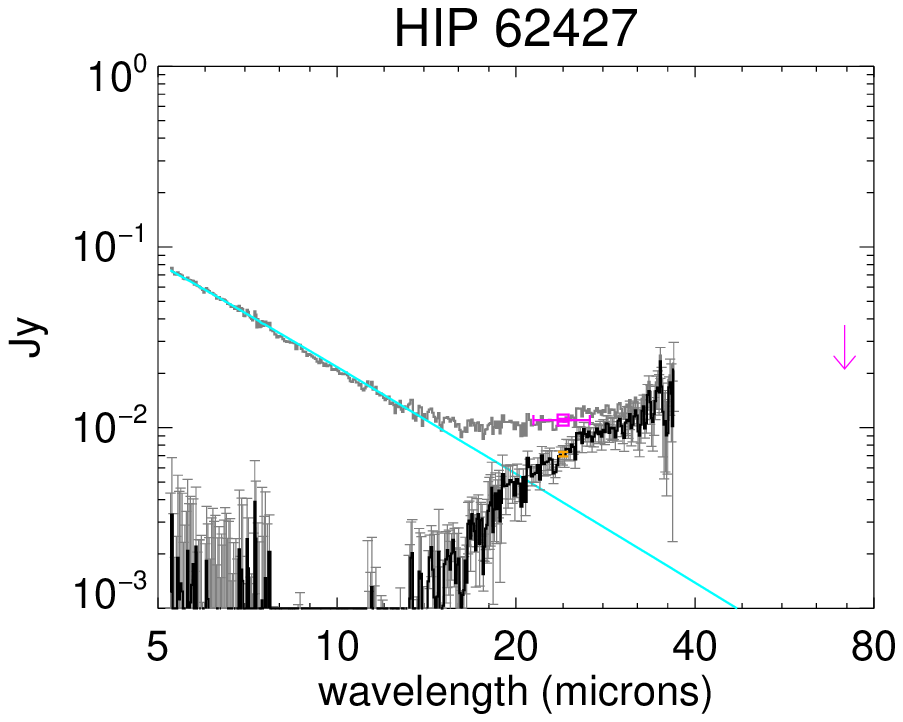} }
\parbox{\stampwidth}{
\includegraphics[width=\stampwidth]{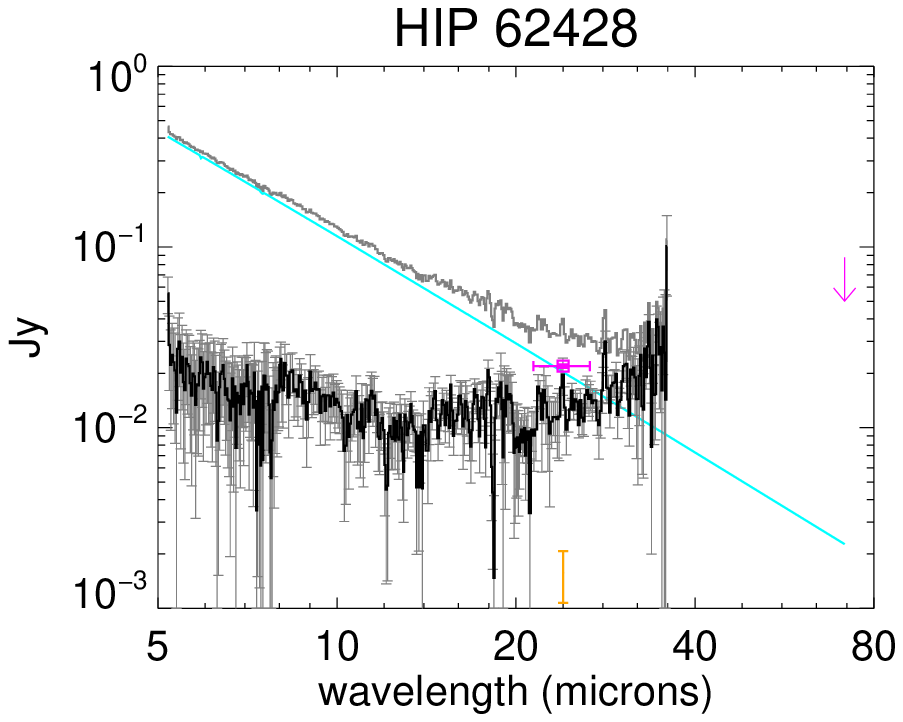} }
\\
\parbox{\stampwidth}{
\includegraphics[width=\stampwidth]{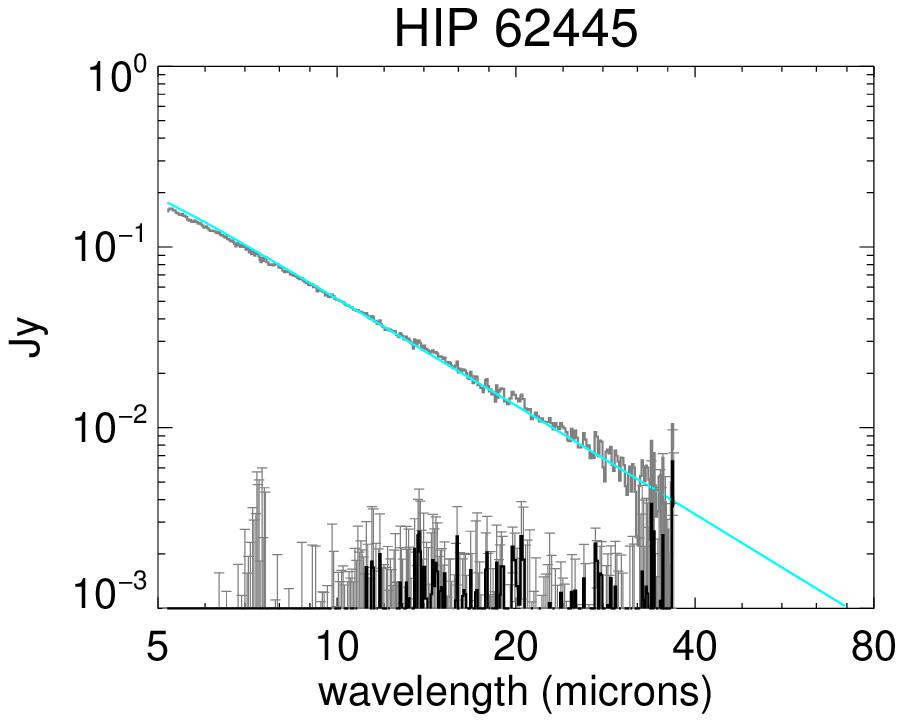} }
\parbox{\stampwidth}{
\includegraphics[width=\stampwidth]{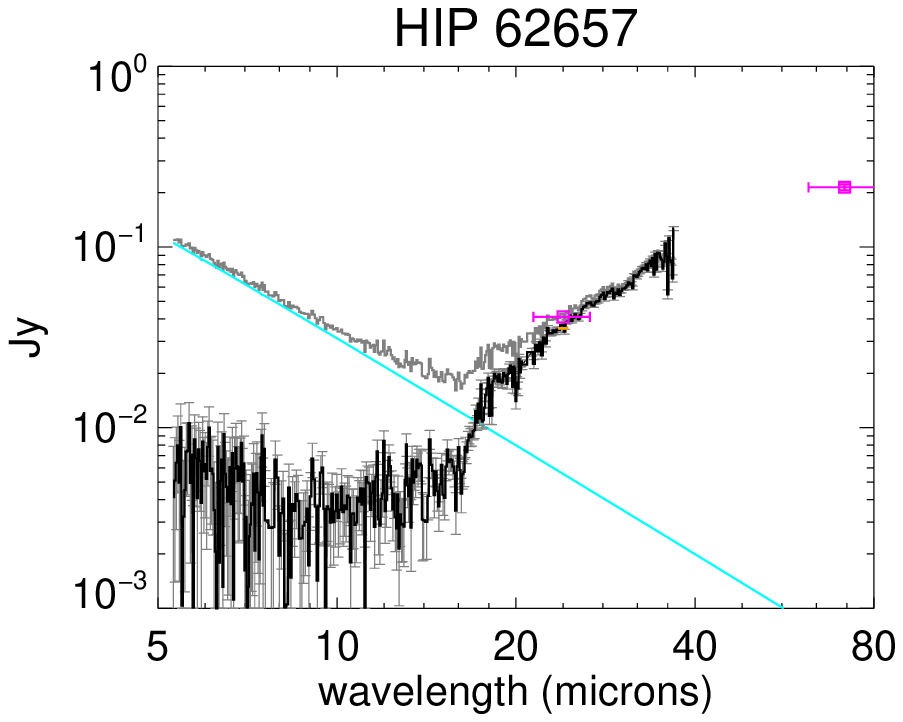} }
\parbox{\stampwidth}{
\includegraphics[width=\stampwidth]{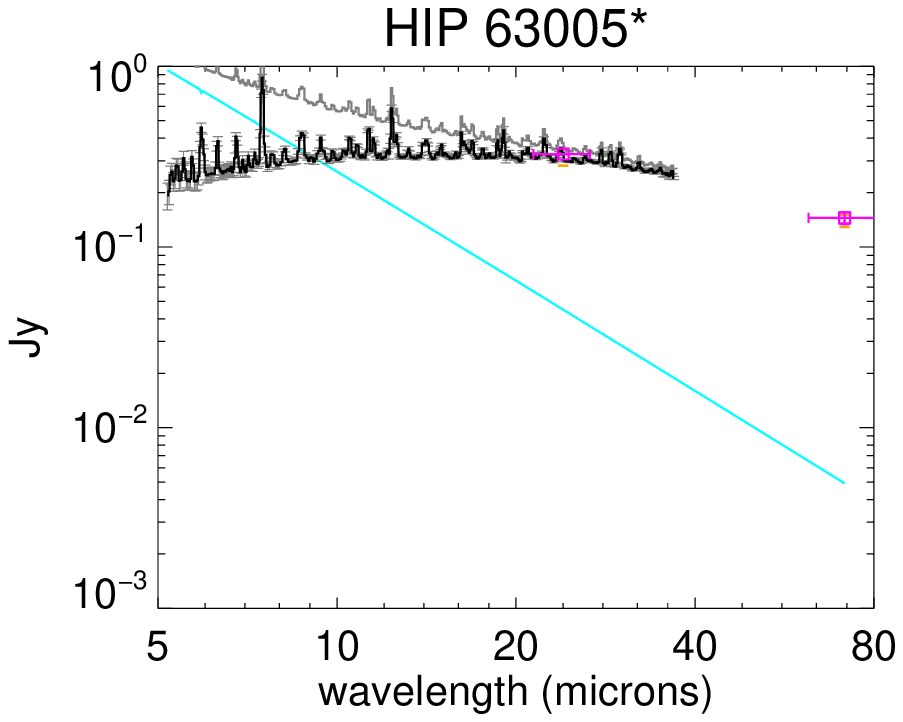} }
\parbox{\stampwidth}{
\includegraphics[width=\stampwidth]{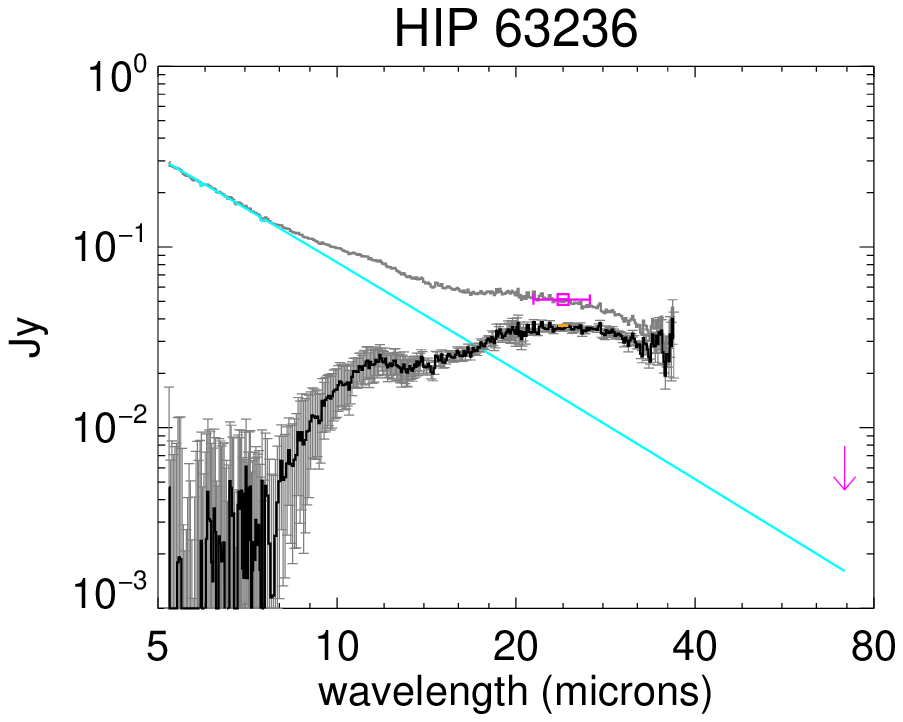} }
\\
\parbox{\stampwidth}{
\includegraphics[width=\stampwidth]{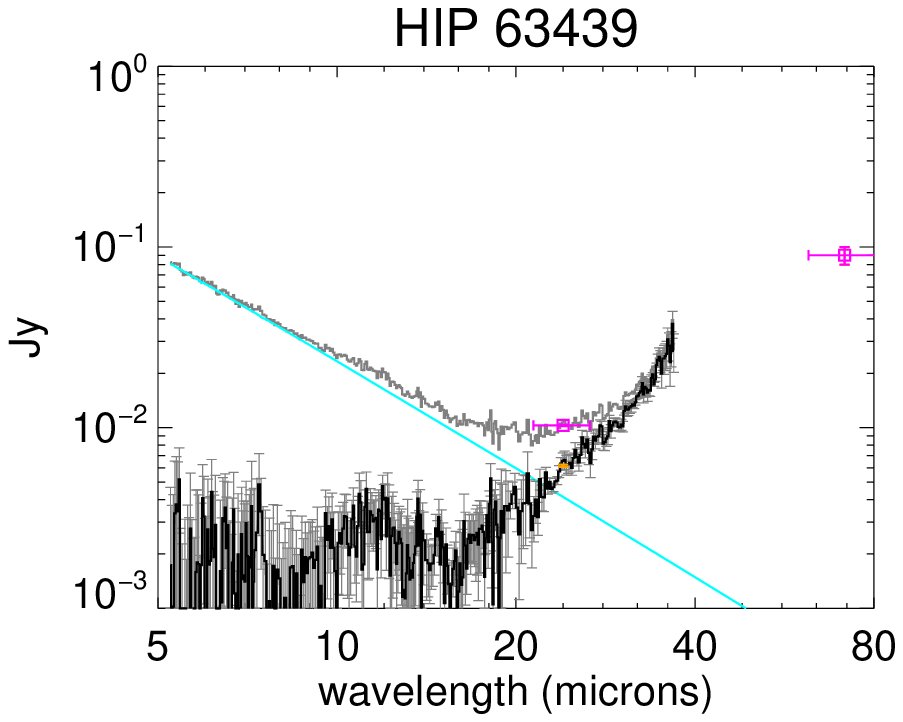} }
\parbox{\stampwidth}{
\includegraphics[width=\stampwidth]{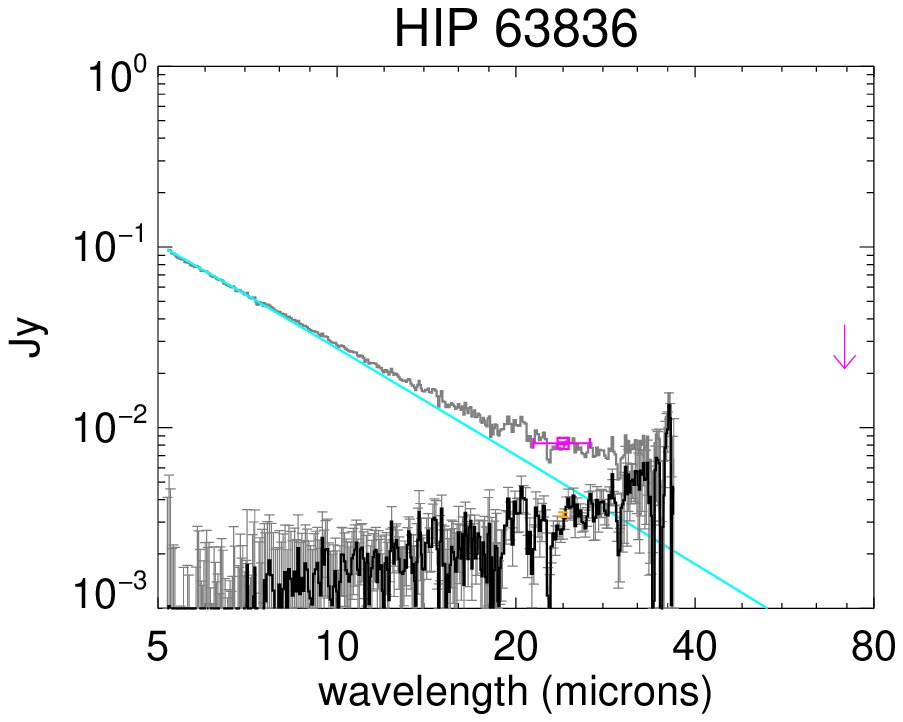} }
\parbox{\stampwidth}{
\includegraphics[width=\stampwidth]{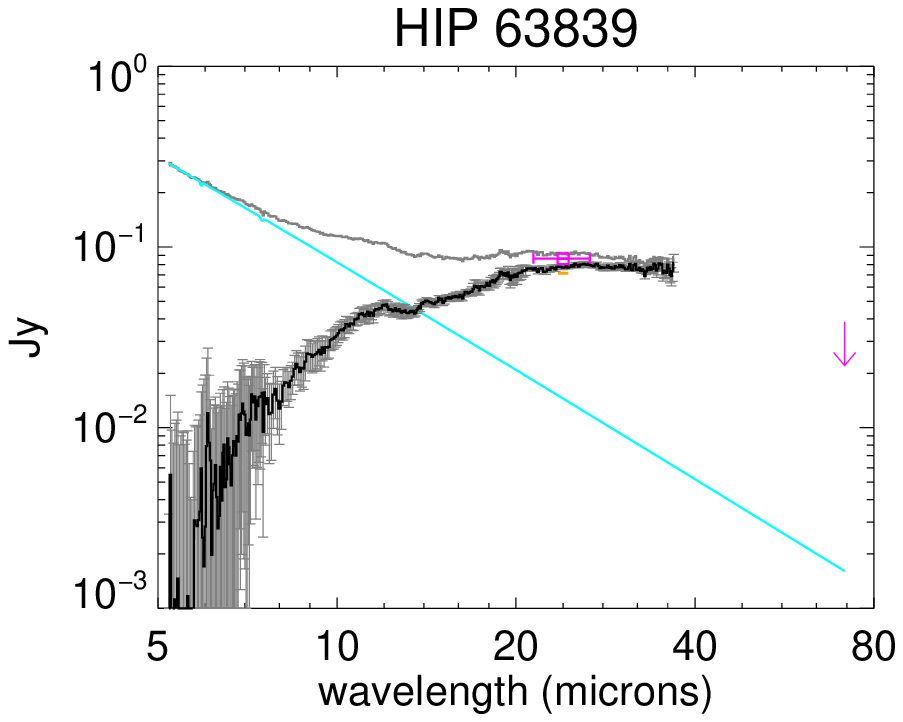} }
\parbox{\stampwidth}{
\includegraphics[width=\stampwidth]{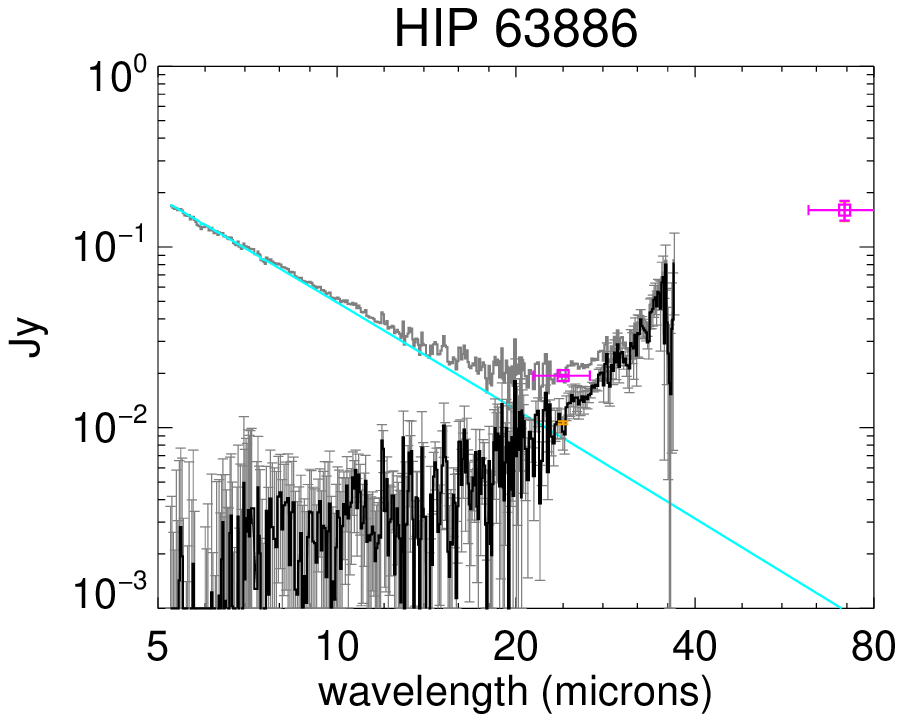} }
\\
\parbox{\stampwidth}{
\includegraphics[width=\stampwidth]{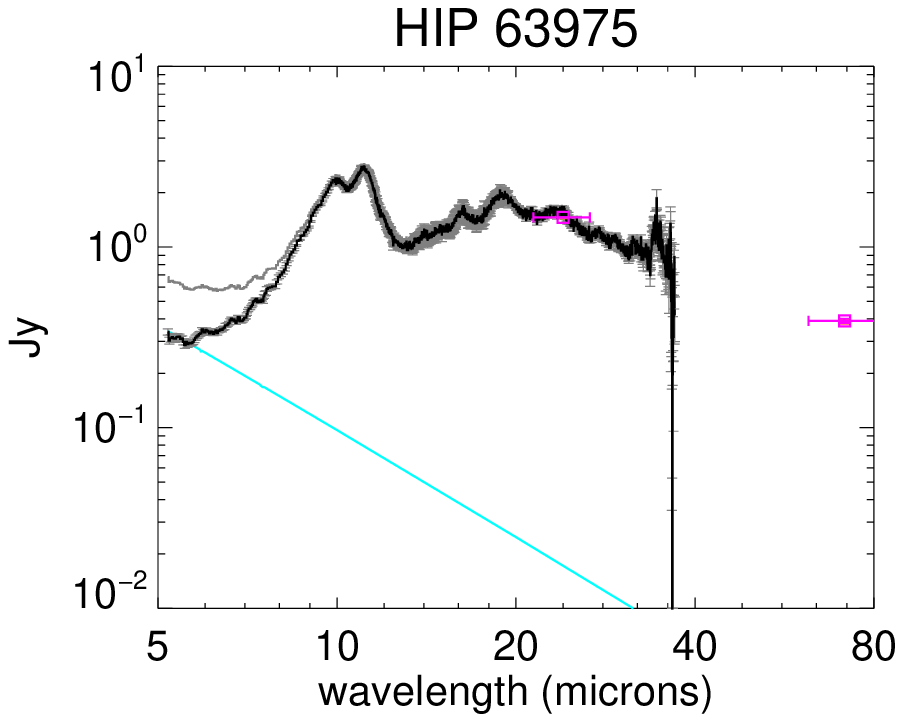} }
\parbox{\stampwidth}{
\includegraphics[width=\stampwidth]{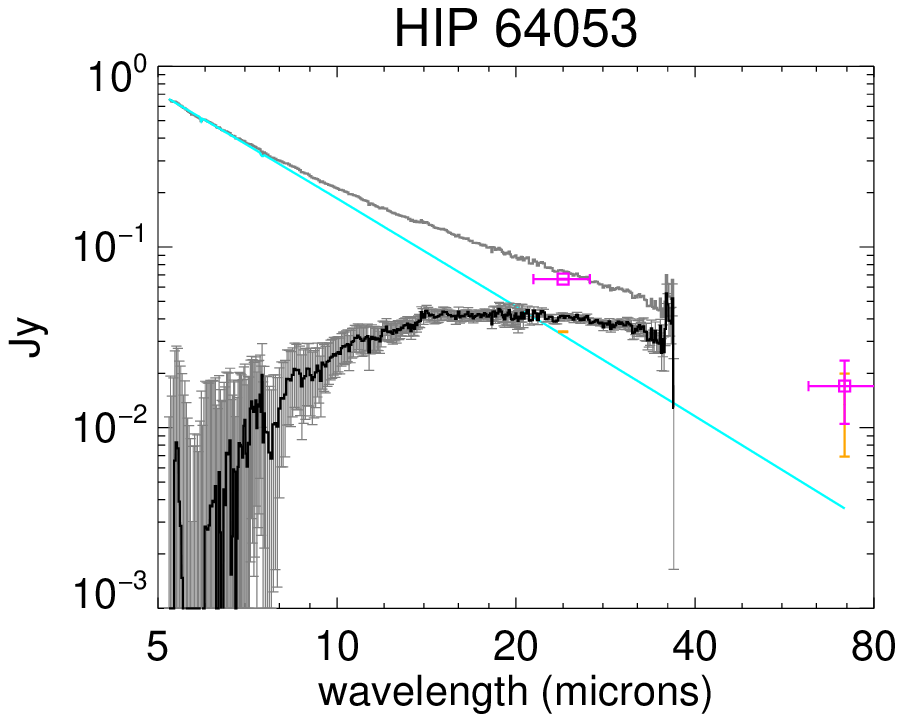} }
\parbox{\stampwidth}{
\includegraphics[width=\stampwidth]{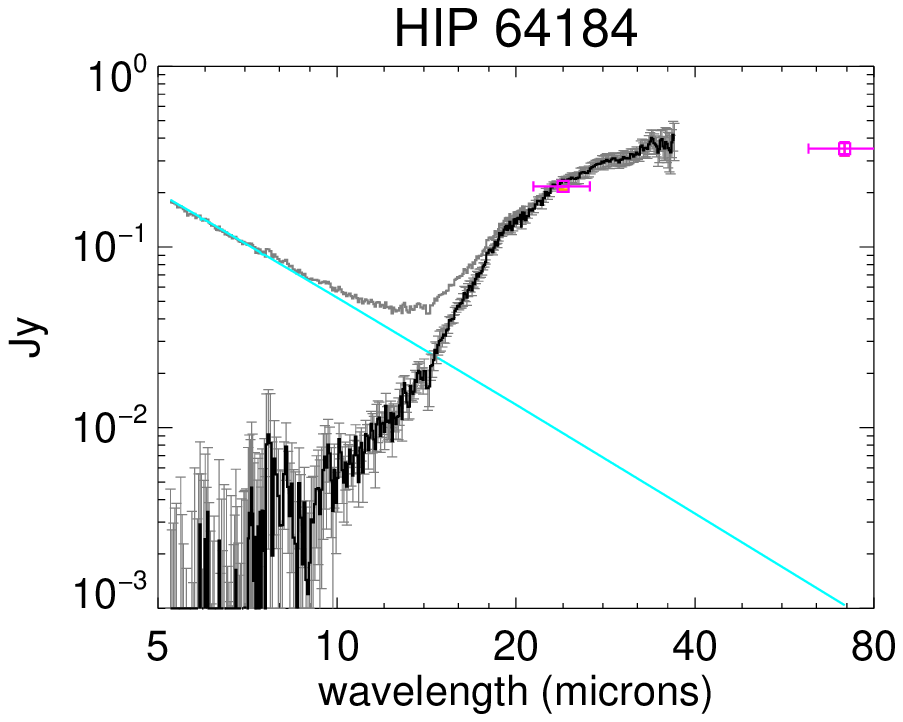} }
\parbox{\stampwidth}{
\includegraphics[width=\stampwidth]{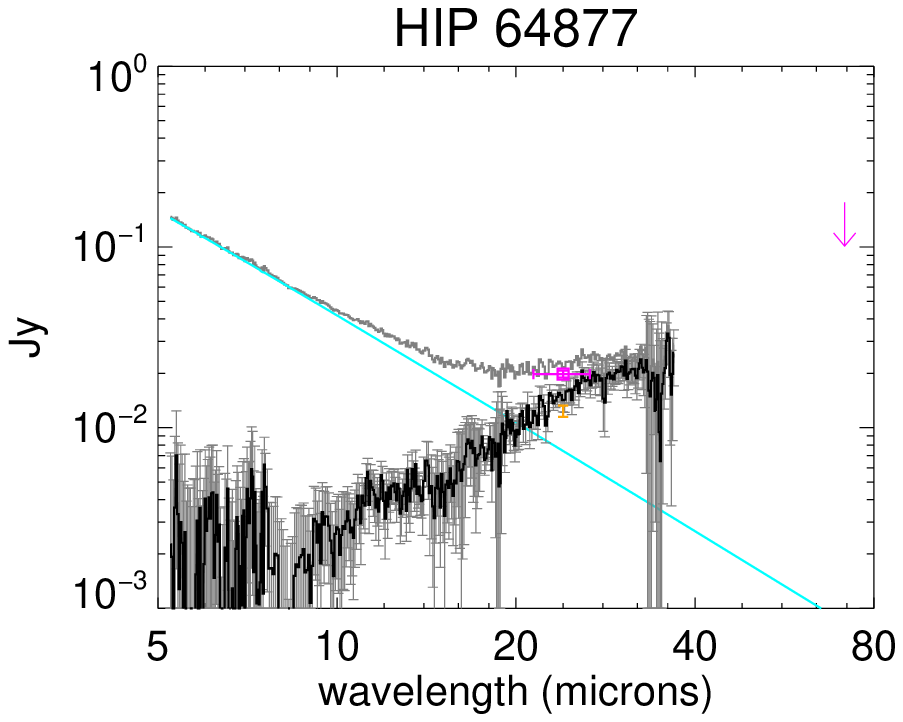} }
\\
\parbox{\stampwidth}{
\includegraphics[width=\stampwidth]{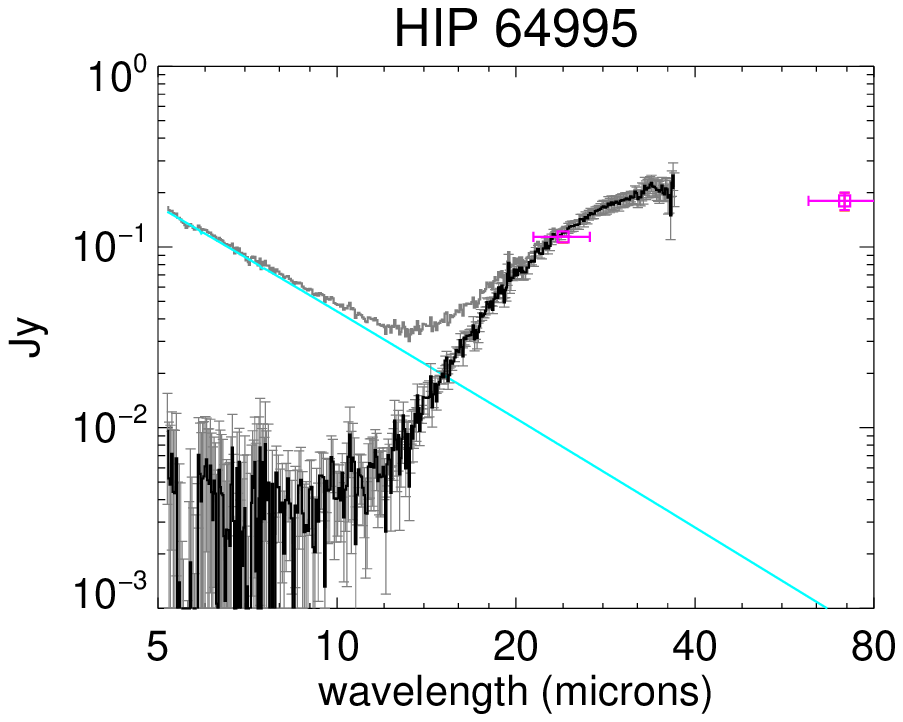} }
\parbox{\stampwidth}{
\includegraphics[width=\stampwidth]{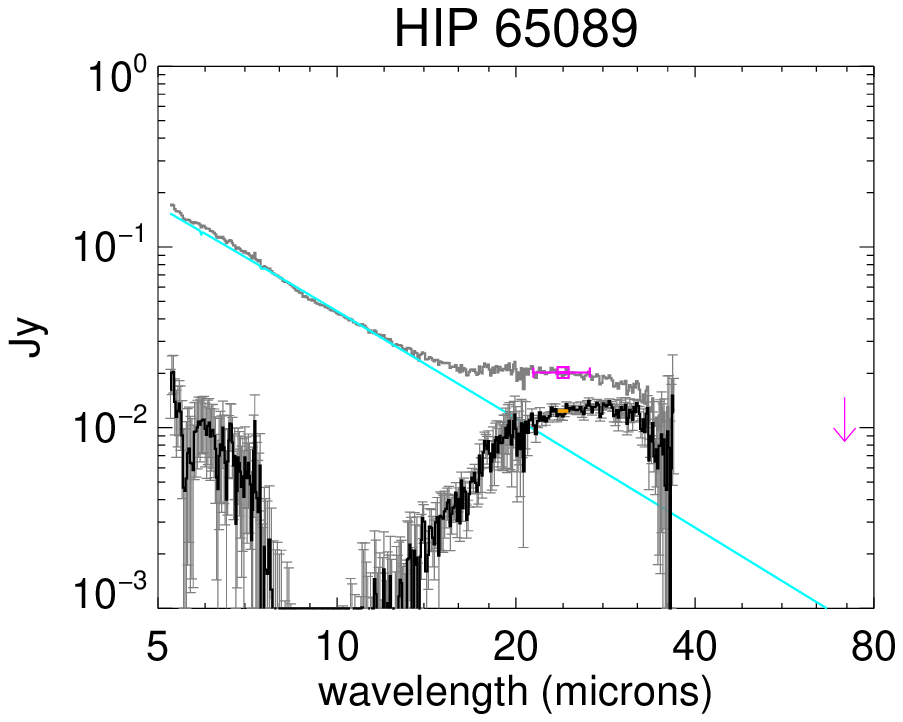} }
\parbox{\stampwidth}{
\includegraphics[width=\stampwidth]{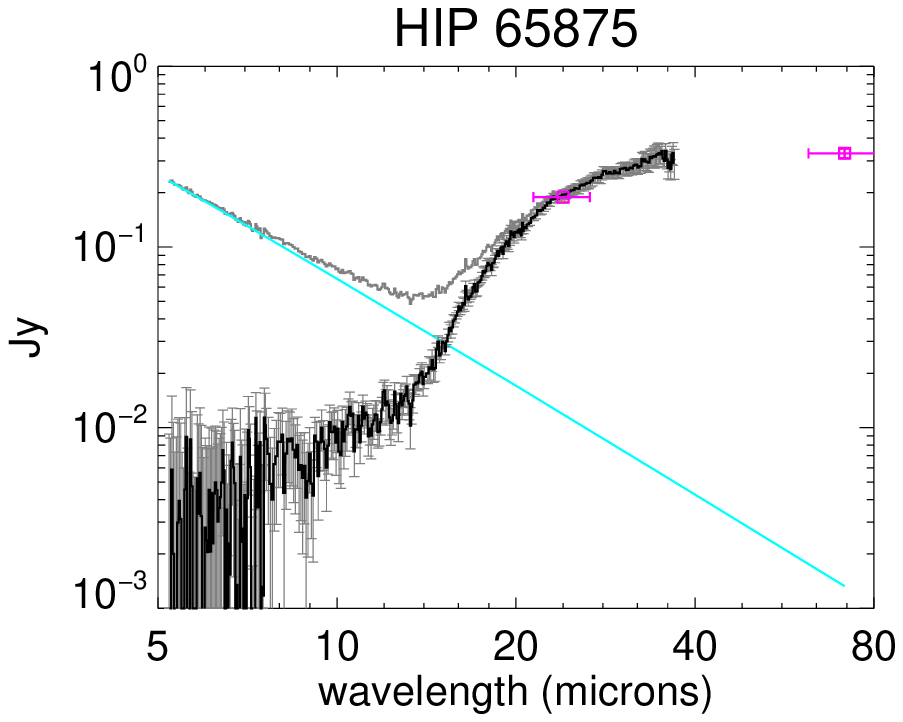} }
\parbox{\stampwidth}{
\includegraphics[width=\stampwidth]{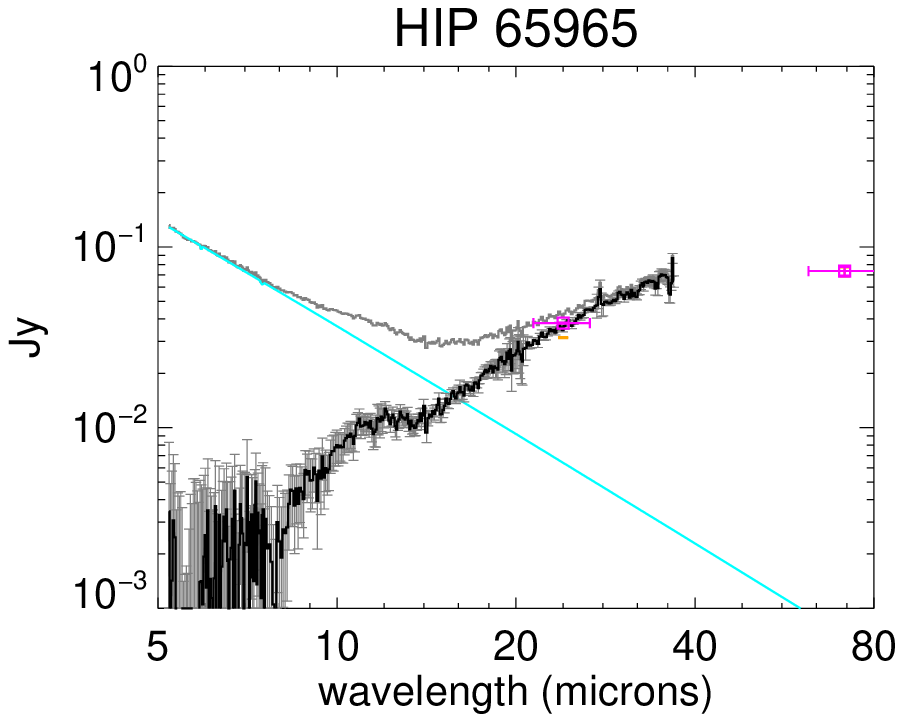} }
\\
\parbox{\stampwidth}{
\includegraphics[width=\stampwidth]{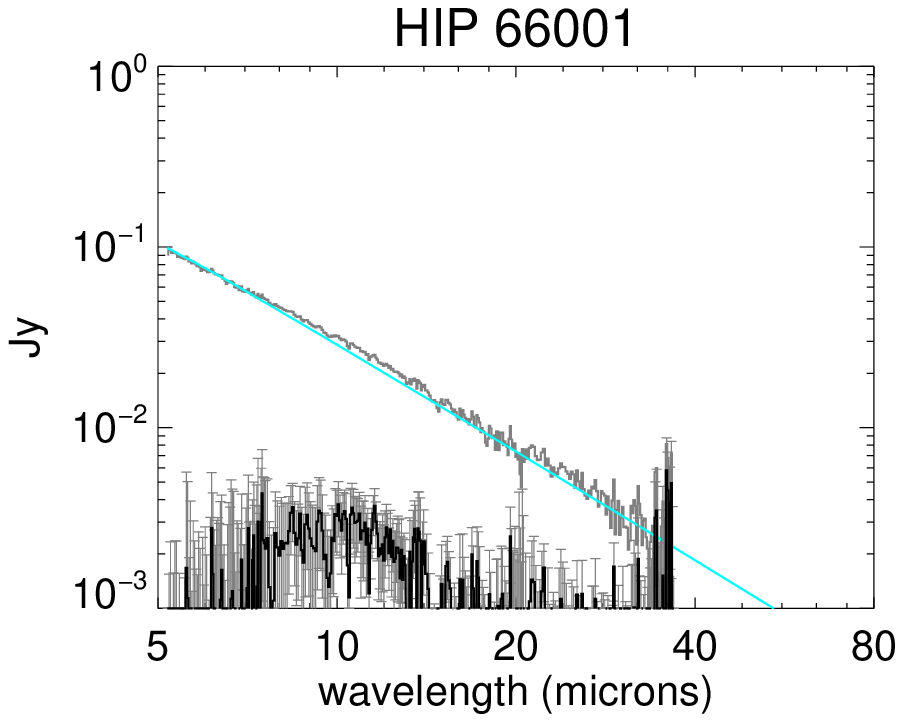} }
\parbox{\stampwidth}{
\includegraphics[width=\stampwidth]{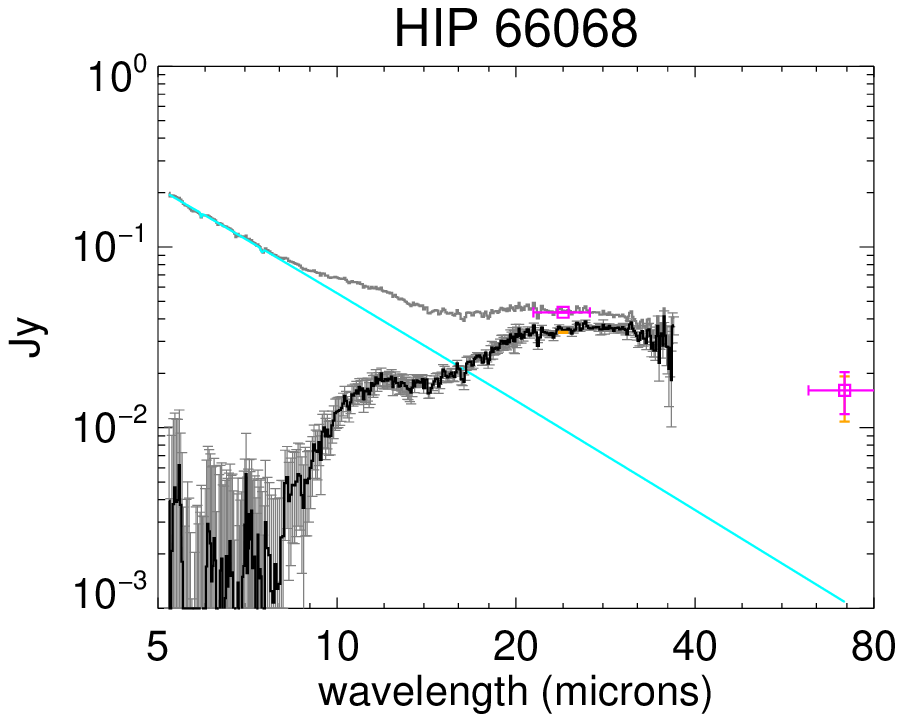} }
\parbox{\stampwidth}{
\includegraphics[width=\stampwidth]{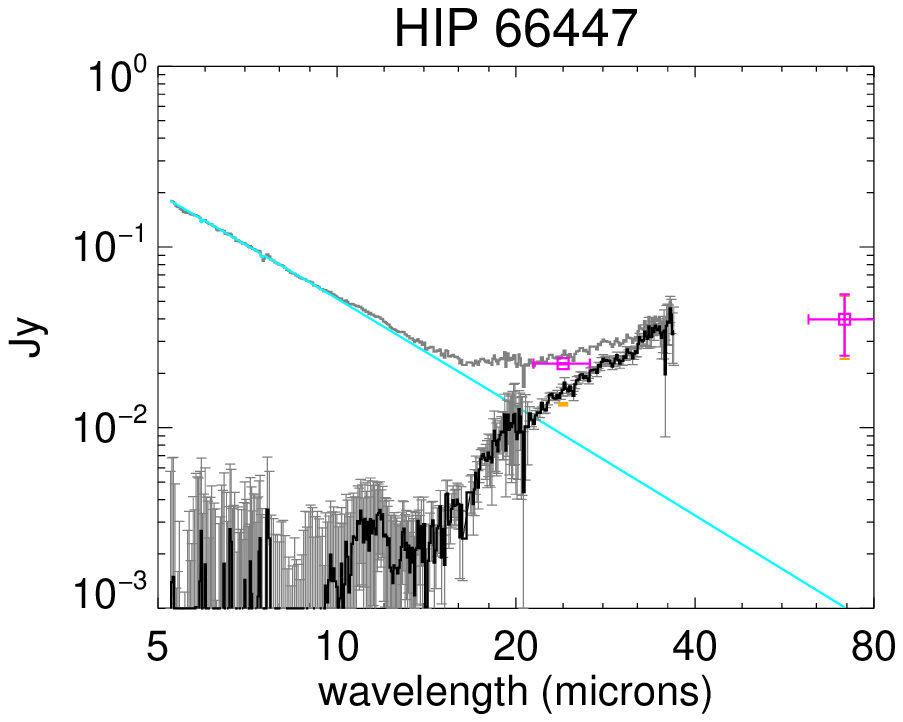} }
\parbox{\stampwidth}{
\includegraphics[width=\stampwidth]{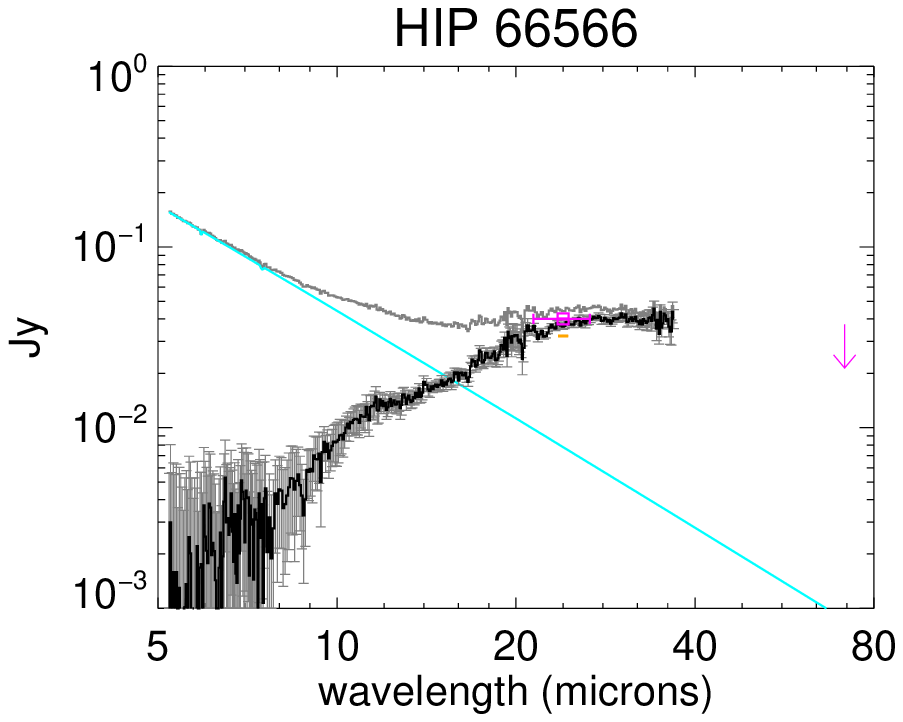} }
\\
\caption{ \label{specfig1}
Continuation Figure \ref{specfig0}, spectra of  objects.}
\end{figure}
\addtocounter{figure}{-1}
\stepcounter{subfig}
\begin{figure}
\parbox{\stampwidth}{
\includegraphics[width=\stampwidth]{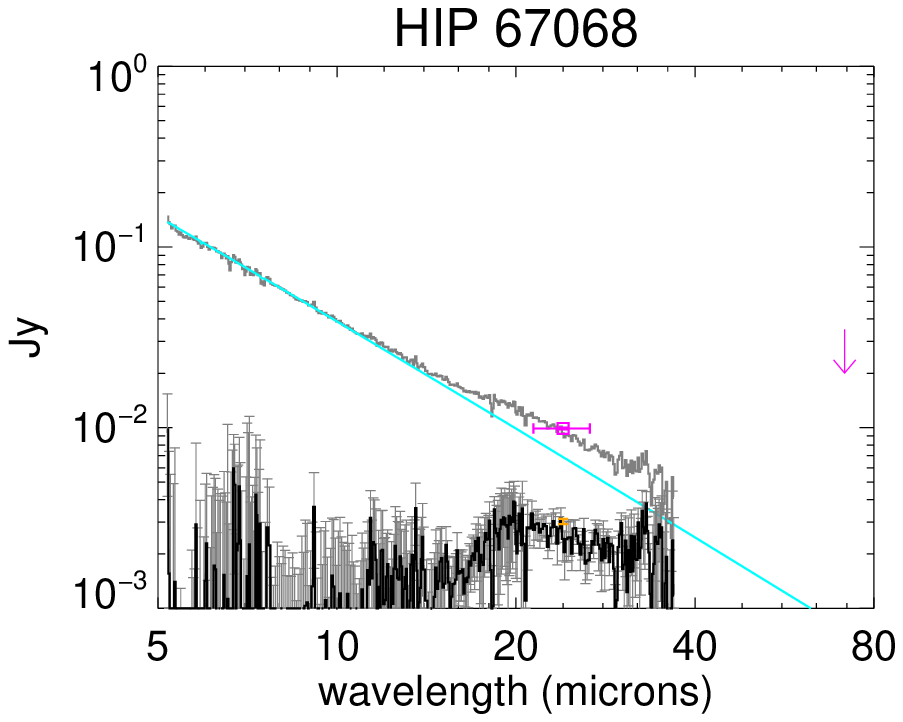} }
\parbox{\stampwidth}{
\includegraphics[width=\stampwidth]{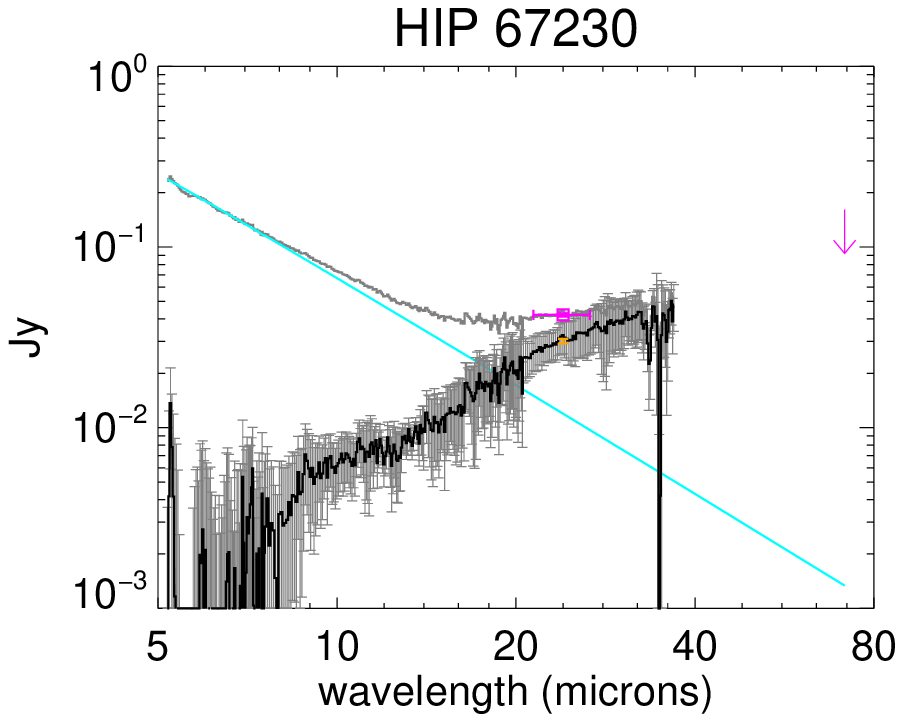} }
\parbox{\stampwidth}{
\includegraphics[width=\stampwidth]{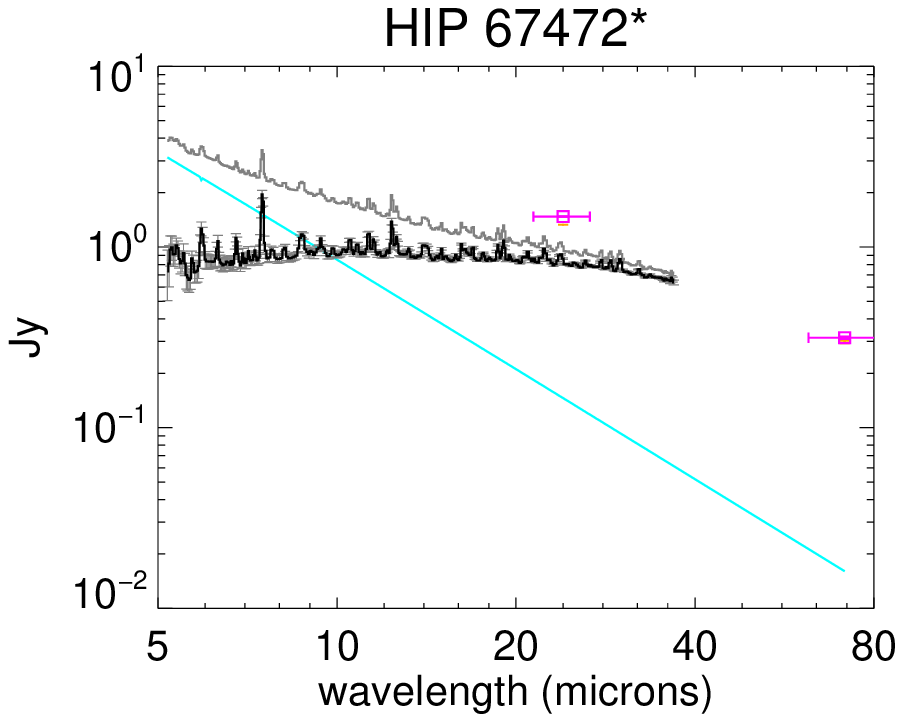} }
\parbox{\stampwidth}{
\includegraphics[width=\stampwidth]{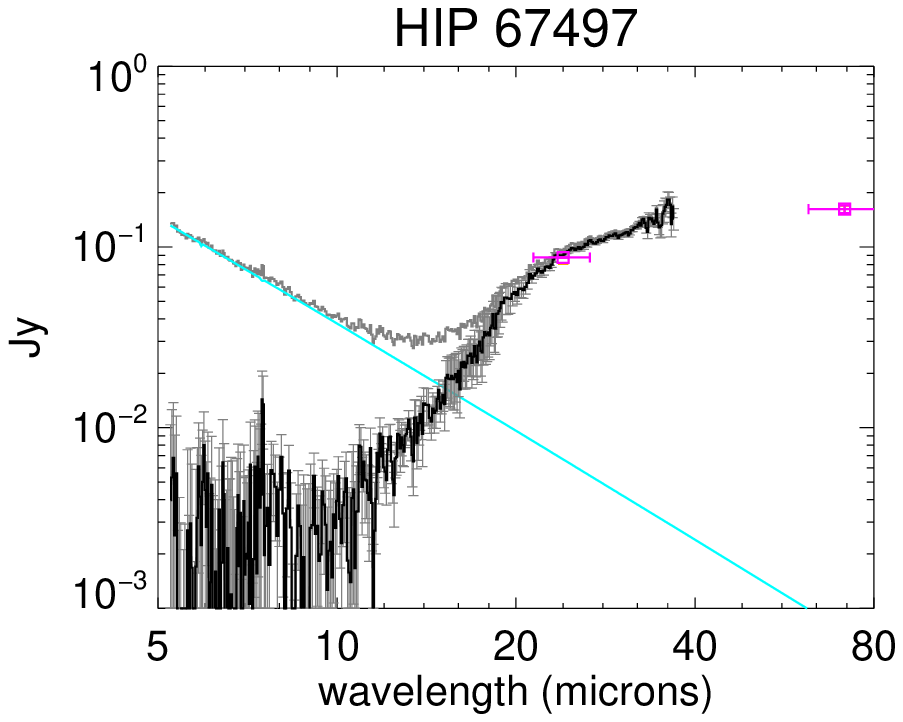} }
\\
\parbox{\stampwidth}{
\includegraphics[width=\stampwidth]{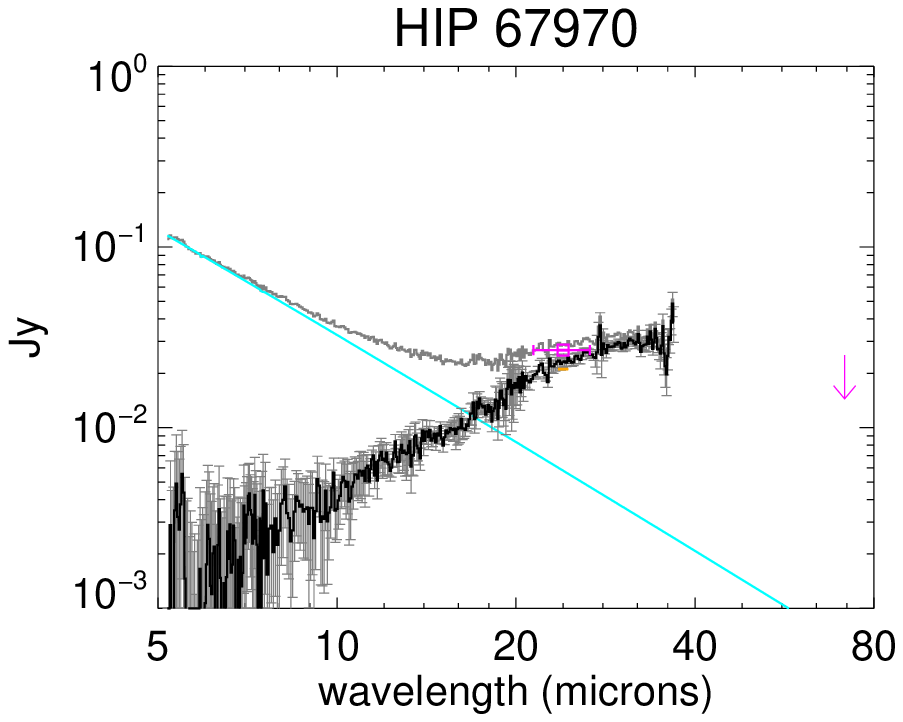} }
\parbox{\stampwidth}{
\includegraphics[width=\stampwidth]{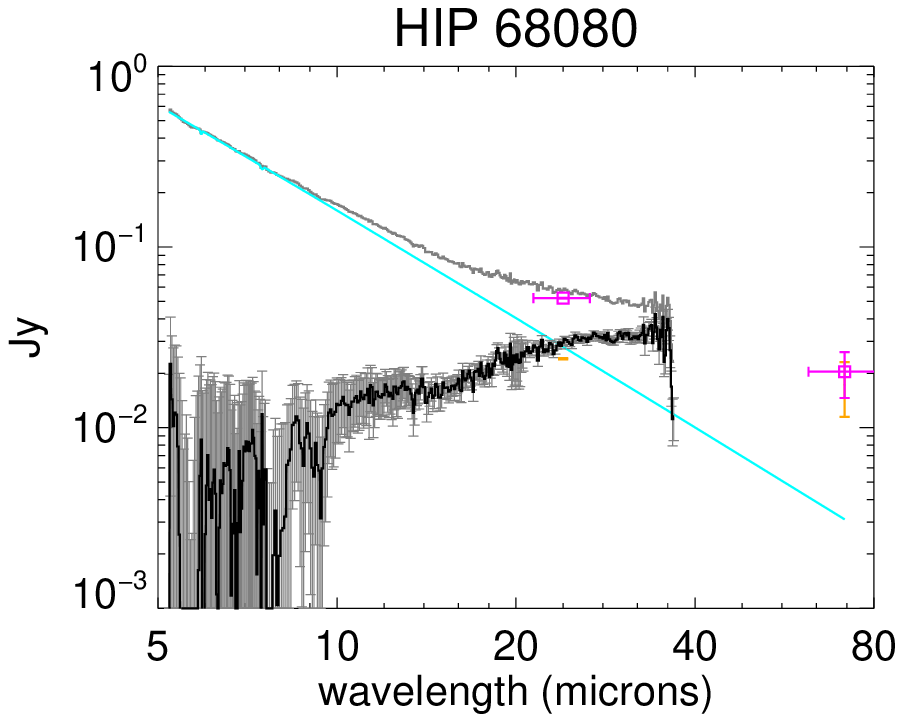} }
\parbox{\stampwidth}{
\includegraphics[width=\stampwidth]{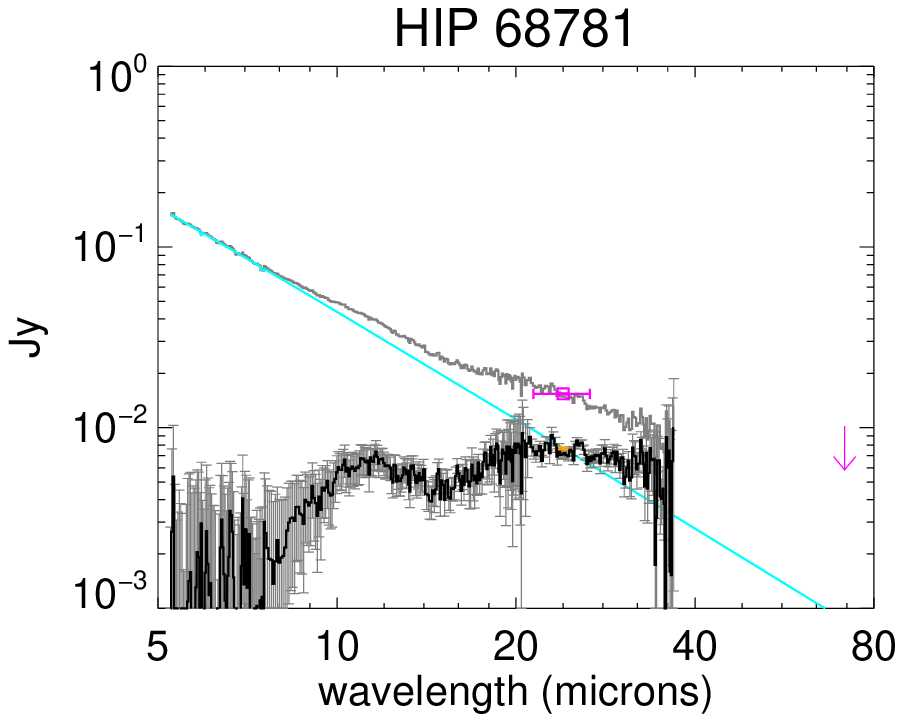} }
\parbox{\stampwidth}{
\includegraphics[width=\stampwidth]{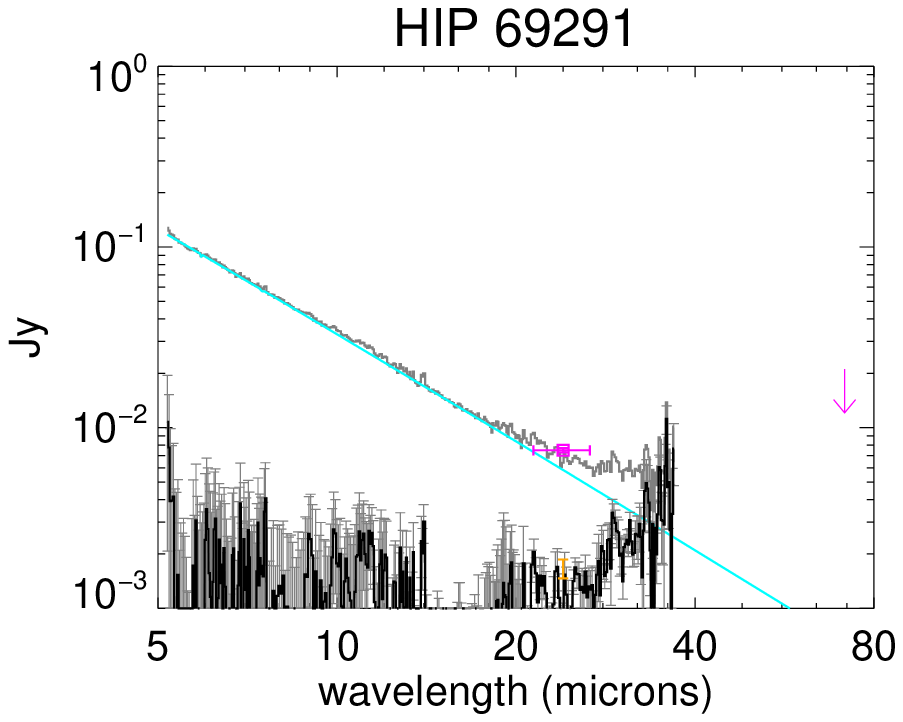} }
\\
\parbox{\stampwidth}{
\includegraphics[width=\stampwidth]{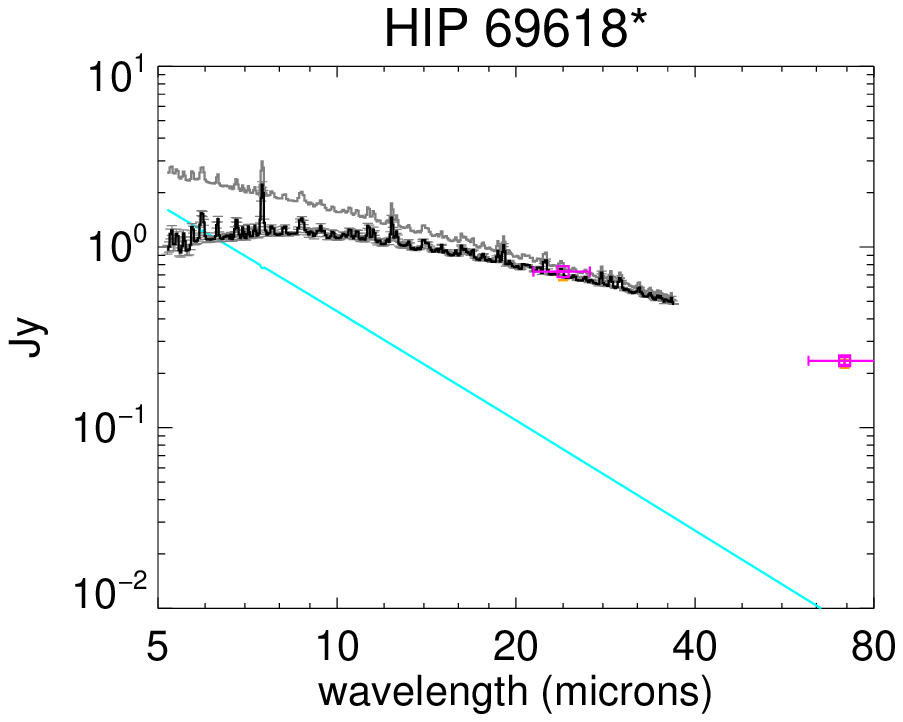} }
\parbox{\stampwidth}{
\includegraphics[width=\stampwidth]{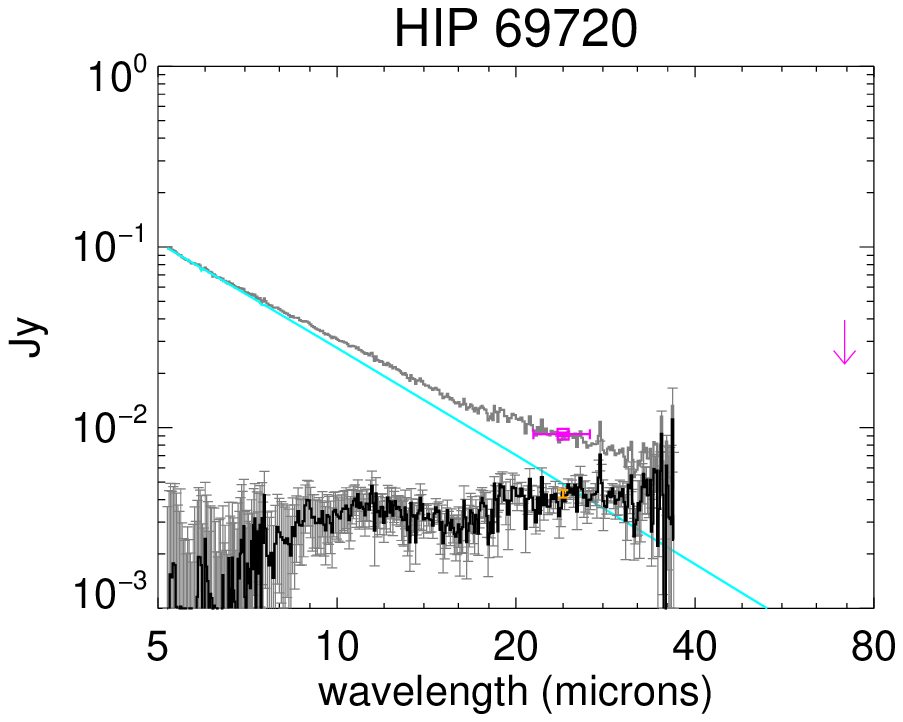} }
\parbox{\stampwidth}{
\includegraphics[width=\stampwidth]{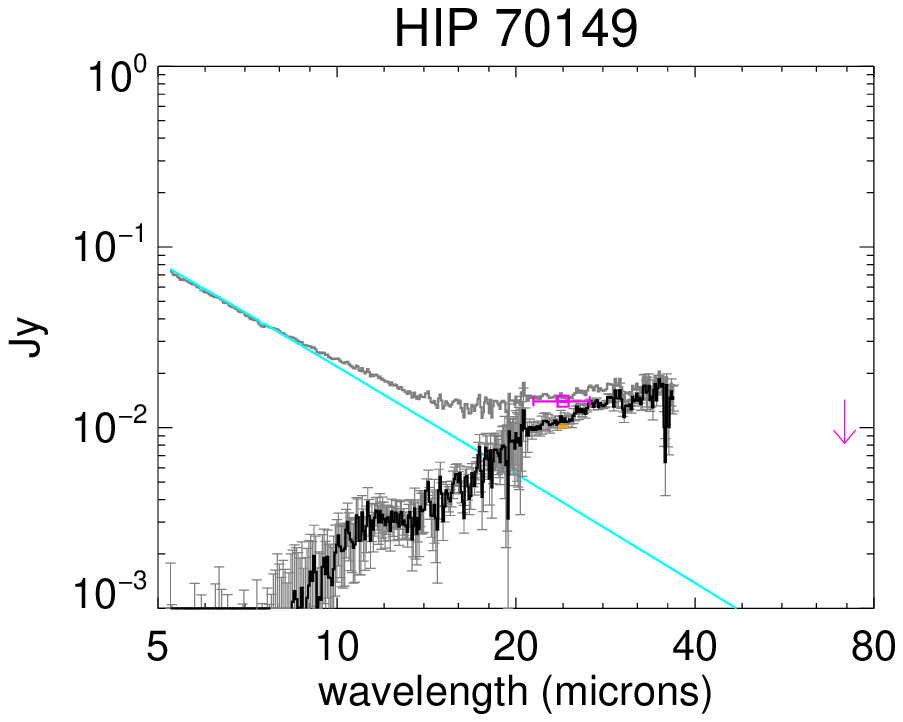} }
\parbox{\stampwidth}{
\includegraphics[width=\stampwidth]{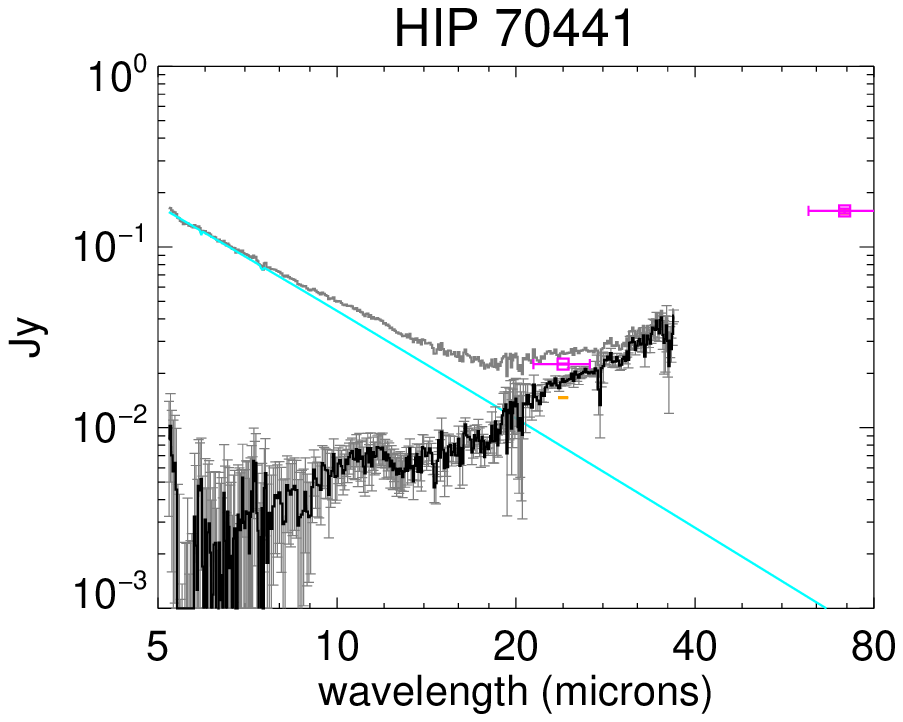} }
\\
\parbox{\stampwidth}{
\includegraphics[width=\stampwidth]{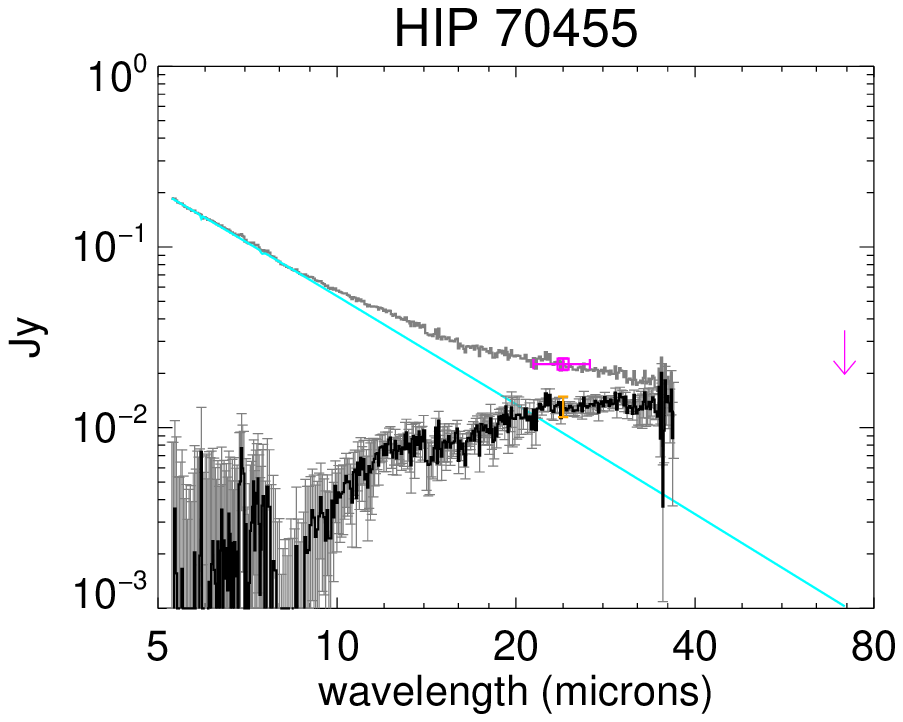} }
\parbox{\stampwidth}{
\includegraphics[width=\stampwidth]{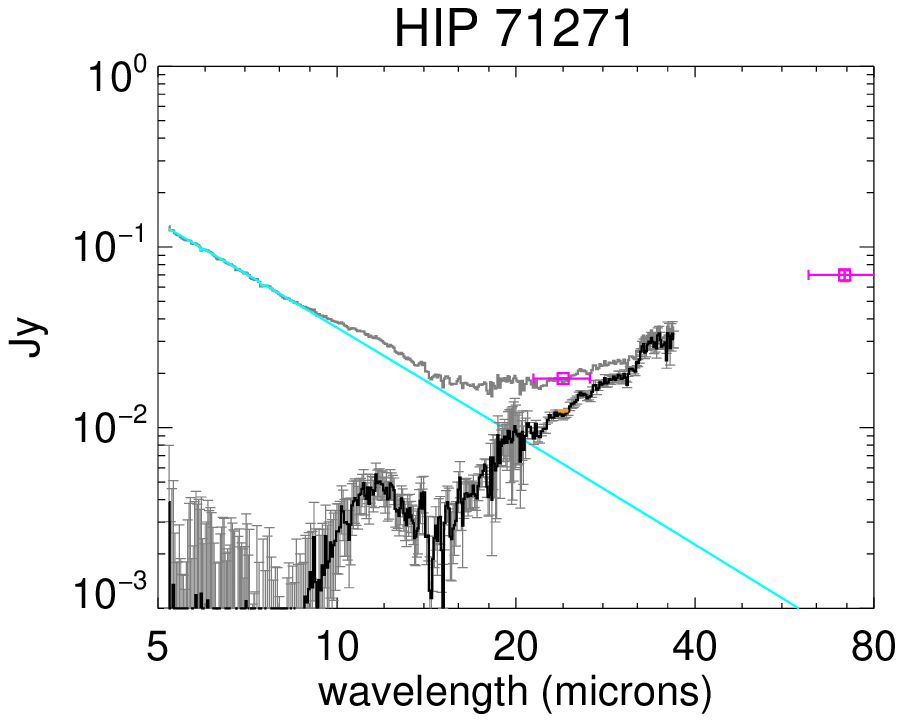} }
\parbox{\stampwidth}{
\includegraphics[width=\stampwidth]{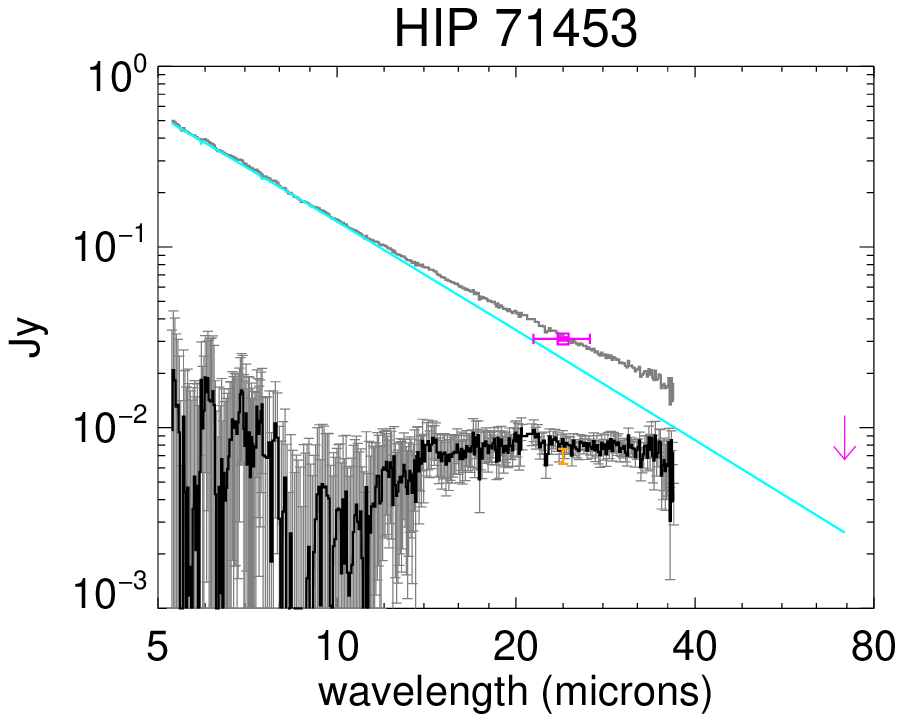} }
\parbox{\stampwidth}{
\includegraphics[width=\stampwidth]{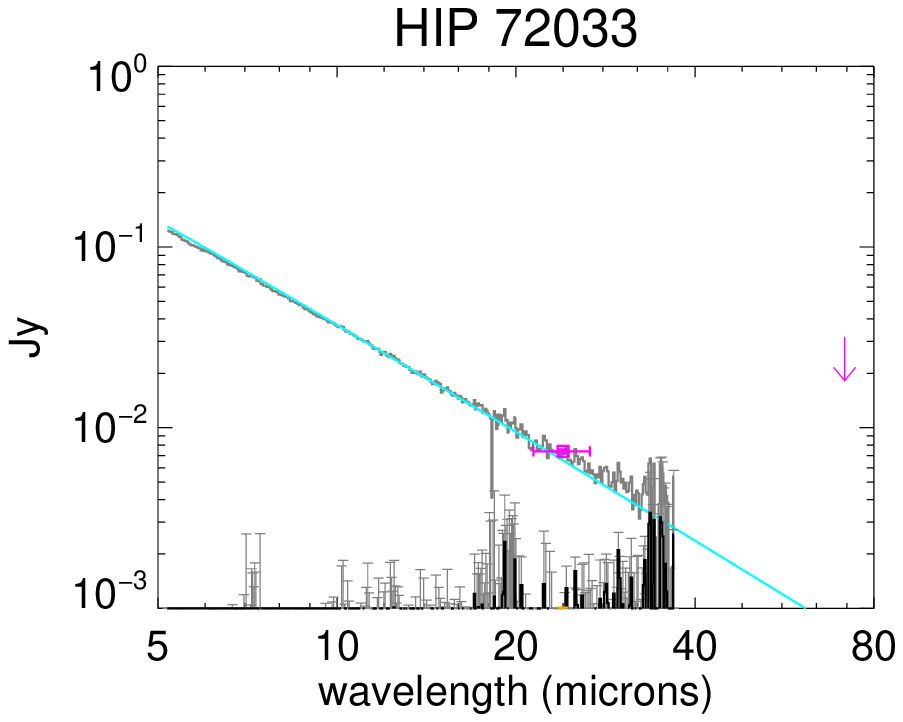} }
\\
\parbox{\stampwidth}{
\includegraphics[width=\stampwidth]{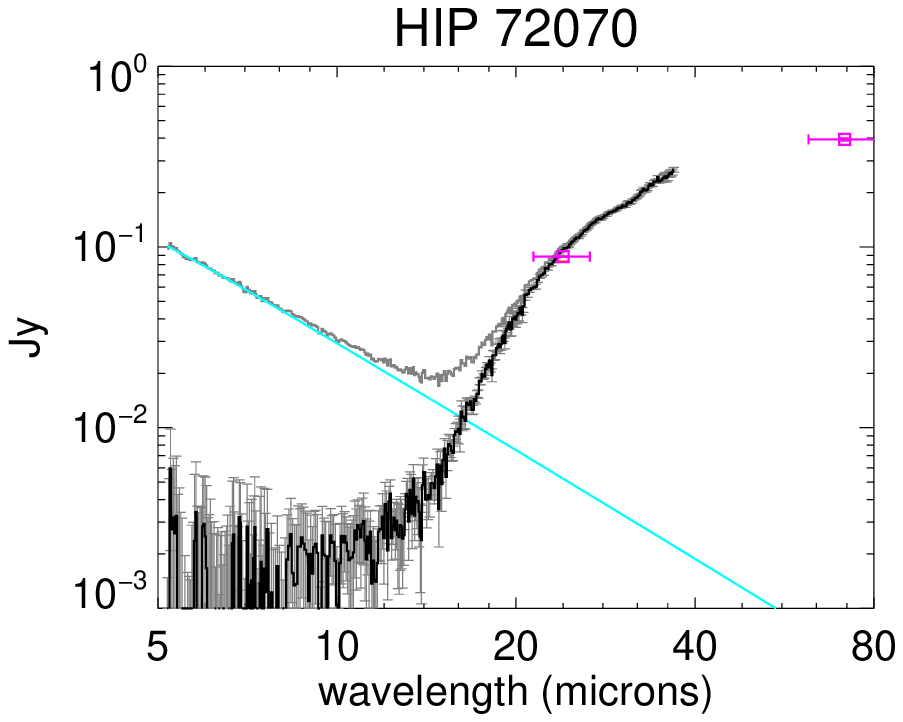} }
\parbox{\stampwidth}{
\includegraphics[width=\stampwidth]{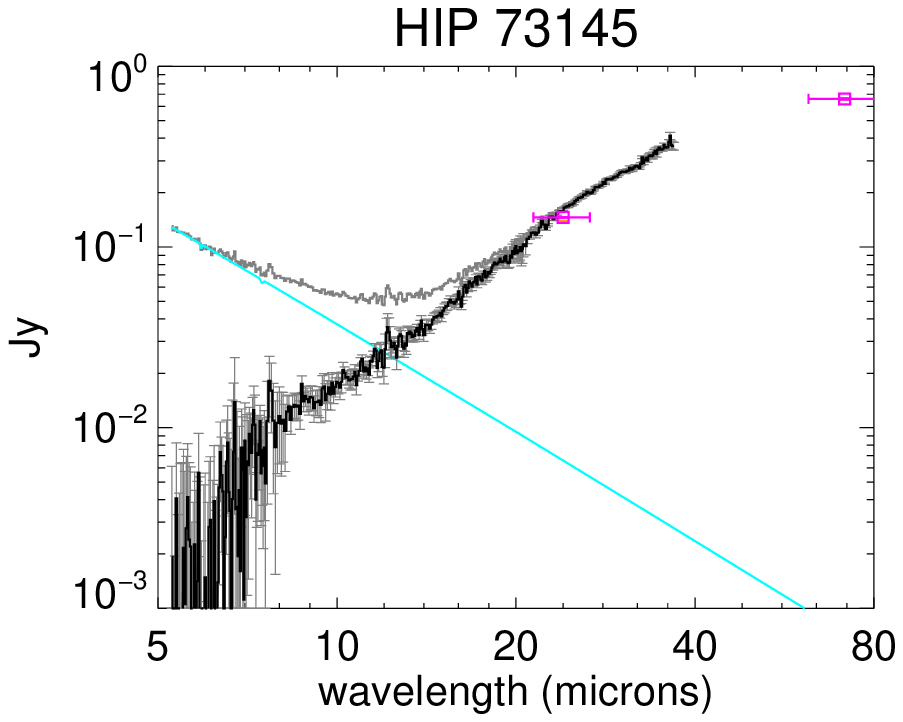} }
\parbox{\stampwidth}{
\includegraphics[width=\stampwidth]{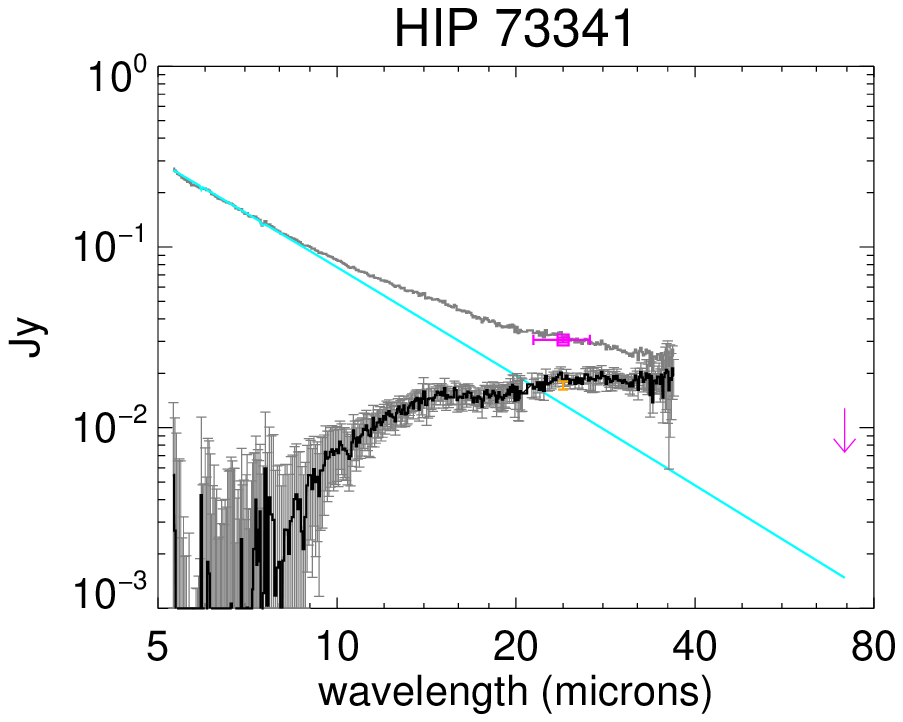} }
\parbox{\stampwidth}{
\includegraphics[width=\stampwidth]{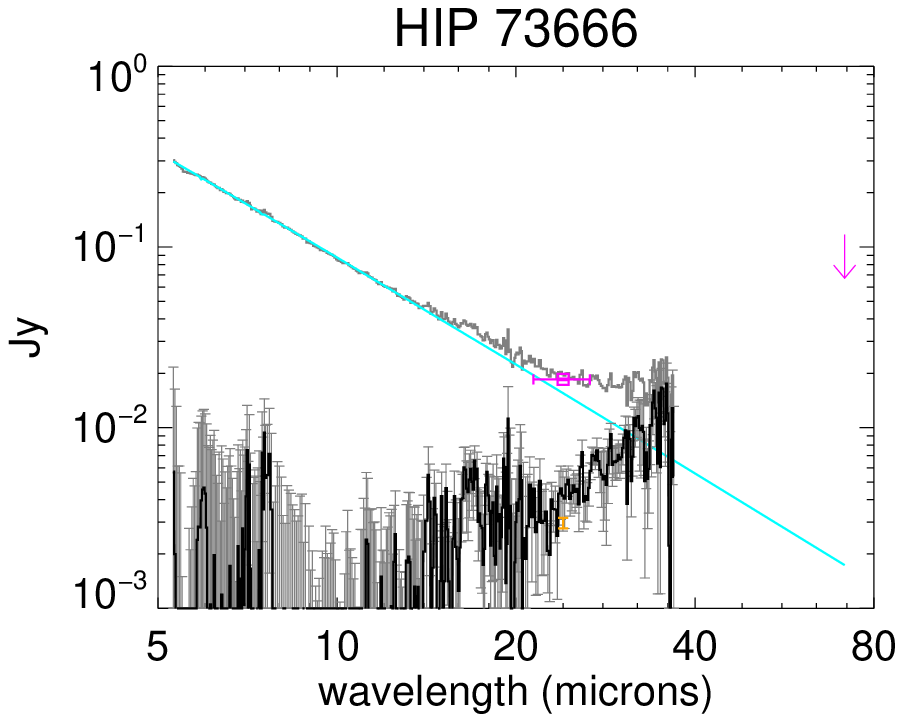} }
\\
\parbox{\stampwidth}{
\includegraphics[width=\stampwidth]{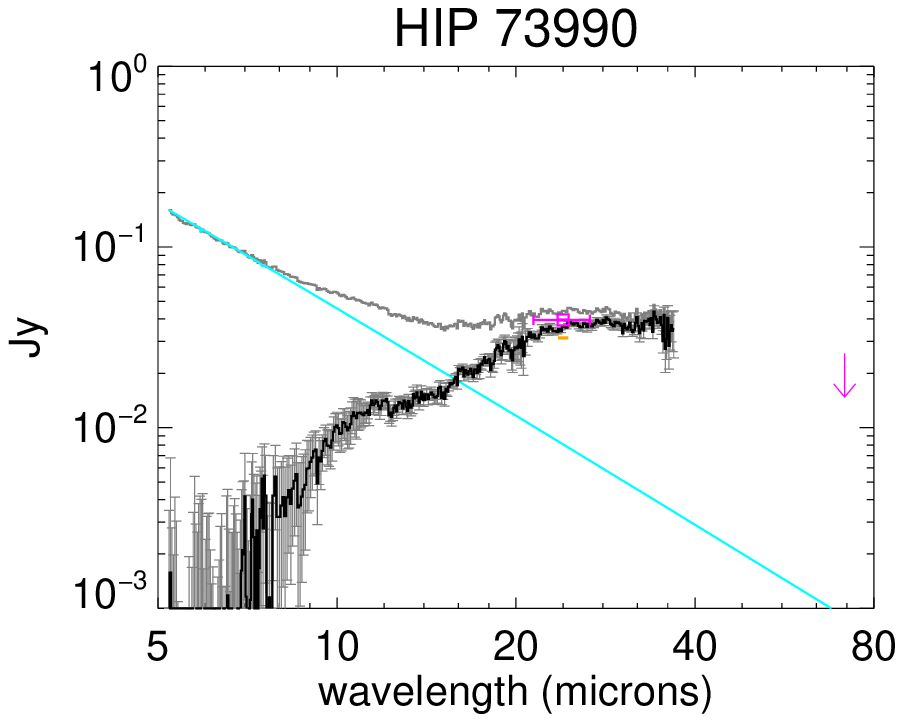} }
\parbox{\stampwidth}{
\includegraphics[width=\stampwidth]{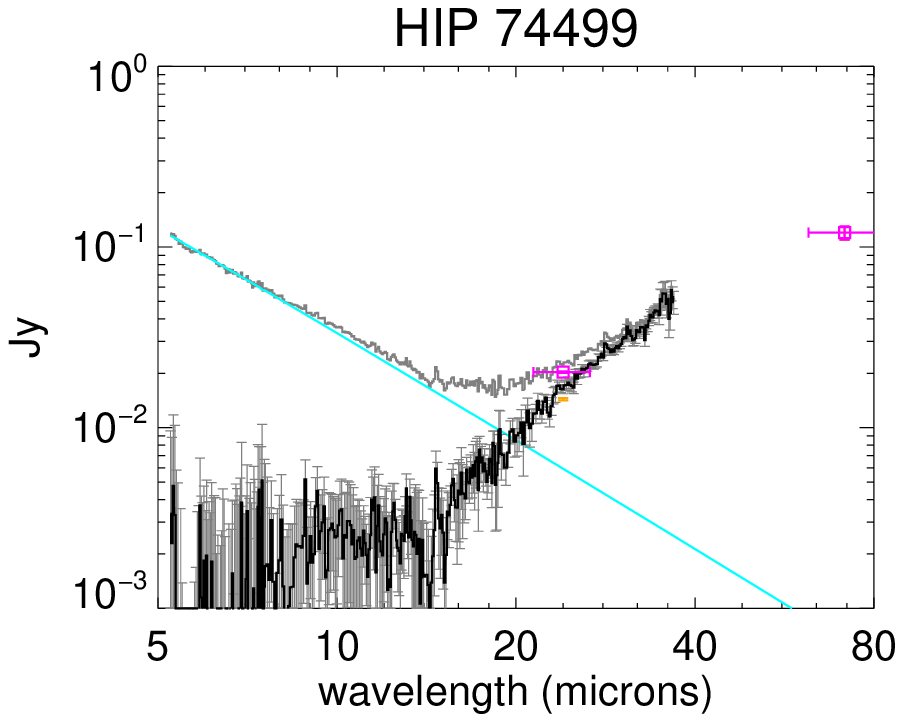} }
\parbox{\stampwidth}{
\includegraphics[width=\stampwidth]{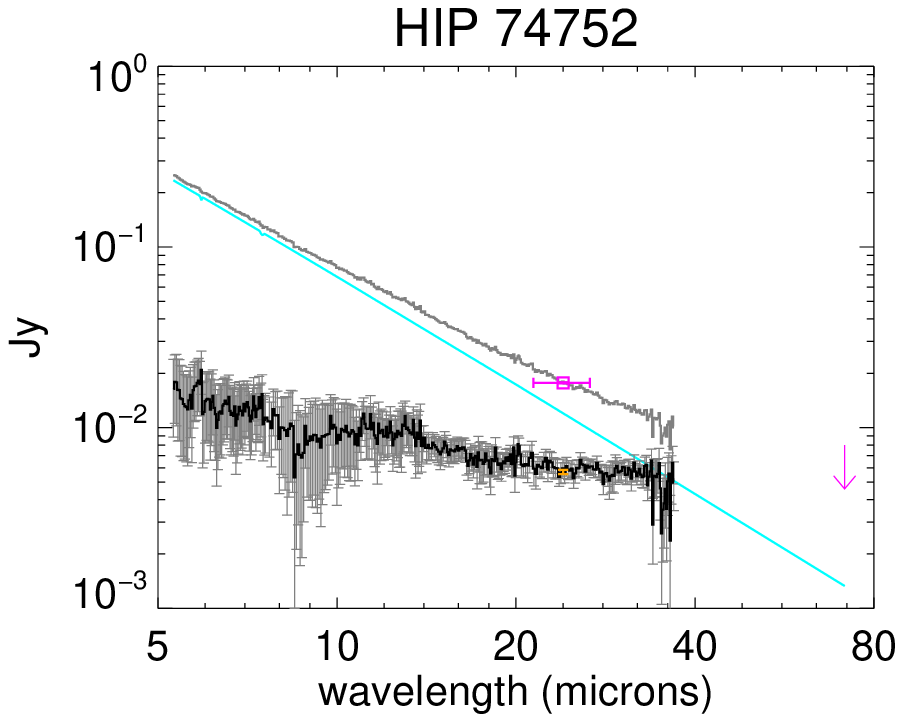} }
\parbox{\stampwidth}{
\includegraphics[width=\stampwidth]{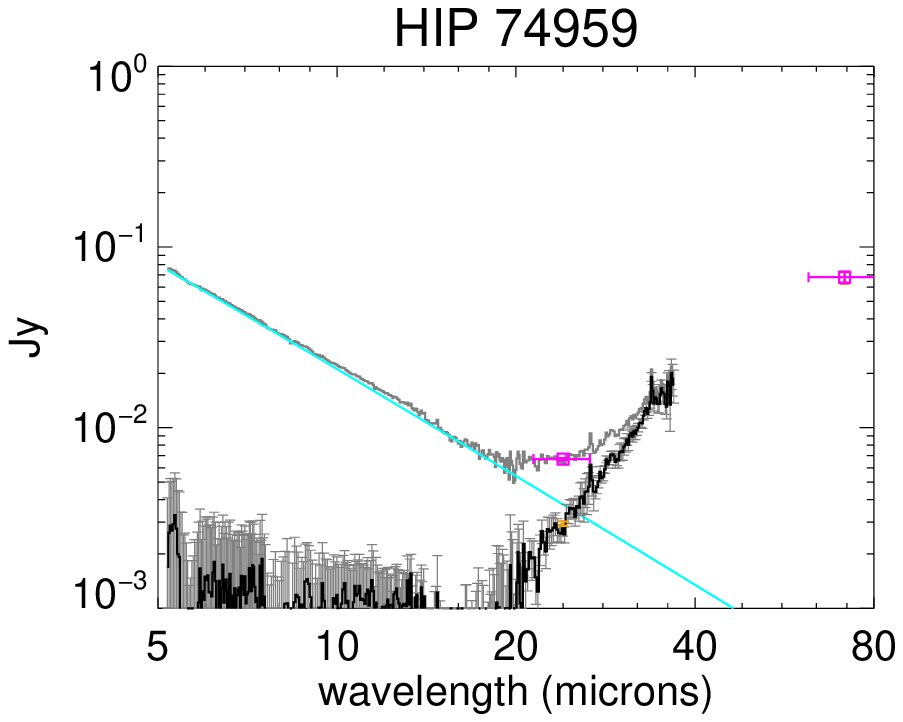} }
\\
\caption{ \label{specfig2}
Continuation Figure \ref{specfig0}, spectra of  objects.}
\end{figure}
\addtocounter{figure}{-1}
\stepcounter{subfig}
\begin{figure}
\parbox{\stampwidth}{
\includegraphics[width=\stampwidth]{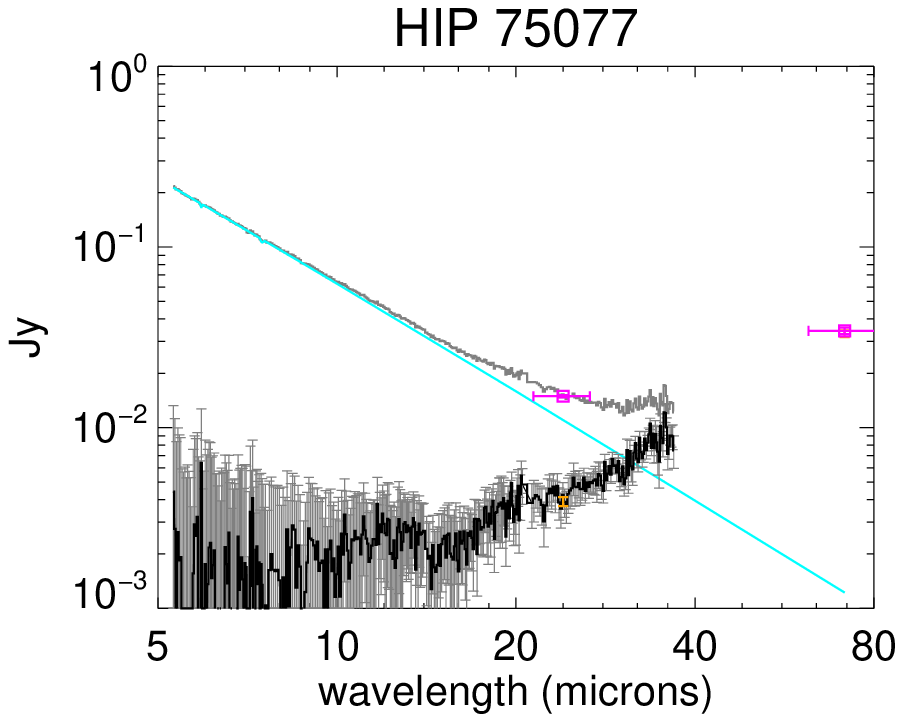} }
\parbox{\stampwidth}{
\includegraphics[width=\stampwidth]{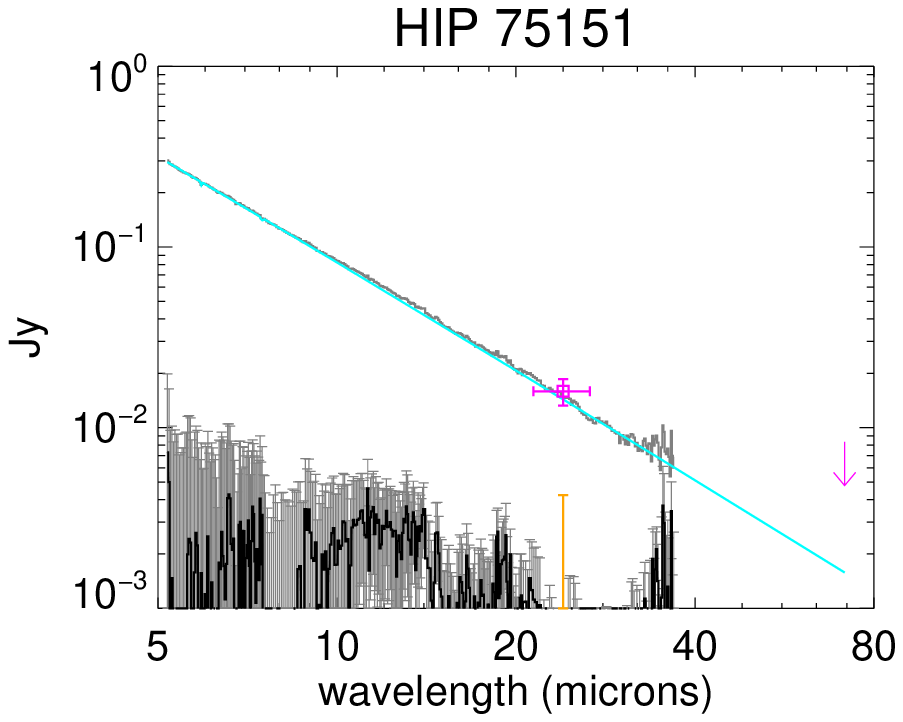} }
\parbox{\stampwidth}{
\includegraphics[width=\stampwidth]{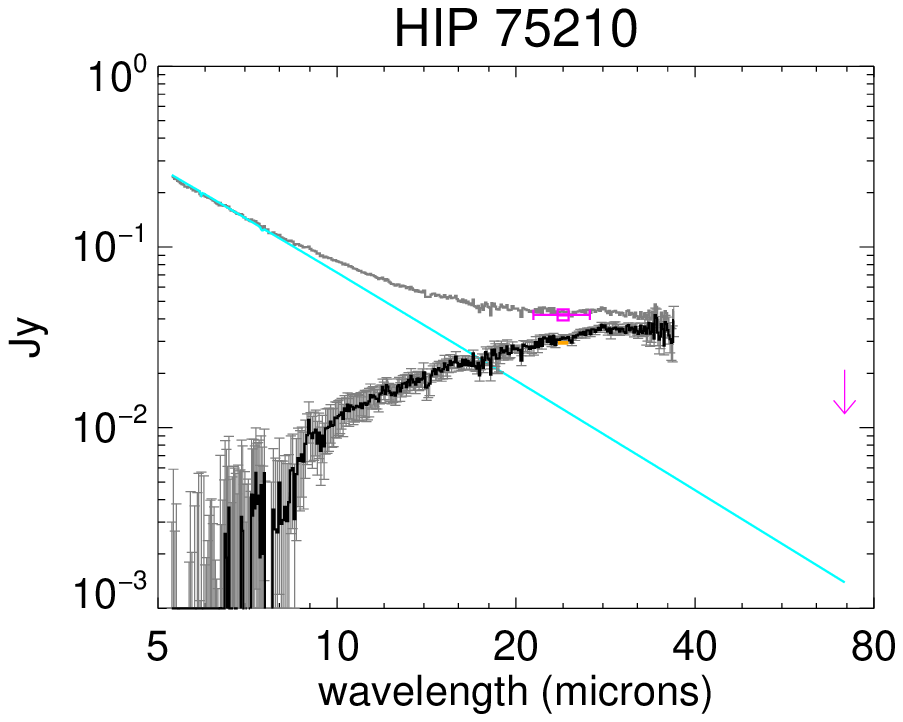} }
\parbox{\stampwidth}{
\includegraphics[width=\stampwidth]{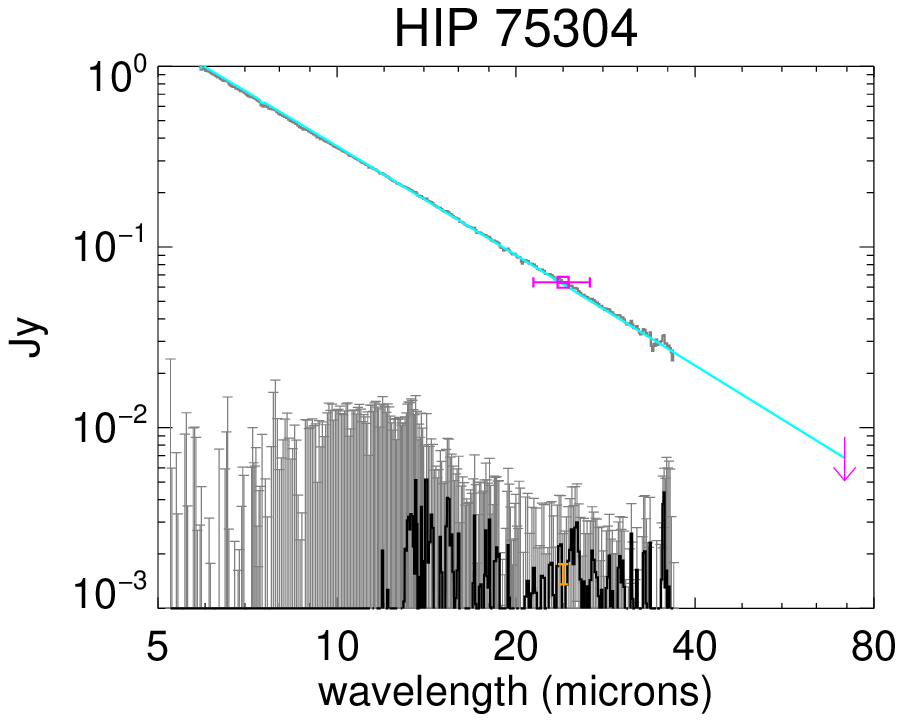} }
\\
\parbox{\stampwidth}{
\includegraphics[width=\stampwidth]{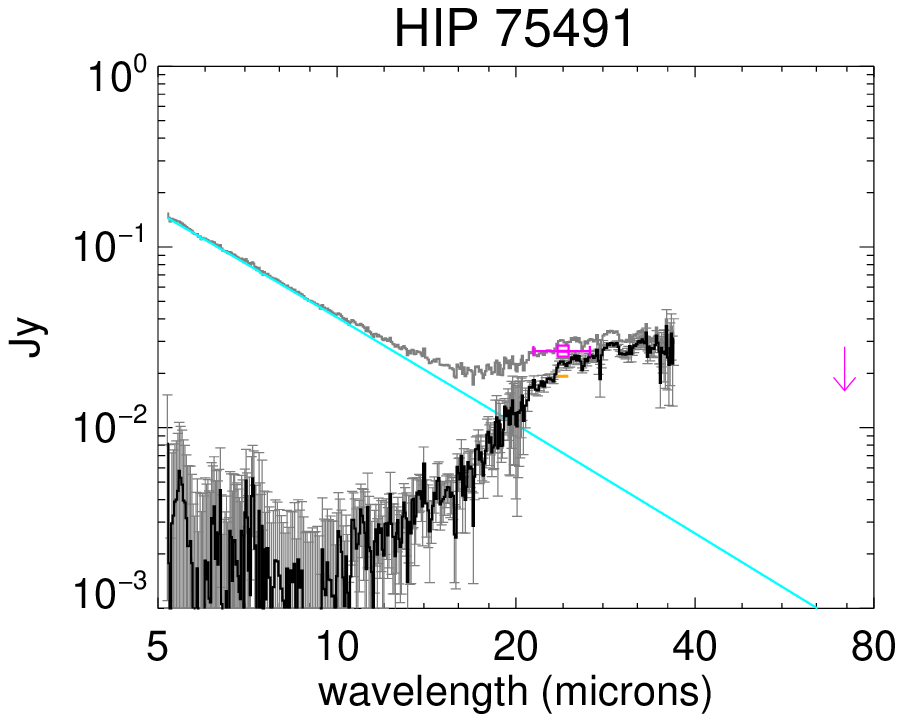} }
\parbox{\stampwidth}{
\includegraphics[width=\stampwidth]{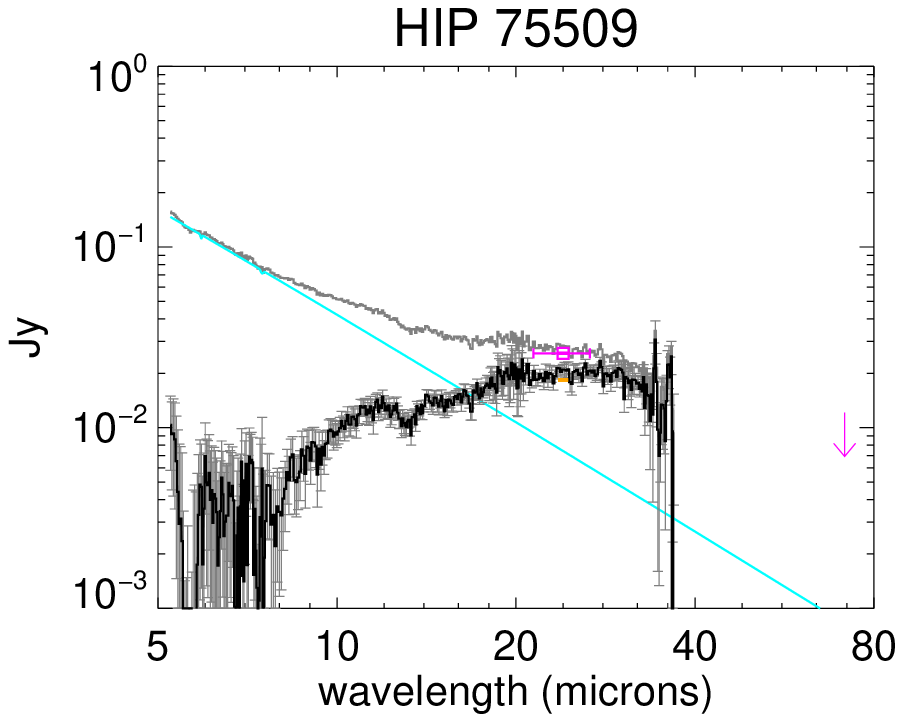} }
\parbox{\stampwidth}{
\includegraphics[width=\stampwidth]{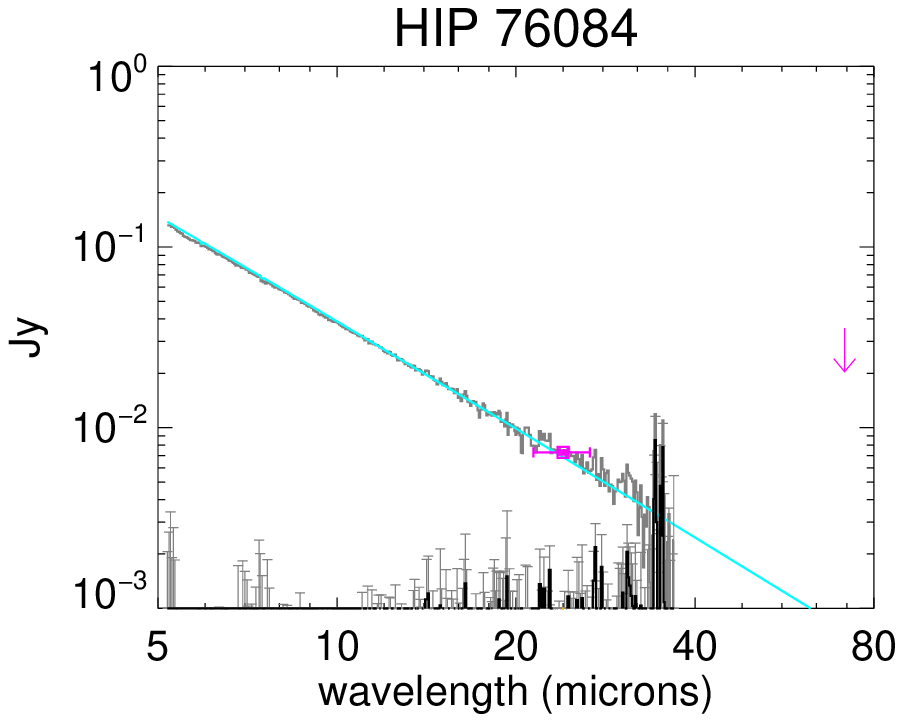} }
\parbox{\stampwidth}{
\includegraphics[width=\stampwidth]{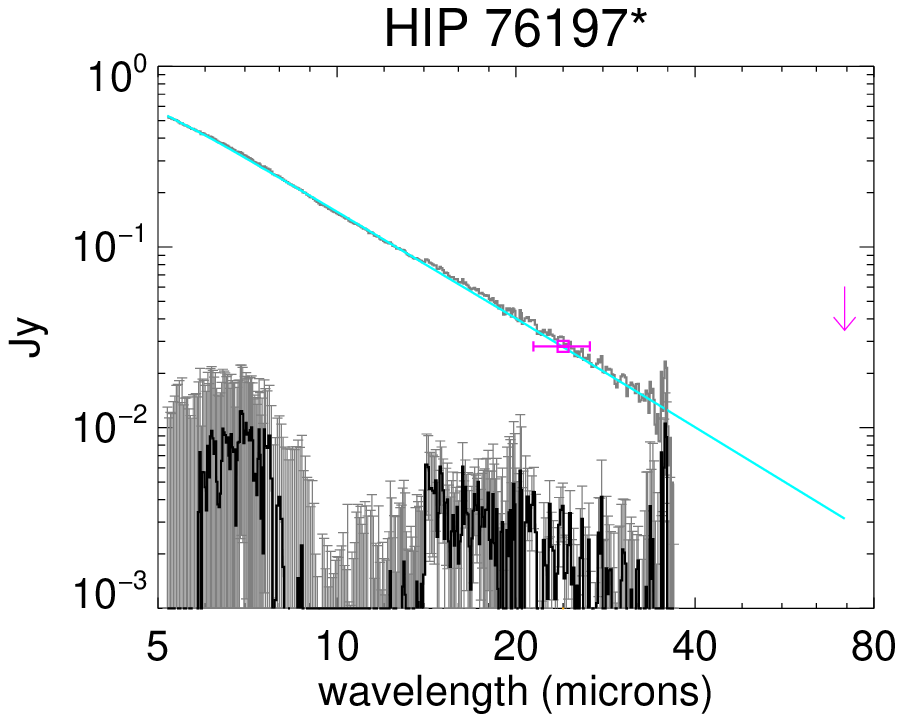} }
\\
\parbox{\stampwidth}{
\includegraphics[width=\stampwidth]{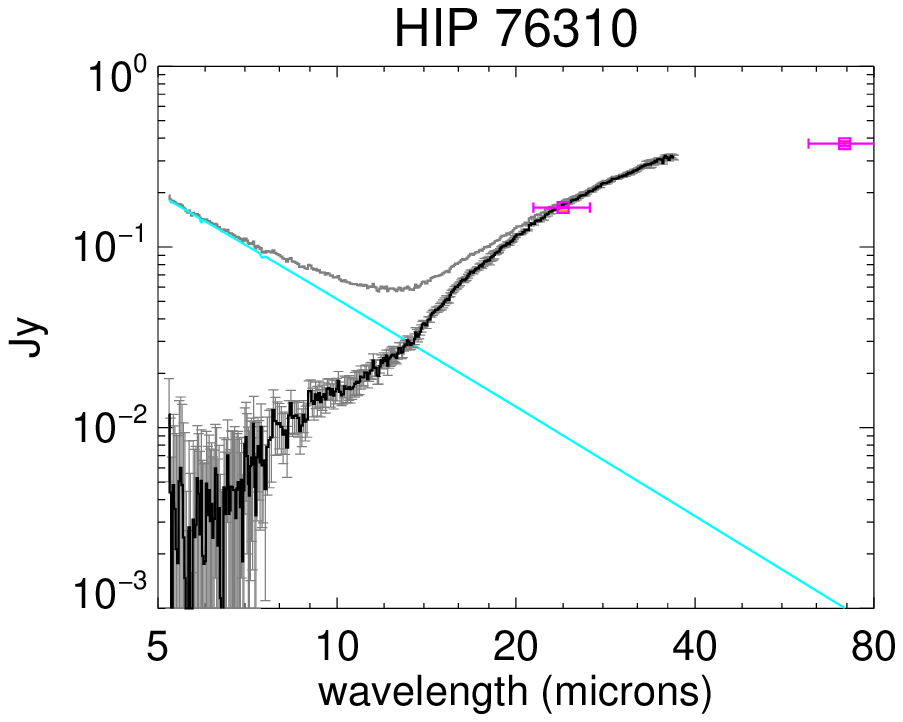} }
\parbox{\stampwidth}{
\includegraphics[width=\stampwidth]{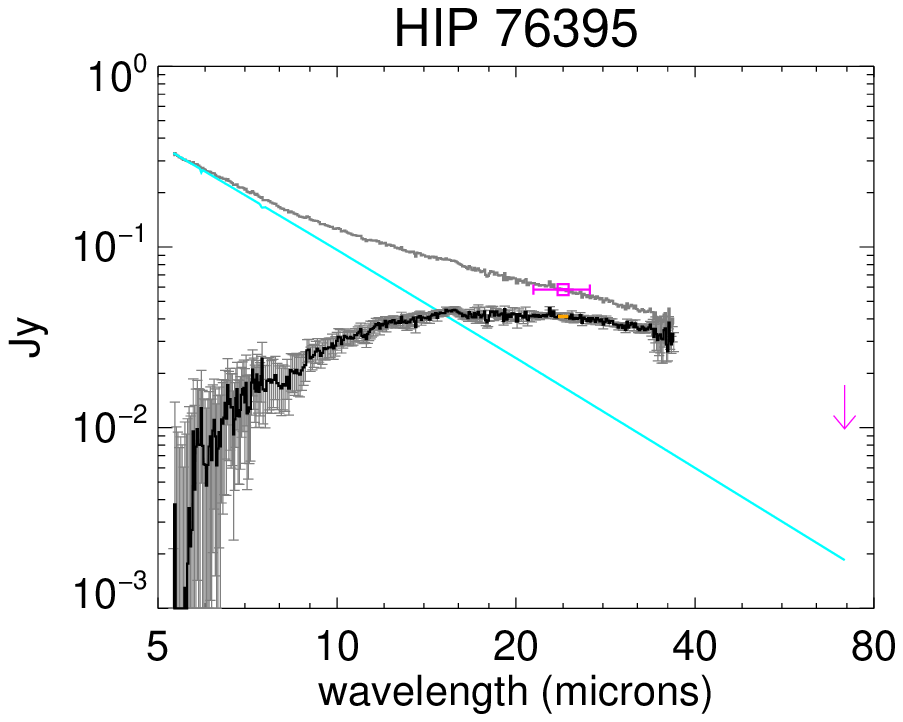} }
\parbox{\stampwidth}{
\includegraphics[width=\stampwidth]{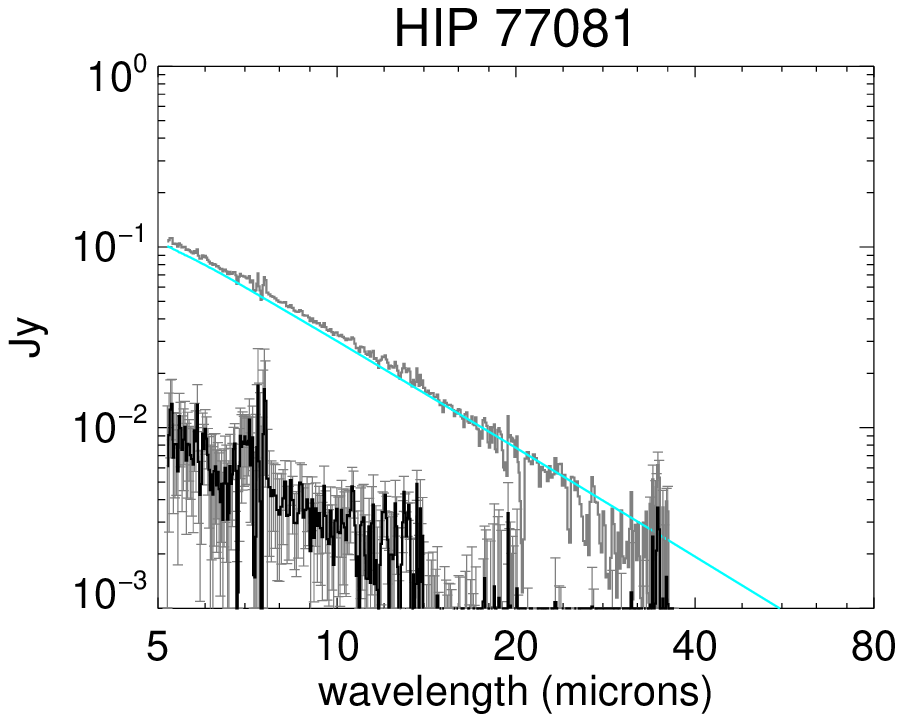} }
\parbox{\stampwidth}{
\includegraphics[width=\stampwidth]{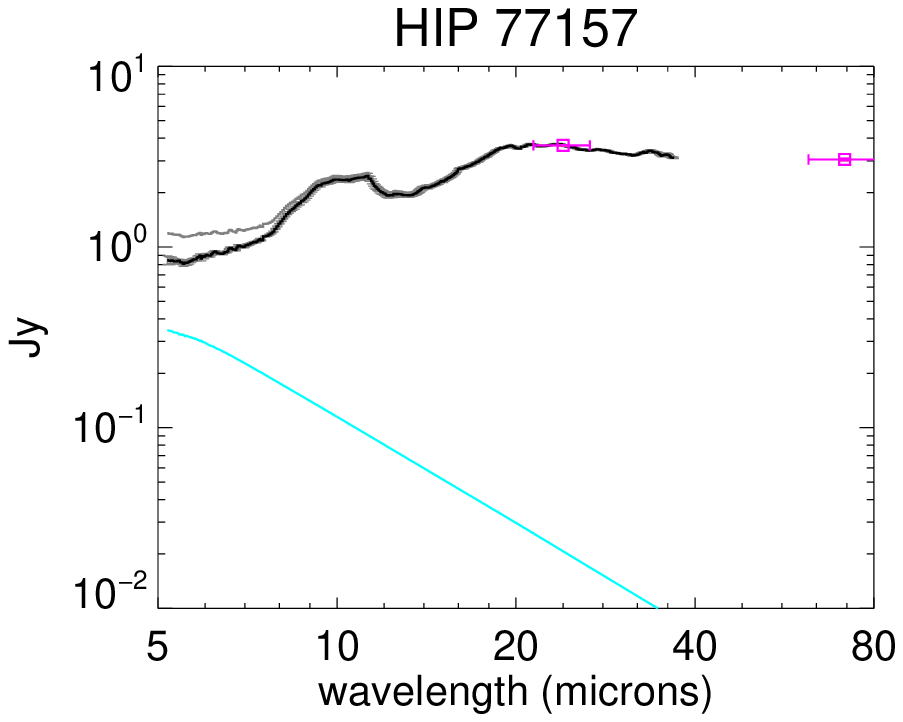} }
\\
\parbox{\stampwidth}{
\includegraphics[width=\stampwidth]{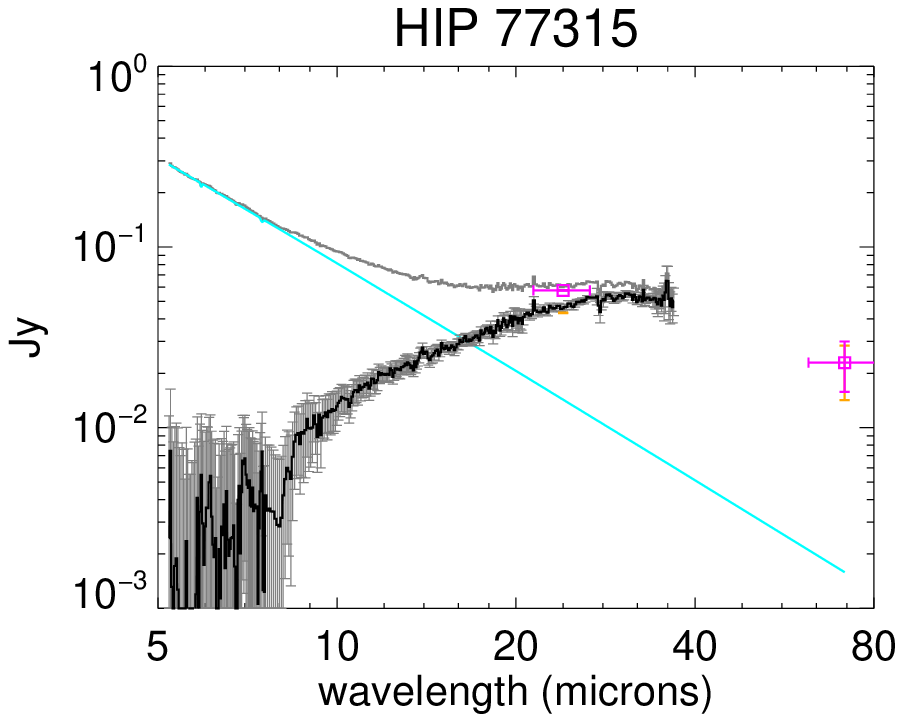} }
\parbox{\stampwidth}{
\includegraphics[width=\stampwidth]{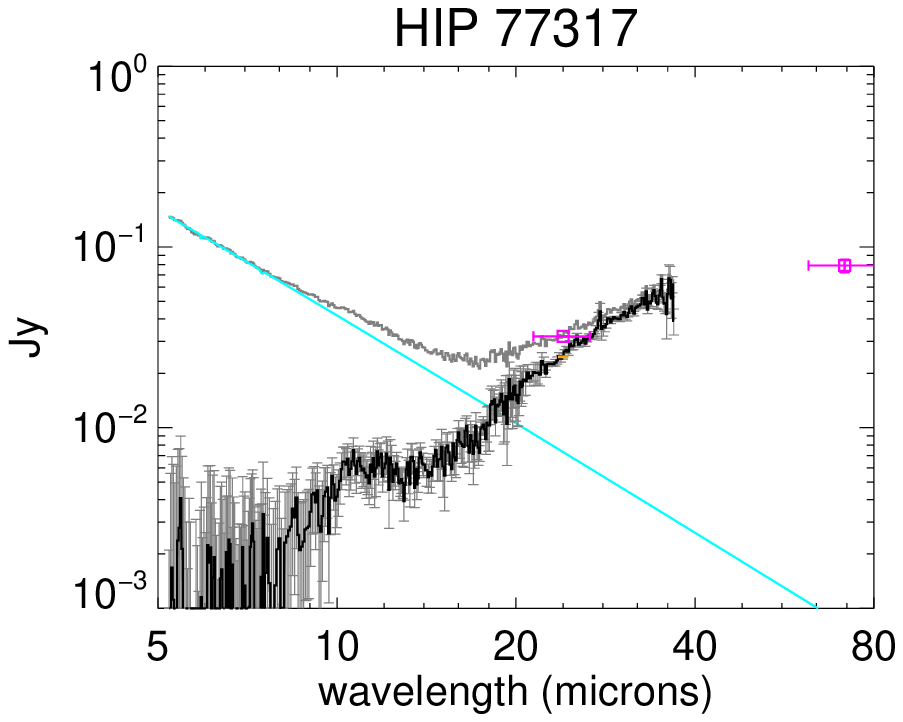} }
\parbox{\stampwidth}{
\includegraphics[width=\stampwidth]{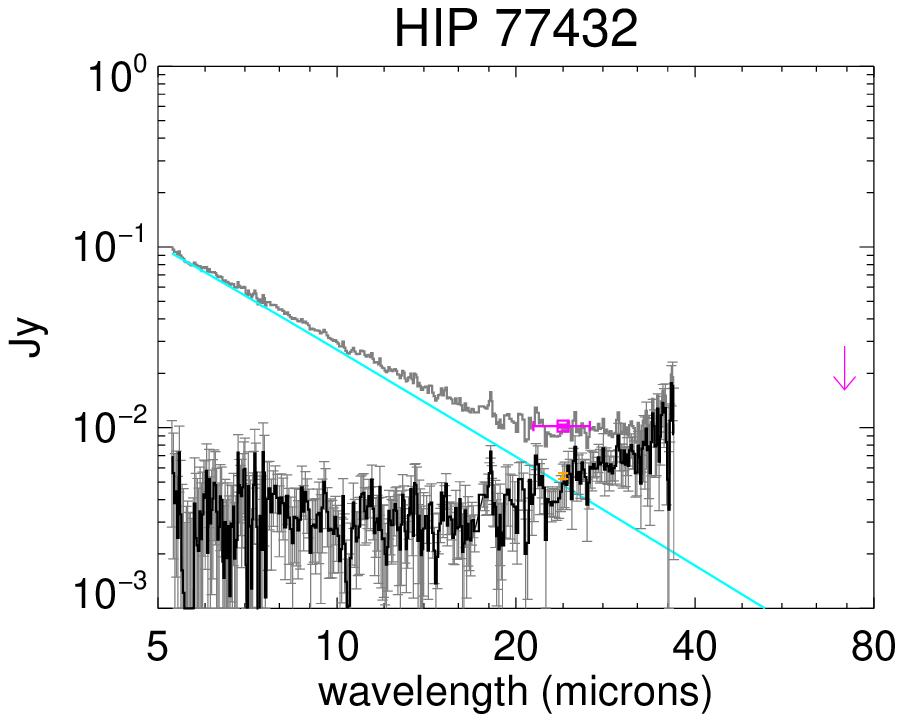} }
\parbox{\stampwidth}{
\includegraphics[width=\stampwidth]{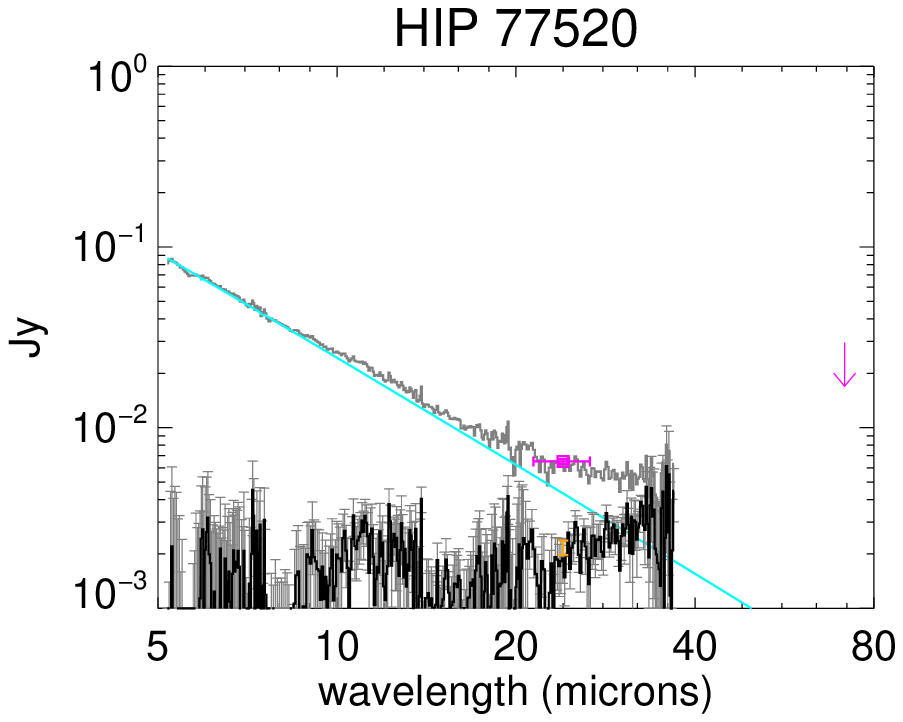} }
\\
\parbox{\stampwidth}{
\includegraphics[width=\stampwidth]{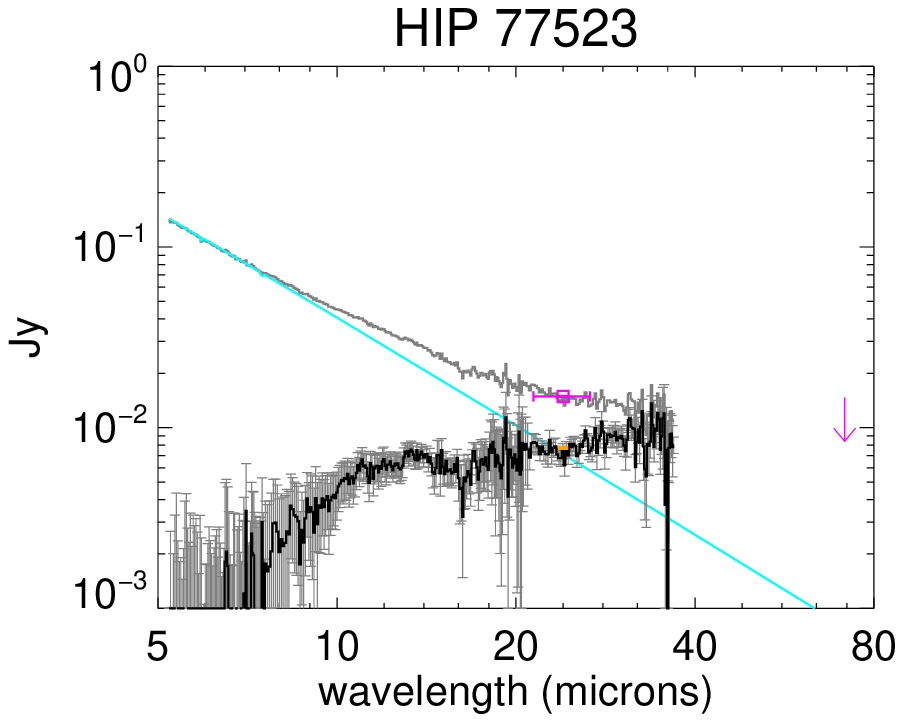} }
\parbox{\stampwidth}{
\includegraphics[width=\stampwidth]{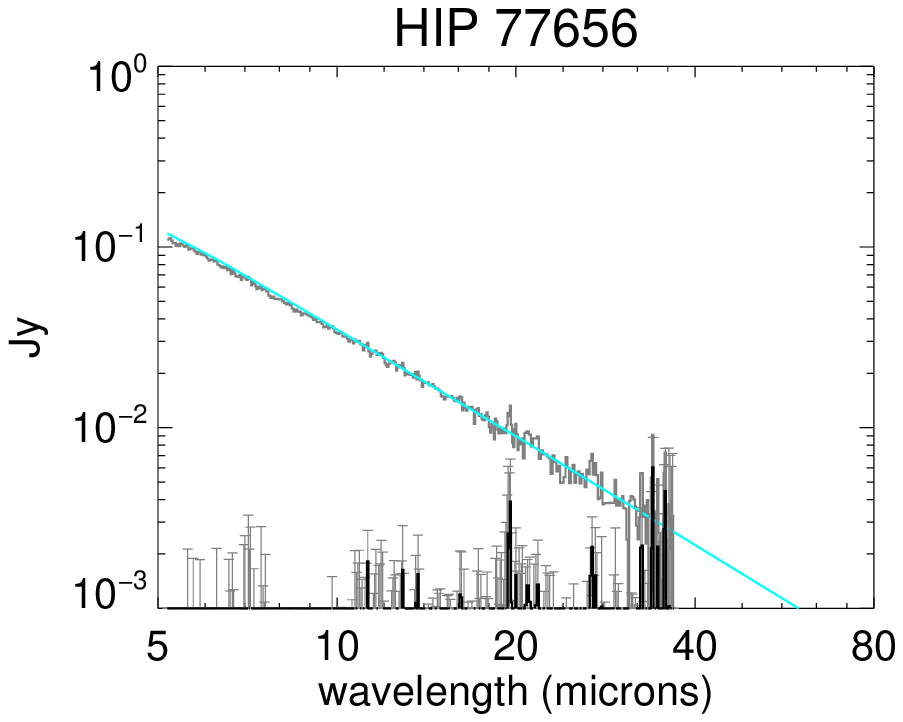} }
\parbox{\stampwidth}{
\includegraphics[width=\stampwidth]{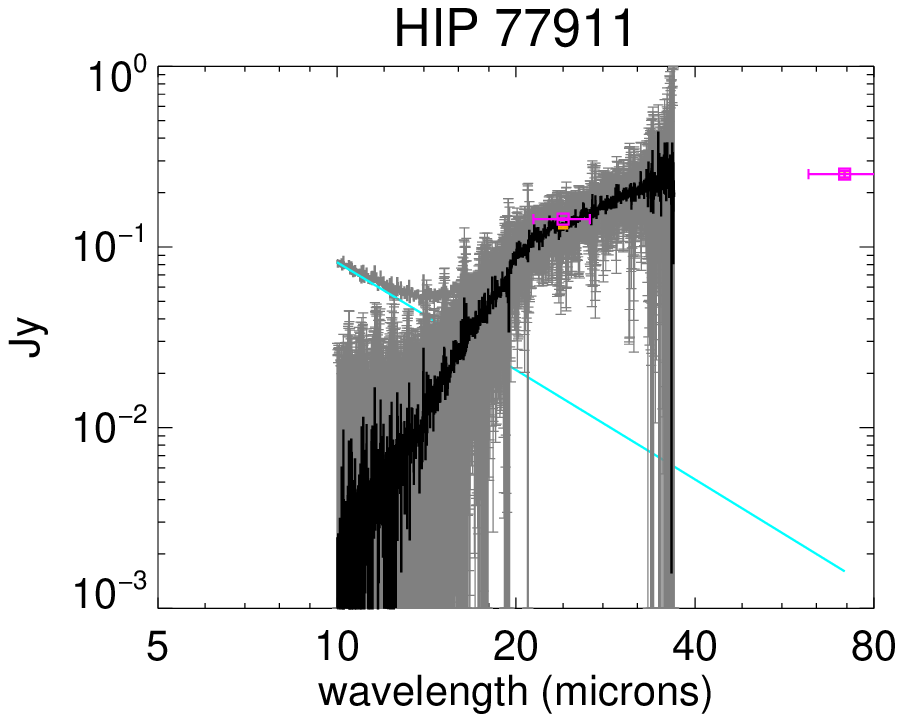} }
\parbox{\stampwidth}{
\includegraphics[width=\stampwidth]{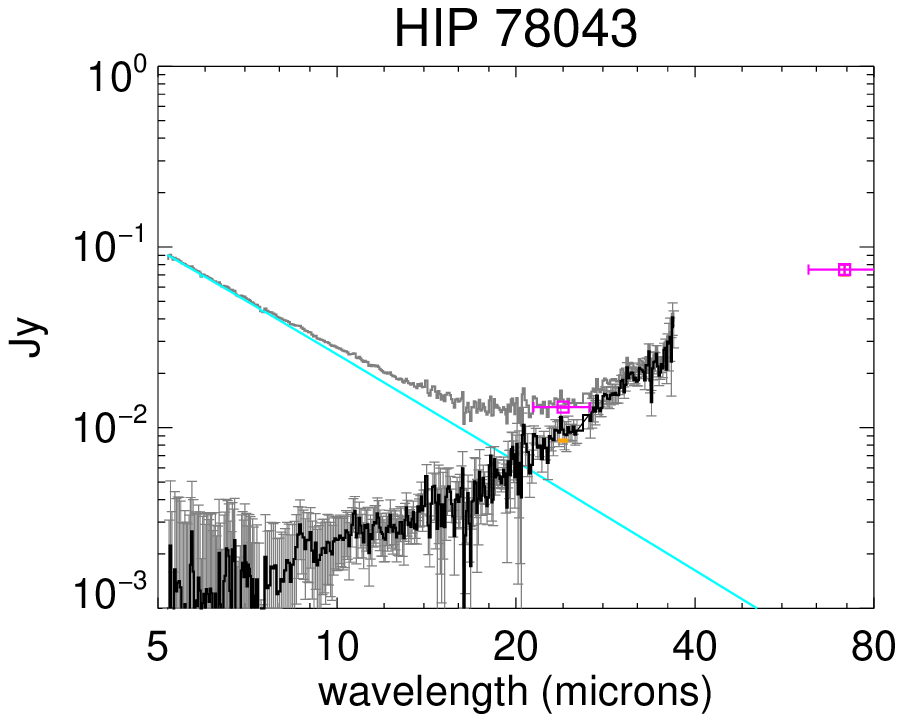} }
\\
\parbox{\stampwidth}{
\includegraphics[width=\stampwidth]{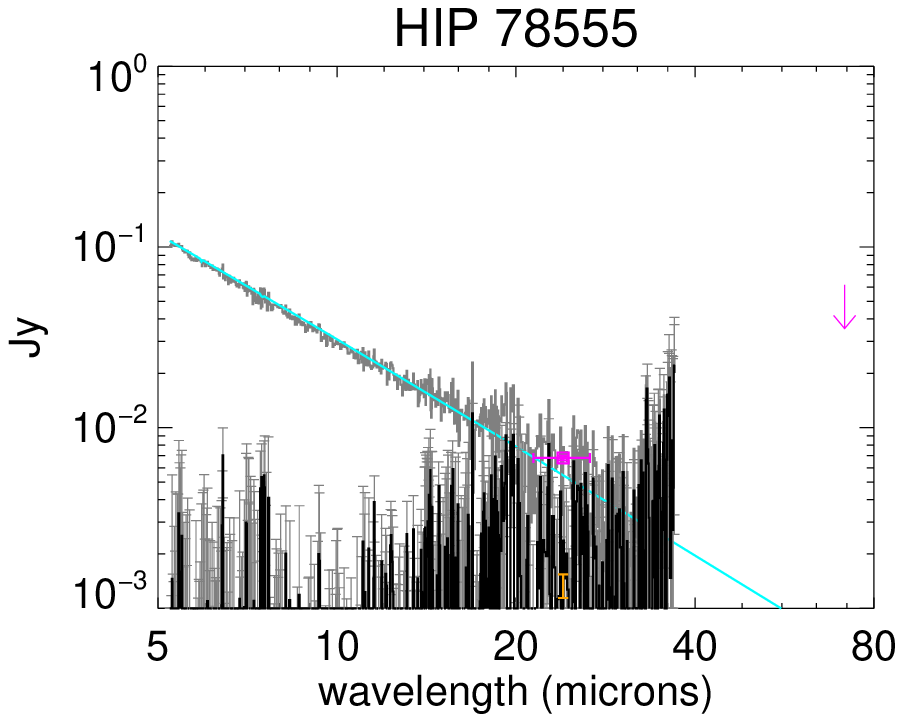} }
\parbox{\stampwidth}{
\includegraphics[width=\stampwidth]{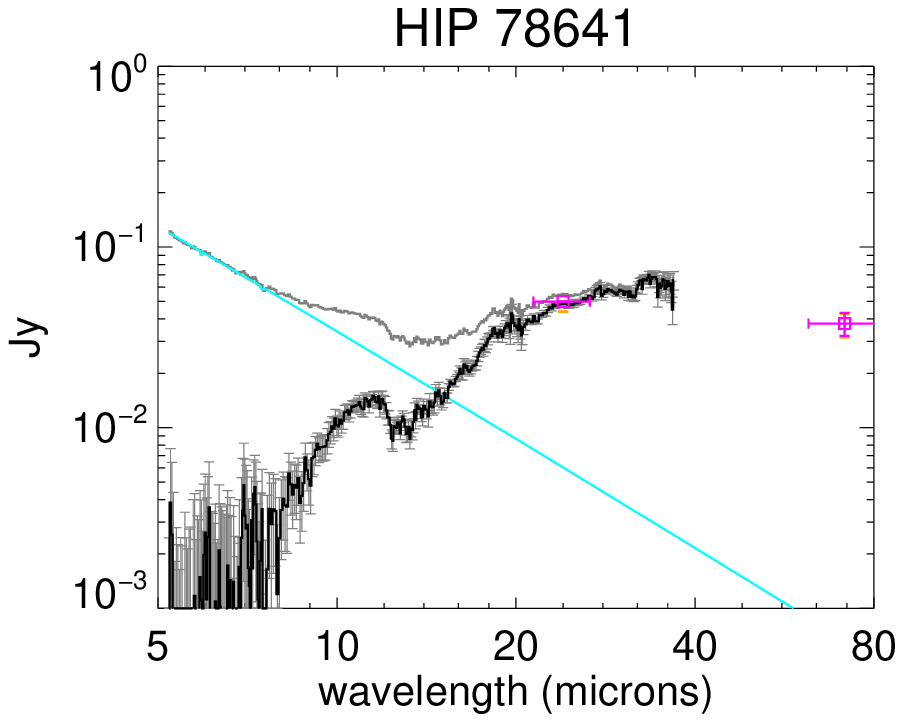} }
\parbox{\stampwidth}{
\includegraphics[width=\stampwidth]{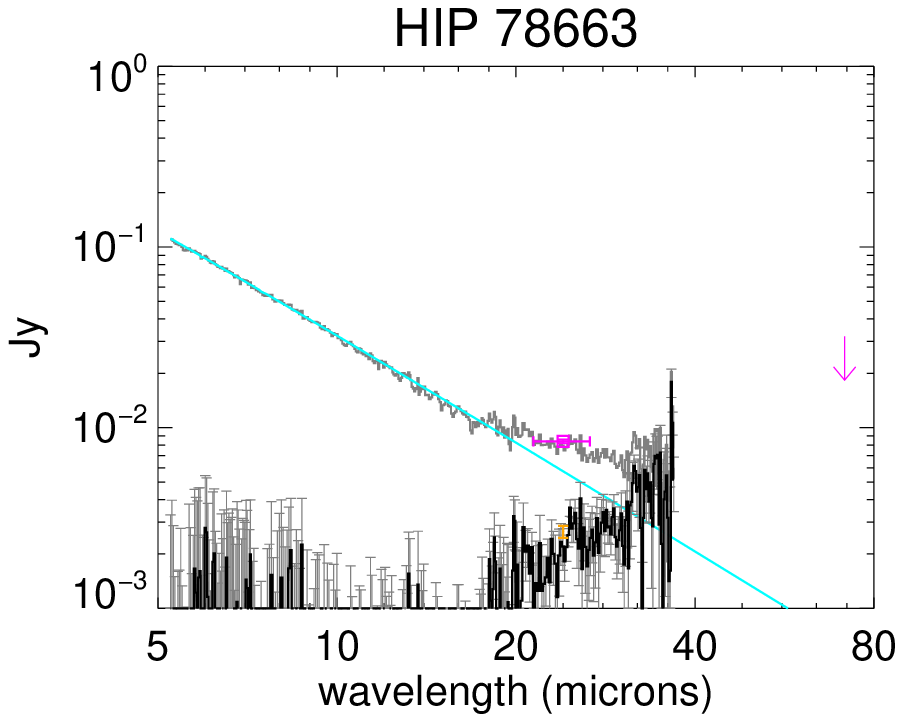} }
\parbox{\stampwidth}{
\includegraphics[width=\stampwidth]{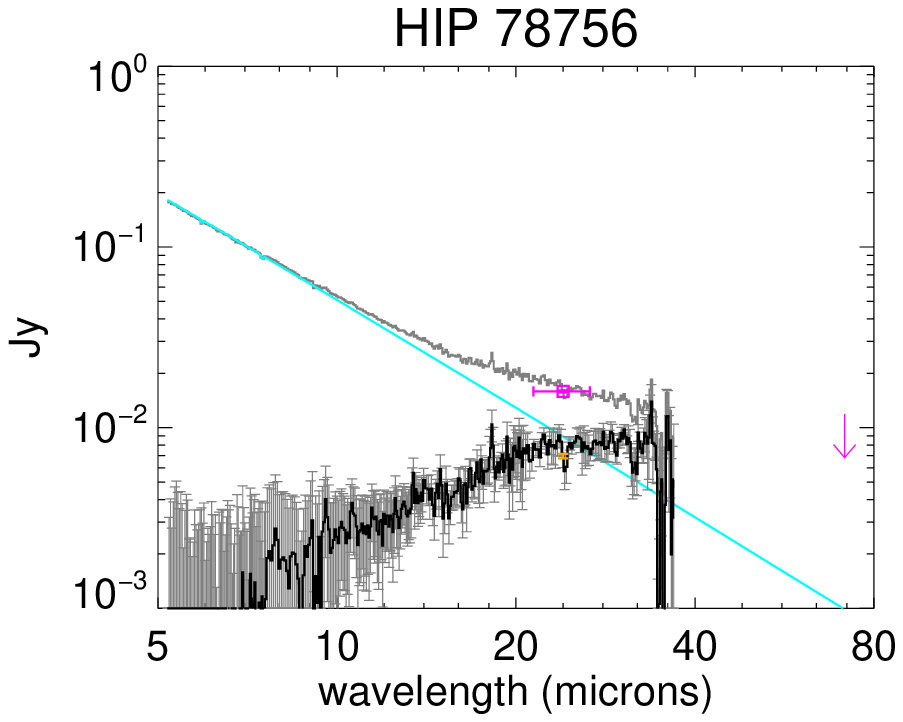} }
\\
\caption{ \label{specfig3}
Continuation Figure \ref{specfig0}, spectra of  objects.}
\end{figure}
\addtocounter{figure}{-1}
\stepcounter{subfig}
\begin{figure}
\parbox{\stampwidth}{
\includegraphics[width=\stampwidth]{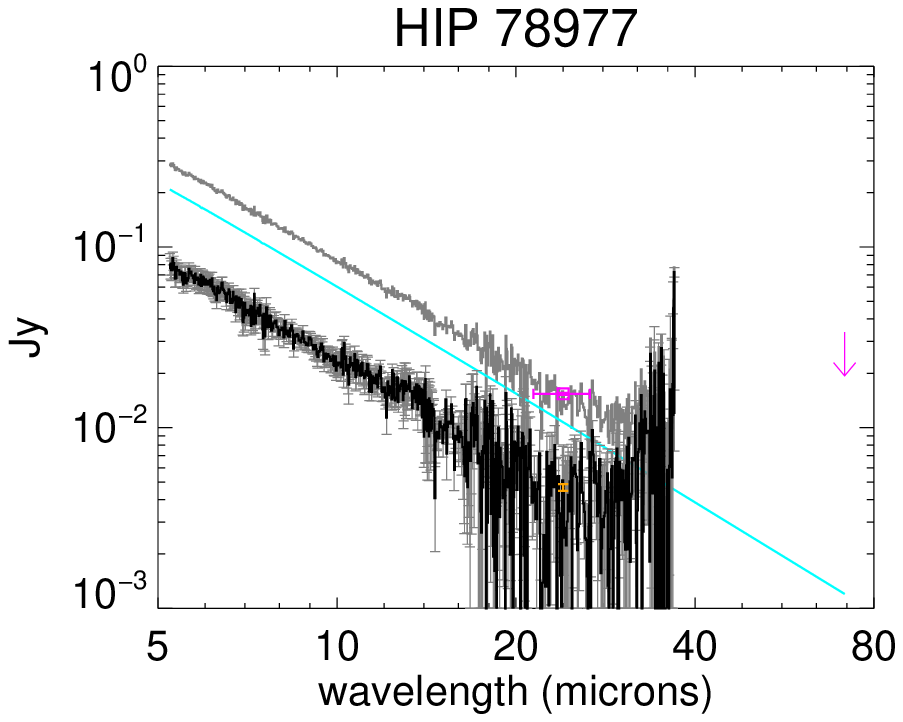} }
\parbox{\stampwidth}{
\includegraphics[width=\stampwidth]{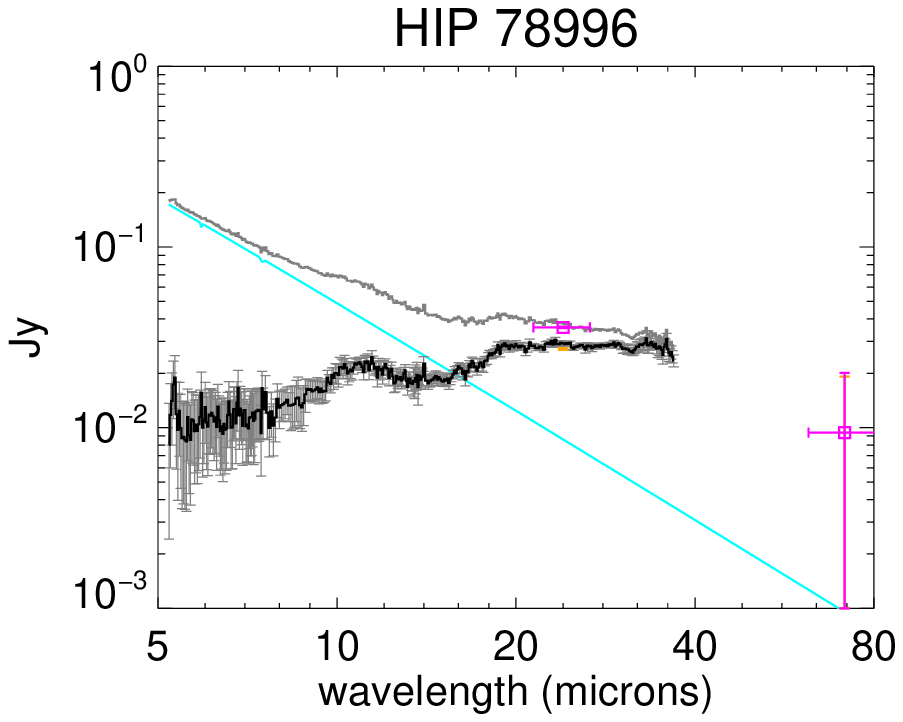} }
\parbox{\stampwidth}{
\includegraphics[width=\stampwidth]{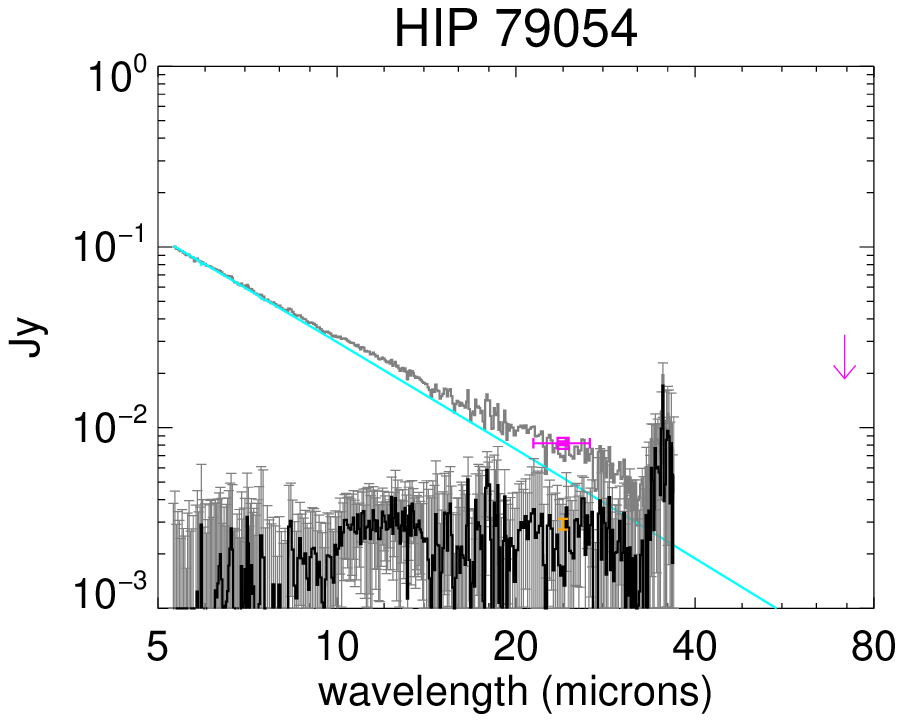} }
\parbox{\stampwidth}{
\includegraphics[width=\stampwidth]{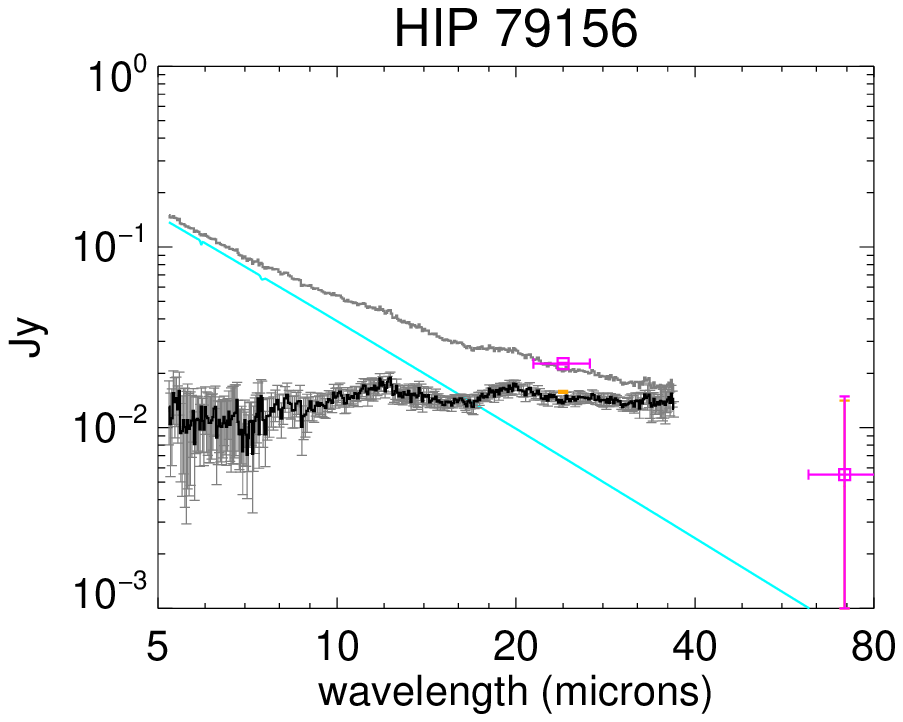} }
\\
\parbox{\stampwidth}{
\includegraphics[width=\stampwidth]{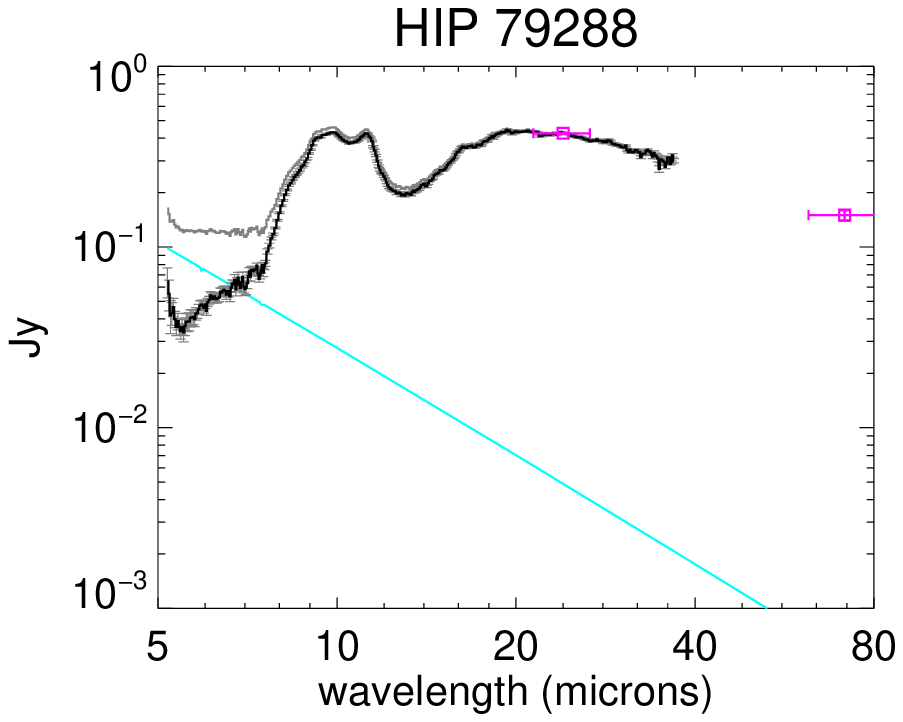} }
\parbox{\stampwidth}{
\includegraphics[width=\stampwidth]{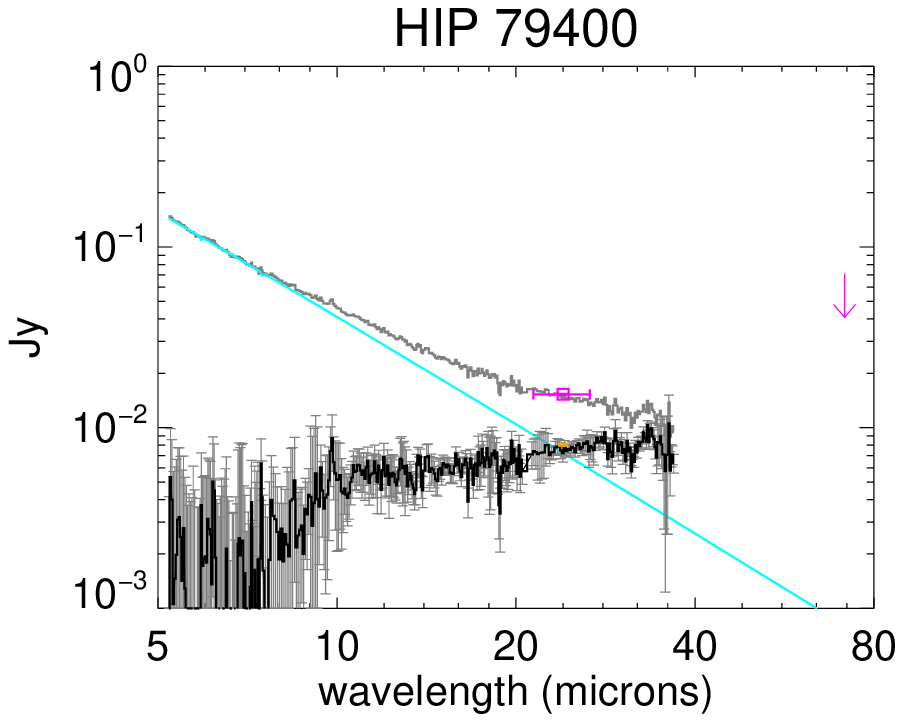} }
\parbox{\stampwidth}{
\includegraphics[width=\stampwidth]{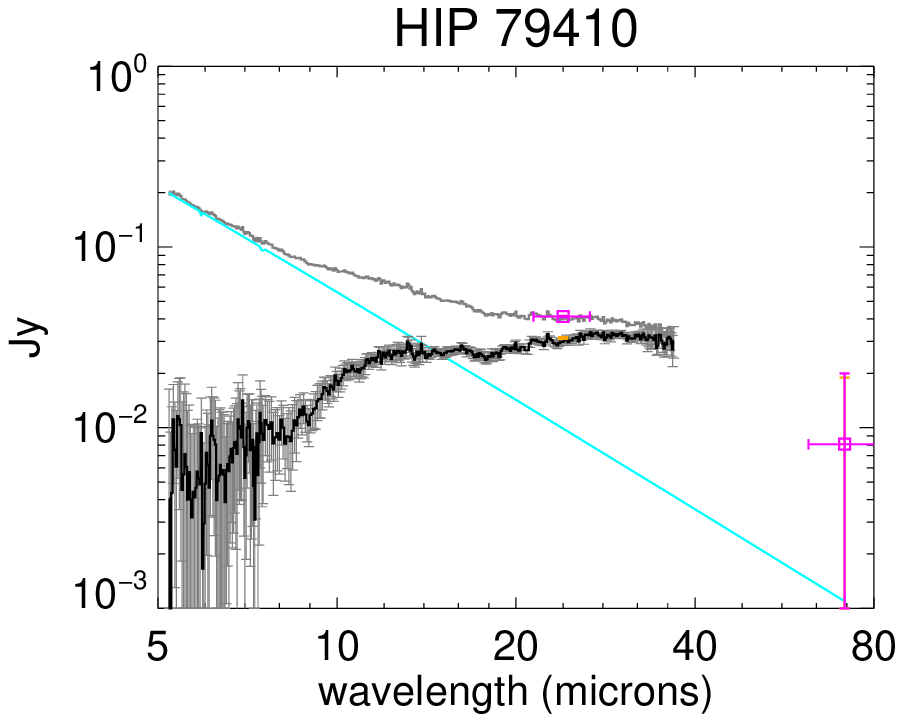} }
\parbox{\stampwidth}{
\includegraphics[width=\stampwidth]{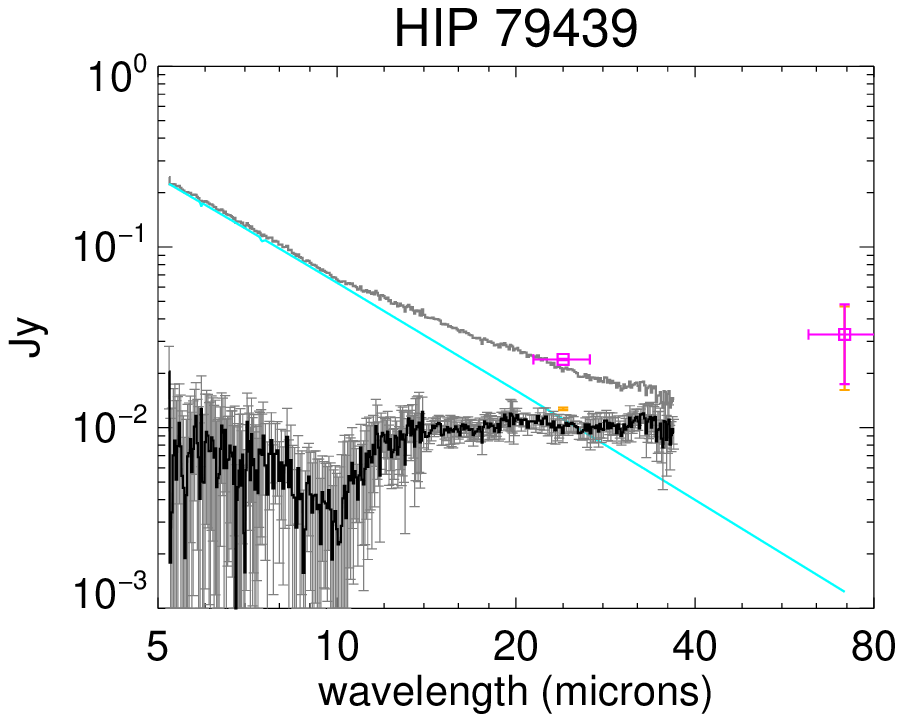} }
\\
\parbox{\stampwidth}{
\includegraphics[width=\stampwidth]{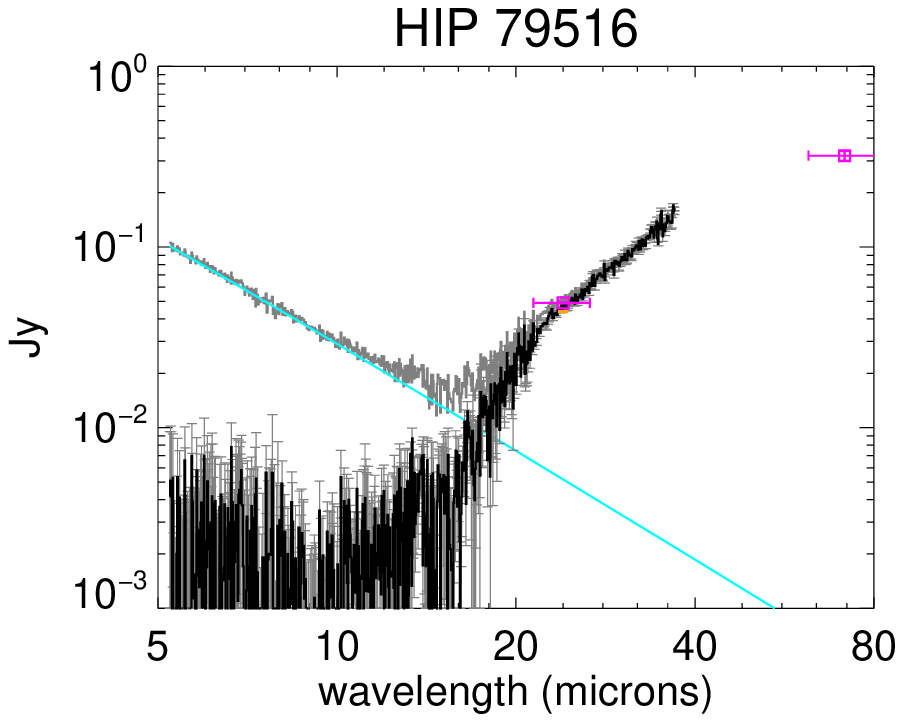} }
\parbox{\stampwidth}{
\includegraphics[width=\stampwidth]{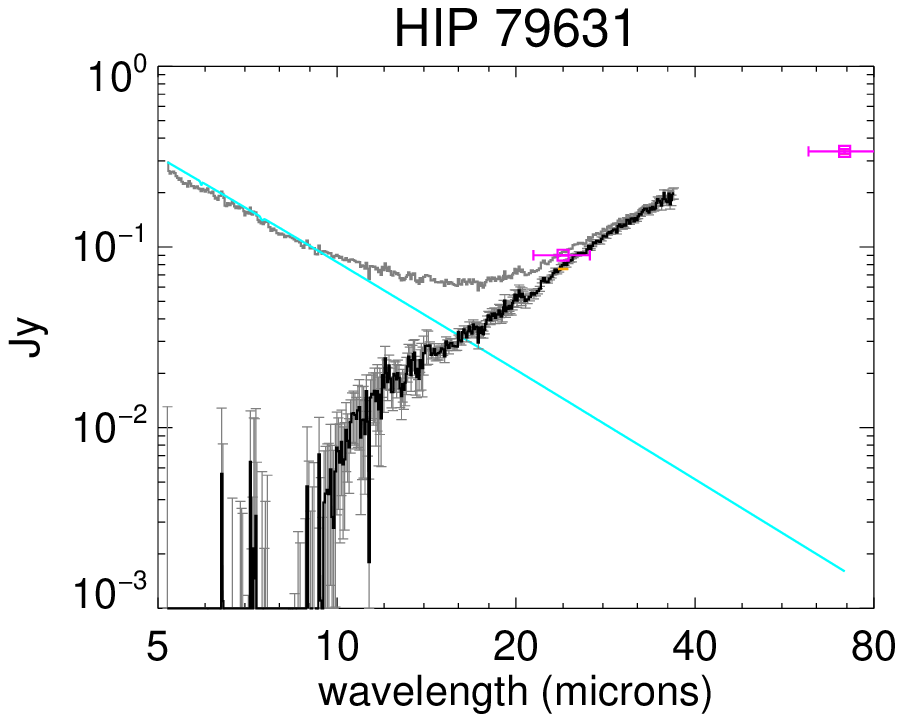} }
\parbox{\stampwidth}{
\includegraphics[width=\stampwidth]{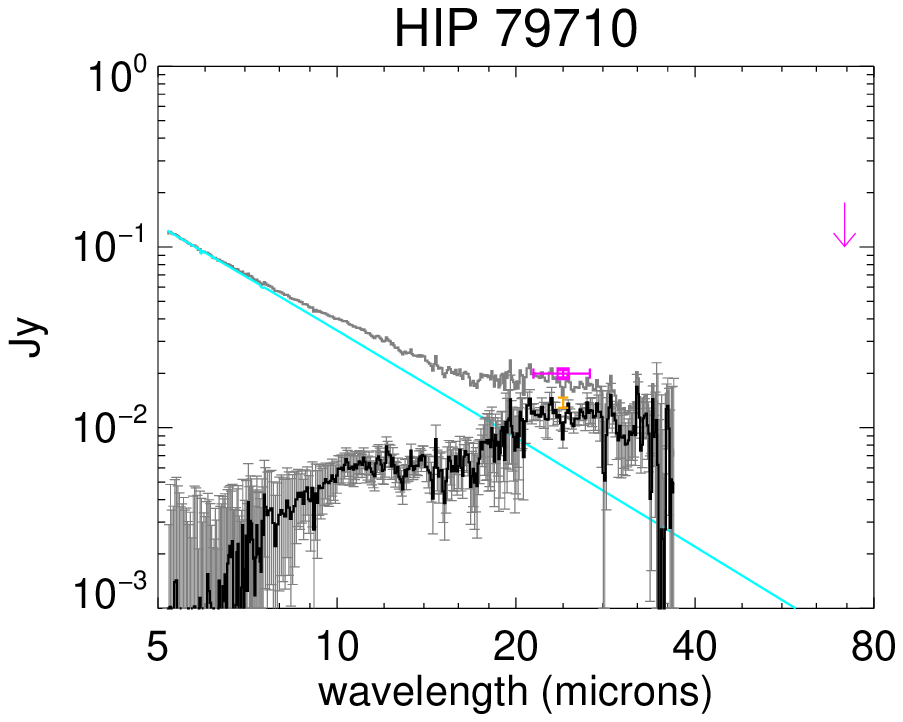} }
\parbox{\stampwidth}{
\includegraphics[width=\stampwidth]{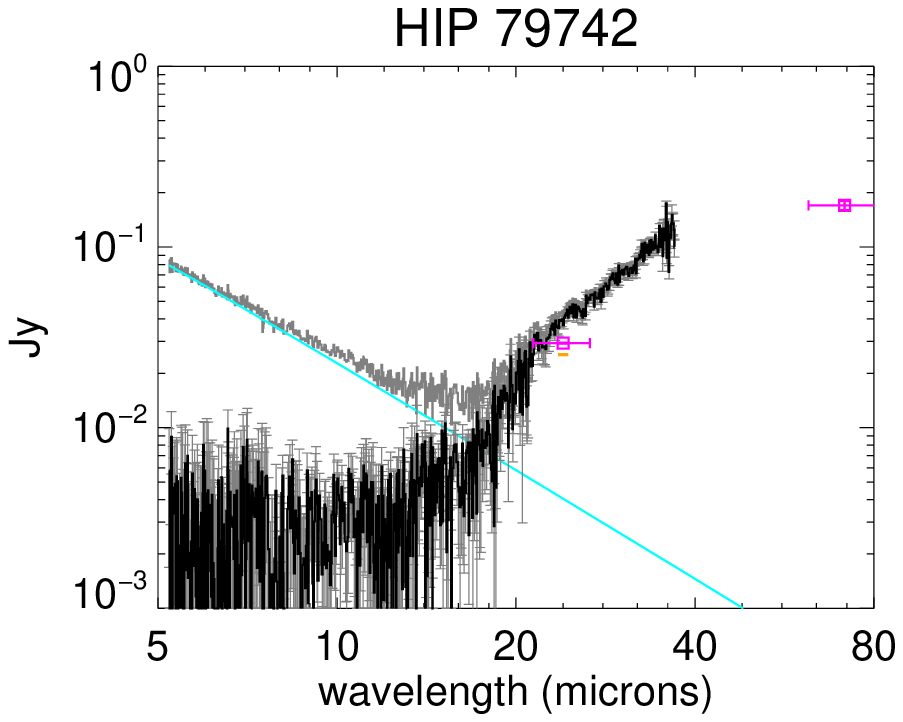} }
\\
\parbox{\stampwidth}{
\includegraphics[width=\stampwidth]{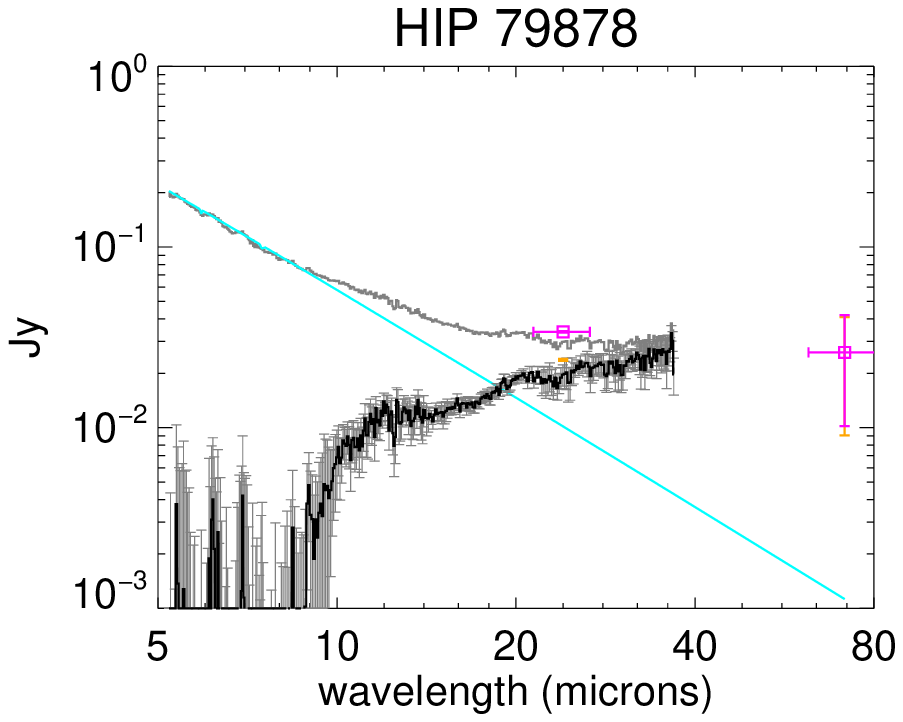} }
\parbox{\stampwidth}{
\includegraphics[width=\stampwidth]{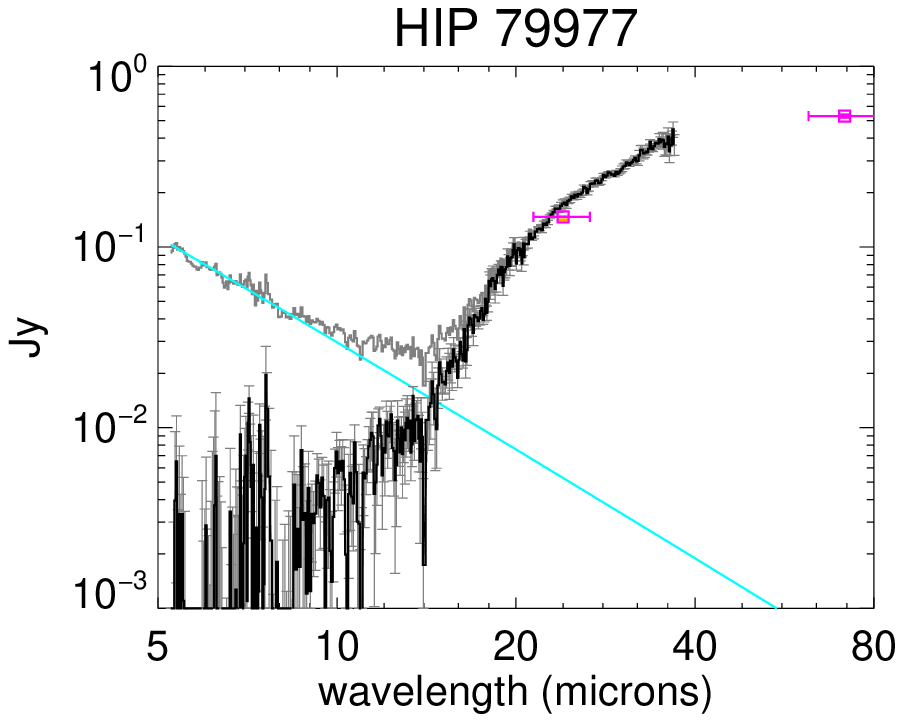} }
\parbox{\stampwidth}{
\includegraphics[width=\stampwidth]{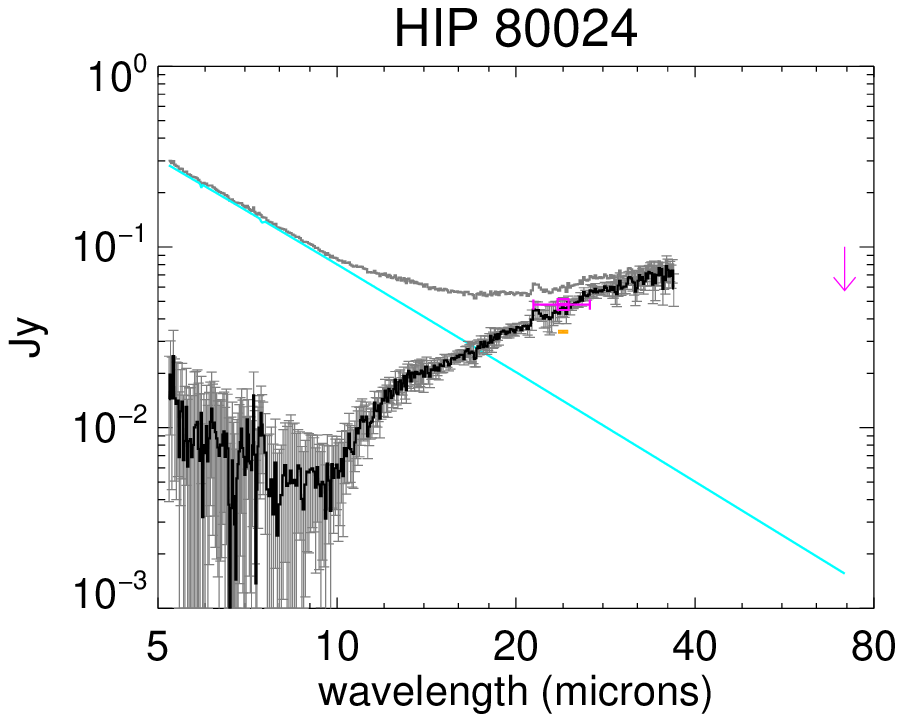} }
\parbox{\stampwidth}{
\includegraphics[width=\stampwidth]{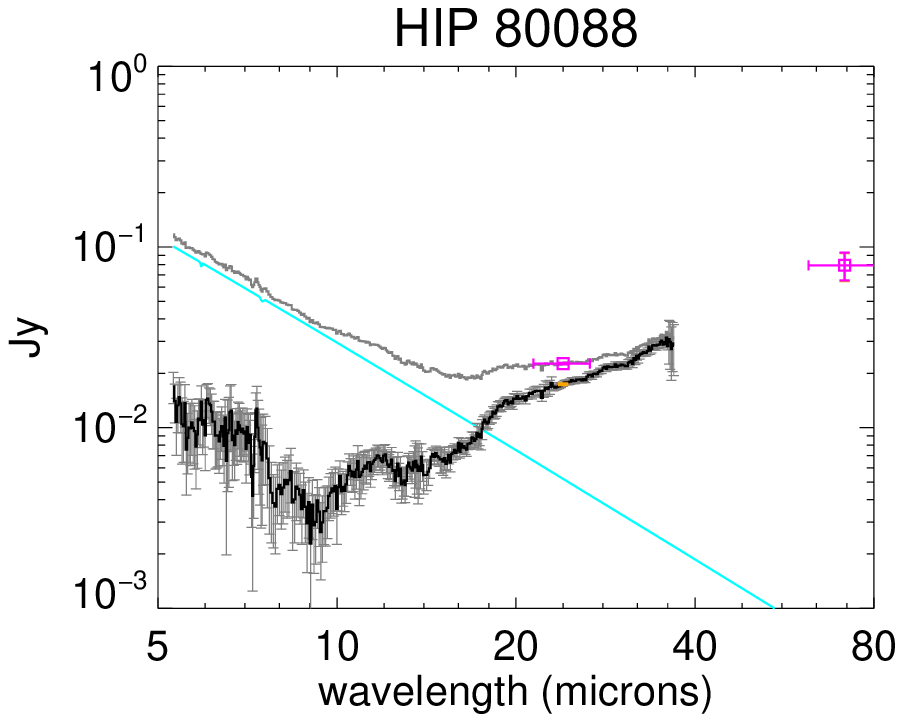} }
\\
\parbox{\stampwidth}{
\includegraphics[width=\stampwidth]{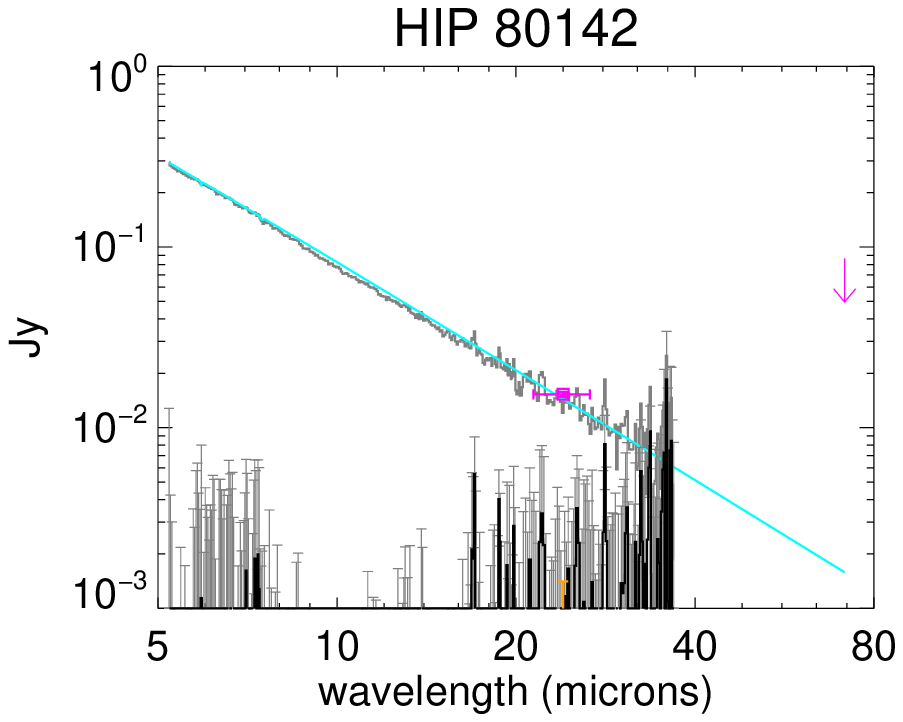} }
\parbox{\stampwidth}{
\includegraphics[width=\stampwidth]{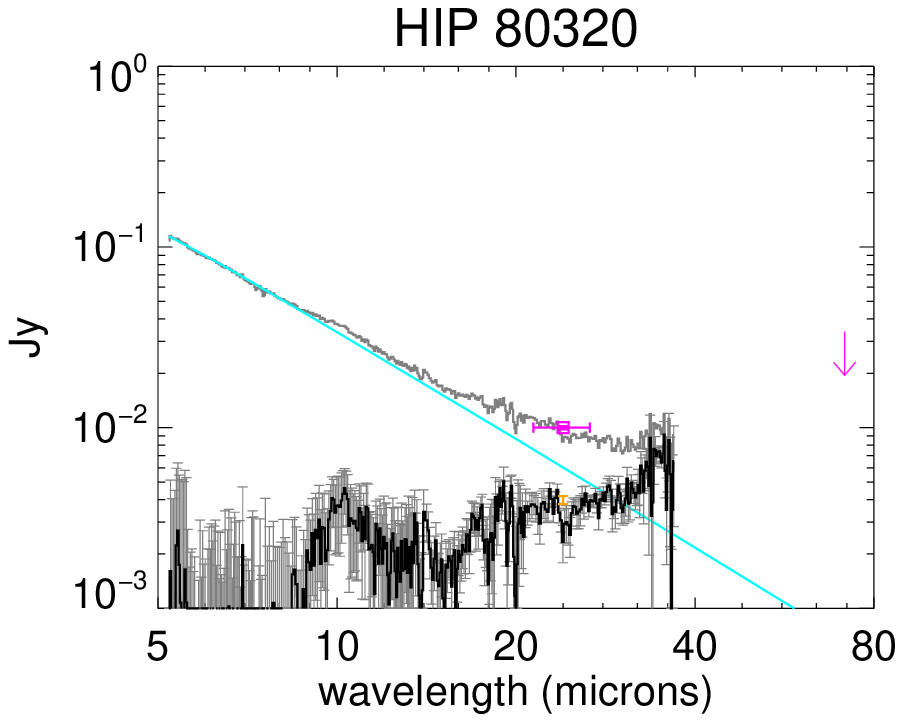} }
\parbox{\stampwidth}{
\includegraphics[width=\stampwidth]{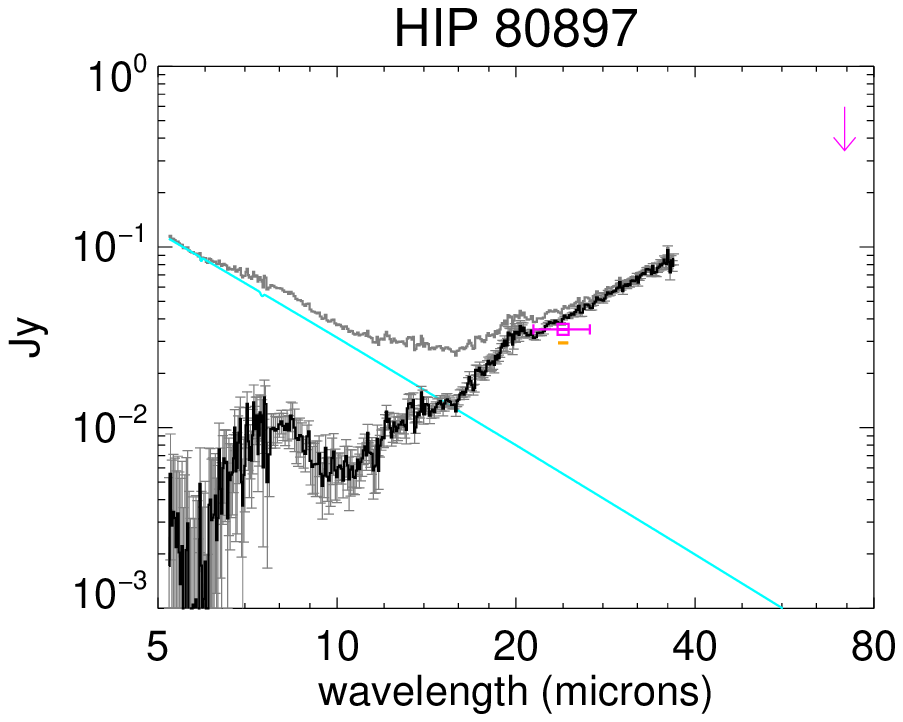} }
\parbox{\stampwidth}{
\includegraphics[width=\stampwidth]{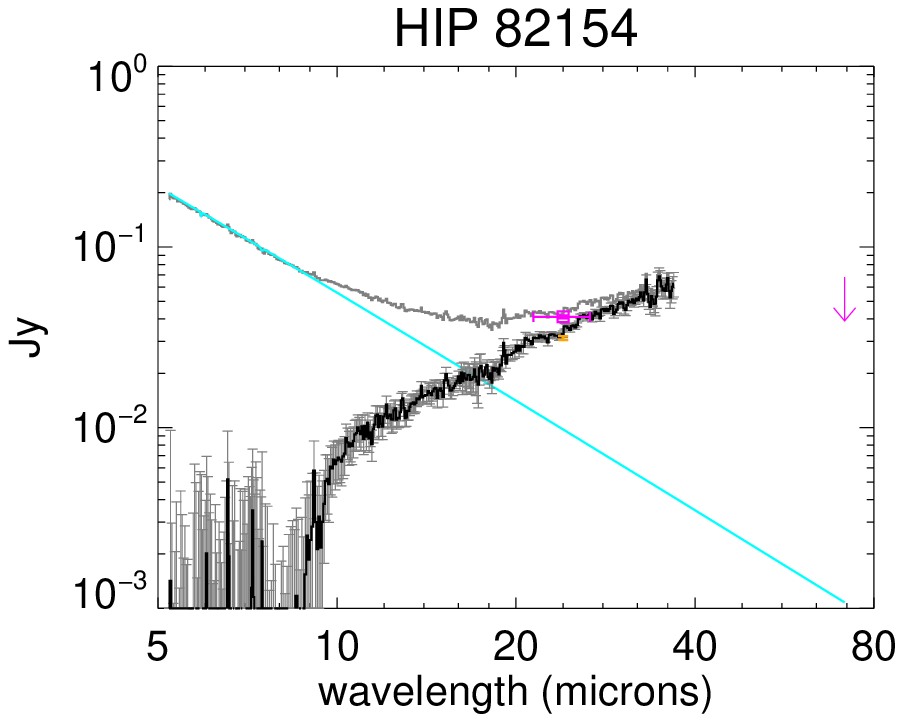} }
\\
\parbox{\stampwidth}{
\includegraphics[width=\stampwidth]{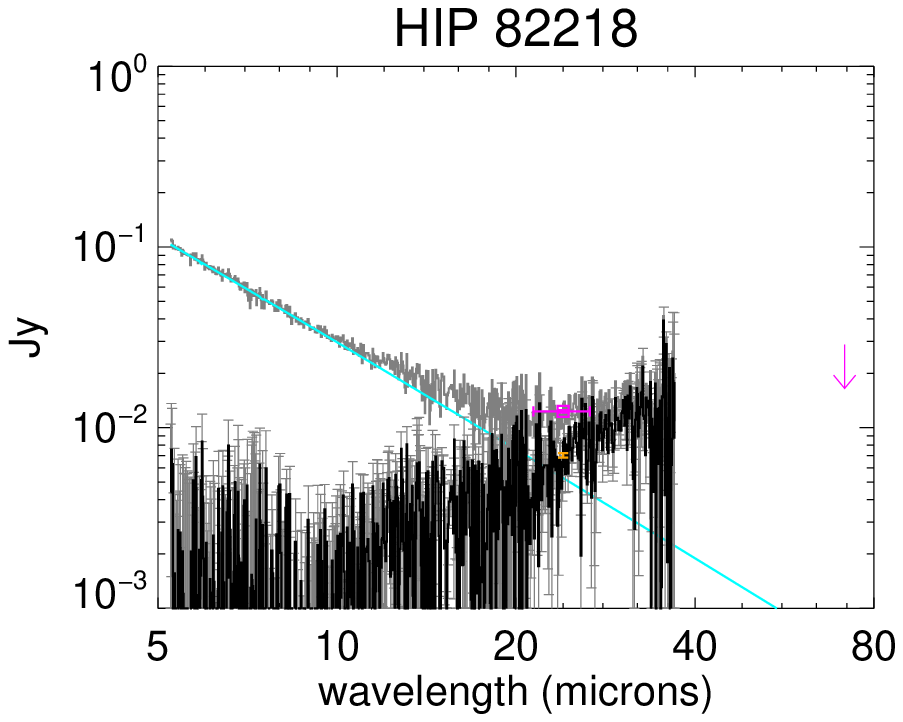} }
\parbox{\stampwidth}{
\includegraphics[width=\stampwidth]{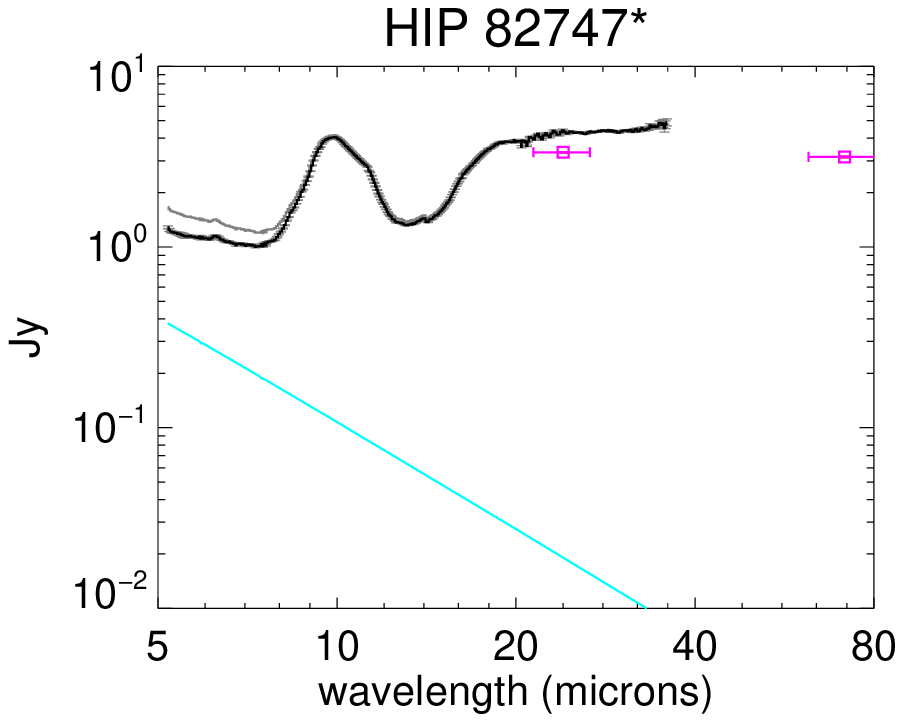} }
\parbox{\stampwidth}{
\includegraphics[width=\stampwidth]{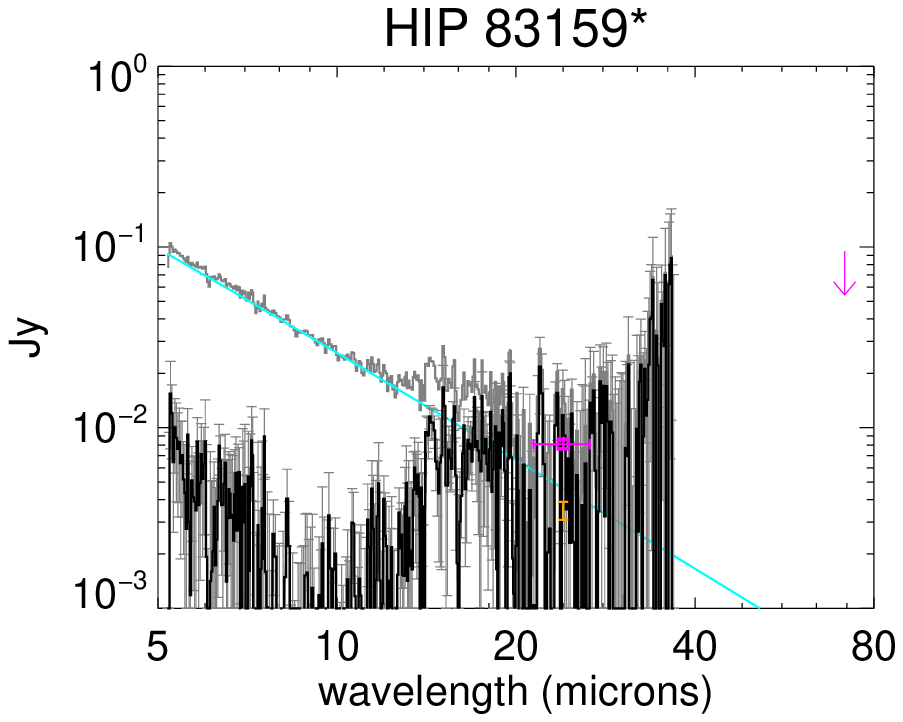} }
\caption{ \label{specfig4}
Continuation Figure \ref{specfig0}, spectra of  objects.}
\end{figure}

\renewcommand{\thefigure}{\arabic{figure}}

\begin{figure}[bt]
  \centerline{\includegraphics[width=4in]{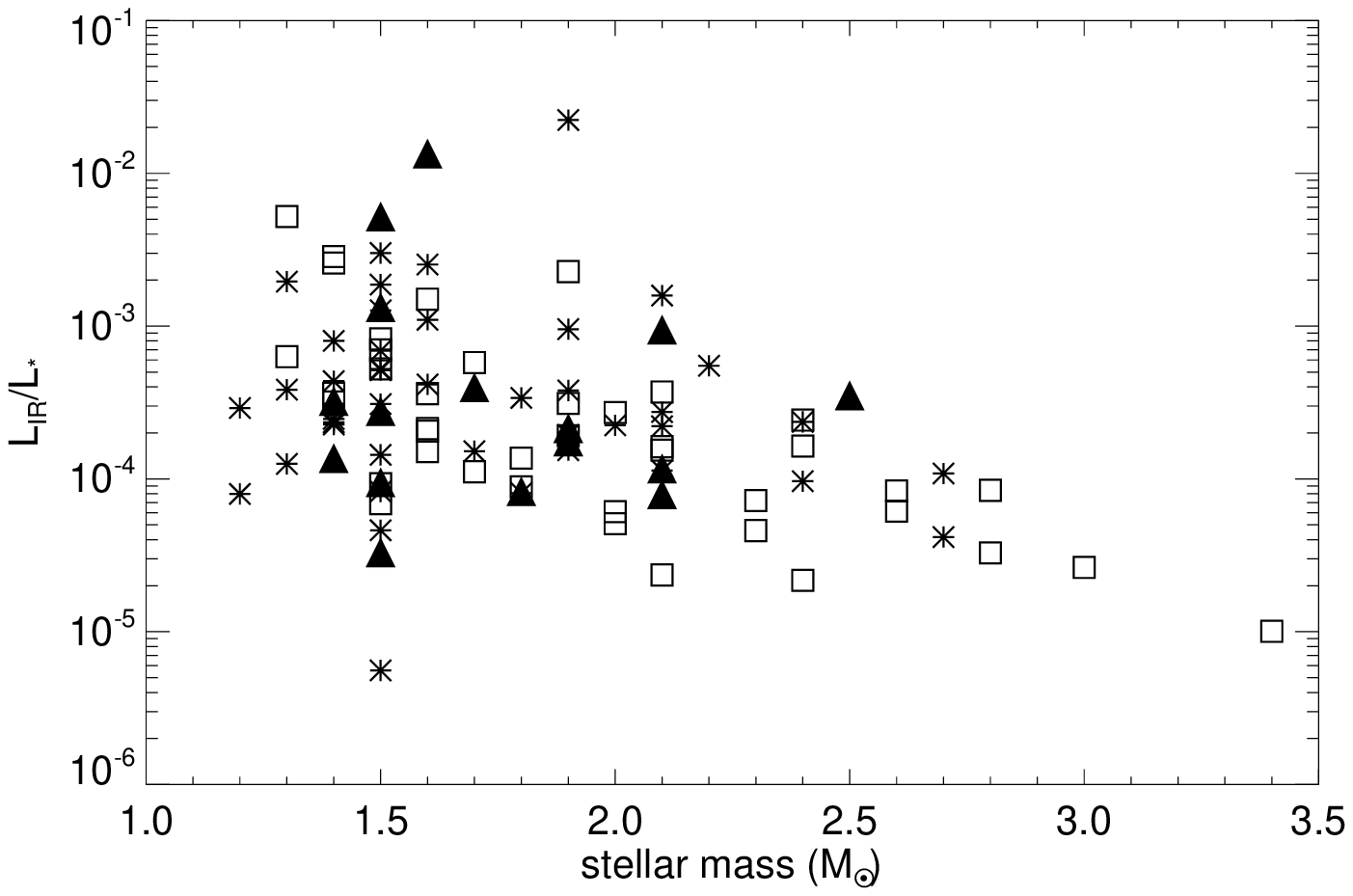}}
\caption{\label{fig:lirlstar}$L_{IR}/L_{*}$ versus stellar mass.
Asterisks are Lower Centaurus Crux,
squares are Upper Centaurus Lupus,
and filled triangles are Upper Scorpius.
}
\end{figure}

In Table \ref{tab:starprops}, we summarize the stellar properties 
of the 119 Sco Cen members whose IRS spectra are discussed here.  
The distances to the sources are taken from 
Hipparcos measurements \citep{2007vanLeeuwen}.  
Of these sources, 
5 are Be stars, and are not analyzed: 
HIP 63005, HIP 67472, HIP 69618, HIP 77859, and HIP 78207.
Another 4 sources are optically thick protoplanetary disks: 
HIP 56354 (HD 100453), HIP 56379 (HD 100546), 
HIP 77157 (HT Lupi), and HIP 82747 
\citep[AK Sco. See e.g.~][]{2006Manoj+,2013Sturm+}.
These sources cannot be adequately modeled 
using simple grain models, so we exclude them from our study.
In Table \ref{tab:diskprops} we tabulate 
the calculated $L\sub{IR}/L_*$ based on the calibrated and 
photosphere-subtracted spectra of the remaining 110 sources.  
In Figure \ref{fig:lirlstar}, we plot $L_{IR}/L_*$ versus stellar mass.  
The downward trend versus stellar mass is contrary to 
what would be expected if disk temperatures or masses
simply scale with stellar mass.  
This can be explained either by decreasing disk mass or decreasing 
dust temperature with increasing stellar mass, 
scenarios that will be discussed in our Results.  

In general, we found that the photosphere models were consistent with the IRS 
observations for all of the stars in our study with the exception of HIP 56673 
(HD 101088) and HIP 78977 (HD 144548). For these two objects, 
this normalization of the IRS spectra produced a Rayleigh-Jeans 
power-law excess at 5-30 $\mu$m excess that could indicate the presence of
a hot dust 
component with $T_{gr}$ $\gg$ 500 K. Visual spectra of HIP 56673B (HD 101088B)
show time-variable H$\alpha$ emission, consistent with accretion observed toward
T Tauri stars \citep{bitner10}. Alternatively, this mismatch in the MIPS 24 $\mu$m flux 
and the IRS spectrum could indicate that these sources possess time variable 
excesses similar to that observed toward ID 8
caused by stochastic grinding events
\citep{2014Meng+}. 

Several sources show little to no excess in the IRS data.  
We omit these non-excess sources based on two criteria: 
the excess significance and the normalized flux ratio.  
These quantities compare the observed versus 
predicted photosphere-only emission over a selected 
wavelength regime.  That is, if $F_{\nu}$ is the 
frequency-dependent flux with uncertainty $\sigma_{\nu}$, then 
\begin{equation}
F(\nu_1,\nu_2) = \frac{\int_{\nu_1}^{\nu_2} F_{\nu}\, d\nu}{\nu_2-\nu_1}
\end{equation}
and the weighted uncertainty is
\begin{equation}
\sigma^2(\nu_1,\nu_2) = \frac{\int_{\nu_1}^{\nu_2} \sigma_{\nu}^2\, d\nu}{\nu_2-\nu_1}.
\end{equation}
We use the subscripts 'obs' and 'pred' to refer to the 
observed and predicted flux, respectively.  
We consider three different passbands: 
$8.5-13$ $\mu$m, $21-26$ $\mu$m, and $30-34$ $\mu$m.  
The flux integrated over each of these passbands are 
F(10 $\mu$m), F(24 $\mu$m), and F(32 $\mu$m), respectively.  

The excess significance is defined to be 
\begin{equation}
\chi = (F\sub{obs}-F\sub{pred})/(\sigma\sub{obs}^2+\sigma\sub{pred}^2)^{1/2}, 
\end{equation}
where $\sigma\sub{obs}$ includes the repeatability error and a  
5\% normalization uncertainty added in quadrature, while 
$\sigma\sub{pred}$ consists of a 3\% normalization uncertainty.  
We calculate the excess significance over our three passbands, 
$\chi_{10}$, $\chi_{24}$, and $\chi_{32}$, and 
list their values in Table \ref{tab:diskprops}. 
We also list values of $\chi\sub{tot}$, which is the 
excess significance calculated over the entire IRS spectrum.  
HIP 56673 and HIP 78977 are listed twice: in the first listing,  
the IRS spectra are normalized to the MIPS 24 micron 
as described in \citet{2014Chen_etal}, and in the second 
listing, marked by an asterisk, 
the spectra are normalized to the photosphere model at 
5-6 microns.  
When these sources are normalized to the photosphere, 
HIP 56673 exhibits a small excess, but HIP 78977 has none.  

The normalized flux ratio is adapted from \citet{carpenter09b}, 
and is defined to be 
\begin{equation}
R_{32/10} = \frac{F\sub{obs}(32\,\mu{\rm m})/F\sub{pred}(32\,\mu{\rm m})}{F\sub{obs}(10\,\mu{\rm m})/F\sub{pred}(10\,\mu{\rm m})}.
\end{equation}
A similar expression is used to calculate $R_{24/10}$.  
In \citet{carpenter09b}, photometric fluxes were used.  Here, 
we integrate the spectrum over the given passband, 
assuming 100\% efficiency.  Since 
$R$ can be considered to be a ratio of the slope of the observed spectrum 
compared to the slope of the predicted spectrum, any error in the 
overall normalization of either spectrum cancels out.  Therefore, the 
error on $R$ is propagated from the repeatability error of the 
observed spectrum alone.  

The excess significance ($\chi$) measures the signal-to-noise of the 
infrared excess at each band pass, while $R$ measures 
the shape of the excess.  In some sources, even though 
the excess significance is formally low, the shape of the
spectrum rises at long wavelengths, indicating that there is 
a notable cold excess.  If $R>1$, then the shape of the 
spectrum indicates a cold excess component, while $R=1$ 
indicates a shape consistent with photospheric emission.  
To determine which spectra to exclude as non-excess sources, 
we use both the $\chi$ and $R$ measures.  
Sources which have $\chi_8$, $\chi_{24}$, $\chi_{30}$, and 
$\chi\sub{tot}$ all less than 3 and whose values of 
$R\sub{24/8}$ and $R\sub{30/8}$ are both less than or 
equal to 1 to within one $\sigma$ are labeled non-excess 
sources and are excluded from further analysis.  
A total of 13 of our sources are non-excess sources, and 
are labeled as such in Table \ref{tab:diskprops}.  
In addition, HIP 78977, when normalized to the photosphere 
model, can be considered a non-excess source.  
This leaves a total of 97 debris disk spectra that
we analyze for their dust properties.  

\setlength{\textwidth}{7in}
\begin{deluxetable}{llcccccccc}
\rotate
\tabletypesize{\footnotesize}
\tablecaption{Stellar Properties \label{tab:starprops}} 
\tablewidth{0pt}
\tablehead{
    \colhead{HIP} &
    \colhead{Name} &
    \colhead{Spectral} &
    \colhead{Distance} &
    \colhead{$T_{\rm eff}$} &
    \colhead{Mass} &
    \colhead{Luminosity}&
    \colhead{$A_V$}&
    \colhead{Program}&
    \colhead{Notes}
    \\
    \omit &
    \omit &
    \colhead{Type} &
    \colhead{(pc)} &
    \colhead{(K)} &
    \colhead{($M_{\odot}$)} &
    \colhead{($L_{\odot}$)} &    
    \omit &
    \omit &
    \omit 
}
\tablewidth{0pt}
\tablecolumns{10}
\startdata
&\\[-3ex]
\cutinhead{Lower Centaurus Crux}
       53524& HD 95086&  A8III  (6) &   90.4 & 7499 &   1.6 &       7.11 &  0.000 & IRS\_DISKS/2 &  (18)   \\   
       55188& HD 98363&  A2V  (6) &  123.6 & 8770 &   1.9 &       11.4 &  0.244 & CCHEN2/40235 &  (18)   \\   
       56354& HD 100453&  A9Ve  (14) &  121.5 & 7447 &   1.6 &       10.4 &  0.214 & IRS\_DISKS/2 &  (18) protoplanetary \\   
       56379& HD 100546&  B9Vne  (6) &   96.9 & 10520 &   2.4 &       26.6 &  0.194 & IRS\_DISKS/2 &  (18) protoplanetary \\   
       56673& HD 101088& F5IV (6) &   93.8 & 6440 &   2.2 &       17.7 &  0.142 & DEBRISII/40651 &  (17) $\lambda^{-2}$ excess \\   
       57524& HD 102458& F9IV (13) &   91.7 & 6115 &   1.2 &       1.89 &  0.178 & DEBRISII/40651 &  (17)   \\   
       57950& HD 103234& F2IV/V (6) &   98.1 & 6890 &   1.5 &       3.90 &  0.066 & DEBRISII/40651 &  (17)   \\   
       58220& HD 103703& F3V (6) &   98.9 & 6740 &   1.5 &       3.35 &  0.096 & YOUNGA/84 &  (17)   \\   
       58528& HD 104231& F5V (6) &  110.5 & 6440 &   1.4 &       3.73 &  0.016 & DEBRISII/40651 &  (17)   \\   
       58720& HD 104600&  B9V  (6) &  105.7 & 11614 &   2.7 &       68.7 &  0.024 & CCHEN2/40235 &  (18)   \\   
       59282& HD 105613&  A3V  (6) &  104.2 & 8551 &   1.8 &       11.6 &  0.192 & CCHEN2/40235 &  (18)   \\   
       59397& HD 105857&  A2V  (6) &  113.0 & 8770 &   1.9 &       14.0 &  0.132 & CCHEN2/40235 &  (18)   \\   
       59481& HD 105994& F3V (7) &  113.1 & 6740 &   1.5 &       4.09 &  0.023 & WARMDISK2/50538 &  (17)   \\   
       59502& HD 106036&  A2V  (6) &  100.7 & 8770 &   1.9 &       13.8 &  0.015 & CCHEN2/40235 &  (18)   \\   
       59693& HD 106389& F6IV (7) &  137.0 & 6360 &   1.3 &       2.32 &  0.184 & DEBRISII/40651 &  (17)   \\   
       59898& HD 106797&  A0V  (6) &   96.0 & 9550 &   2.1 &       32.9 &  0.022 & CCHEN2/40235 &  (18)   \\   
       59960& HD 106906& F5V (6) &   92.1 & 6440 &   1.5 &       5.06 &  0.000 & IRS\_DISKS/2 &  (17)   \\   
       60183& HD 107301&  B9V  (6) &   93.9 & 10814 &   2.4 &       36.8 &  0.067 & CCHEN2/40235 &  (18)   \\   
       60348& HD 107649& F5V (7) &   93.7 & 6440 &   1.4 &       2.13 &  0.022 & YOUNGA/84 &  (17)   \\   
       60561& HD 107947&  A0V  (6) &   91.1 & 9550 &   2.1 &       17.5 &  0.000 & CCHEN2/40235 &  (18)   \\   
       60710& HD 108257&  B3Vn  (2) &  137.4 & 17298 &   5.4 &       809. &  0.070 & IRS\_DISKS/2 &  (18)   \\  
       61049& HD 108857& F7V (6) &   97.0 & 6280 &   1.4 &       3.13 &  0.143 & CCHEN2/40235 &  (17)   \\   
       61087& HD 108904& F6V (6) &   97.5 & 6360 &   1.5 &       4.97 &  0.050 & CCHEN2/40235 &  (17)   \\   
       61684& HD 109832&  A9V  (6) &  111.9 & 7447 &   1.6 &       7.14 &  0.261 & CCHEN2/40235 &  (18)   \\   
       61782& HD 110058&  A0V  (7) &  107.4 & 9550 &   2.1 &       10.2 &  0.436 & IRS\_DISKS/2 &  (18)   \\   
       62134& HD 110634& F2V (7) &  115.6 & 6890 &   1.5 &       3.74 &  0.026 & YOUNGA/84 &  (17)   \\  
       62427& HD 111103& F8 (1) &  142.7 & 6200 &   1.4 &       3.17 &  0.000 & DEBRISII/40651 &  (17)   \\   
       62445& HD 111170& G4.5IVe (13) &  130.5 & 5728 &   1.6 &       4.17 &  0.660 & RUBBLE/148 &  (13)   \\  
       62657& HD 111520& F5/6V (7) &  108.6 & 6400 &   1.3 &       2.60 &  0.028 & CCHEN/241 &  (17)   \\  
       63005& HD 112091&  B5V(e)  (2) &  124.8 & 16634 &   5.0 &       541. &  0.200 & CCHEN2/40235 &  (18) classical Be \\   
       63236& HD 112383&  A2IV/V  (6) &  110.7 & 8770 &   1.9 &       20.0 &  0.000 & CCHEN2/40235 &  (18)   \\   
       63439& HD 112810& F3/5IV/V (7) &  143.3 & 6590 &   1.4 &       3.52 &  0.000 & DEBRISII/40651 &  (17)   \\   
       63836& HD 113524& F6/8 (7) &  107.4 & 6280 &   1.3 &       2.31 &  0.000 & DEBRISII/40651 &  (17)   \\   
       63839& HD 113457&  A0V  (6) &   99.4 & 9550 &   2.1 &       20.3 &  0.000 & CCHEN2/40235 &  (18)   \\   
       63886& HD 113556& F2V (6) &  106.7 & 6890 &   1.5 &       4.91 &  0.024 & IRS\_DISKS/2 &  (17)   \\   
       63975& HD 113766& F3/5V (7) &  122.5 & 6590 &   1.9 &       11.9 &  0.000 & IRS\_DISKS/2 &  (17)   \\   
       64053& HD 113902&  B8/9V  (7) &  100.1 & 11695 &   2.7 &       77.6 &  0.070 & CCHEN2/40235 &  (18)   \\   
       64184& HD 114082& F3V (6) &   85.5 & 6740 &   1.5 &       3.18 &  0.078 & IRS\_DISKS/2 &  (17)   \\   
       64877& HD 115361& F5V (6) &  125.0 & 6440 &   1.5 &       5.02 &  0.000 & CCHEN2/40235 &  (17)   \\   
       64995& HD 115600& F2IV/V (6) &  110.5 & 6890 &   1.5 &       4.79 &  0.000 & IRS\_DISKS/2 &  (17)   \\   
       65089& HD 115820&  A7/8V  (7) &   96.5 & 7656 &   1.7 &       4.83 &  0.026 & CCHEN2/40235 &  (18)   \\   
       65875& HD 117214& F6V (6) &  110.3 & 6360 &   1.6 &       5.64 &  0.000 & IRS\_DISKS/2 &  (17)   \\   
       65965& HD 117484&  B9.5V  (7) &  147.3 & 10593 &   2.4 &       25.9 &  0.100 & CCHEN/40235 &  (18)   \\   
       66001& HD 117524& G2.5IV (13) &  152.4 & 5834 &   1.2 &       2.14 &  0.110 & RUBBLE/148 &  (13)   \\  
       66068& HD 117665&  A1/2V  (7) &  147.9 & 8974 &   1.9 &       24.4 &  0.000 & CCHEN/40235 &  (18)   \\   
       66566& HD 118588&  A1V  (7) &  126.4 & 9204 &   2.0 &       14.9 &  0.097 & CCHEN/40235 &  (18)   \\   
       67068& HD 119511& F3V (7) &   91.6 & 6740 &   1.5 &       2.70 &  0.000 & WARMDISK2/50538 &  (17)   \\   
       67230& HD 119718& F5V (6) &  131.8 & 6440 &   1.8 &       8.67 &  0.037 & CCHEN2/40235 &  (17)   \\
\cutinhead{Upper Centaurus Lupus}
       66447& HD 118379&  A3IV/V  (7) &  121.7 & 8551 &   1.8 &       13.3 &  0.184 & TD\_GTO/50485 &  (18)   \\   
       67472& HD 120324&  B2V:e  (8) &  155.0 & 20512 &   7.3 &   6.53e+03 &  0.287 & CCHEN2/40235 &  (18) classical Be \\   
       67497& HD 120326& F0V (7) &  107.4 & 7200 &   1.6 &       4.45 &  0.158 & DEBRISII/40651 &  (17)   \\   
       67970& HD 121189& F3V (7) &  118.8 & 6740 &   1.5 &       3.85 &  0.070 & CCHEN2/40235 &  (17)   \\   
       68080& HD 121336&  A1Vn (11) &  139.9 & 9204 &   2.0 &       64.9 &  0.082 & CCHEN2/40235 &  (18)   \\   
       68781& HD 122705&  A4V  (5) &  112.9 & 8279 &   1.8 &       8.89 &  0.000 & TD\_GTO/50485 &  (18)   \\   
       69291& HD 123889& F2V (9) &  132.3 & 6890 &   1.5 &       5.02 &  0.043 & WARMDISK2/50638 &  (17)   \\   
       69618& HD 124367&  B4Vne  (2) &  147.7 & 16982 &   5.2 &   1.01e+03 &  0.409 & CCHEN2/40235 &  (18) classical Be \\   
       69720& HD 124619& F0V (6) &  133.3 & 7200 &   1.6 &       5.00 &  0.228 & DEBRISII/40651 &  (17)   \\   
       70149& HD 125541&  A9V  (7) &  113.3 & 7447 &   1.6 &       3.18 &  0.146 & TD\_GTO/50485 &  (18)   \\   
       70441& HD 126062&  A1V  (7) &  110.4 & 9204 &   2.0 &       11.4 &  0.018 & CCHEN2/40235 &  (18)   \\   
       70455& HD 126135&  B8V  (7) &  165.0 & 11967 &   2.8 &       70.3 &  0.096 & GOWERNER2005/20132 &  (18)   \\   
       71271& HD 127750&  A0V  (7) &  175.7 & 9550 &   2.1 &       26.9 &  0.040 & TD\_GTO/50485 &  (18)   \\   
       71453& HD 128207&  B8V  (9) &  147.5 & 13490 &   3.4 &       210. &  0.010 & GOWERNER2005/20132 &  (18)   \\   
       72033& HD 129490& F7IV/V (7) &  155.8 & 6280 &   1.5 &       5.46 &  0.297 & DEBRISII/40651 &  (17)   \\  
       72070& HD 129590& G1V (13) &  132.6 & 5945 &   1.3 &       2.84 &  0.047 & CCHEN2/42035 &  (17)   \\   
       73145& HD 131835&  A2IV  (9) &  122.7 & 8770 &   1.9 &       10.5 &  0.187 & CCHEN2/40235 &  (18)   \\   
       73341& HD 132238&  B8V  (9) &  162.6 & 12359 &   3.0 &       112. &  0.040 & GOWERNER2005/20132 &  (18)   \\   
       73666& HD 133075& F3IV (9) &  151.5 & 6740 &   2.1 &       16.5 &  0.364 & DEBRISII/40651 &  (17)   \\   
       73990& HD 133803&  A9V  (9) &  124.8 & 7447 &   1.6 &       8.04 &  0.211 & TD\_GTO/50485 &  (18)   \\   
       74499& HD 134888& F3/5V (9) &   89.9 & 6590 &   1.5 &       2.06 &  0.058 & DEBRISII/40651 &  (17)   \\   
       74752& HD 135454&  B9.5V  (7) &  173.3 & 10351 &   2.3 &       66.2 &  0.023 & GOWERNER2005/20132 &  (18)   \\   
       74959& HD 135953& F5V (9) &  133.2 & 6440 &   1.3 &       2.68 &  0.080 & TD\_GTO/50485 &  (17)   \\   
       75077& HD 136246&  A1V  (9) &  131.6 & 9204 &   2.0 &       22.7 &  0.114 & GOWERNER2005/20132 &  (18)   \\   
       75151& HD 136347&  B9IVSi(SrCr)  (11) &  143.3 & 11641 &   2.7 &       68.2 &  0.031 & GOWERNER2005/20132 &  (18)   \\  
       75210& HD 136482&  B8/9V  (9) &  136.2 & 11324 &   2.6 &       54.2 &  0.033 & GOWERNER2005/20132 &  (18)   \\   
       75304& HD 136664&  B4V  (2) &  159.2 & 16711 &   5.0 &   1.27e+03 &  0.052 & IREXT/20294 &  (18)   \\  
       75491& HD 137057& F3V (9) &  168.6 & 6740 &   1.9 &       9.59 &  0.051 & CCHEN2/40235 &  (17)   \\   
       75509& HD 137119&  A2V  (9) &  107.2 & 8770 &   1.9 &       9.02 &  0.067 & CCHEN2/40235 &  (18)   \\   
       76084& HD 138296& F2V (9) &  142.7 & 6890 &   1.7 &       6.64 &  0.191 & DEBRISII/40651 &  (17)   \\  
       76395& HD 138923&  B8V  (3) &  106.5 & 11967 &   2.8 &       53.6 &  0.041 & GOWERNER2005/20132 &  (18)   \\   
       77081& HD 140374& G7.5IV (13) &  200.8 & 5521 &   1.4 &       2.24 &  0.150 & RUBBLE/148 &  (13)   \\  
       77157& HT Lupi& K3Ve (15) &  141.2 & 4730 &   1.1 &       5.09 &  1.138 & CCHEN2/40235 &  (17) protoplanetary \\   
       77315& HD 140817&  A0V  (9) &  147.3 & 9550 &   2.1 &       41.5 &  0.105 & CCHEN2/40235 &  (18)   \\   
       77317& HD 140840&  B9/A0V  (9) &  125.8 & 10593 &   2.4 &       21.5 &  0.029 & CCHEN2/40235 &  (18)   \\   
       77432& HD 141011& F5V (7) &   96.3 & 6440 &   1.4 &       1.90 &  0.000 & DEBRISII/40651 &  (17)   \\   
       77520& HD 141254& F3V (9) &  100.8 & 6740 &   1.5 &       1.87 &  0.194 & WARMDISK2/50538 &  (17)   \\   
       77523& HD 141327&  B9V  (9) &  195.3 & 10304 &   2.3 &       48.3 &  0.139 & TD\_GTO/50485 &  (18)   \\   
       77656& HD 141521& G5IV (13) &  140.1 & 5702 &   1.4 &       2.57 &  0.520 & RUBBLE/148 &  (13)   \\  
       78043& HD 142446& F3V (9) &  144.3 & 6740 &   1.5 &       4.66 &  0.124 & TD\_GTO/50485 &  (17)   \\   
       78555& HD 143538& F0V (9) &  106.3 & 7200 &   1.6 &       3.55 &  0.178 & CCHEN/241 &  (17)   \\  
       78641& HD 143675&  A5IV/V  (9) &  113.4 & 8072 &   1.7 &       6.32 &  0.049 & CCHEN2/40235 &  (18)   \\   
       78756& HD 143939&  B9III  (10) &  144.7 & 10740 &   2.4 &       40.6 &  0.000 & TD\_GTO/50485 &  (18)   \\  
       79400& HD 145357&  A5V  (7) &  146.8 & 8072 &   1.7 &       12.8 &  0.342 & TD\_GTO/50485 &  (18)   \\   
       79516& HD 145560& F5V (7) &  133.7 & 6440 &   1.4 &       3.84 &  0.001 & CCHEN/241 &  (17)   \\   
       79631& HD 145880&  B9.5V  (9) &  127.9 & 10593 &   2.4 &       35.3 &  0.355 & TD\_GTO/50485 &  (18)   \\   
       79710& HD 145972& F0V (7) &  127.4 & 7200 &   1.6 &       5.78 &  0.094 & TD\_GTO/50485 &  (17)   \\   
       79742& HD 146181& F6V (19) &  146.2 & 6360 &   1.4 &       3.66 &  0.000 & CCHEN/241 &  (17)   \\   
       80142& HD 147001&  B7V  (7) &  137.2 & 11912 &   2.8 &       78.5 &  0.166 & TD\_GTO/50485 &  (18)   \\  
       80897& HD 148657&  A0V  (9) &  165.6 & 9550 &   2.1 &       21.3 &  0.864 & TD\_GTO/50485 &  (18)   \\  
       82154& HD 151109&  B8V  (4) &  222.7 & 11298 &   2.6 &       117. &  0.074 & TD\_GTO/50485 &  (18)   \\   
       82747& AK Sco& F5V (9) &  102.8 & 6440 &   1.5 &       4.72 &  1.098 & IRSDISKS/2 &  (17) protoplanetary \\   
       83159& HD 153232& F5V (9) &  146.6 & 6440 &   1.5 &       4.17 &  0.000 & CCHEN/241 &  (17)   \\  
\cutinhead{Upper Scorpius}
       76310& HD 138813& A0V (16) &  150.8 & 9750 &   2.1 &       30.5 &  0.155 & JMCARP/30091 &  (20)   \\   
       77859& HD 142184& B2V (12) & 130.9 & \nodata & \nodata& \nodata & \nodata & JMCARP/30091 & classical Be \\
       77911& HD 142315& B9V (19) &  147.7 & 10000 &   2.5 &       50.2 &  0.202 & JMCARP/30091 &  (20)   \\   
       78207& HD 142983& B8Ia/Iab (12) & 143.5 & \nodata & \nodata& \nodata & \nodata & JMCARP/30091 & classical Be \\
       78663& HD 143811& F5V (9) &  144.3 & 6440 &   1.5 &       4.86 &  0.093 & DEBRISII/40651 &  (17)   \\   
       78977& HD 144548& F7V (15) &  116.7 & 6280 &   1.5 &       5.02 &  0.365 & CCHEN/241 &  (17) $\lambda^{-2}$ excess \\   
       78996& HD 144587& A9V (16) &  108.5 & 8750 &   1.4 &       11.4 &  0.967 & JMCARP/30091 &  (20)   \\   
       79054& HD 144729& F0V (12) &  138.9 & 7200 &   1.5 &       5.47 &  0.585 & CCHEN/241 &  (17)   \\  
       79156& HD 144981& A0V (16) &  170.4 & 9750 &   1.9 &       31.6 &  0.574 & JMCARP/30091 &  (20)   \\   
       79288& HD 145263& F0V (12) &  149.9 & 7200 &   1.6 &       6.40 &  0.399 & DEBRISII/40651 &  (17)   \\  
       79410& HD 145554& B9V (16) &  140.4 & 10000 &   1.9 &       32.5 &  0.577 & JMCARP/30091 &  (20)   \\   
       79439& HD 145631& B9V (16) &  131.8 & 10000 &   1.8 &       32.7 &  0.666 & JMCARP/30091 &  (20)   \\   
       79878& HD 146606& A0V (16) &  129.4 & 10000 &   2.1 &       27.1 &  0.016 & JMCARP/30091 &  (20)   \\   
       79977& HD 146897& F2/3V (12) &  122.7 & 6815 &   1.5 &       3.66 &  0.341 & IRS\_DISKS/2 &  (17)   \\   
       80024& HD 147010& B9II (16) &  163.4 & 10500 &   2.1 &       73.7 &  0.772 & JMCARP/30091 &  (20)   \\   
       80088& HD 147137& A9V (19) &  139.1 & 9000 &   1.7 &       12.7 &  1.147 & JMCARP/30091 &  (20)   \\   
       80320& HD 147594& G3IV (15) &  142.0 & 5830 &   1.4 &       3.39 &  0.031 & WARMDISK2/50538 &  (17)   \\   
       82218& HD 151376& F2/3V (12) &  135.7 & 6815 &   1.5 &       4.48 &  0.300 & CCHEN/241 &  (17)   \\   

\enddata
\vspace{-5ex}
\tablerefs{
References:
(1)  \citet{1920AnHar..95....1C},
(2)  \citet{1969ApJ...157..313H},
(3)  \citet{1970MmRAS..72..233H},
(4)  \citet{1971AJ.....76..237S},
(5)  \citet{1972AJ.....77..474G},
(6)  \citet{1975mcts.book.....H},
(7)  \citet{1978mcts.book.....H},
(8)  \citet{1978rmsa.book.....M},
(9)  \citet{1982mcts.book.....H},
(10) \citet{1983A&AS...51..143G},
(11) \citet{1984ApJS...55..657C},
(12) \citet{1988mcts.book.....H},
(13) \citet{2002Mamajek_etal},
(14) \citet{2003AJ....126.2971V},
(15) \citet{2006A&A...460..695T},
(16) \citet{Preibisch2008},
(17) \citet{2011Chen_etal},
(18) \citet{2012Chen_etal},
(19) \citet{2012PecautMamajekBubar},
(20) \citet{2014Chen_etal}
}
\end{deluxetable}
\setlength{\textwidth}{6.5in}

\begin{deluxetable}{ccccccccccl}
\tabletypesize{\footnotesize}
\tablecaption{Inferred disk properties\label{tab:diskprops}} 
\tablewidth{0pt}
\tablehead{
    \colhead{HIP} &
    \colhead{HD name} &
    \colhead{$L_{\rm IR}/L_*$}&
    \multicolumn{4}{c}{excess significance ($\chi$)} &
    \colhead{$R_{24/10}$} &
    \colhead{$R_{32/10}$} &
    \colhead{$a_{\rm min}$} &
    note 
    \\
    \omit &
    \omit &
    \omit &
    $\chi_{10}$ &
    $\chi_{24}$ &
    $\chi_{32}$ &
    total &
    \omit &
    \omit &
    \colhead{($\mu$m)} &
    \omit
}
\tablewidth{0pt}
\tablecolumns{11}
\startdata
&\\[-3ex]
\cutinhead{Lower Centaurus Crux}
53524 & HD 95086 &  1.10e-03 &   1.85 &   7.10 &   9.75 &   2.13 & $  3.21\pm  0.30$ & $ 12.53\pm  1.08$ &  1.8 &   \\
55188 & HD 98363 &  9.56e-04 &   6.47 &  16.55 &   7.98 &   5.19 & $  9.97\pm  0.31$ & $ 23.44\pm  2.65$ &  2.3 &   \\
56673 & HD 101088 &  5.50e-04 &   2.86 &   4.37 &   2.90 &   2.91 & $  1.12\pm  0.02$ & $  1.11\pm  0.07$ &  3.2 & $\lambda^{-2}$ excess \\
56673* & HD 101088 &  4.42e-05 &   0.23 &   2.00 &   1.30 &   0.28 & $  1.12\pm  0.02$ & $  1.11\pm  0.07$ &  3.2 & *\\
57524 & HD 102458 &  2.91e-04 &   1.27 &   2.56 &   2.13 &   1.19 & $  1.40\pm  0.19$ & $  1.79\pm  0.41$ &  0.4 &   \\
57950 & HD 103234 &  1.44e-04 &   0.96 &   7.04 &   4.65 &   0.97 & $  1.76\pm  0.07$ & $  2.45\pm  0.30$ &  1.1 &   \\
58220 & HD 103703 &  6.91e-04 &   4.40 &   6.30 &   4.24 &   3.50 & $  2.19\pm  0.23$ & $  3.17\pm  0.60$ &  0.9 &   \\
58528 & HD 104231 &  2.35e-04 &   2.47 &   7.48 &   3.55 &   1.15 & $  1.98\pm  0.12$ & $  2.57\pm  0.47$ &  1.1 &   \\
58720 & HD 104600 &  1.09e-04 &   3.25 &  15.10 &  17.02 &   2.44 & $  3.72\pm  0.06$ & $  8.04\pm  0.16$ &  8.6 &   \\
59282 & HD 105613 &  8.07e-05 &   2.40 &   9.26 &   4.78 &   0.94 & $  1.91\pm  0.06$ & $  2.58\pm  0.33$ &  2.5 &   \\
59397 & HD 105857 &  1.74e-04 &   2.32 &  12.06 &   9.20 &   1.95 & $  3.01\pm  0.13$ & $  4.39\pm  0.34$ &  2.8 &   \\
59481 & HD 105994 &  8.28e-05 &   0.17 &   1.86 &   1.16 &   0.42 & $  1.19\pm  0.09$ & $  1.39\pm  0.34$ &  1.1 &   \\
59502 & HD 106036 &  3.79e-04 &   5.49 &  13.80 &   5.17 &   3.38 & $  4.71\pm  0.18$ & $  8.45\pm  1.45$ &  2.8 &   \\
59693 & HD 106389 &  3.83e-04 &   3.48 &   4.58 &   2.27 &   1.57 & $  1.67\pm  0.18$ & $  1.87\pm  0.47$ &  0.7 &   \\
59898 & HD 106797 &  2.21e-04 &   3.85 &  16.05 &  17.48 &   3.11 & $  4.31\pm  0.06$ & $  8.23\pm  0.14$ &  5.5 &   \\
59960 & HD 106906 &  1.27e-03 &   0.44 &  13.60 &  16.09 &   2.49 & $  7.11\pm  0.35$ & $ 23.54\pm  1.04$ &  1.4 &   \\
60183 & HD 107301 &  9.65e-05 &   1.60 &  15.08 &  14.85 &   1.74 & $  4.06\pm  0.14$ & $  8.46\pm  0.40$ &  5.4 &   \\
60348 & HD 107649 &  2.27e-04 &   1.50 &   6.69 &   3.92 &   1.07 & $  1.83\pm  0.11$ & $  3.03\pm  0.51$ &  0.4 &   \\
60561 & HD 107947 &  1.13e-04 &   2.33 &  11.31 &  10.33 &   1.49 & $  2.44\pm  0.08$ & $  4.33\pm  0.27$ &  3.1 &   \\
60710 & HD 108257 & -2.13e-06 &  -0.43 &  -0.11 &   0.25 &  -0.21 & $  1.02\pm  0.07$ & $  1.11\pm  0.31$ & 44.7 & no excess \\
61049 & HD 108857 &  4.36e-04 &   3.46 &  12.94 &  10.42 &   1.86 & $  3.13\pm  0.09$ & $  4.16\pm  0.25$ &  0.9 &   \\
61087 & HD 108904 &  5.24e-04 &   2.43 &   5.69 &   1.95 &   1.88 & $  2.93\pm  0.35$ & $  3.58\pm  1.38$ &  1.4 &   \\
61684 & HD 109832 &  4.16e-04 &   1.14 &  13.01 &  15.03 &   1.79 & $  3.93\pm  0.13$ & $ 10.36\pm  0.39$ &  1.8 &   \\
61782 & HD 110058 &  1.58e-03 &   3.69 &  17.41 &  18.00 &   6.63 & $ 23.13\pm  1.24$ & $ 72.52\pm  3.83$ &  1.9 &   \\
62134 & HD 110634 &  4.61e-05 &   0.08 &   1.59 &   0.27 &   0.15 & $  1.59\pm  0.36$ & $  1.34\pm  1.24$ &  1.0 &   \\
62427 & HD 111103 &  2.46e-04 &  -0.57 &   7.19 &   5.17 &   0.57 & $  2.76\pm  0.22$ & $  6.38\pm  1.02$ &  0.9 &   \\
62445 & HD 111170 & -1.57e-04 &  -0.12 &   0.31 &  -0.01 &  -0.58 & $  1.04\pm  0.10$ & $  1.01\pm  0.39$ &  1.1 & no excess \\
62657 & HD 111520 &  1.96e-03 &   1.81 &  14.82 &  17.50 &   3.49 & $  5.97\pm  0.23$ & $ 20.37\pm  0.70$ &  0.8 &   \\
63236 & HD 112383 &  1.54e-04 &   2.57 &  12.60 &  10.61 &   1.73 & $  2.71\pm  0.12$ & $  3.76\pm  0.26$ &  3.9 &   \\
63439 & HD 112810 &  8.00e-04 &   1.38 &   6.25 &  10.92 &   1.18 & $  2.00\pm  0.16$ & $  6.48\pm  0.47$ &  1.0 &   \\
63836 & HD 113524 &  1.26e-04 &   1.00 &   4.34 &   2.42 &   0.47 & $  1.46\pm  0.09$ & $  2.56\pm  0.66$ &  0.7 &   \\
63839 & HD 113457 &  2.74e-04 &   5.07 &  15.37 &  16.07 &   3.34 & $  4.13\pm  0.14$ & $  7.12\pm  0.26$ &  3.6 &   \\
63886 & HD 113556 &  5.15e-04 &   0.84 &   4.68 &   4.88 &   0.75 & $  2.05\pm  0.22$ & $  6.74\pm  1.15$ &  1.3 &   \\
63975 & HD 113766 &  2.23e-02 &  17.40 &  18.78 &  14.00 &  13.12 & $  4.00\pm  0.11$ & $  4.93\pm  0.27$ &  2.5 &   \\
64053 & HD 113902 &  4.17e-05 &   2.27 &   9.93 &   8.92 &   1.28 & $  1.89\pm  0.04$ & $  2.54\pm  0.14$ &  9.6 &   \\
64184 & HD 114082 &  3.01e-03 &   1.83 &  13.19 &   9.41 &   3.82 & $ 19.43\pm  1.17$ & $ 56.92\pm  5.46$ &  0.8 &   \\
64877 & HD 115361 &  3.09e-04 &   1.22 &   5.85 &   1.86 &   1.04 & $  2.59\pm  0.25$ & $  5.43\pm  2.41$ &  1.4 &   \\
64995 & HD 115600 &  1.87e-03 &   1.57 &  12.22 &   8.98 &   3.49 & $ 12.76\pm  0.88$ & $ 40.36\pm  4.16$ &  1.3 &   \\
65089 & HD 115820 &  1.52e-04 &  -0.32 &   9.60 &   6.66 &   1.21 & $  2.46\pm  0.09$ & $  3.65\pm  0.36$ &  1.2 &   \\
65875 & HD 117214 &  2.53e-03 &   2.14 &  13.25 &   9.56 &   4.05 & $ 13.32\pm  0.72$ & $ 37.70\pm  3.41$ &  1.5 &   \\
65965 & HD 117484 &  2.37e-04 &   3.10 &  14.72 &  14.76 &   2.92 & $  4.93\pm  0.18$ & $ 13.81\pm  0.64$ &  3.9 &   \\
66001 & HD 117524 &  7.98e-05 &   1.31 &   0.88 &  -0.01 &   0.28 & $  1.04\pm  0.13$ & $  0.91\pm  0.42$ &  0.6 & no excess \\
66068 & HD 117665 &  1.94e-04 &   3.18 &  13.91 &  11.39 &   2.17 & $  3.45\pm  0.10$ & $  5.71\pm  0.34$ &  4.6 &   \\
66566 & HD 118588 &  2.24e-04 &   3.00 &  14.26 &  11.26 &   2.69 & $  4.32\pm  0.16$ & $  8.16\pm  0.55$ &  2.9 &   \\
67068 & HD 119511 &  5.58e-06 &   0.28 &   3.79 &   2.30 &   0.03 & $  1.33\pm  0.07$ & $  1.52\pm  0.23$ &  0.7 &   \\
67230 & HD 119718 &  3.39e-04 &   1.51 &   4.08 &   3.30 &   1.05 & $  2.99\pm  0.49$ & $  6.06\pm  1.56$ &  2.0 &   \\
\cutinhead{Upper Centaurus Lupus}
66447 & HD 118379 &  1.37e-04 &   0.45 &  10.08 &  10.05 &   0.80 & $  2.44\pm  0.10$ & $  6.29\pm  0.46$ &  2.8 &   \\
67497 & HD 120326 &  1.50e-03 &   1.31 &  15.53 &  15.65 &   3.50 & $ 11.52\pm  0.70$ & $ 31.39\pm  2.00$ &  1.1 &   \\
67970 & HD 121189 &  5.25e-04 &   2.13 &  11.04 &  12.42 &   2.10 & $  3.93\pm  0.23$ & $  8.62\pm  0.52$ &  1.0 &   \\
68080 & HD 121336 &  6.10e-05 &   1.27 &   8.88 &   9.60 &   0.83 & $  1.81\pm  0.04$ & $  2.83\pm  0.14$ & 10.9 &   \\
68781 & HD 122705 &  8.87e-05 &   1.93 &   7.62 &   4.35 &   0.96 & $  1.72\pm  0.06$ & $  2.18\pm  0.27$ &  2.0 &   \\
69291 & HD 123889 &  6.89e-05 &   0.62 &   2.22 &   2.62 &   0.38 & $  1.14\pm  0.06$ & $  1.71\pm  0.28$ &  1.4 &   \\
69720 & HD 124619 &  1.51e-04 &   1.97 &   5.28 &   3.49 &   0.98 & $  1.59\pm  0.11$ & $  2.18\pm  0.36$ &  1.3 &   \\
70149 & HD 125541 &  2.06e-04 &   1.57 &   9.20 &   7.94 &   1.22 & $  3.28\pm  0.21$ & $  6.80\pm  0.68$ &  0.8 &   \\
70441 & HD 126062 &  2.72e-04 &   1.99 &   9.09 &   9.59 &   1.57 & $  2.68\pm  0.15$ & $  6.52\pm  0.51$ &  2.2 &   \\
70455 & HD 126135 &  3.28e-05 &   1.35 &   7.62 &   6.47 &   1.01 & $  2.08\pm  0.12$ & $  3.25\pm  0.33$ &  8.5 &   \\
71271 & HD 127750 &  1.62e-04 &   1.32 &  10.74 &  12.09 &   1.10 & $  2.51\pm  0.08$ & $  7.01\pm  0.37$ &  4.6 &   \\
71453 & HD 128207 &  1.01e-05 &   0.34 &   4.17 &   5.05 &   0.52 & $  1.29\pm  0.04$ & $  1.53\pm  0.08$ & 19.5 &   \\
72033 & HD 129490 & -1.06e-04 &  -0.20 &   0.45 &   0.64 &  -0.48 & $  1.06\pm  0.09$ & $  1.23\pm  0.34$ &  1.5 & no excess \\
72070 & HD 129590 &  5.22e-03 &   1.15 &  18.03 &  18.66 &   5.45 & $ 15.21\pm  0.43$ & $ 60.57\pm  1.80$ &  0.9 &   \\
73145 & HD 131835 &  2.27e-03 &   5.10 &  17.74 &  17.89 &   6.89 & $ 14.32\pm  0.69$ & $ 49.81\pm  2.48$ &  2.2 &   \\
73341 & HD 132238 &  2.64e-05 &   1.63 &   7.60 &   6.90 &   0.93 & $  2.03\pm  0.11$ & $  3.06\pm  0.28$ & 12.2 &   \\
73666 & HD 133075 &  2.36e-05 &  -0.13 &   2.35 &   2.85 &   0.13 & $  1.25\pm  0.08$ & $  1.99\pm  0.33$ &  3.1 &   \\
73990 & HD 133803 &  3.62e-04 &   2.95 &  14.05 &  12.14 &   2.30 & $  4.12\pm  0.16$ & $  7.35\pm  0.44$ &  2.0 &   \\
74499 & HD 134888 &  8.25e-04 &   1.10 &   8.84 &  12.42 &   1.27 & $  3.14\pm  0.22$ & $ 10.34\pm  0.67$ &  0.3 &   \\
74752 & HD 135454 &  7.22e-05 &   1.70 &   4.89 &   5.94 &   1.61 & $  1.29\pm  0.08$ & $  1.61\pm  0.11$ &  9.7 &   \\
74959 & HD 135953 &  6.31e-04 &   0.80 &   4.85 &  10.84 &   0.85 & $  1.60\pm  0.11$ & $  5.59\pm  0.35$ &  0.8 &   \\
75077 & HD 136246 &  5.09e-05 &   0.57 &   3.55 &   6.70 &   0.41 & $  1.31\pm  0.07$ & $  2.07\pm  0.13$ &  4.1 &   \\
75151 & HD 136347 &  5.01e-06 &   0.49 &   0.58 &   0.21 &   0.18 & $  1.01\pm  0.04$ & $  0.99\pm  0.10$ &  8.5 & no excess \\
75210 & HD 136482 &  6.14e-05 &   2.53 &  11.56 &  10.42 &   1.62 & $  2.78\pm  0.10$ & $  5.03\pm  0.32$ &  7.1 &   \\
75304 & HD 136664 & -3.90e-06 &  -0.26 &   0.32 &   0.21 &  -0.42 & $  1.04\pm  0.04$ & $  1.04\pm  0.05$ & 73.8 & no excess \\
75491 & HD 137057 &  3.12e-04 &   0.67 &  11.41 &  10.90 &   1.35 & $  3.45\pm  0.16$ & $  7.43\pm  0.51$ &  2.0 &   \\
75509 & HD 137119 &  1.92e-04 &   3.38 &   9.02 &   4.78 &   2.13 & $  2.77\pm  0.17$ & $  4.03\pm  0.64$ &  1.9 &   \\
76084 & HD 138296 & -7.25e-05 &  -0.31 &   0.50 &   0.10 &  -0.44 & $  1.07\pm  0.09$ & $  1.06\pm  0.35$ &  1.6 & no excess \\
76395 & HD 138923 &  8.42e-05 &   4.39 &  12.14 &  12.72 &   2.77 & $  2.48\pm  0.07$ & $  3.61\pm  0.14$ &  6.6 &   \\
77081 & HD 140374 &  3.63e-04 &   1.36 &  -0.52 &  -0.45 &   0.96 & $  0.85\pm  0.11$ & $  0.68\pm  0.47$ &  0.1 & no excess \\
77315 & HD 140817 &  1.54e-04 &   2.69 &  12.88 &  13.92 &   2.15 & $  3.39\pm  0.11$ & $  6.34\pm  0.26$ &  6.9 &   \\
77317 & HD 140840 &  1.64e-04 &   1.82 &  10.72 &  14.39 &   1.79 & $  3.63\pm  0.19$ & $ 10.90\pm  0.48$ &  3.3 &   \\
77432 & HD 141011 &  3.19e-04 &   1.65 &   3.85 &   3.14 &   1.30 & $  1.74\pm  0.21$ & $  3.26\pm  0.76$ &  0.2 &   \\
77520 & HD 141254 &  9.31e-05 &   1.18 &   2.47 &   2.58 &   0.47 & $  1.28\pm  0.13$ & $  2.07\pm  0.43$ &  0.2 &   \\
77523 & HD 141327 &  4.59e-05 &   1.86 &   6.65 &   4.79 &   0.92 & $  1.77\pm  0.10$ & $  2.87\pm  0.39$ &  7.2 &   \\
77656 & HD 141521 & -1.73e-04 &  -0.41 &  -0.06 &  -0.03 &  -0.61 & $  1.02\pm  0.15$ & $  1.00\pm  0.33$ &  0.7 & no excess \\
78043 & HD 142446 &  7.00e-04 &   1.60 &   5.90 &   9.73 &   1.46 & $  2.54\pm  0.25$ & $  7.80\pm  0.60$ &  1.3 &   \\
78555 & HD 143538 & -7.40e-05 &  -0.52 &   0.52 &   0.34 &  -0.21 & $  1.29\pm  0.47$ & $  1.55\pm  1.51$ &  0.9 & no excess \\
78641 & HD 143675 &  5.77e-04 &   4.11 &  15.67 &  14.25 &   3.61 & $  6.15\pm  0.18$ & $ 14.16\pm  0.66$ &  1.5 &   \\
78756 & HD 143939 &  2.17e-05 &   0.82 &   5.42 &   4.14 &   0.52 & $  1.73\pm  0.11$ & $  2.53\pm  0.36$ &  5.9 &   \\
79400 & HD 145357 &  1.11e-04 &   1.47 &   5.35 &   8.60 &   0.95 & $  1.75\pm  0.16$ & $  2.63\pm  0.20$ &  2.9 &   \\
79516 & HD 145560 &  2.85e-03 &   0.59 &  14.18 &  15.53 &   2.82 & $  8.20\pm  0.56$ & $ 33.35\pm  2.30$ &  1.1 &   \\
79631 & HD 145880 &  2.43e-04 &   1.20 &  15.19 &  17.19 &   1.73 & $  5.18\pm  0.26$ & $ 17.28\pm  0.88$ &  5.2 &   \\
79710 & HD 145972 &  2.14e-04 &   2.39 &   7.07 &   3.32 &   1.26 & $  2.42\pm  0.19$ & $  3.27\pm  0.72$ &  1.5 &   \\
79742 & HD 146181 &  2.59e-03 &   1.24 &   8.71 &  14.00 &   2.62 & $  8.12\pm  0.98$ & $ 32.95\pm  3.14$ &  1.1 &   \\
80142 & HD 147001 & -1.42e-05 &  -0.92 &  -0.03 &   0.20 &  -0.47 & $  1.06\pm  0.27$ & $  1.19\pm  0.54$ &  9.4 & no excess \\
80897 & HD 148657 &  3.71e-04 &   3.05 &  15.32 &  16.48 &   3.81 & $  5.96\pm  0.25$ & $ 17.90\pm  0.79$ &  3.7 &   \\
82154 & HD 151109 &  8.35e-05 &   1.94 &  13.05 &  12.35 &   1.75 & $  3.73\pm  0.12$ & $  9.06\pm  0.51$ & 14.6 &   \\
83159 & HD 153232 &  5.23e-04 &   0.32 &   0.46 &   0.34 &   0.39 & $  1.81\pm  1.80$ & $  3.91\pm  9.26$ &  1.1 & no excess \\
\cutinhead{Upper Scorpius}
76310 & HD 138813 &  9.50e-04 &   4.36 &  18.00 &  17.65 &   6.55 & $ 12.95\pm  0.39$ & $ 39.23\pm  1.38$ &  5.2 &   \\
77911 & HD 142315 &  3.52e-04 &   0.11 &   2.58 &   1.75 &   0.75 & $  8.83\pm  4.33$ & $ 25.22\pm 16.45$ &  7.0 &   \\
78663 & HD 143811 &  3.30e-05 &  -0.15 &   2.48 &   2.20 &   0.06 & $  1.34\pm  0.12$ & $  2.18\pm  0.53$ &  1.3 &   \\
78977 & HD 144548 &  1.33e-03 &   4.72 &   1.60 &   1.38 &   4.52 & $  0.97\pm  0.15$ & $  1.50\pm  0.53$ &  1.4 & $\lambda^{-2}$ excess \\
78977* & HD 144548 &  9.55e-05 &   0.21 &  -0.13 &   0.90 &   0.10 & $  0.97\pm  0.15$ & $  1.50\pm  0.53$ &  1.4 & no excess*\\
78996 & HD 144587 &  3.24e-04 &   4.70 &  13.91 &  11.09 &   3.37 & $  2.93\pm  0.10$ & $  4.80\pm  0.31$ &  3.1 &   \\
79054 & HD 144729 &  9.49e-05 &   1.15 &   1.09 &   0.55 &   0.51 & $  1.33\pm  0.37$ & $  1.50\pm  1.07$ &  1.5 & no excess \\
79156 & HD 144981 &  2.15e-04 &   4.66 &  10.36 &   8.08 &   3.09 & $  2.16\pm  0.10$ & $  3.27\pm  0.28$ &  5.9 &   \\
79288 & HD 145263 &  1.35e-02 &  18.25 &  19.36 &  19.13 &  16.29 & $  5.84\pm  0.08$ & $  8.87\pm  0.15$ &  1.6 &   \\
79410 & HD 145554 &  1.76e-04 &   4.38 &  12.65 &  10.87 &   2.68 & $  2.86\pm  0.11$ & $  4.95\pm  0.32$ &  6.0 &   \\
79439 & HD 145631 &  8.33e-05 &   1.16 &   6.31 &   5.77 &   1.20 & $  1.72\pm  0.11$ & $  2.47\pm  0.26$ &  6.4 &   \\
79878 & HD 146606 &  7.98e-05 &   1.72 &   6.34 &   5.92 &   1.01 & $  2.47\pm  0.22$ & $  4.50\pm  0.59$ &  4.6 &   \\
79977 & HD 146897 &  5.21e-03 &   1.71 &  15.88 &  15.25 &   4.94 & $ 24.15\pm  1.87$ & $ 89.38\pm  7.21$ &  1.0 &   \\
80024 & HD 147010 &  1.17e-04 &   1.70 &   7.70 &   7.76 &   2.12 & $  3.54\pm  0.30$ & $  8.22\pm  0.86$ & 11.6 &   \\
80088 & HD 147137 &  3.98e-04 &   2.41 &  11.77 &  14.09 &   2.90 & $  3.44\pm  0.16$ & $  7.91\pm  0.36$ &  2.9 &   \\
80320 & HD 147594 &  1.37e-04 &   1.09 &   4.65 &   3.22 &   0.45 & $  1.44\pm  0.09$ & $  2.22\pm  0.38$ &  1.0 &   \\
82218 & HD 151376 &  2.77e-04 &   0.52 &   3.21 &   2.78 &   0.55 & $  2.07\pm  0.35$ & $  4.75\pm  1.39$ &  1.2 &   \\

\enddata
\tablenotetext{*}{HIP 56673 and HIP 78977 are listed twice, the marked listing 
indicating that the spectrum was normalized to the photosphere model 
rather than the MIPS 24 micron measurement.}
\end{deluxetable}

\clearpage

\section{Debris Disk Modeling}

Our modeling of the dust in the debris disks under study goes beyond 
a simple blackbody model.  This is because the spectral range
covered by IRS includes various silicate features.  In order to 
model the IRS spectra in better detail, we include grain properties 
such as size, temperature, and composition and generate model 
spectra using Mie theory.  

\subsection{Optical Properties of Grains}

We assume that the dust is composed of amorphous 
silicates of olivine and pyroxene composition, 
the optical constants for which are adopted from 
\citet{Dorschner1995} and \citet{Jaeger1994}, assuming 
a $\mbox{Mg}:\mbox{Fe}$ ratio of 1 for both species\footnote{
Optical constants for all silicate species are available at 
http://www.astro.uni-jena.de/Laboratory/OCDB/newsilicates.html.}.

In order to calculate equilibrium temperatures, we need the optical 
constants at short wavelengths as well.  For $\lambda \lesssim 2\mu$m, 
we use the optical constants for astronomical silicates
from \citet{DraineLee1984}.

The minimum grain size, $a_{\rm min}$, is estimated by assuming 
that radiation pressure removes the smallest grains if 
$\beta(=F_{\rm rad}/F_{\rm grav})>0.5$, so that 
\begin{equation}\label{eq:amin}
a_{\rm min} > 
\frac{6 L_* \langle Q_{\rm pr}(a) \rangle}{16\pi G M_* c \rho_s}
\end{equation}
\citep{1988Artymowicz},
where $L_*$ and $M_*$ are the stellar luminosity and mass, 
$\rho_s$ is the density of an individual grain, and 
$\langle Q_{\rm pr}(a) \rangle$ is the spectrum-averaged radiation pressure
coupling coefficient, given by 
\(
\langle Q_{\rm pr}(a) \rangle = 
\int Q_{\rm pr}(a,\lambda) F_{\lambda} \, d\lambda/
(\int F_{\lambda} \, d\lambda).
\)
These values are tabulated in \citet{2011Chen_etal} 
for F- and G-type stars, and \citet{2012Chen_etal} for B- and A-type stars.  
In Table \ref{tab:starprops} we
list the stellar mass and luminosity assumed for 
modeling the dust for each source, and in Table \ref{tab:diskprops}, 
we list the estimated minimum grain size.

\citet{2011Chen_etal,2012Chen_etal} estimated the color temperature 
of the dust from the the ratio of 24 to 70 micron MIPS photometric
excess, and the grain distance calculated assuming the 
grains had a temperature equal to the color temperature 
and a size equal to an average grain size of 
$\langle a \rangle = 5a_{\rm min}/3$.  
The IRS spectra provide more detailed information on the temperature, 
size, and composition of the dust grains than the MIPS photometry.  
In this work we allow the grain size to be a free parameter.  
The distribution of grain sizes depends on the model type 
implemented, as described in \S\ref{sec:fittypes}, 
and $a_{\rm min}$ is treated as a lower limit on grain size.  

For computational simplicity, 
we use Mie theory to calculate the optical constants for 
scattering and absorption of light on particles of different sizes.
The main element missing from a Mie-theory treatment would
be grain porosity, which could result in underestimating 
the grain sizes and $\beta$ values
\citep{1998Lisse+,2007Kolokolova+,2001Kolokolova+}.
We use the Oxford IDL routines\footnote{
http://www-atm.physics.ox.ac.uk/code/mie/index.html}
to calcuate the optical constants.  
This generates $Q\upp{ext}(\lambda,a)$ and $Q\upp{sca}(\lambda,a)$, 
the extinction and scattering efficiencies, respectively, 
as a function of wavelength $\lambda$ and grain radius $a$.  
Then the absorption efficiency is $Q\upp{abs} = Q\upp{ext}-Q\upp{sca}$.  
Scattered stellar light does not contribute significantly at 
the Spitzer IRS wavelengths, 
so we consider only the thermal component of emission 
in modeling the spectra.

\subsection{Grain Temperatures and Distances}

The dust grains are assumed to be heated by stellar irradiation 
and in thermal equilibrium.  The amount of radiation absorbed 
by a grain is the incident stellar flux modified by the 
absorption efficiency $Q\sub{abs}$ scaled by the cross-section 
of the grain.  
The absorption efficiencies of olivine and pyroxene are 
$Q\upp{abs}_{o}$ and $Q\upp{abs}_{p}$, respectively.  
The fractional composition of olivine is $f_o$, so the 
fractional composition of pyroxene is $(1-f_o)$. 
The stellar flux at a distance $r$ from the star is 
\(F_{\nu,*} = \pi (R_*/r)^2 I_{\nu,*}(T\sub{eff})\).  
Then the total power absorbed by a grain is 
\begin{equation}
P\sub{abs}= \int_0^{\infty} 
\pi^2 a^2 Q\upp{abs} \left(\frac{R_*}{r}\right)^2 I_{\nu,*}(T\sub{eff})
\, d\nu
\end{equation}
where $I_{\nu,*}$ is the stellar spectrum, for which we use the 
best fit Kurucz models.  

The total emergent power of a grain of 
radius $a$ and temperature $T\sub{gr}$ is 
\begin{equation}
P\sub{emit} = \int_0^{\infty}
4 \pi^2 a^2 Q\upp{abs} B_{\nu}(T\sub{gr}) \, d\nu
\end{equation}
where $B_{\nu}$ is the Planck function.

To find the equilibrium temperature of the grain, we set 
$P\sub{emit}=P\sub{abs}$ and find that  
\begin{equation}\label{eq:tequil}
\int_0^{\infty}
Q\upp{abs} B_{\nu}(T\sub{gr}) \, d\nu = 
\frac{R_*^2}{4r^2}
\int_0^{\infty} 
Q\upp{abs} I_{\nu,*}(T\sub{eff})\, d\nu
\end{equation}
The absorption efficiency, 
$Q\sub{abs}$, is itself a function of $a$ and $\nu$,
and also depends on the composition,
as determined from Mie theory and the optical constants of 
the constituents, namely oliving and pyroxene.
We set $Q\upp{abs}$ equal to the the absorption efficiencies of 
olivine ($Q\upp{abs}_{o}$) or pyroxene ($Q\upp{abs}_{p}$), 
depending on the grain composition.  
We assume a segregated spheres distribution for the composition of grains 
where each grain is either pure olivine or pure pyroxene, and 
$f_o$ and $f_p$ are the mass fraction of grains consisting of 
olivine or pyroxene, respectively.
The distance of the grains derived from the grain temperatures
is sensitive to the composition.  For instance, highly reflective
grains, such as ices, absorb energy less efficiently, therefore
for the same equilibrium temperature
they will be at smaller stellocentric distances compared to more
absorptive grains.
For olivine and pyroxene, we find that the differences in distances
are not significant.  We also find that including a wider
range of compositions, such as crystalline silicates, 
does not significantly improve fits to the mid-IR spectra 
enough to justify the additional free parameters.
Thus, the scope of this work is limited to olivine and
pyroxene compositions only.  

To facilitate calculations of equilibrium temperatures, we 
tabulate values of $r$ versus $T\sub{gr}$ as a function of 
$a$ and composition and interpolate on the grid.  
The equilibrium temperature of a grain of a given size is a proxy 
for its distance from the star.  

We assume that the disks are optically thin, so the 
total spectrum of the disk is the summation of the emission 
of all the grains in the disk.  
We define $n(a,r)$ to be the number density of grains 
between distance $r$ and $r+\delta r$ and grain size $a$ and $a+\delta a$
grain size distribution as a function 
of $a$ and $r$ so that the total number of grains is 
$N=\int\int n(a,r)\, 2\pi r \, dr da $
and the total mass of the disk is 
\begin{equation}\label{eq:massintegral}
M = \int\int \frac{4\pi a^3}{3} \rho_d \, n(a,r)\, 2\pi r \, dr da 
\end{equation}
where $\rho_d$ is the bulk density of the dust grains
(3.3 g/cm$^3$).  
The emitted spectrum of a single grain of radius 
$a$ at a distance $r$ from the star is 
\[
\pi \frac{a^2}{d^2} Q\sub{abs}(a,\lambda) B_{\nu}[T\sub{gr}(T\sub{eff},a,r)]
\]
where $d$ is the distance between the observer and the star.  
Thus, the total integrated spectrum for a distribution of 
particles in a disk is 
\begin{equation}\label{eq:spectrumintegral}
F_{\nu} = \int\int 
\pi \frac{a^2}{d^2} Q\sub{abs}(a,\lambda) B_{\nu}[T\sub{gr}(T\sub{eff},a,r)]
n(a,r)\, 2\pi r \, dr da.
\end{equation}

\subsection{Fitting the Spectra}
\label{sec:fittypes}

We fit our spectra to two different dust distribution models, 
allowing grain size, temperature/distance, and composition 
to be free parameters in order 
to understand the properties of the debris disks.  The model types 
were 
1) a single uniform grain size with a single temperature and 
composition, and 
2) a two grain model 
where each population has a uniform grain size, temperature, 
and composition.  In each case, the best 
fitting size and temperature parameters are determined by 
minimizing the reduced $\chi_{\nu}^2$.  
This was implemented using the IDL routine MPFITFUN.  
The detailed description of each model follows, including 
a listing of the free parameters for each model type.  

\begin{enumerate}
\item {\em Single grain model:} 
This model assumes a population of grains of a uniform 
single size and temperature, as if they were distributed 
in a ring (or a shell) of uniform radius around the star.  
The free parameters for this model are 
the grain radius $a$, temperature $T\sub{gr}$, 
and composition $(f_o,f_p)$.  
The total number of particles $N$ sets the overall normalization
of the spectrum.  Since the grain size distribution and 
stellocentric distance are delta functions, 
then the brightness of this model disk assuming a distance of 
$d$ is 
\begin{equation}
F_{\nu}= N\frac{\pi a^2}{d^2}
\left[ f_o Q\upp{abs}_o(a,\lambda) + f_p Q\upp{abs}_p(a,\lambda)\right]
B_{\nu}(T\sub{gr}).
\end{equation}
This mass of this disk is then 
\begin{equation}
M\sub{disk} = N \frac{4\pi a^3}{3} \rho_d.
\end{equation}

Our assumption is that all the grains are at the same temperature.  
If the absorption efficiencies were to differ greatly between 
olivine and pyroxene, then the grains would not be co-located at the 
same distance from the star.  For the grain sizes and temperature 
ranges explored, the absorption efficiencies 
are similar enough 
that the dust grains are effectively co-located.  For typical temperatures 
and grain sizes in our fits, the calculated distances between pure olivine 
and pure pyroxene grains differs by $\sim10\%$.  

\item {\em Two-grain model:}\\ 
This model is a superposition of two uniform grain populations, 
each with its own independent grain size, temperature, and 
composition.  
That is, one ring of grains with 
grain radius $a_1$, temperature $T_1$, composition 
$(f_{o1},f_{p1})$, and number $N_1$, 
plus a second ring with parameters 
$a_2$, $T_2$, $(f_{o2},f_{p2})$, and $N_2$. 
Then the brightness of this disk model is 
\begin{eqnarray}
F_{\nu}&=& N_1\frac{\pi a_1^2}{d_1^2}
\left[f_{o,1} Q\upp{abs}_o(a_1,\nu) + f_{p,1} Q\upp{abs}_p(a_1,\nu) \right]
(a_1,\nu) B_{\nu}(T_1)
\nonumber\\ &&\quad 
+N_2\frac{\pi a_2^2}{d_2^2}
\left[f_{o,2} Q\upp{abs}_o(a_2,\nu) + f_{p,2} Q\upp{abs}_p(a_2,\nu) \right]
B_{\nu}(T_2).
\end{eqnarray}
This mass of this disk is then 
\begin{equation}
M\sub{disk} = \left(N_1 \frac{4\pi a_1^3}{3} 
                  + N_2 \frac{4\pi a_2^3}{3}\right) \rho_d
\end{equation}

\end{enumerate}

\section{Results}

For each fit, the grain temperature, grain size, and amorphous silicate composition are allowed to be free parameters.
That is, the single grain model has 4 free parameters
(temperature, grain size, olivine:pyroxene ratio, and total mass),
and the two-grain model has 8 free parameters.
Once we have carried out a least-squares fit to the excess spectrum of 
the single-grain and two-grain population models, 
as described in section \ref{sec:fittypes}, we need to 
determine which model best describes each object.
If the reduced $\chi_{\nu}^2$ value is less than 2, then we 
declare the fit to be reasonable.  The resulting 
fits for the 96 sources analyzed are shown in Tables 
\ref{tab:fits1}-\ref{tab:fits0}.
These tables also list the derived stellocentric distance of each grain
population, $r\sub{gr}$, which is calculated from 
the fitted grain properties and assumed stellar properties,
according to Equation (\ref{eq:tequil}).
In Figures \ref{fitfig0}-\ref{fitfig4}, we show these fits to the 
infrared excess spectra.  In each plot, the
photosphere-subtracted 
spectrum is plotted, along with the best fit single-grain 
and two-grain models to the excess spectrum.  

\setlength{\stampwidth}{0.24\textwidth}

\renewcommand{\thefigure}{\arabic{figure}\alph{subfig}}
\setcounter{subfig}{1}

\begin{figure}
\parbox{\stampwidth}{
\includegraphics[width=\stampwidth]{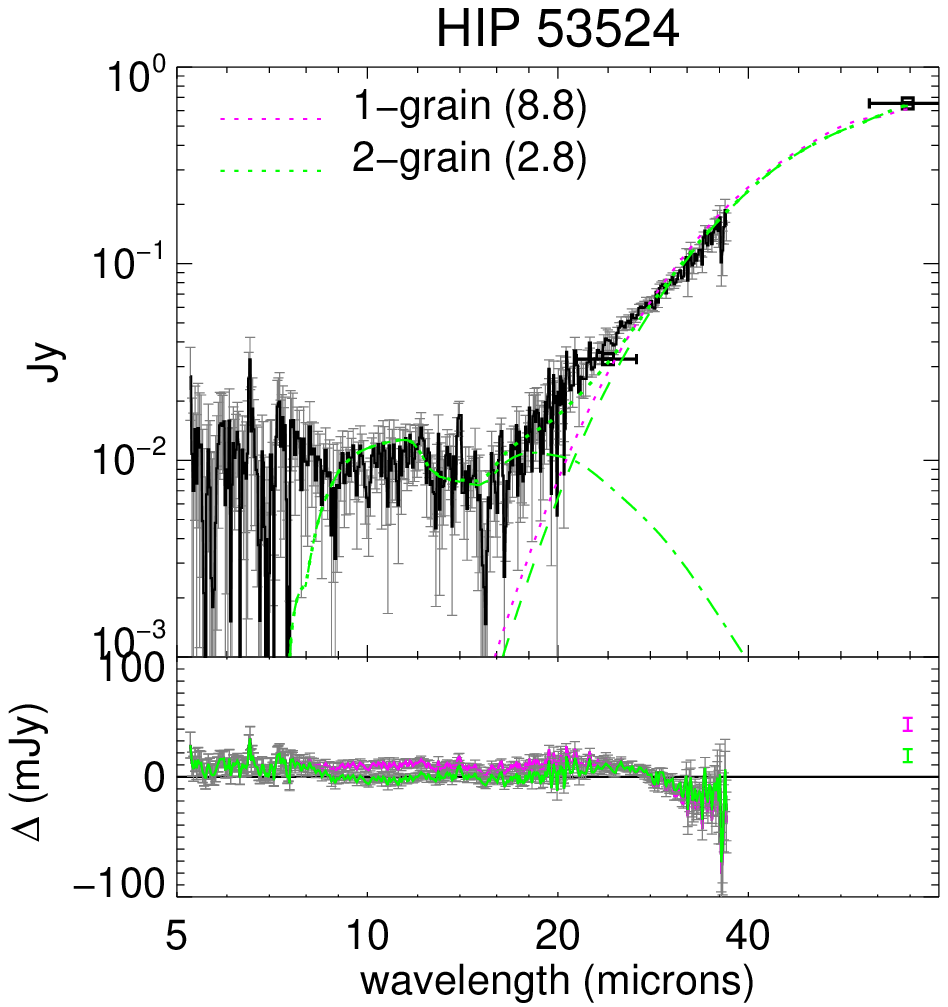} }
\parbox{\stampwidth}{
\includegraphics[width=\stampwidth]{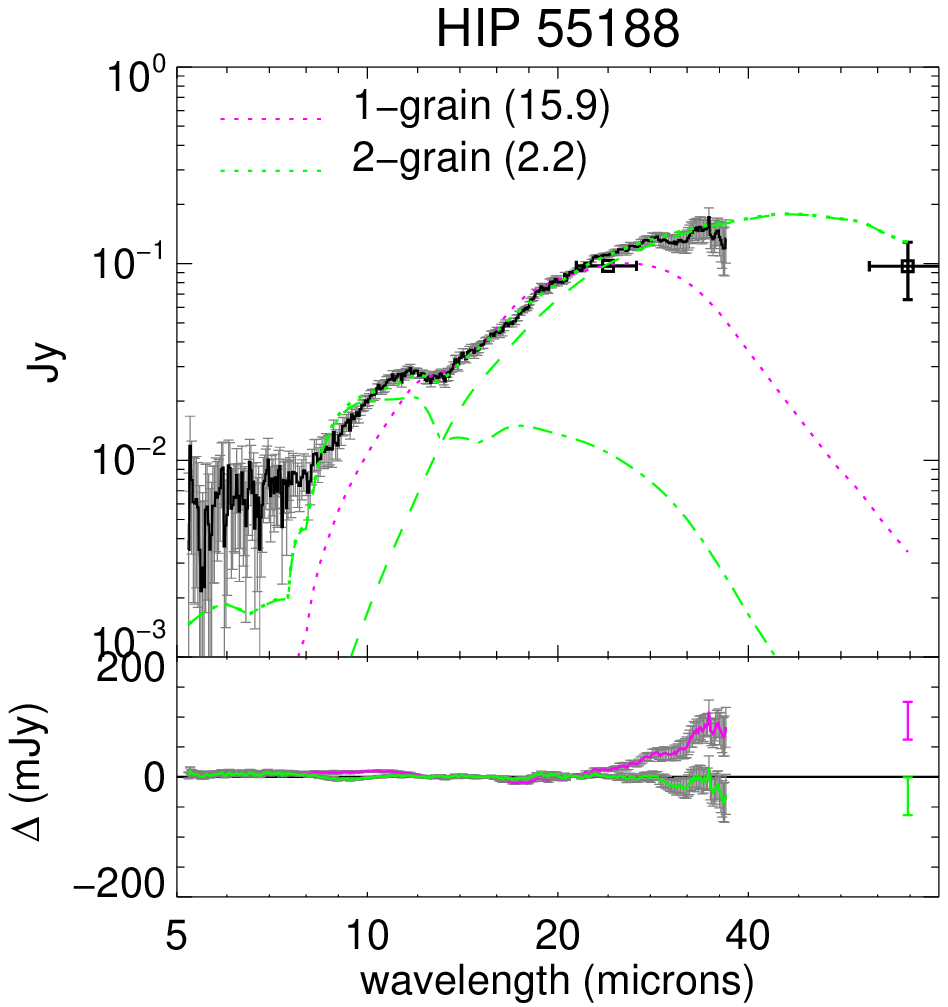} }
\parbox{\stampwidth}{
\includegraphics[width=\stampwidth]{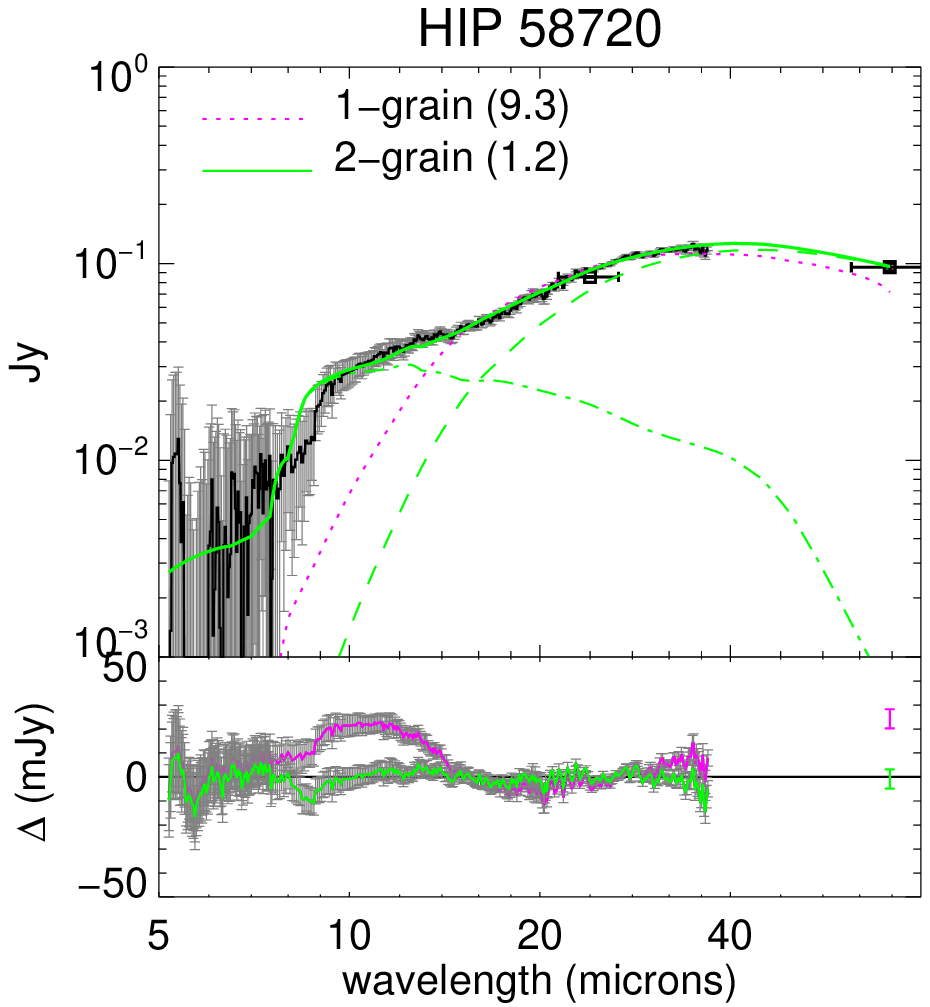} }
\parbox{\stampwidth}{
\includegraphics[width=\stampwidth]{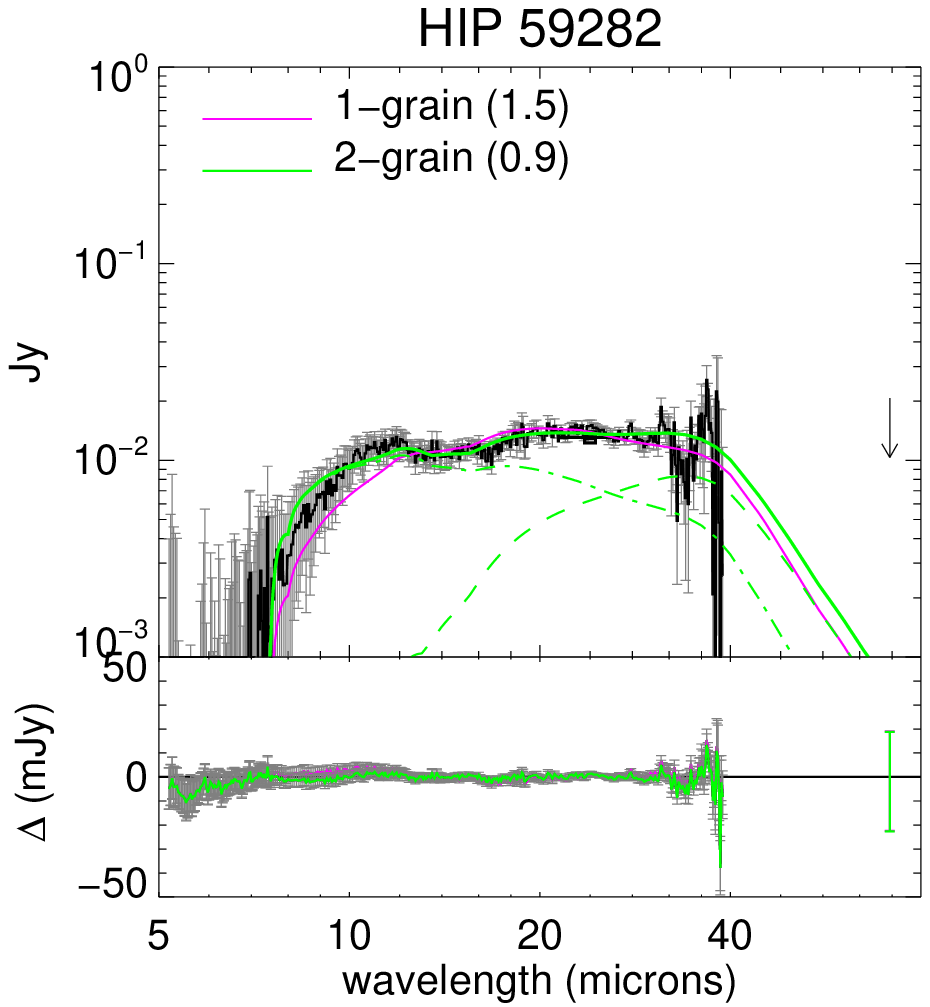} }
\\
\parbox{\stampwidth}{
\includegraphics[width=\stampwidth]{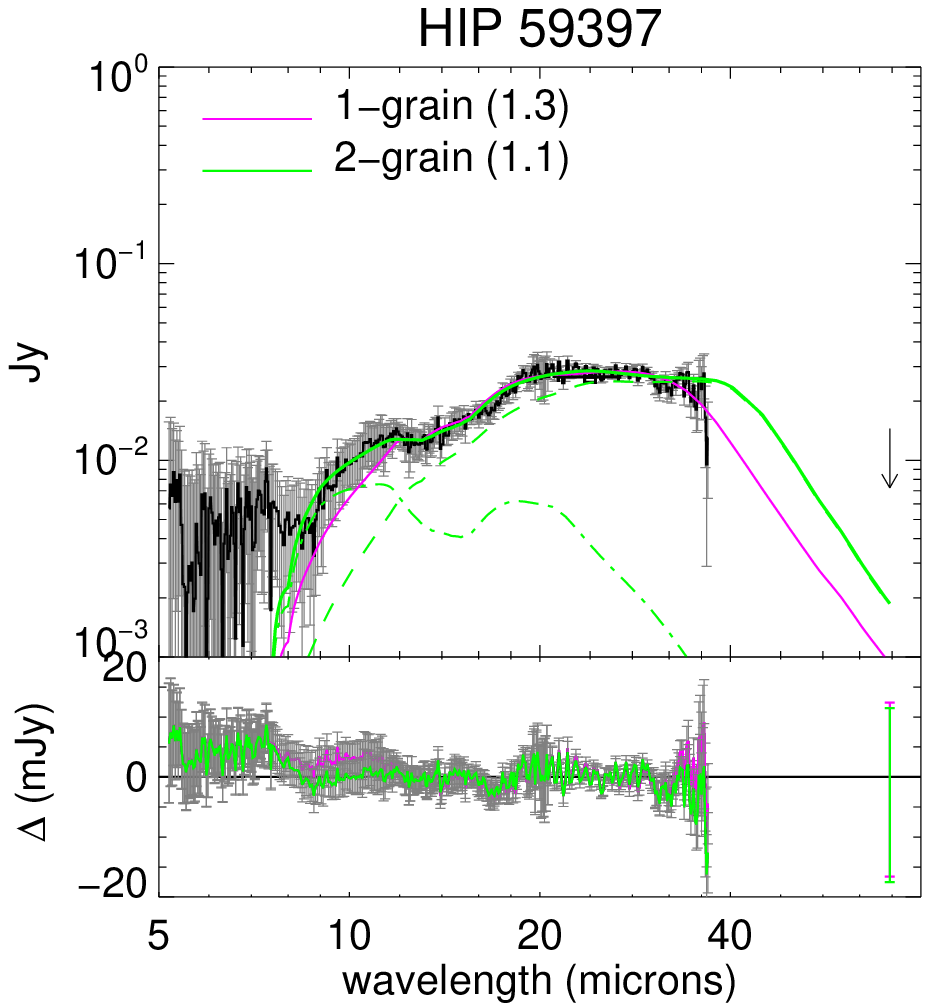} }
\parbox{\stampwidth}{
\includegraphics[width=\stampwidth]{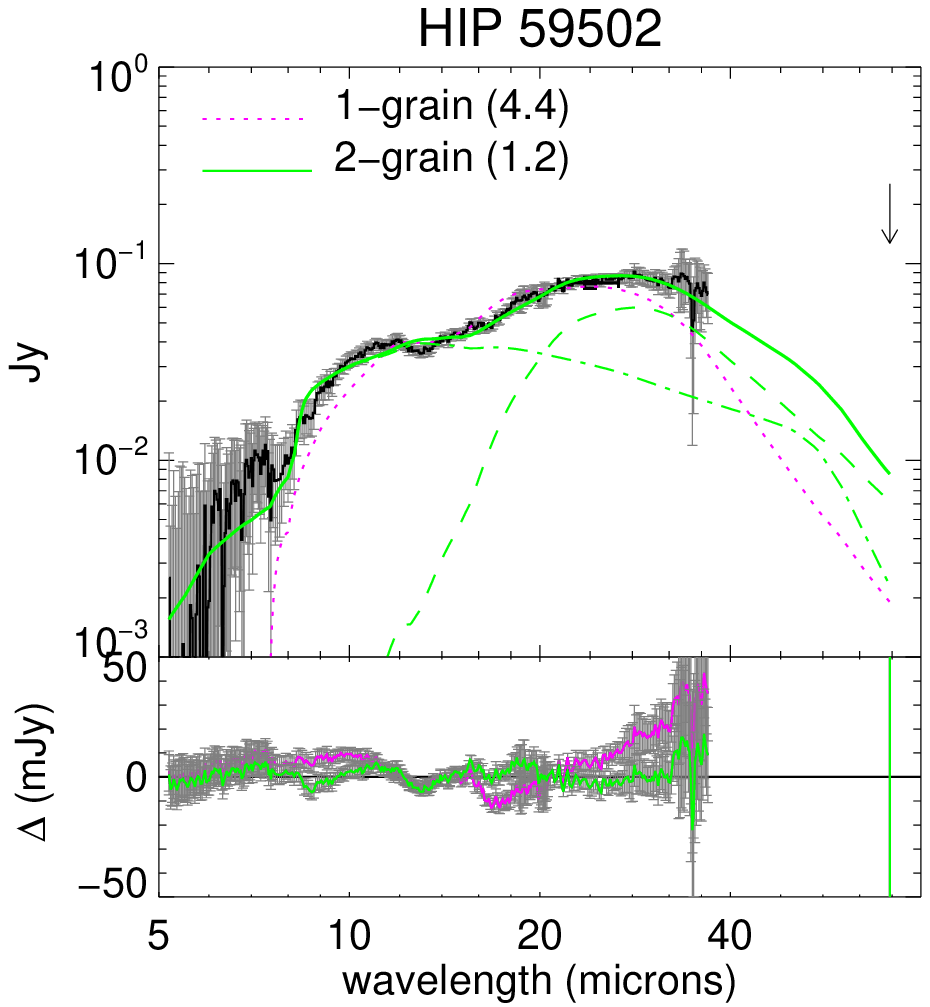} }
\parbox{\stampwidth}{
\includegraphics[width=\stampwidth]{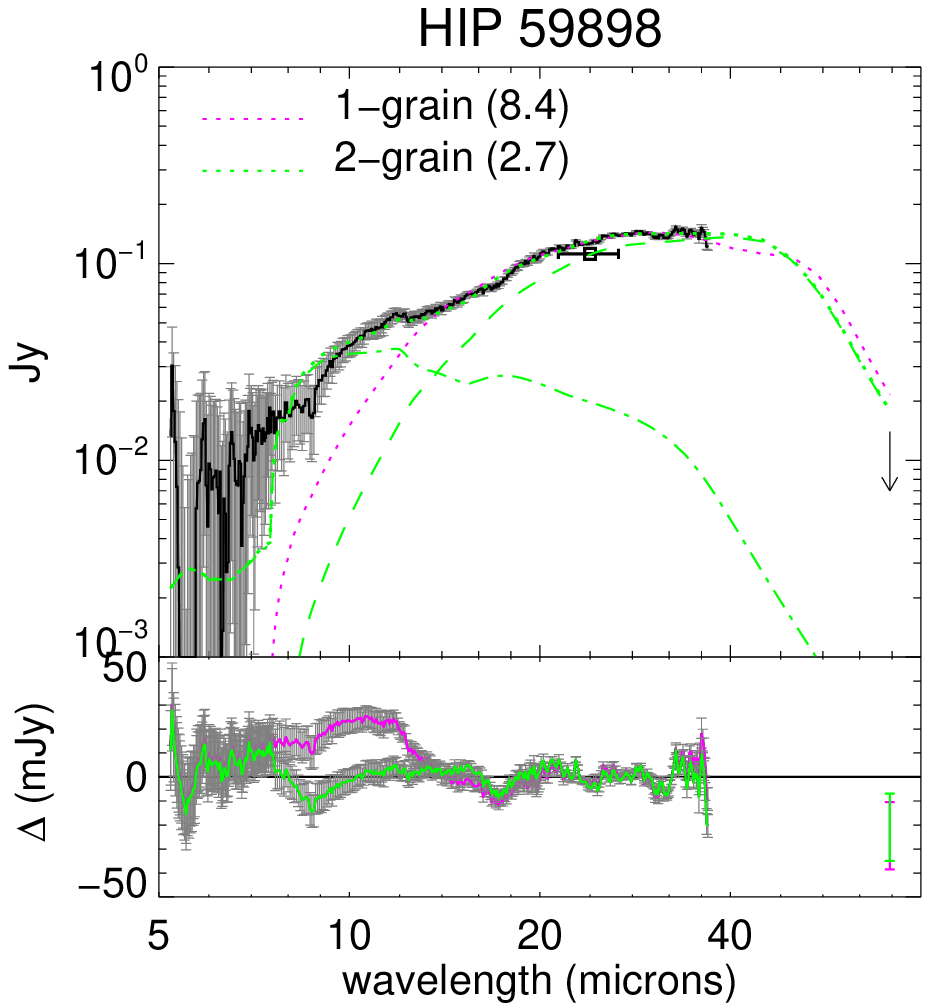} }
\parbox{\stampwidth}{
\includegraphics[width=\stampwidth]{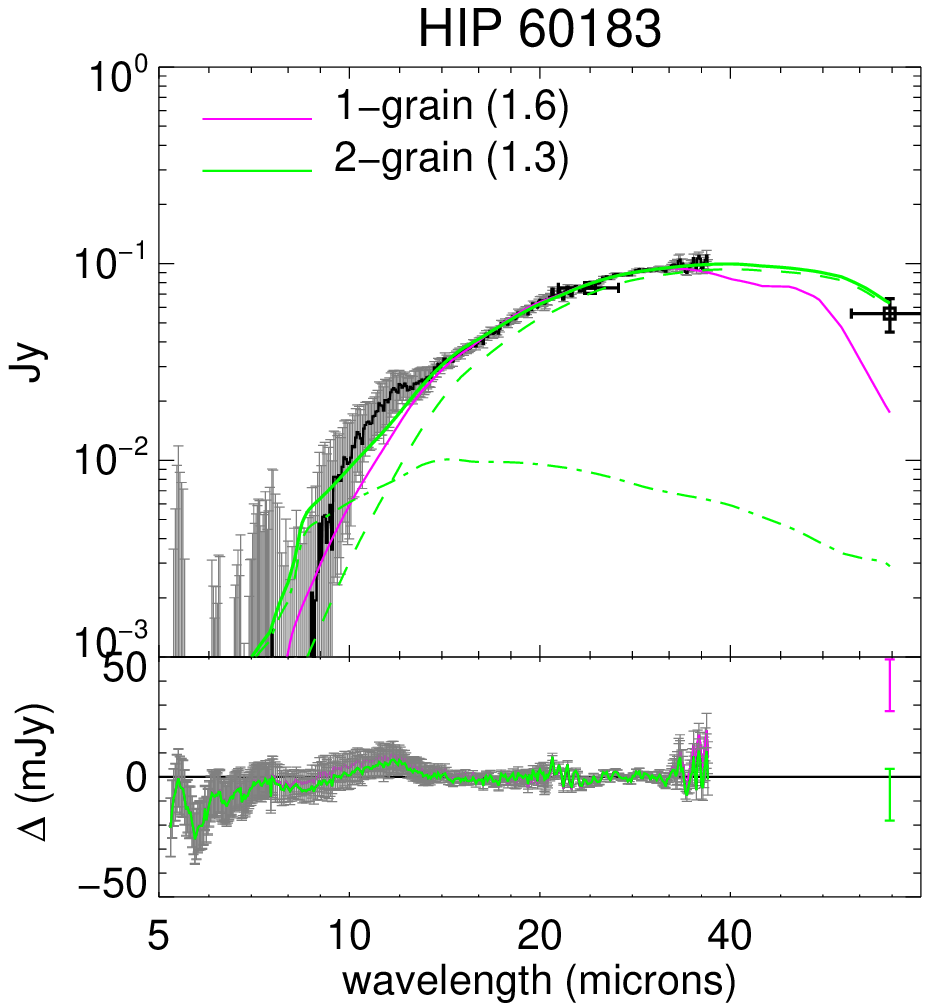} }
\\
\parbox{\stampwidth}{
\includegraphics[width=\stampwidth]{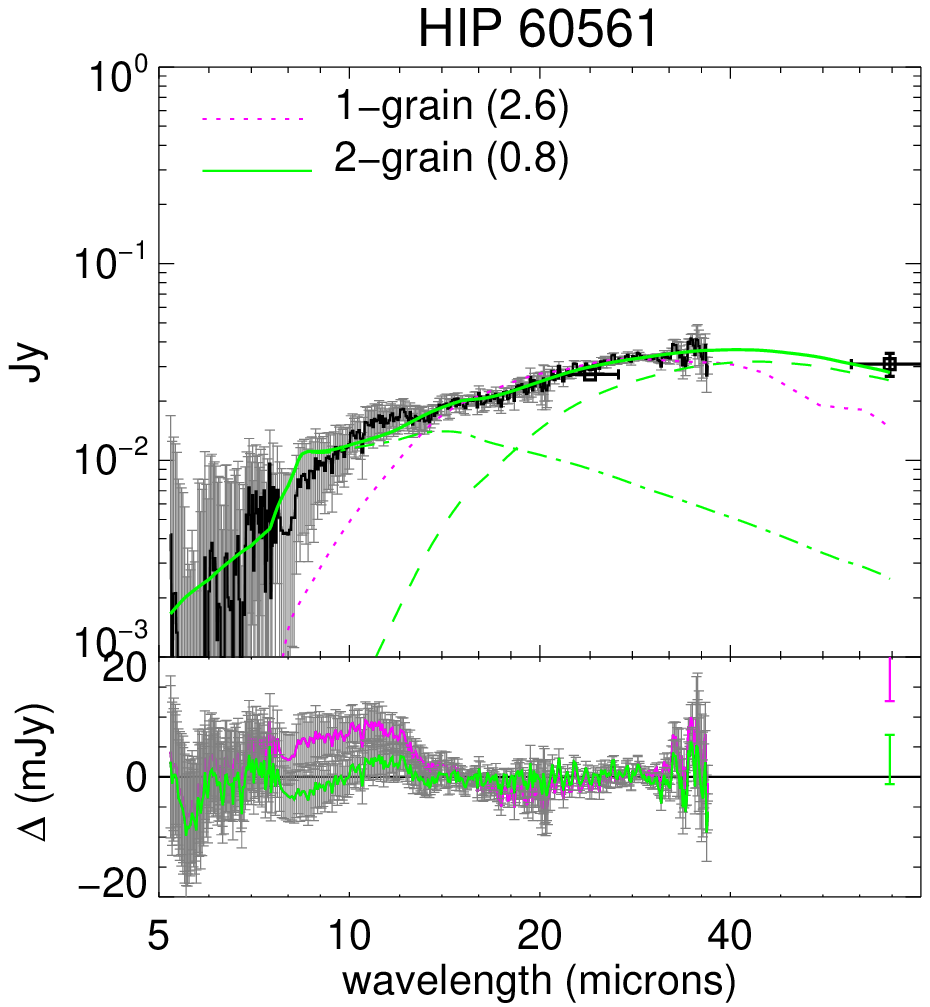} }
\parbox{\stampwidth}{
\includegraphics[width=\stampwidth]{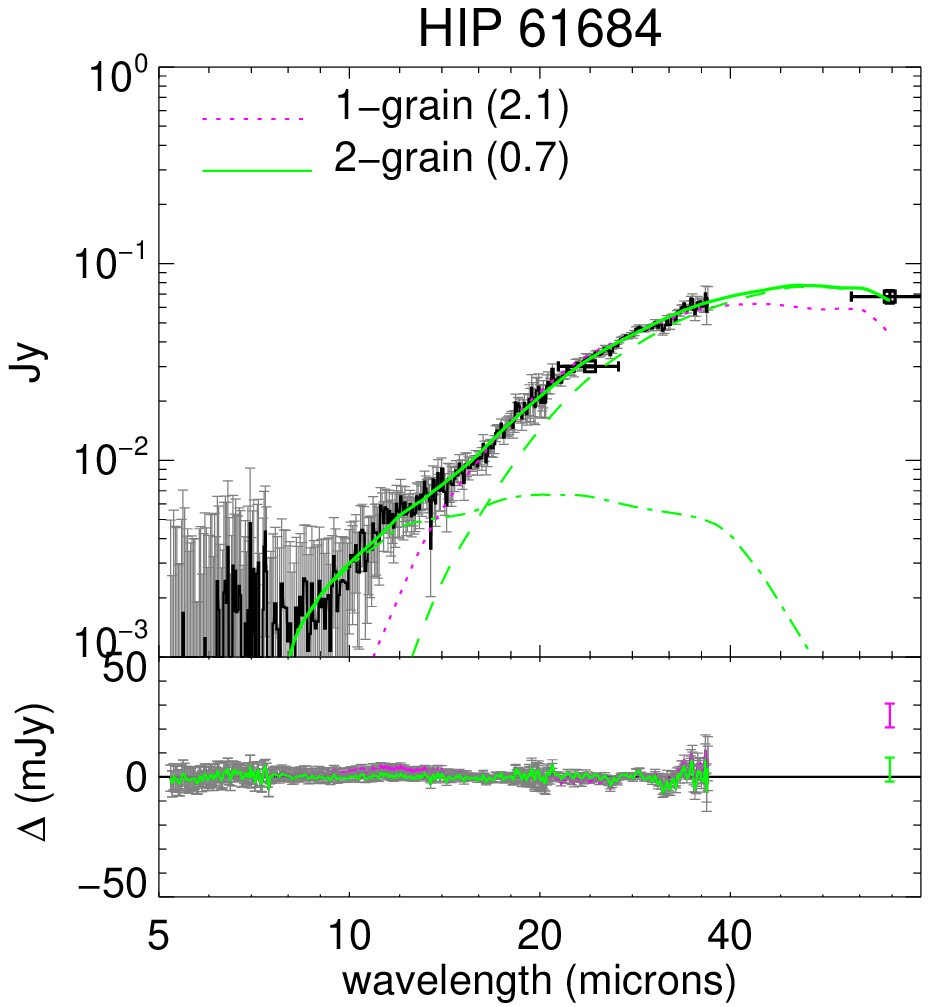} }
\parbox{\stampwidth}{
\includegraphics[width=\stampwidth]{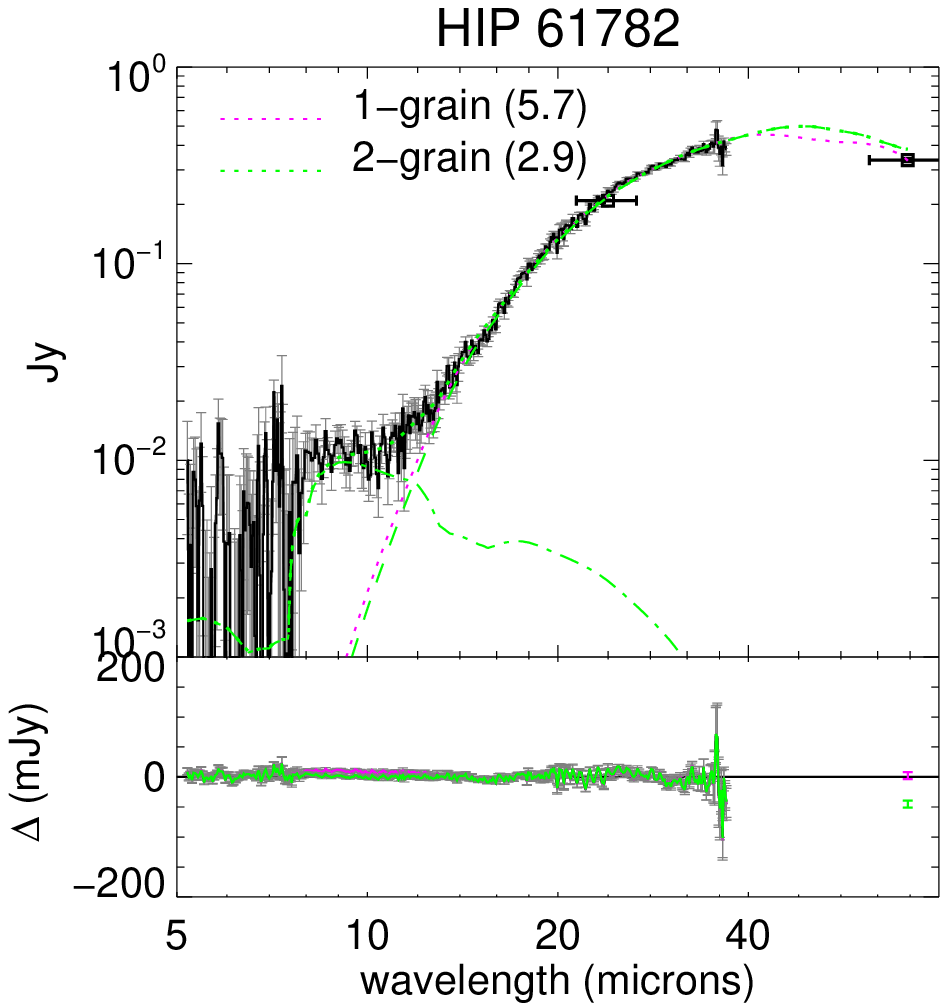} }
\parbox{\stampwidth}{
\includegraphics[width=\stampwidth]{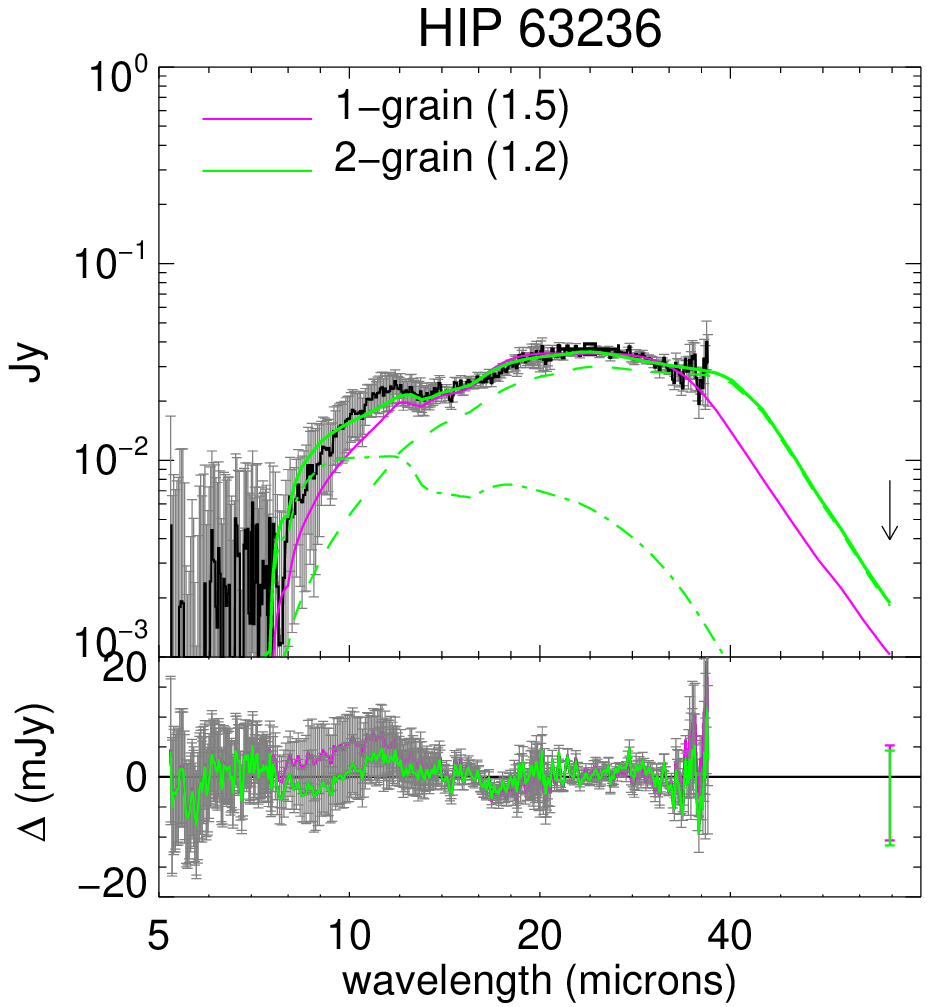} }
\\
\parbox{\stampwidth}{
\includegraphics[width=\stampwidth]{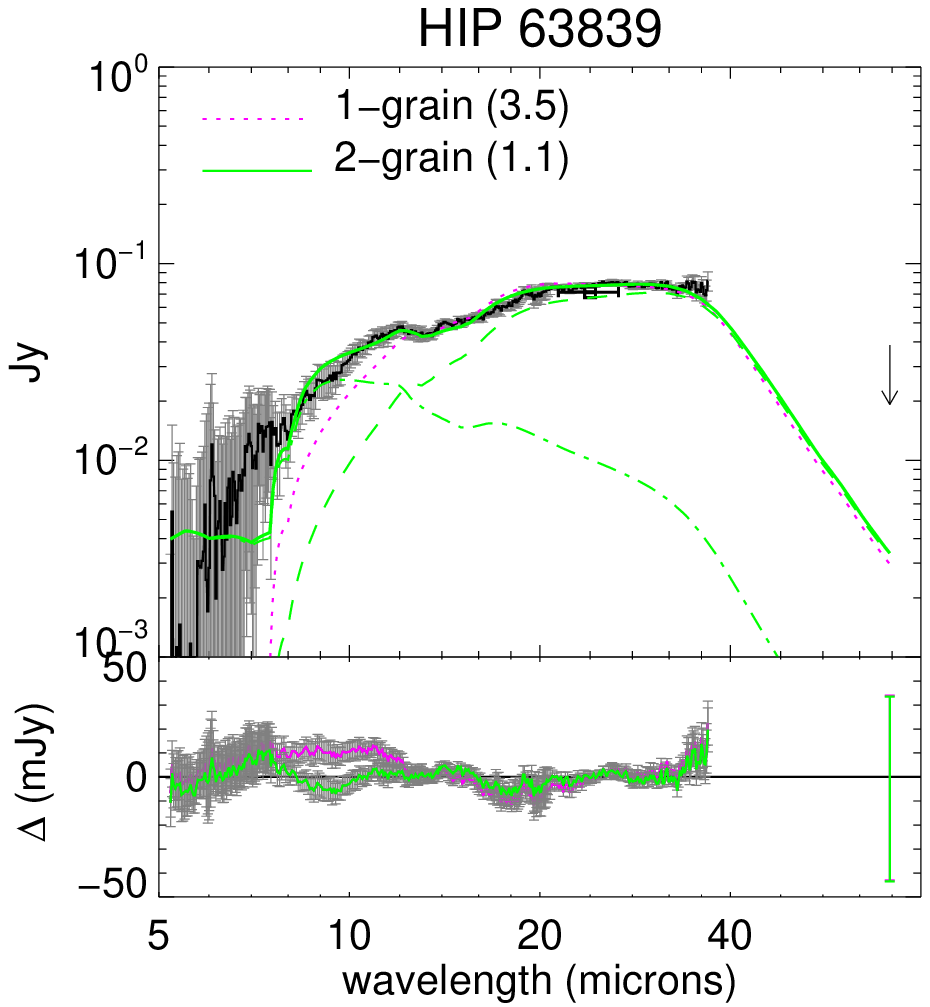} }
\parbox{\stampwidth}{
\includegraphics[width=\stampwidth]{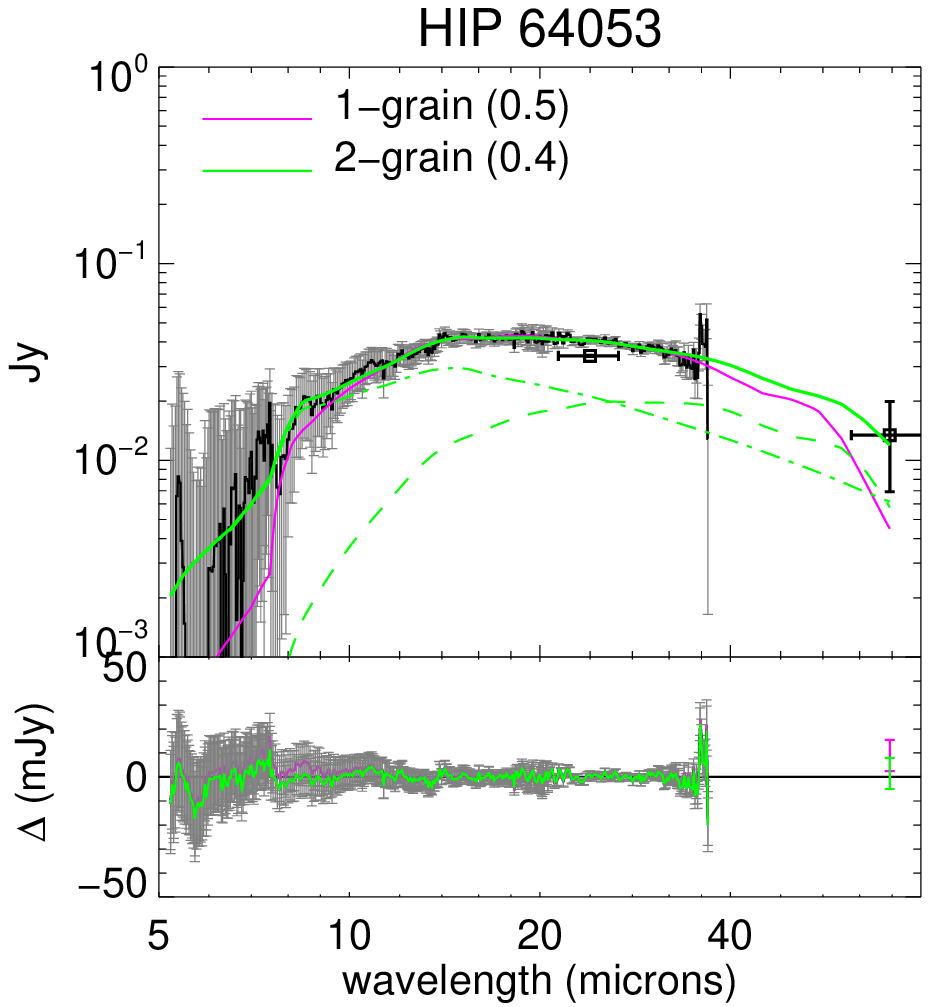} }
\parbox{\stampwidth}{
\includegraphics[width=\stampwidth]{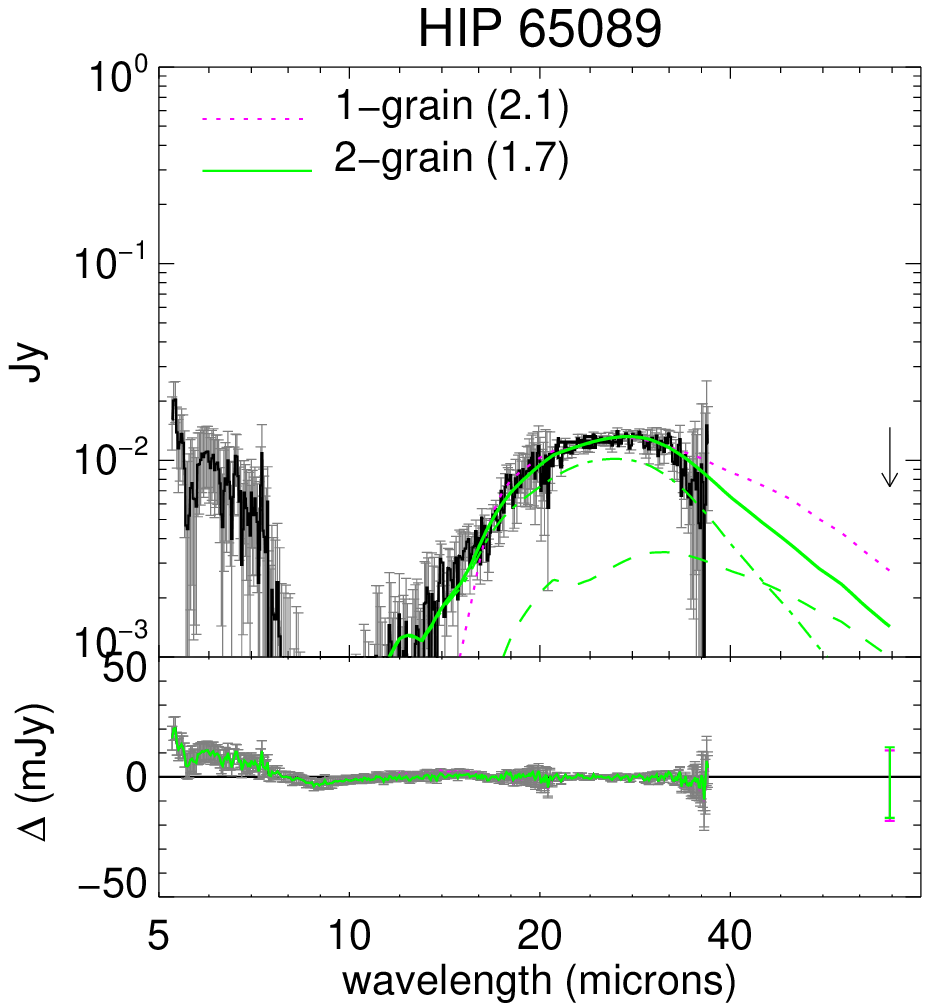} }
\parbox{\stampwidth}{
\includegraphics[width=\stampwidth]{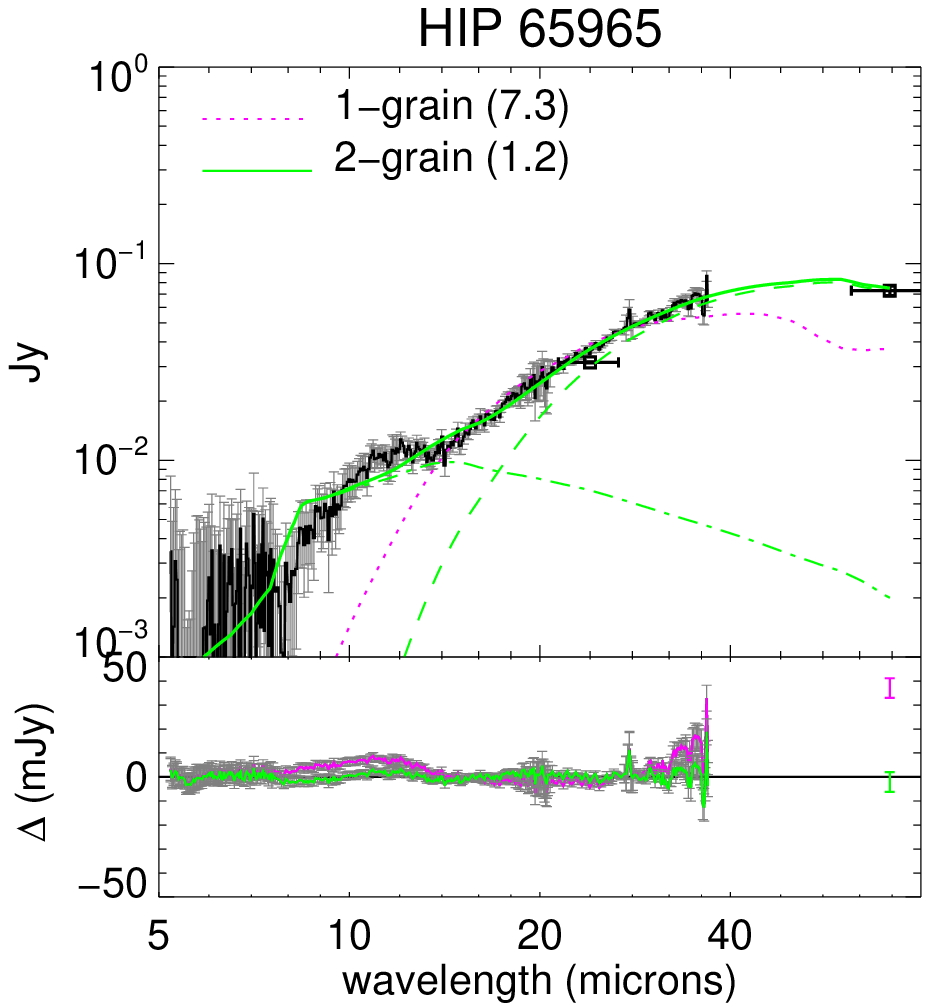} }
\\
\parbox{\stampwidth}{
\includegraphics[width=\stampwidth]{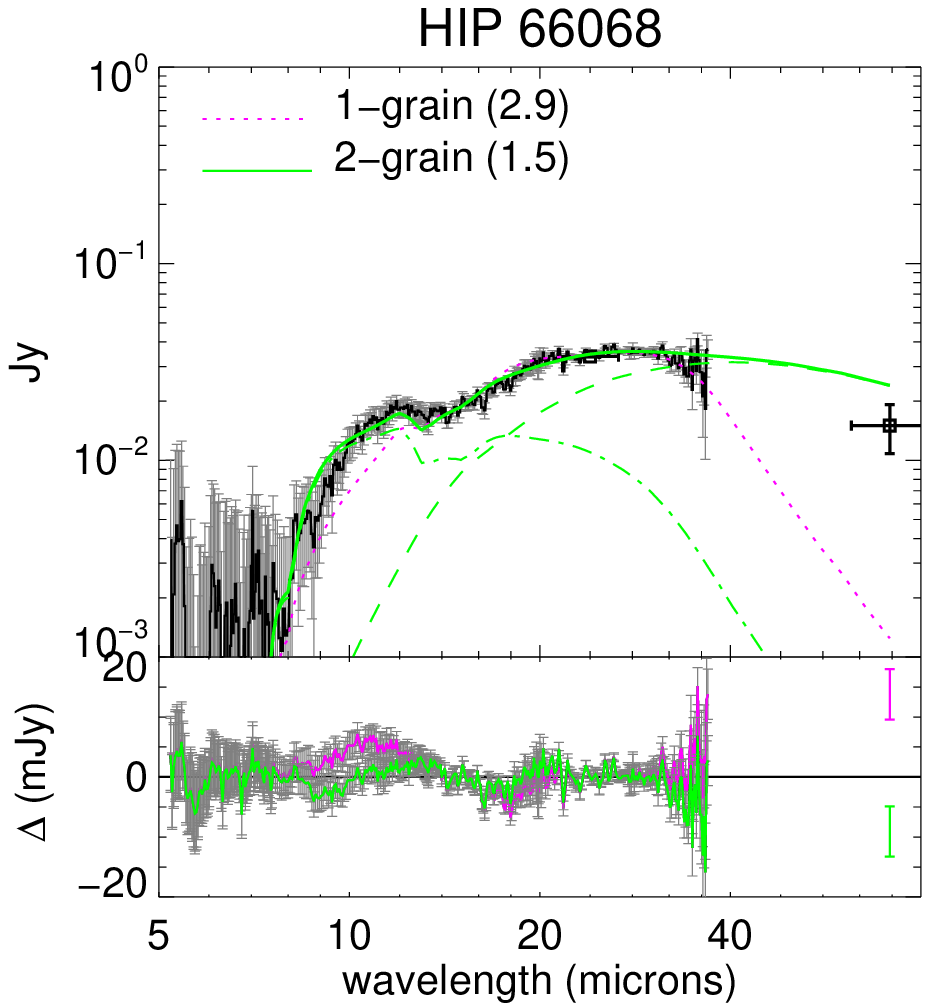} }
\parbox{\stampwidth}{
\includegraphics[width=\stampwidth]{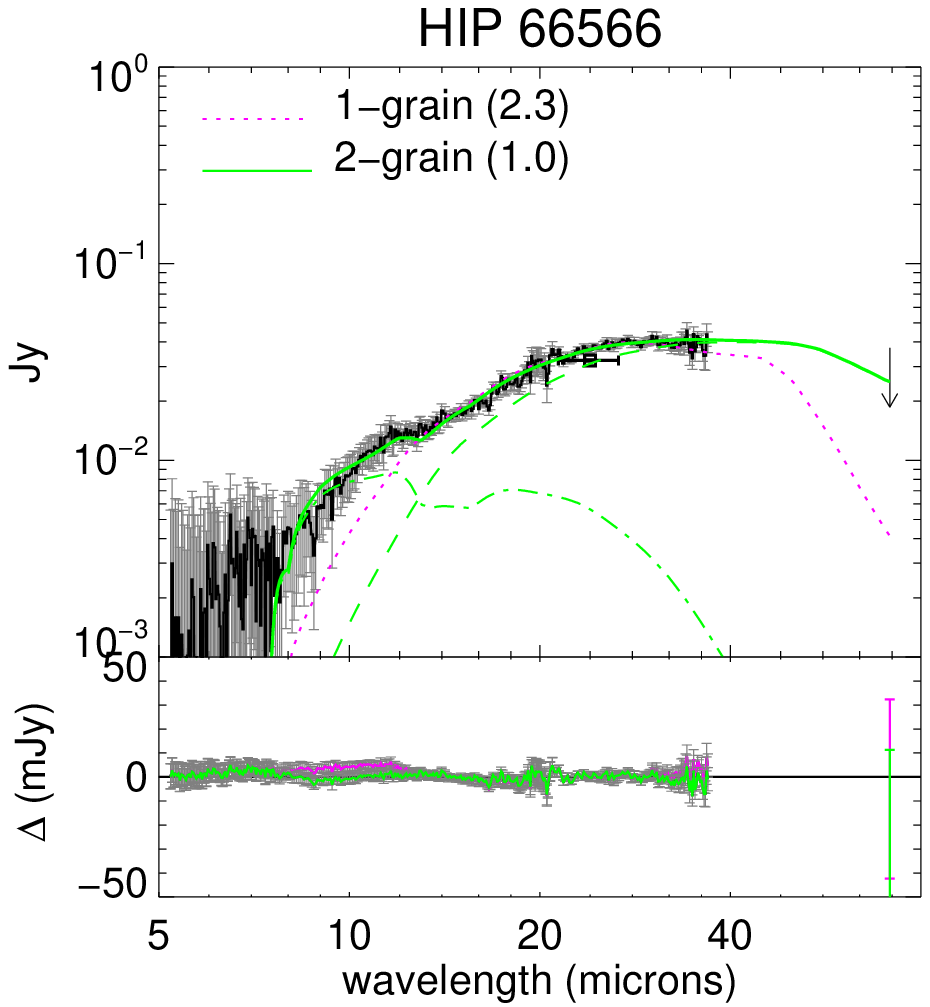} }
\parbox{\stampwidth}{
\includegraphics[width=\stampwidth]{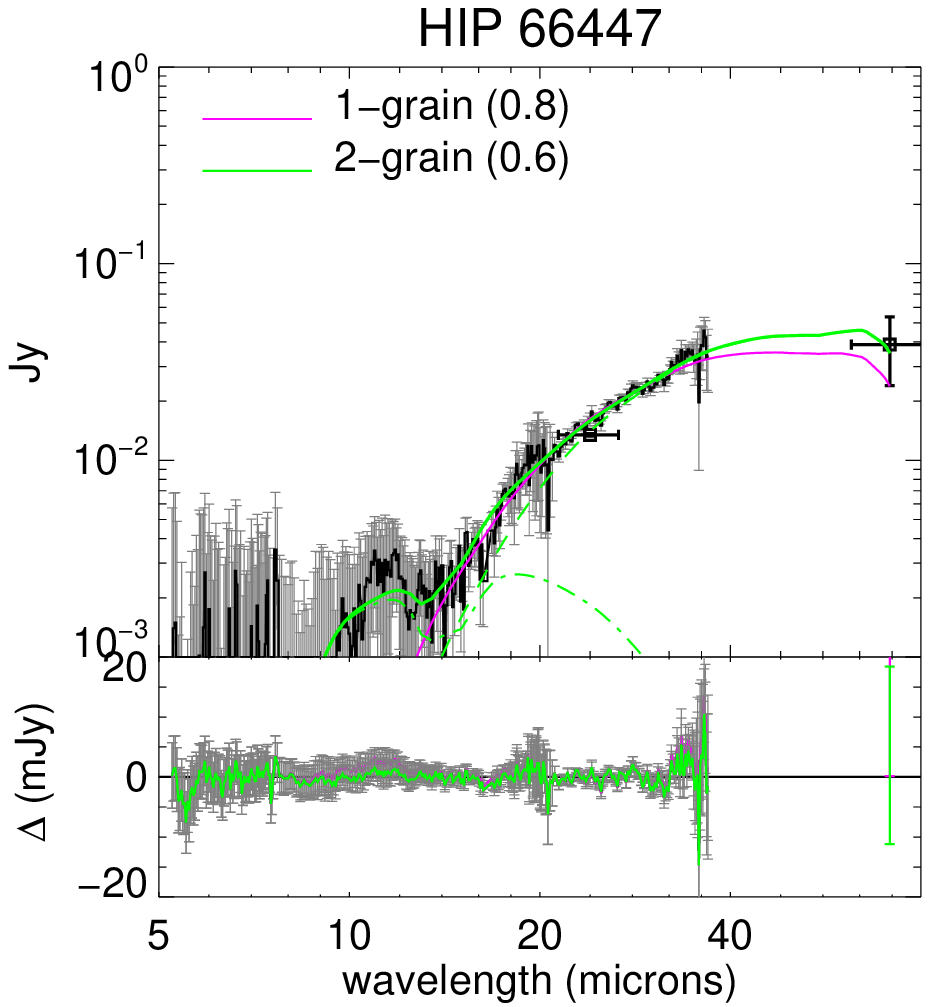} }
\parbox{\stampwidth}{
\includegraphics[width=\stampwidth]{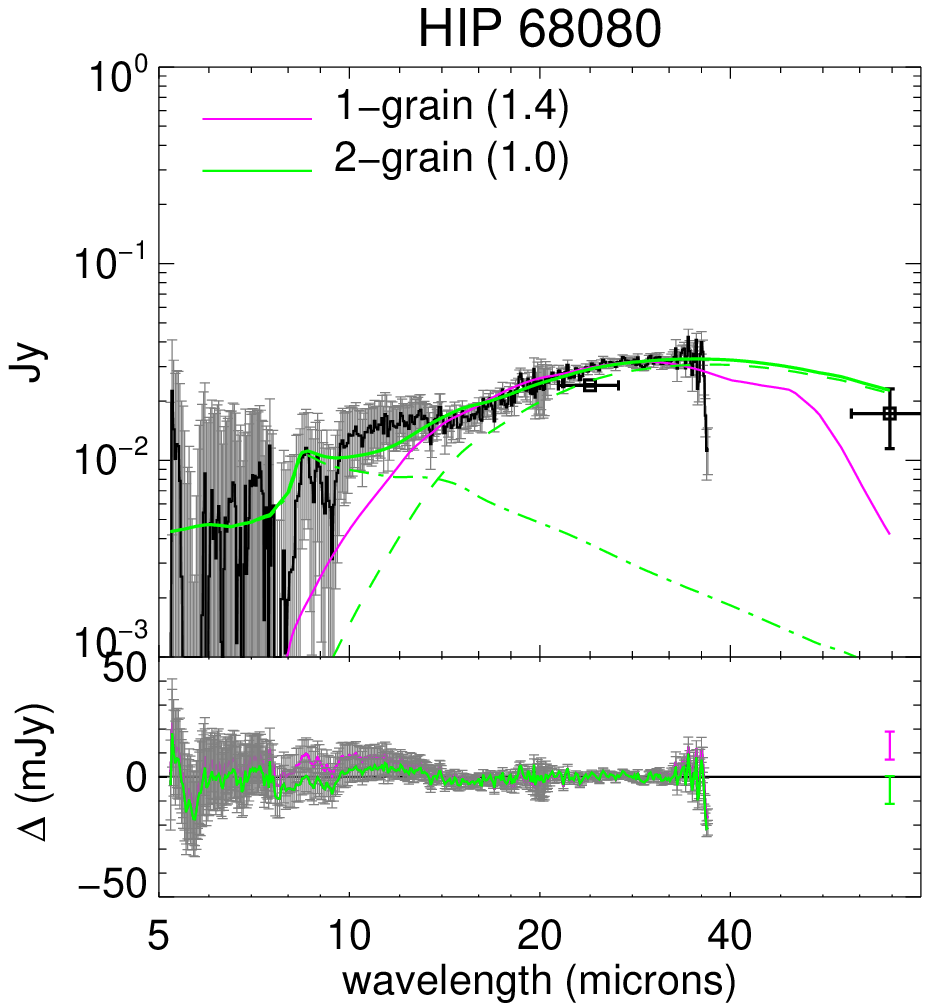} }
\\
\vspace{-2ex}
\caption{ \label{fitfig0}
\small \protect\input{fitcaption}
}
\end{figure}
\addtocounter{figure}{-1}
\stepcounter{subfig}
\begin{figure}
\parbox{\stampwidth}{
\includegraphics[width=\stampwidth]{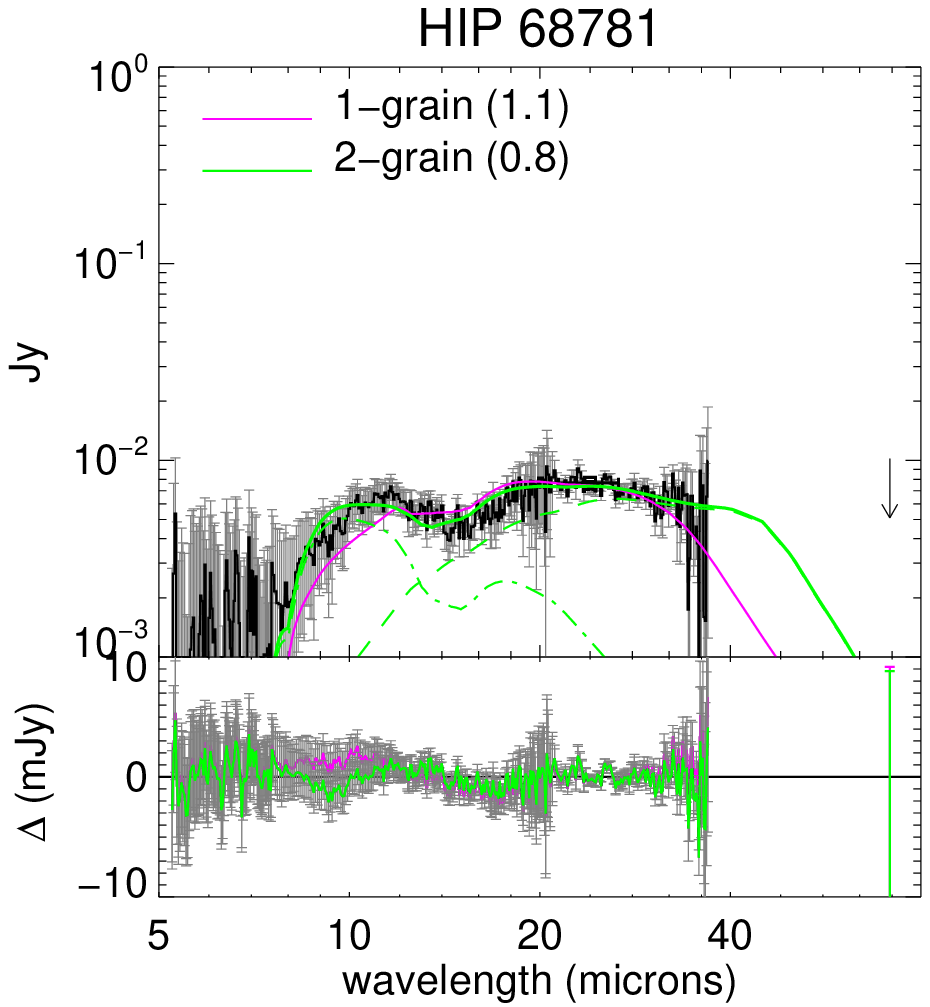} }
\parbox{\stampwidth}{
\includegraphics[width=\stampwidth]{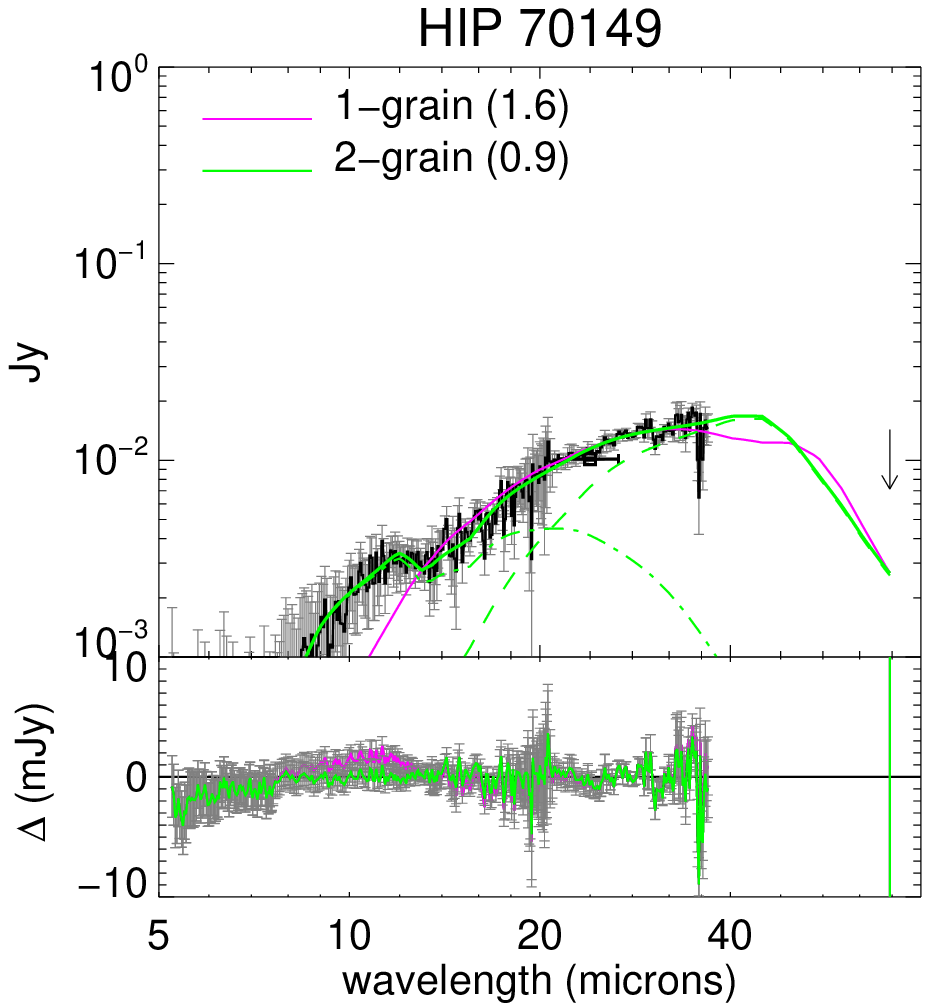} }
\parbox{\stampwidth}{
\includegraphics[width=\stampwidth]{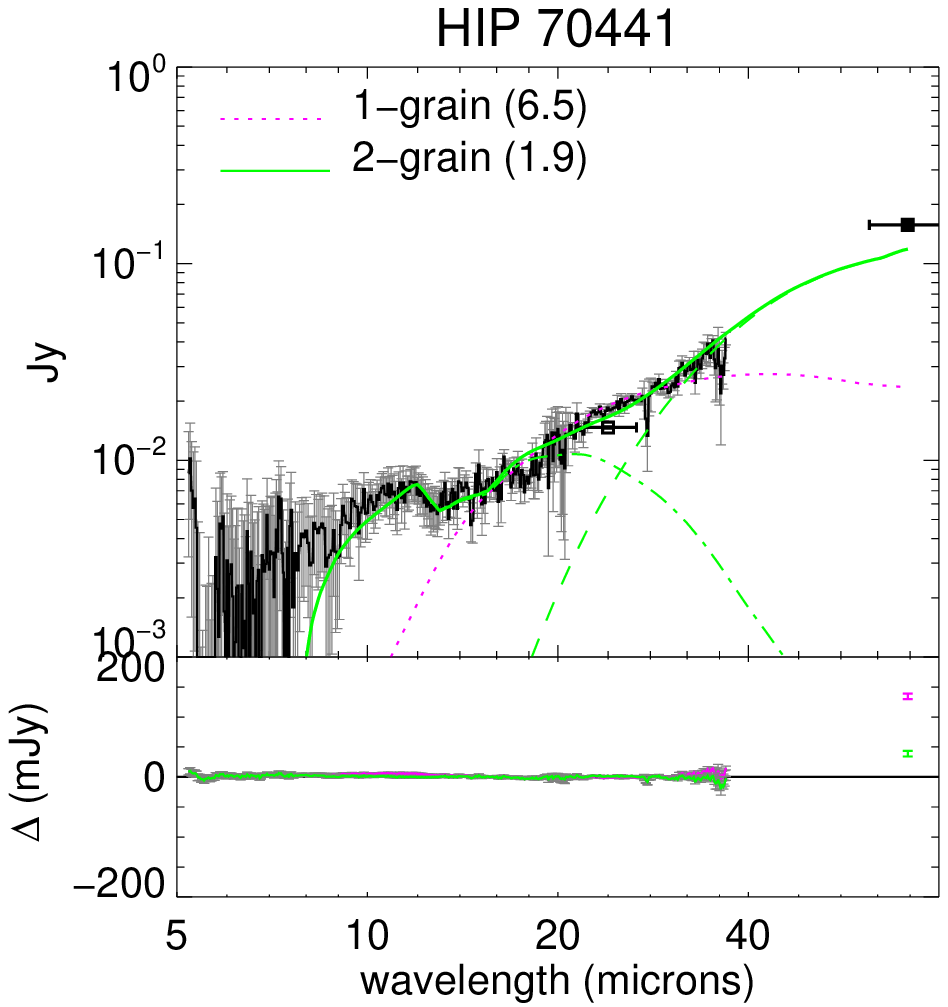} }
\parbox{\stampwidth}{
\includegraphics[width=\stampwidth]{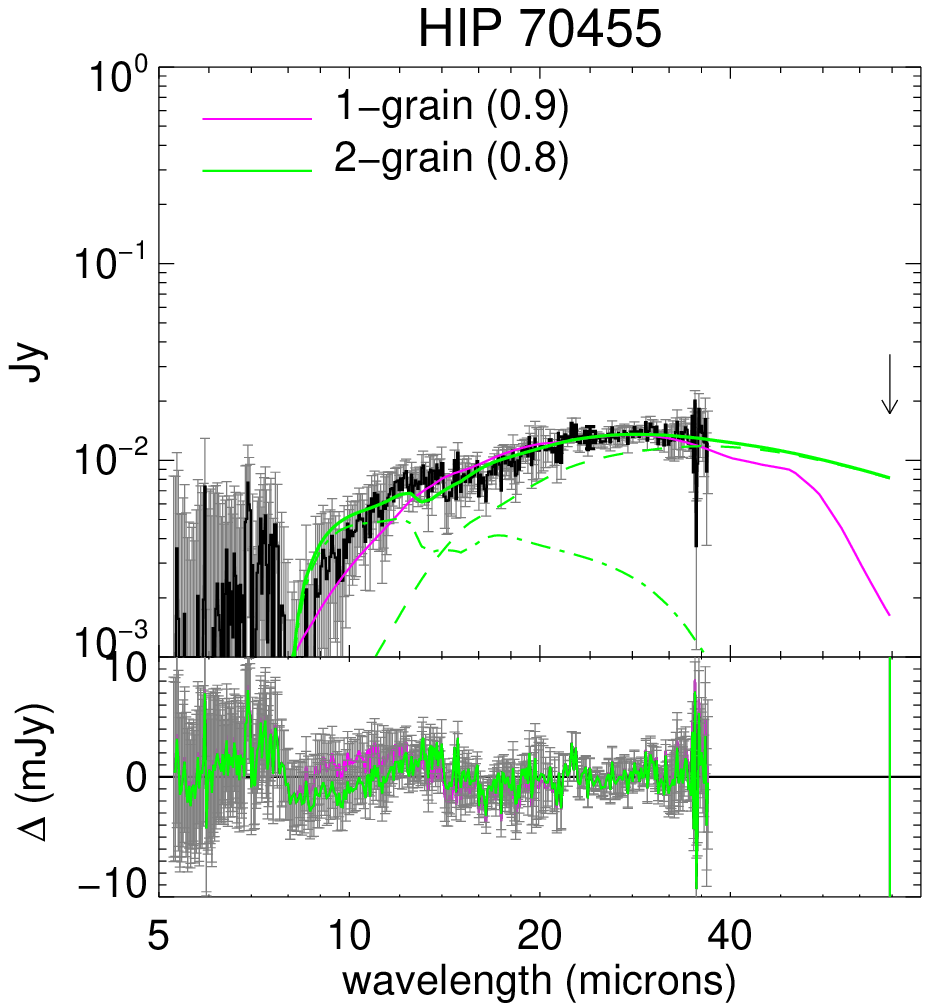} }
\\
\parbox{\stampwidth}{
\includegraphics[width=\stampwidth]{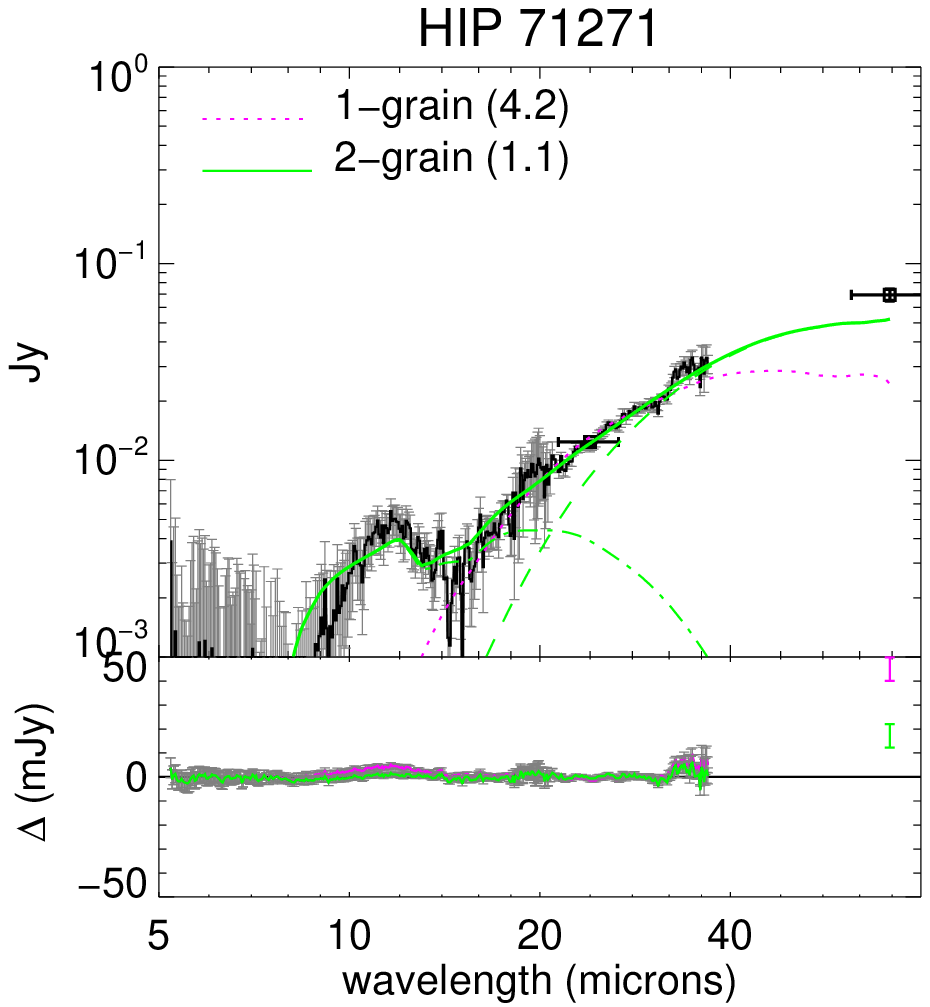} }
\parbox{\stampwidth}{
\includegraphics[width=\stampwidth]{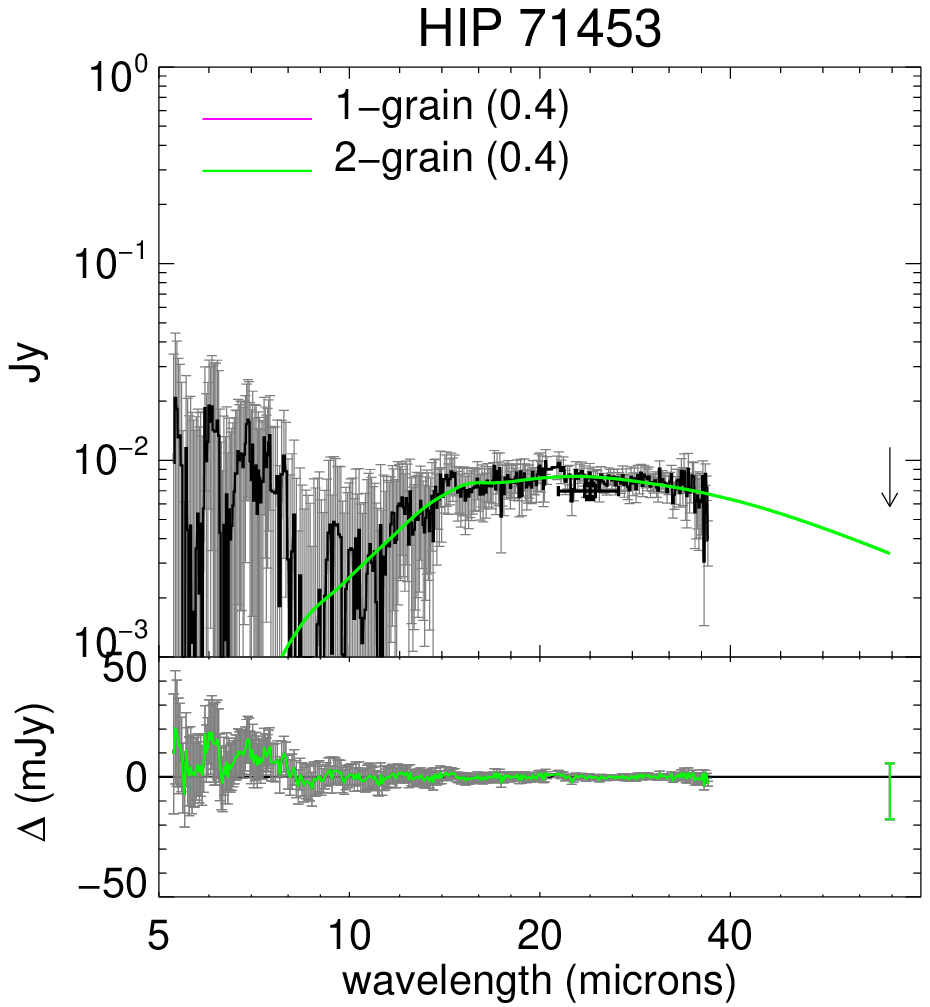} }
\parbox{\stampwidth}{
\includegraphics[width=\stampwidth]{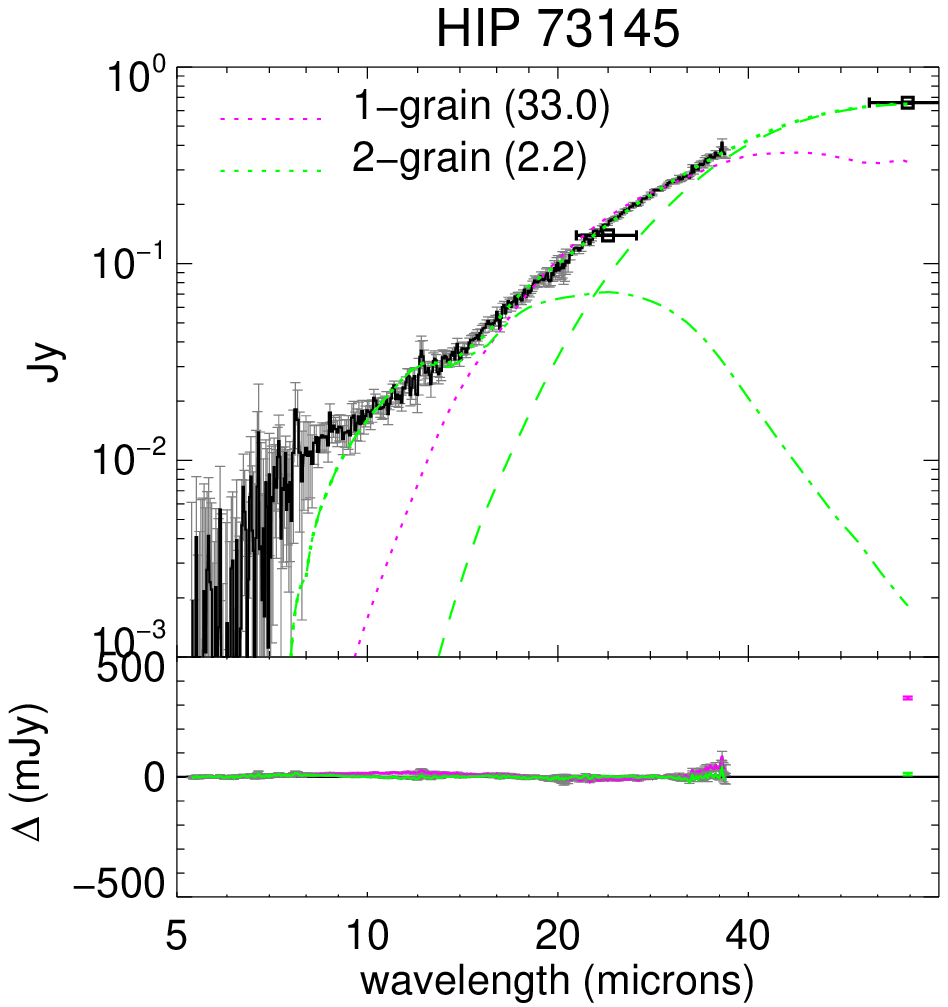} }
\parbox{\stampwidth}{
\includegraphics[width=\stampwidth]{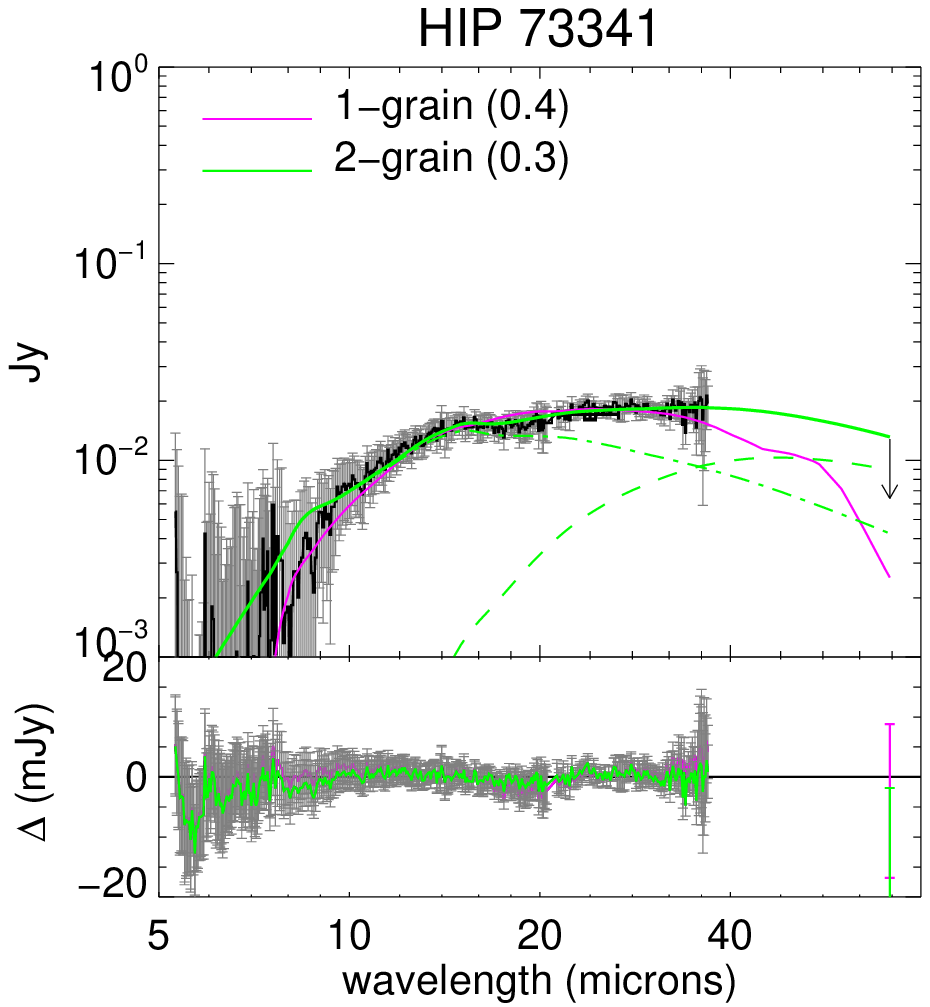} }
\\
\parbox{\stampwidth}{
\includegraphics[width=\stampwidth]{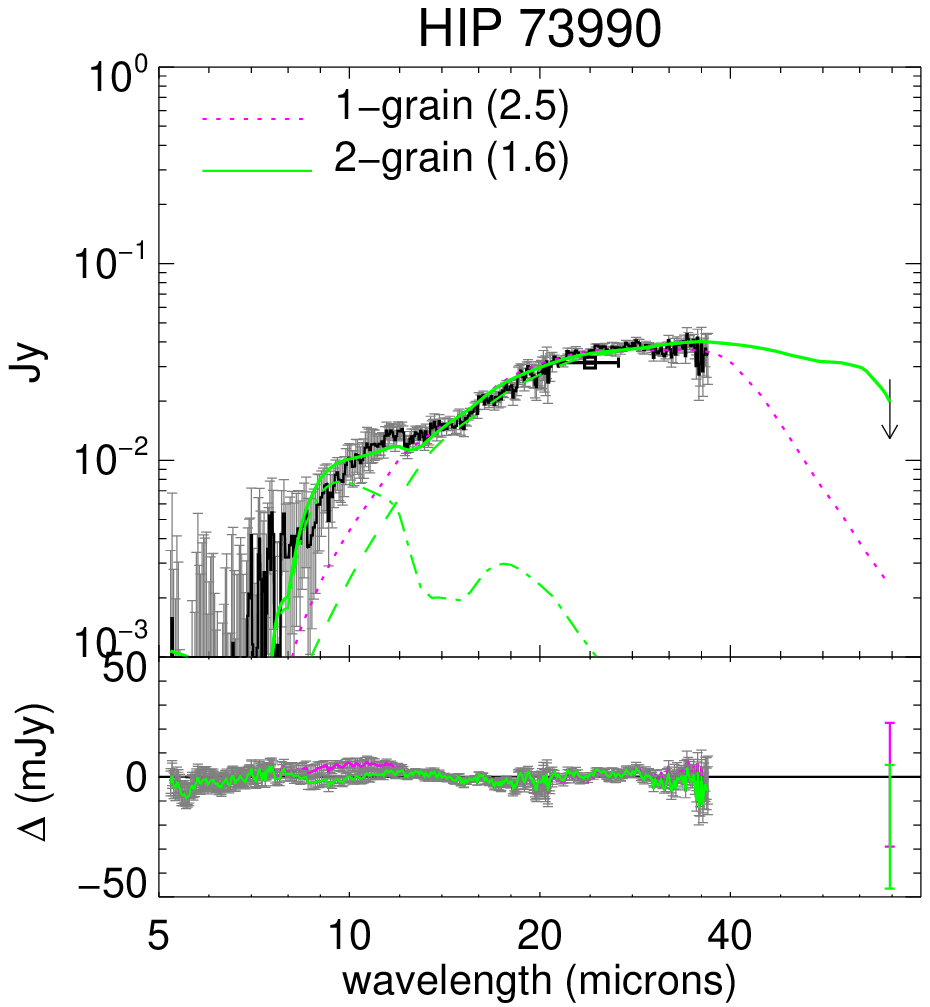} }
\parbox{\stampwidth}{
\includegraphics[width=\stampwidth]{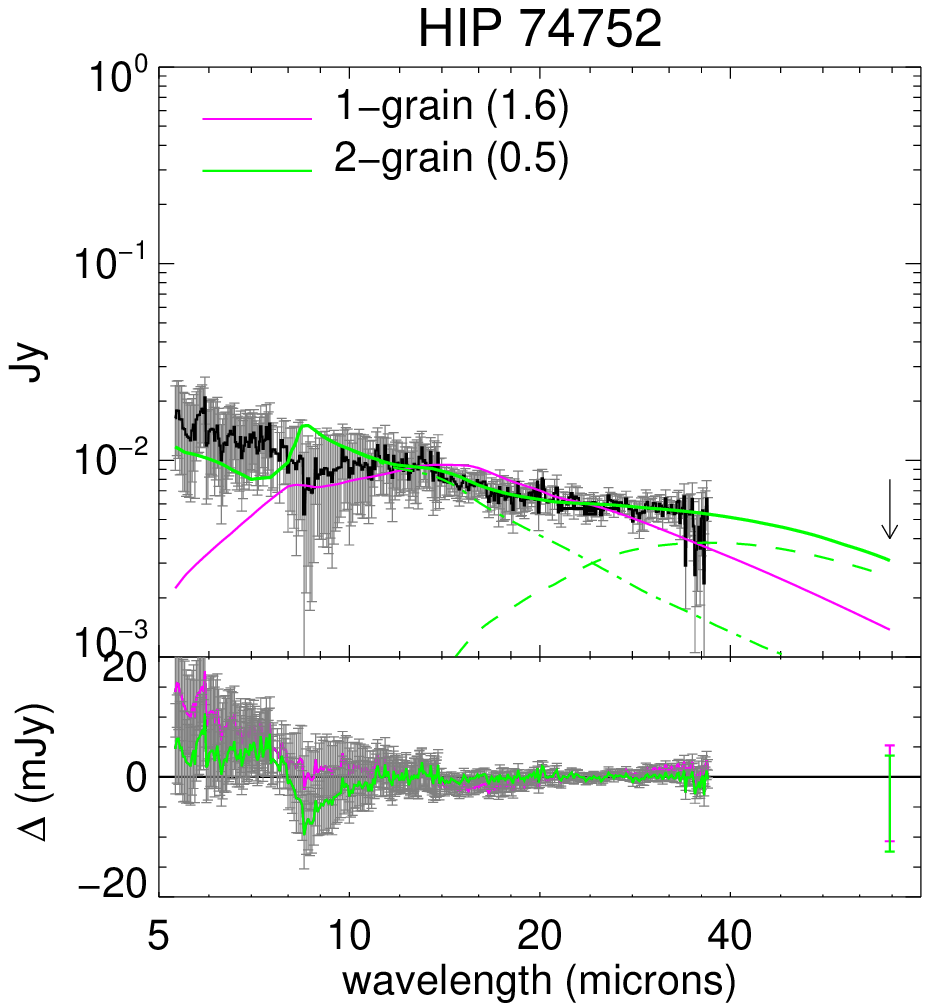} }
\parbox{\stampwidth}{
\includegraphics[width=\stampwidth]{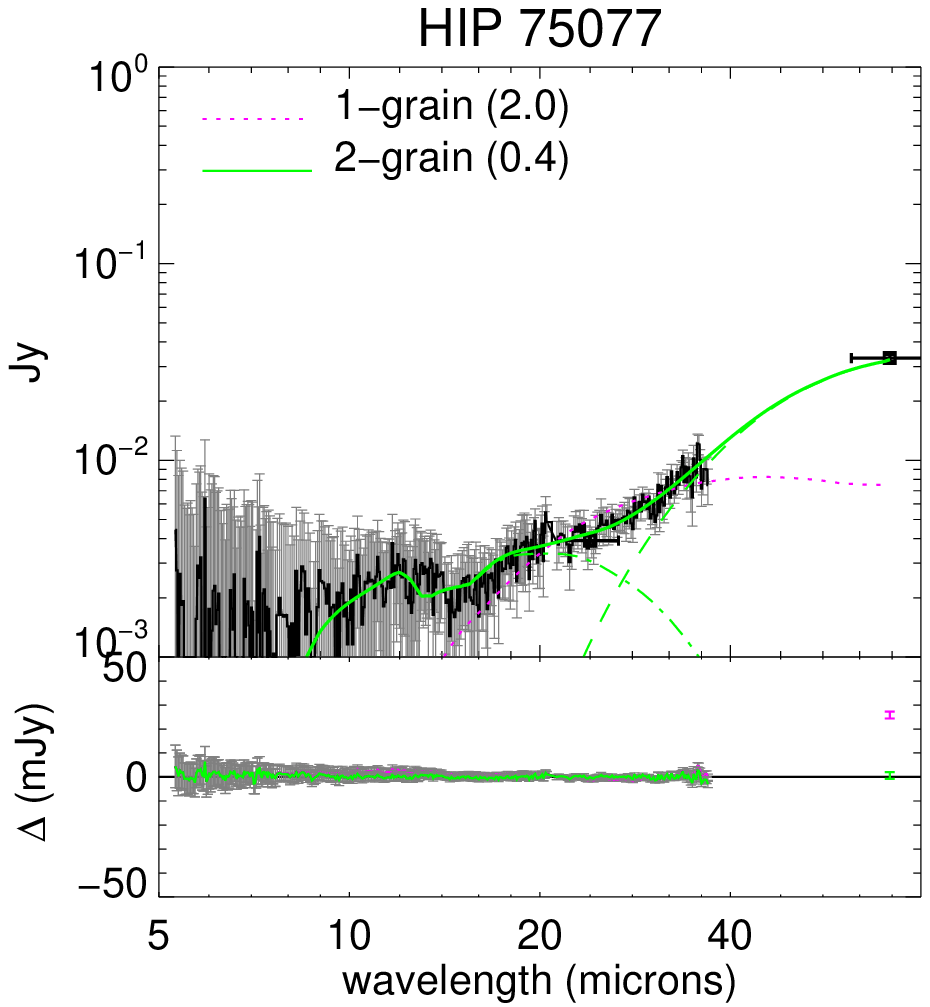} }
\parbox{\stampwidth}{
\includegraphics[width=\stampwidth]{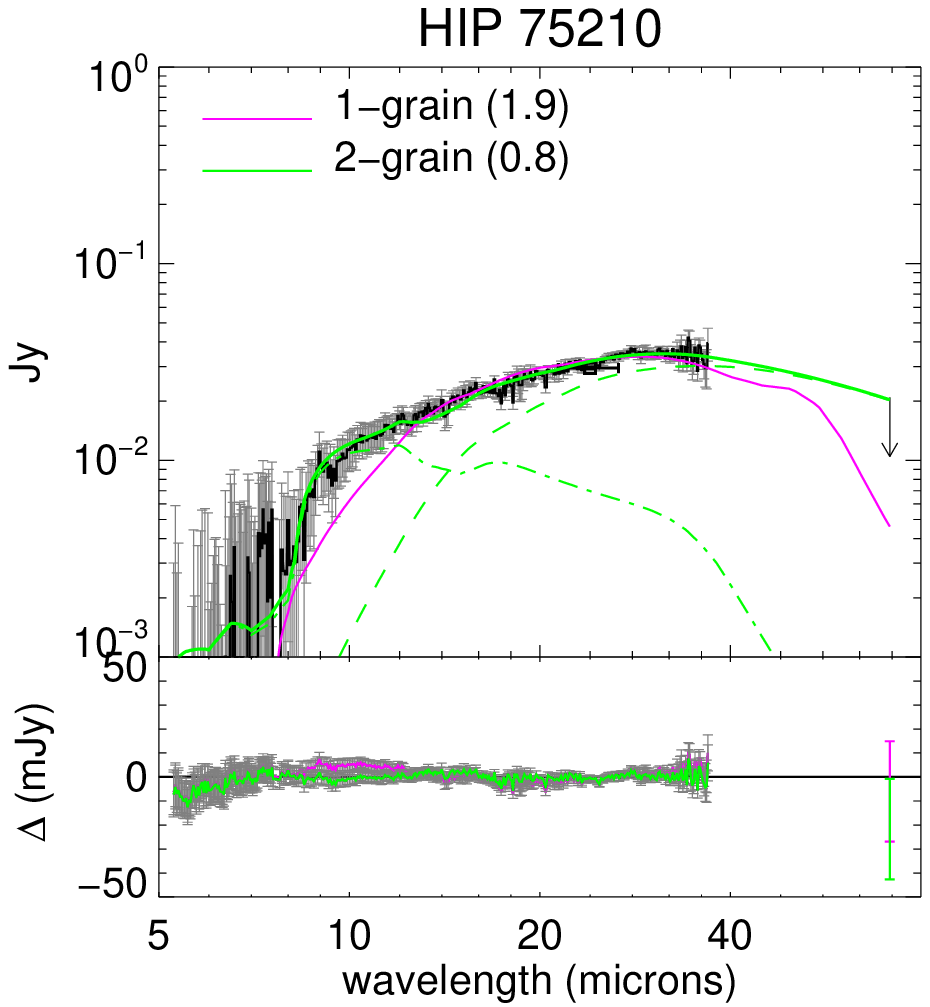} }
\\
\parbox{\stampwidth}{
\includegraphics[width=\stampwidth]{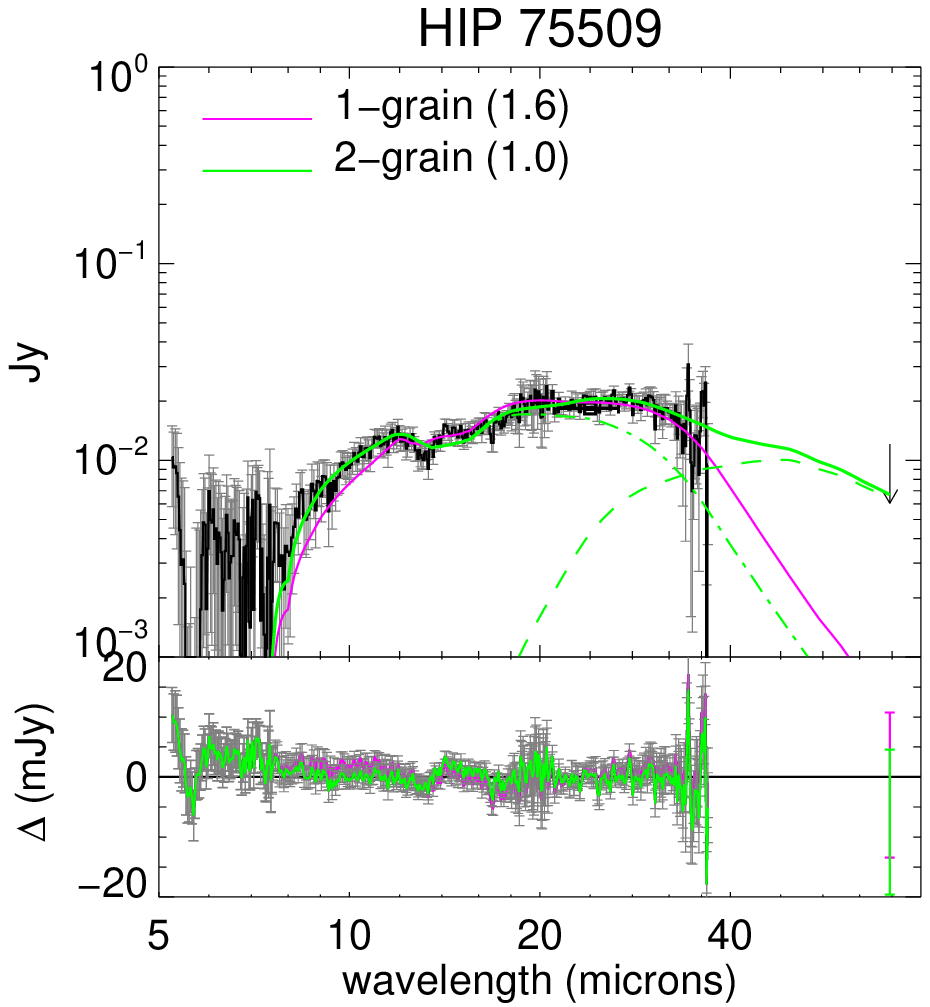} }
\parbox{\stampwidth}{
\includegraphics[width=\stampwidth]{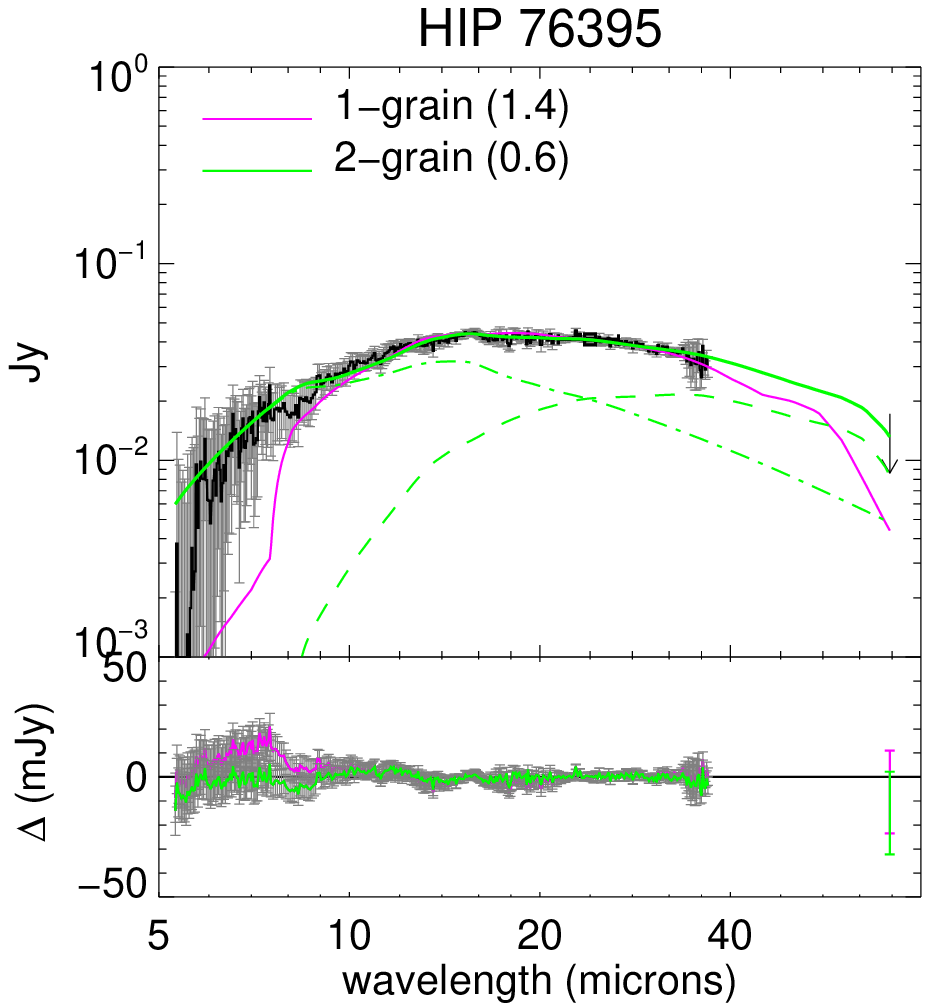} }
\parbox{\stampwidth}{
\includegraphics[width=\stampwidth]{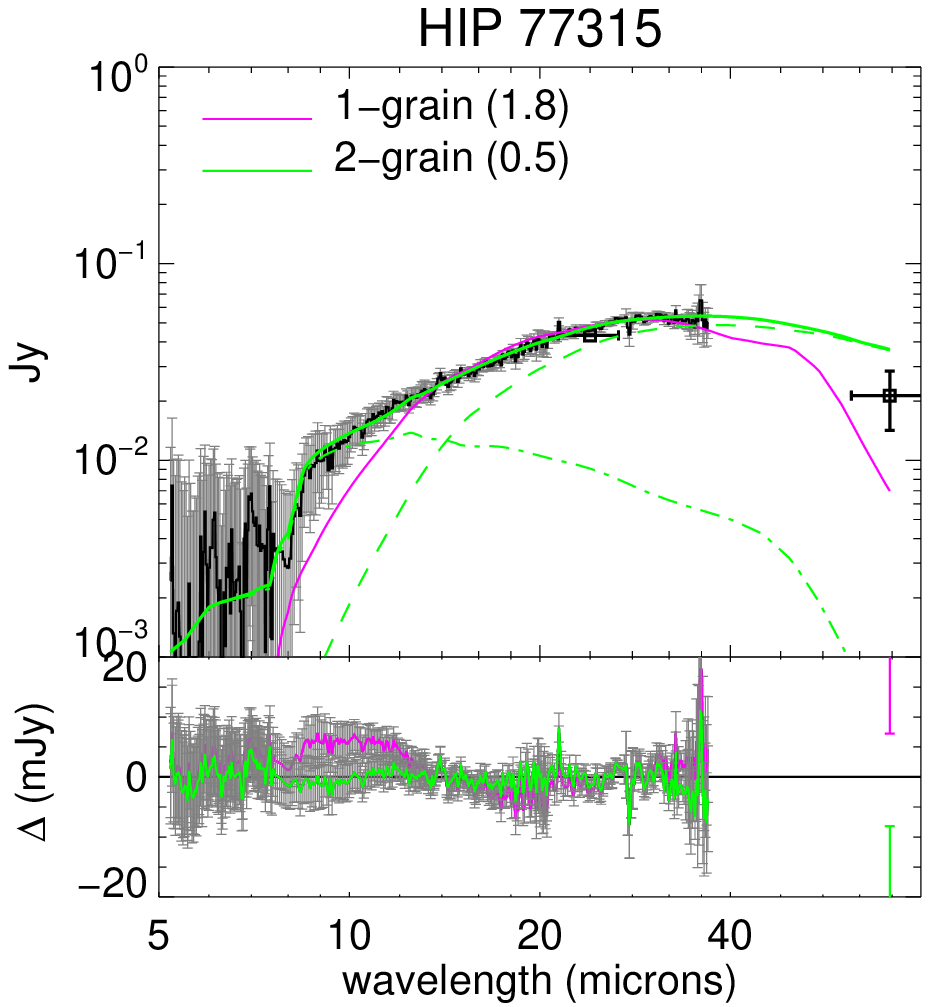} }
\parbox{\stampwidth}{
\includegraphics[width=\stampwidth]{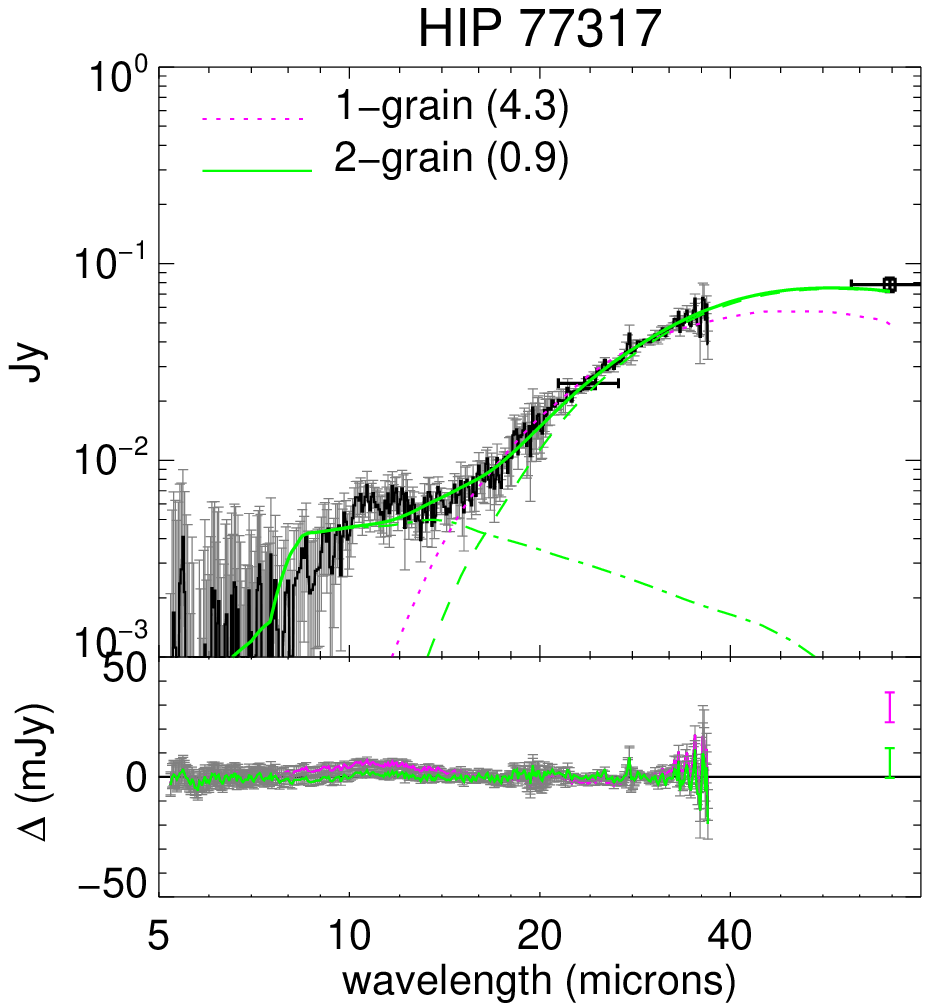} }
\\
\parbox{\stampwidth}{
\includegraphics[width=\stampwidth]{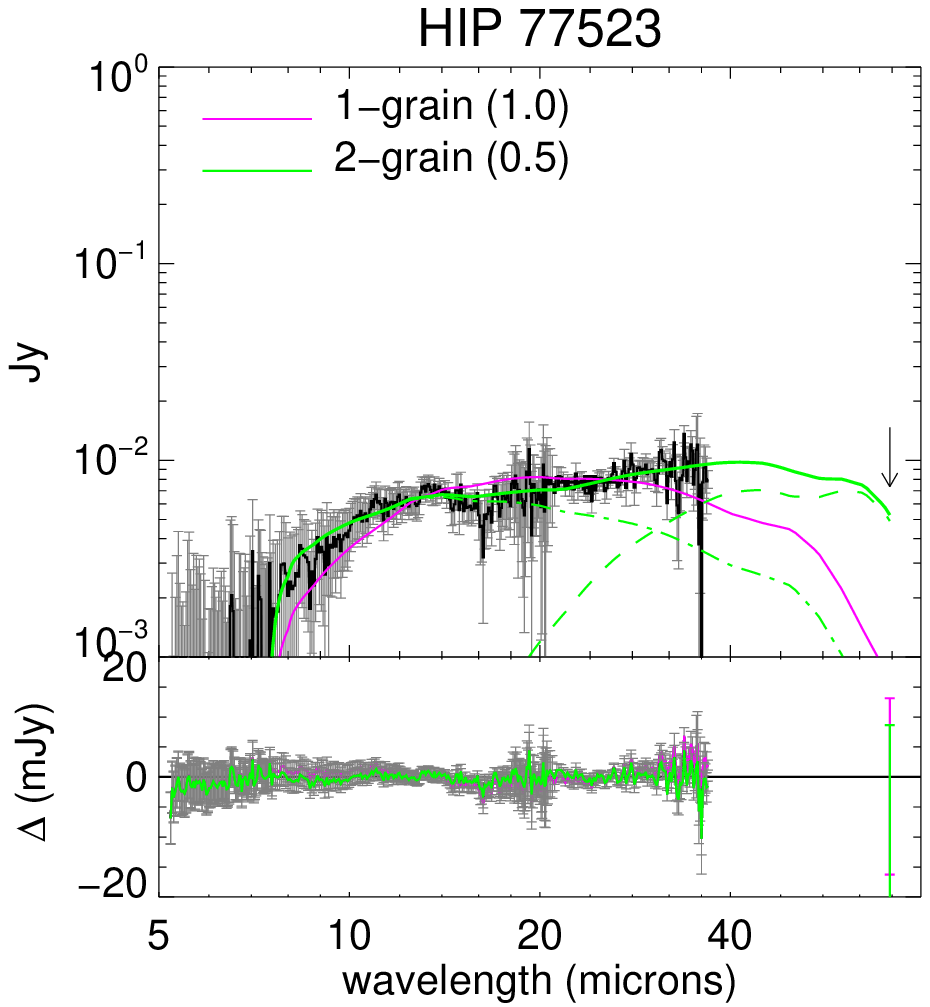} }
\parbox{\stampwidth}{
\includegraphics[width=\stampwidth]{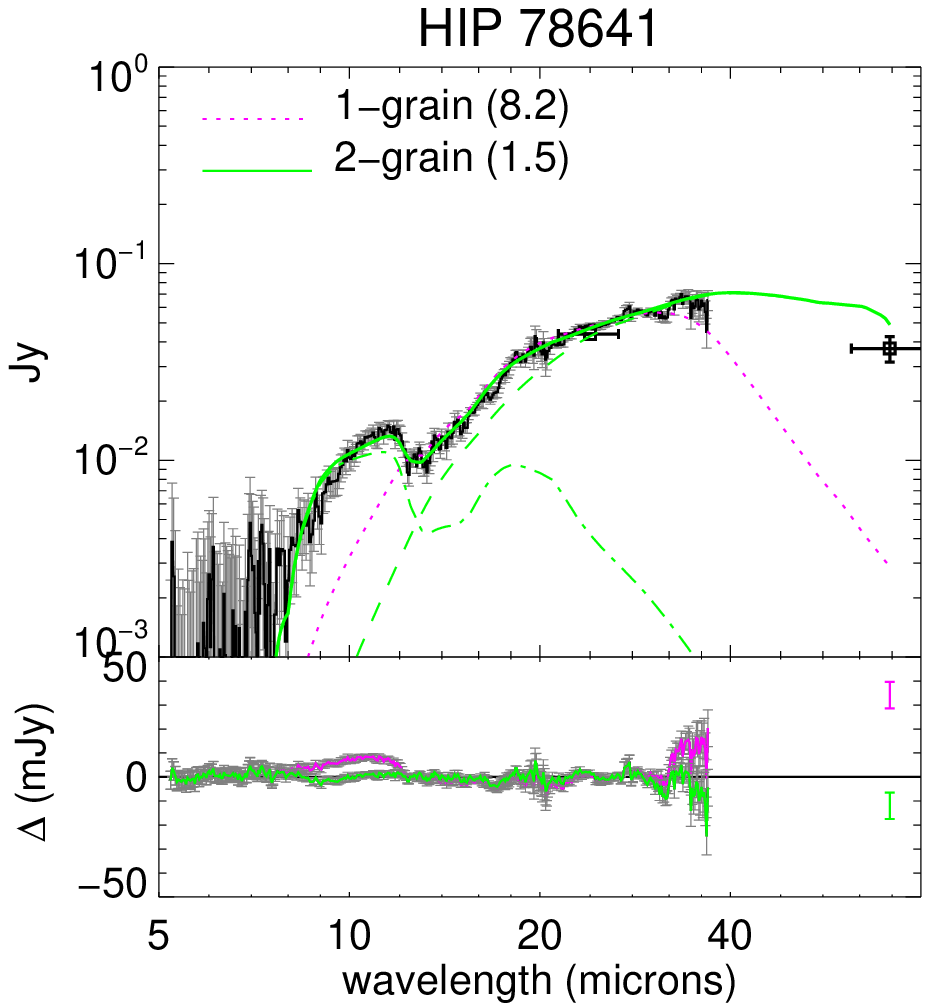} }
\parbox{\stampwidth}{
\includegraphics[width=\stampwidth]{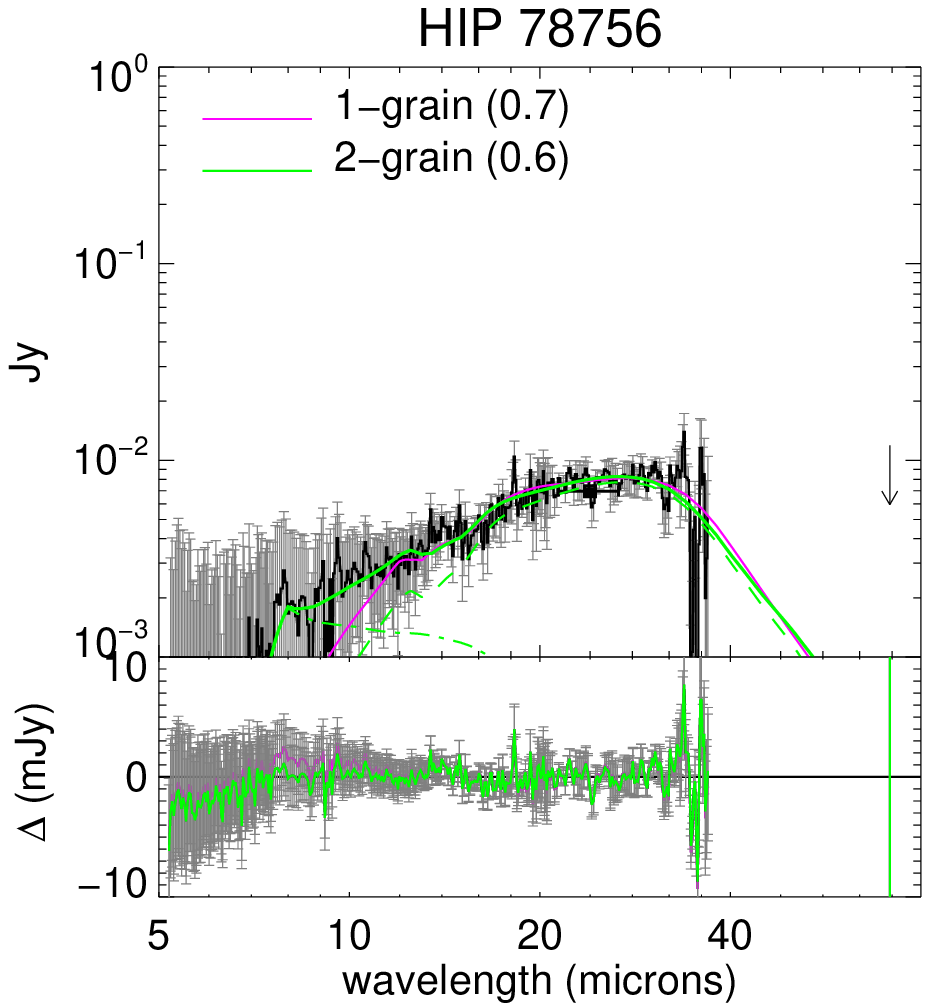} }
\parbox{\stampwidth}{
\includegraphics[width=\stampwidth]{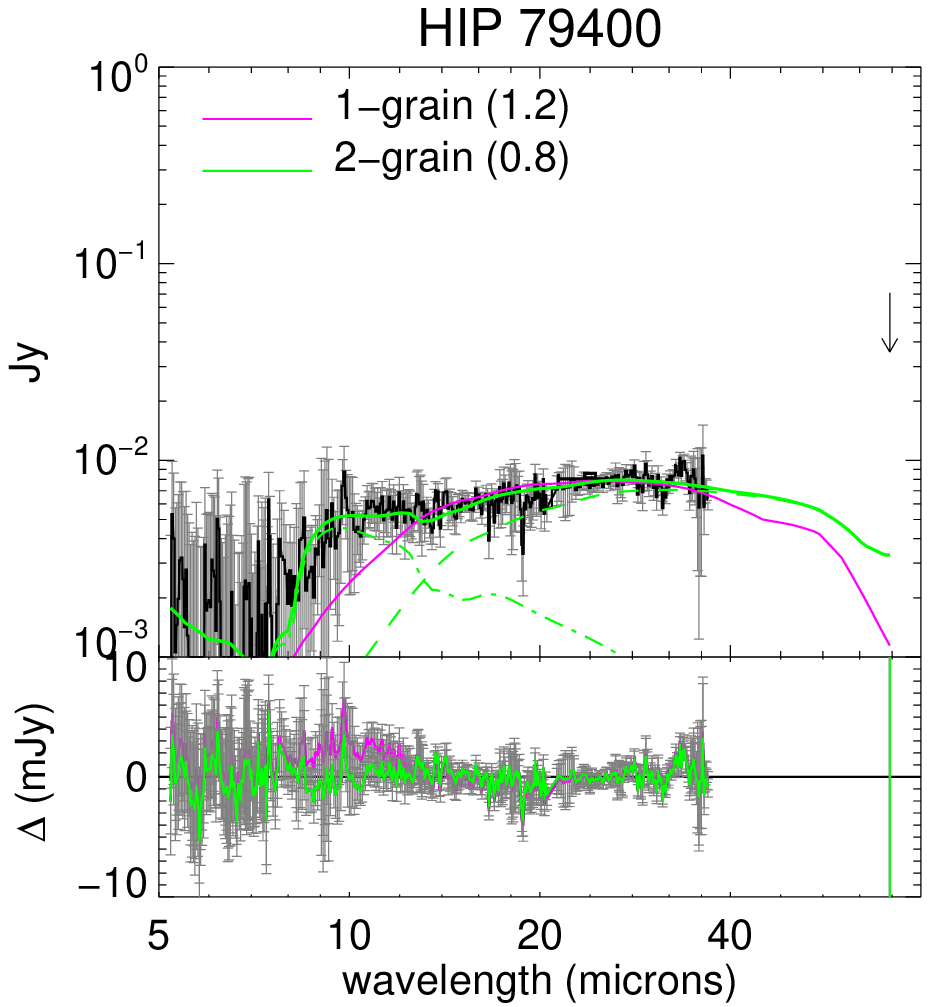} }
\\
\caption{ \label{fitfig1}
Continuation Figure \ref{fitfig0}.}
\end{figure}
\addtocounter{figure}{-1}
\stepcounter{subfig}
\begin{figure}
\parbox{\stampwidth}{
\includegraphics[width=\stampwidth]{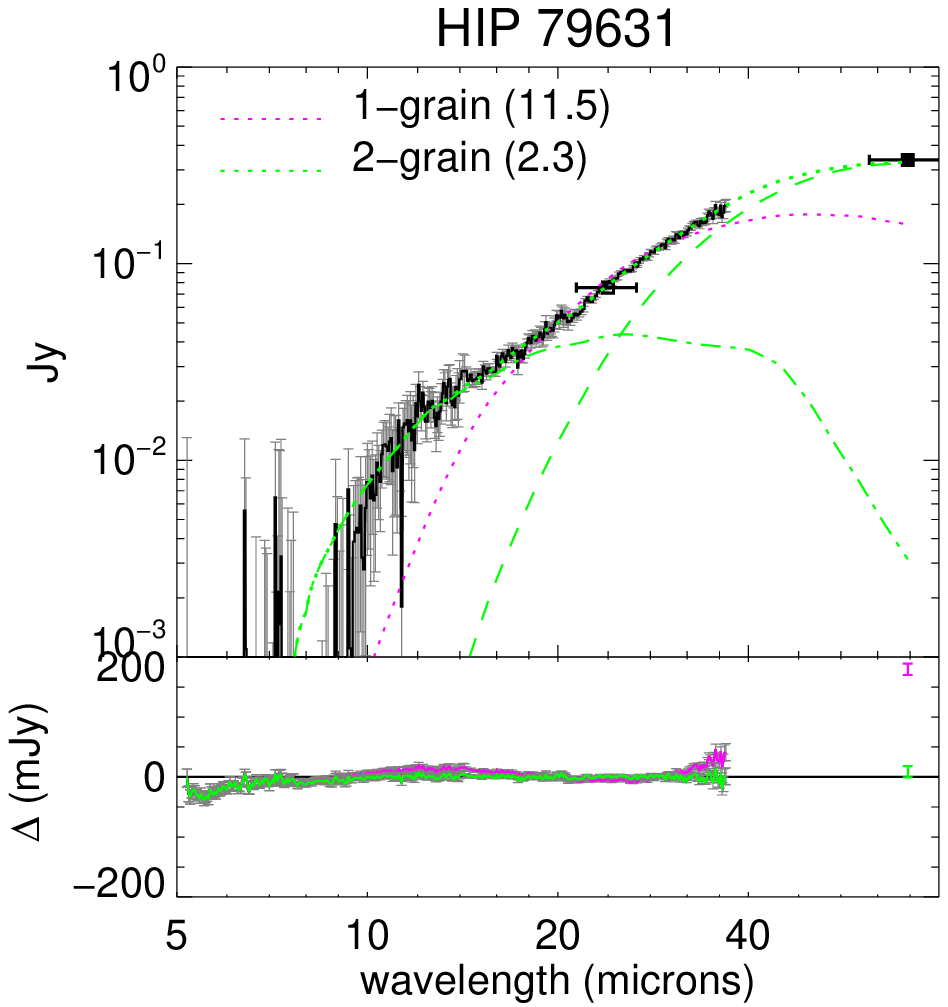} }
\parbox{\stampwidth}{
\includegraphics[width=\stampwidth]{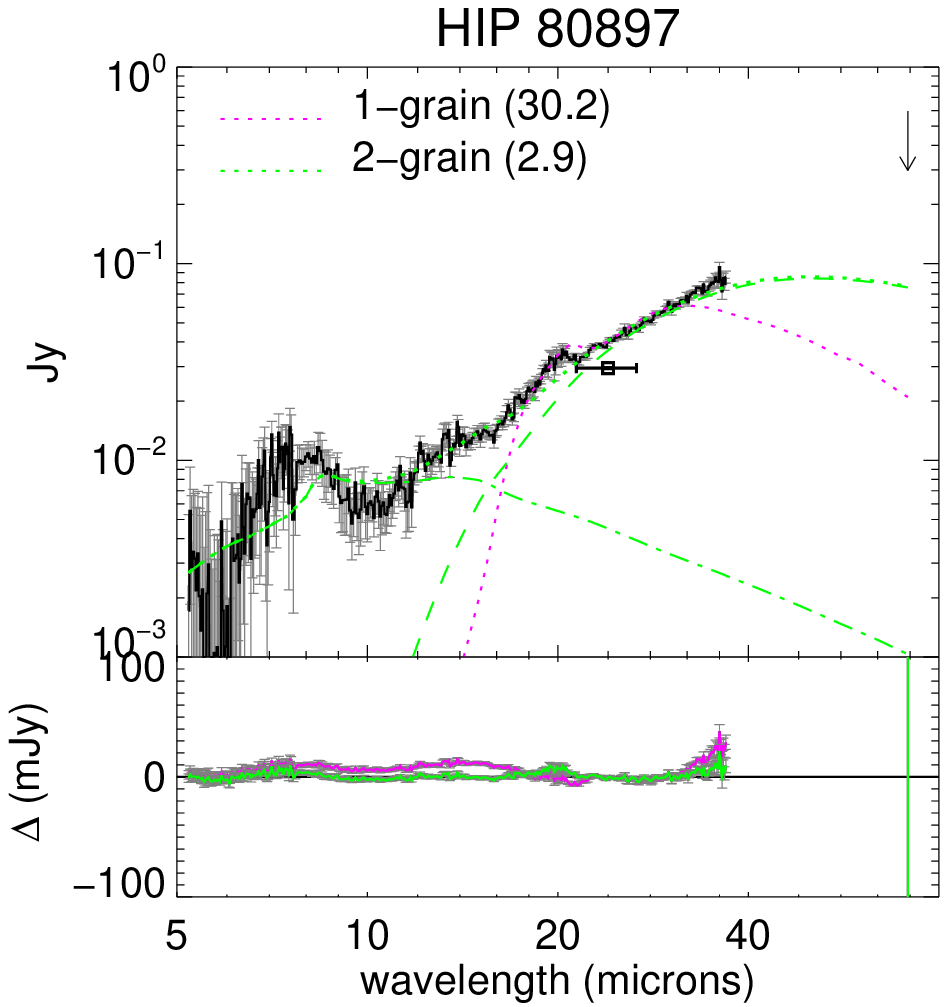} }
\parbox{\stampwidth}{
\includegraphics[width=\stampwidth]{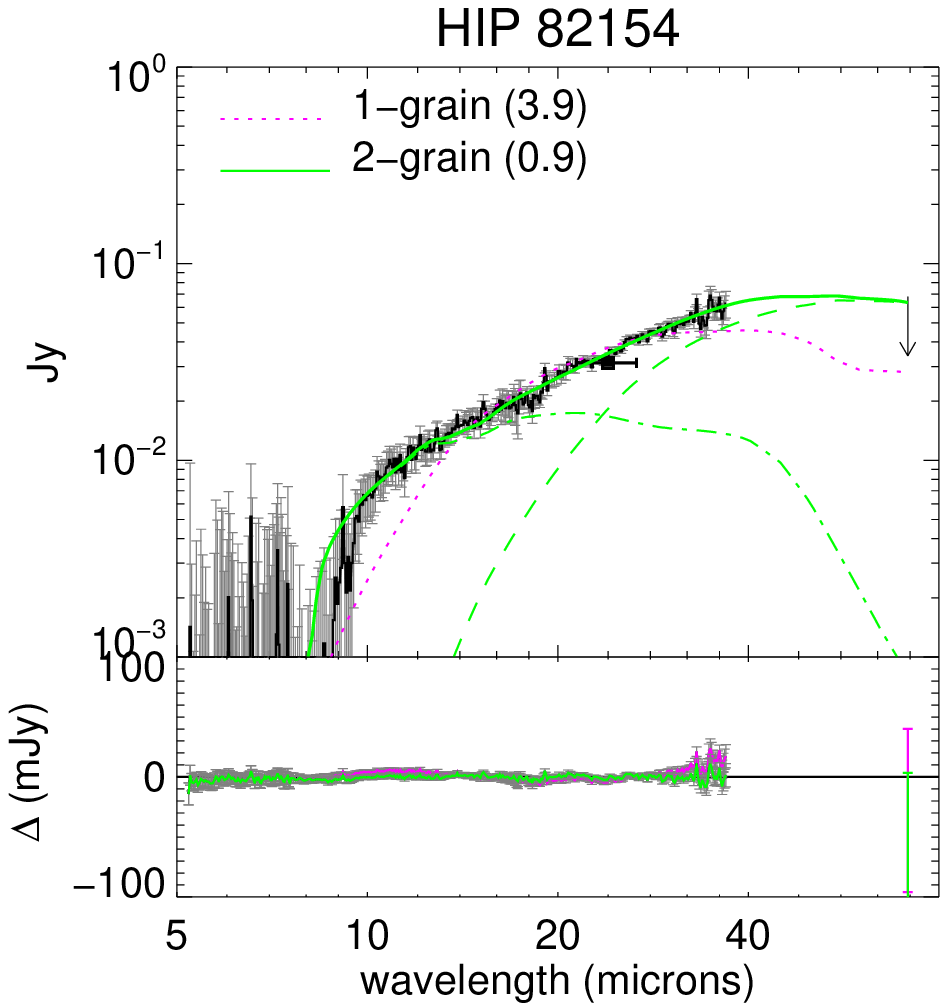} }
\parbox{\stampwidth}{
\includegraphics[width=\stampwidth]{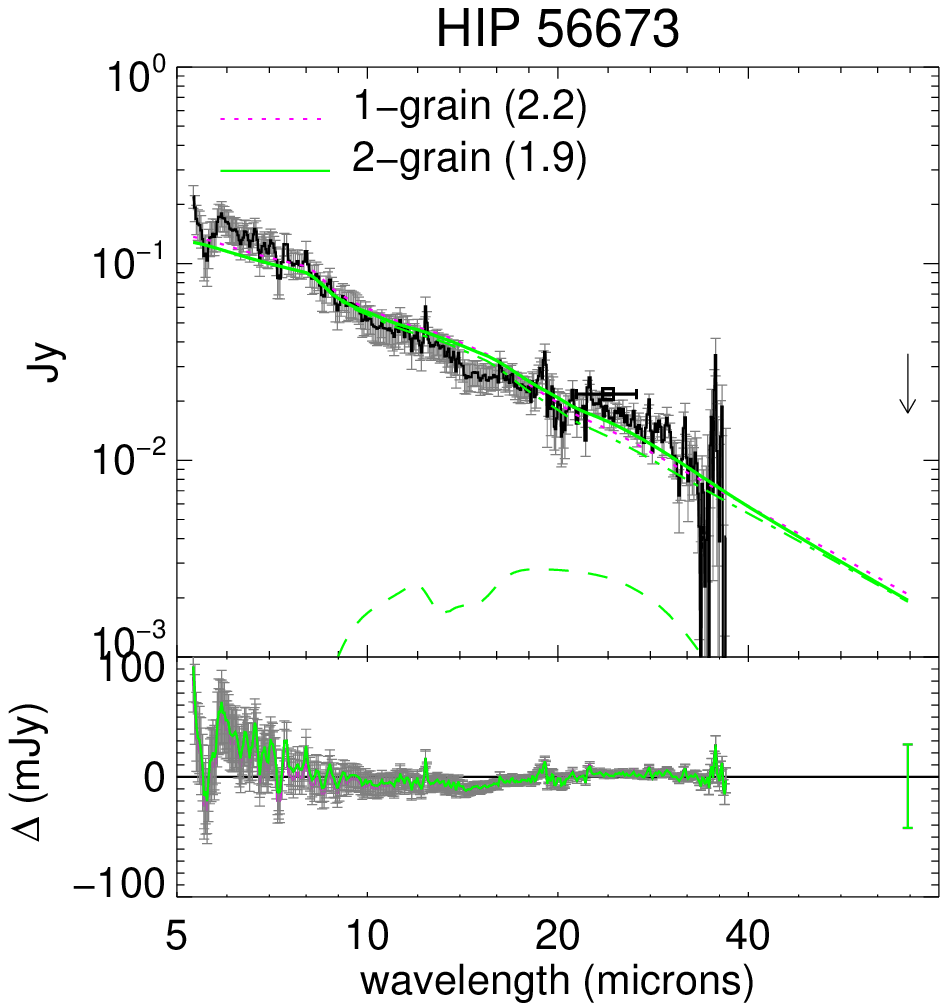} }
\\
\parbox{\stampwidth}{
\includegraphics[width=\stampwidth]{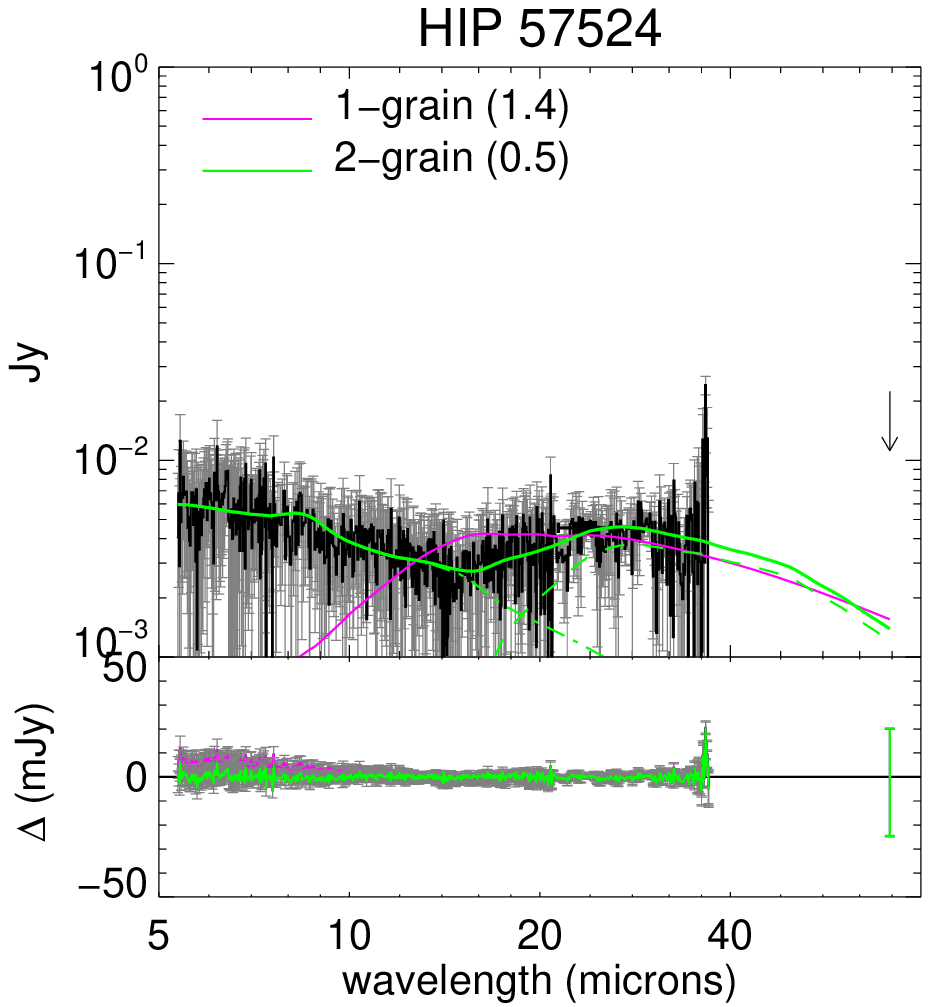} }
\parbox{\stampwidth}{
\includegraphics[width=\stampwidth]{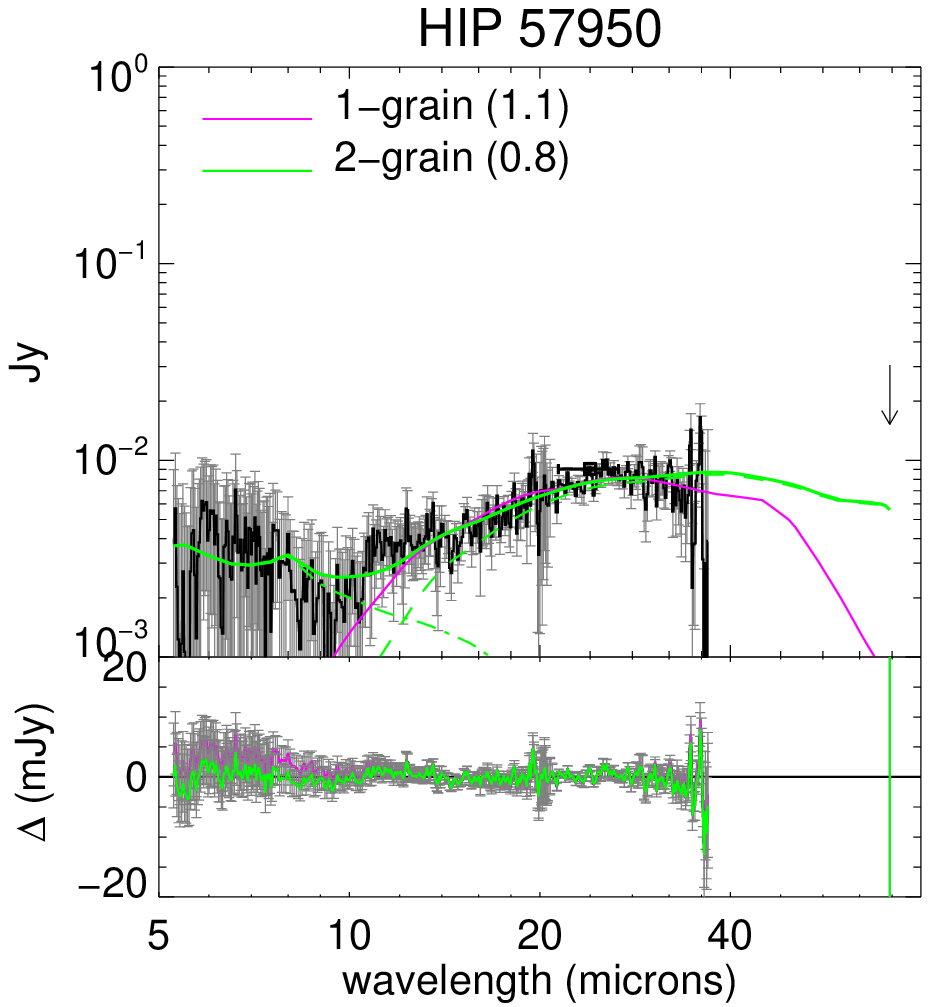} }
\parbox{\stampwidth}{
\includegraphics[width=\stampwidth]{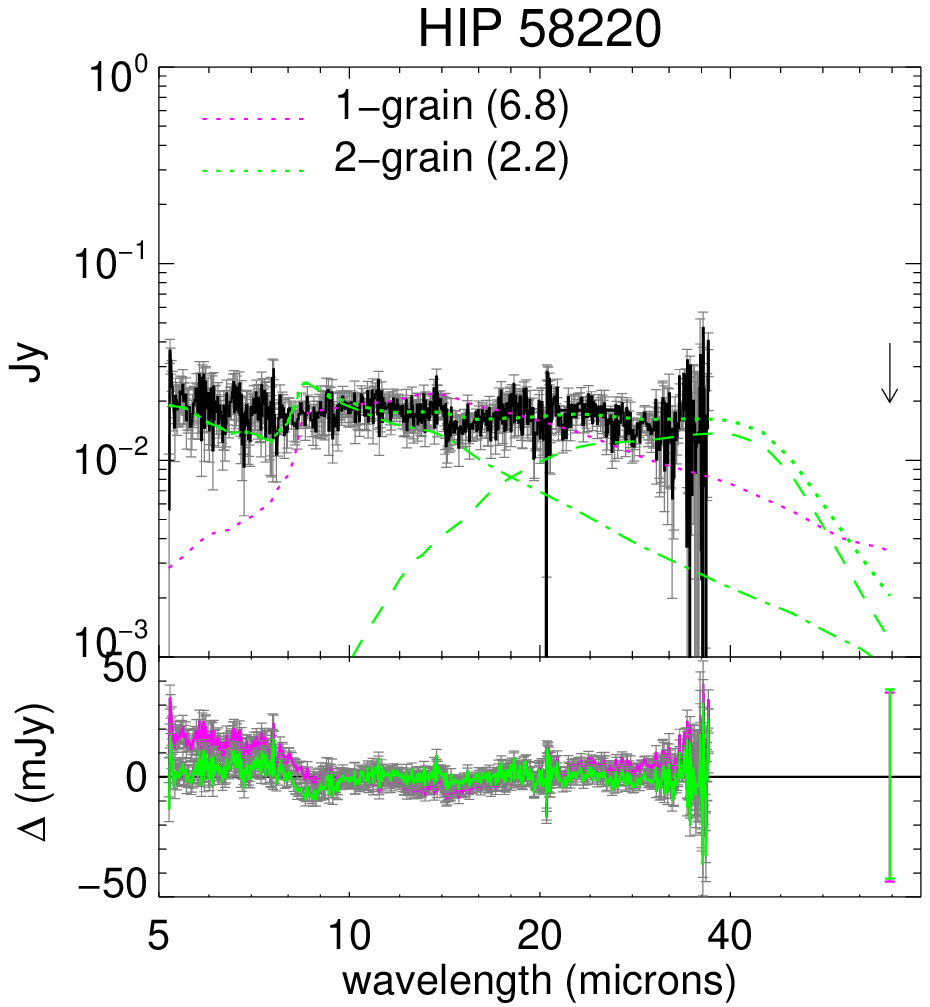} }
\parbox{\stampwidth}{
\includegraphics[width=\stampwidth]{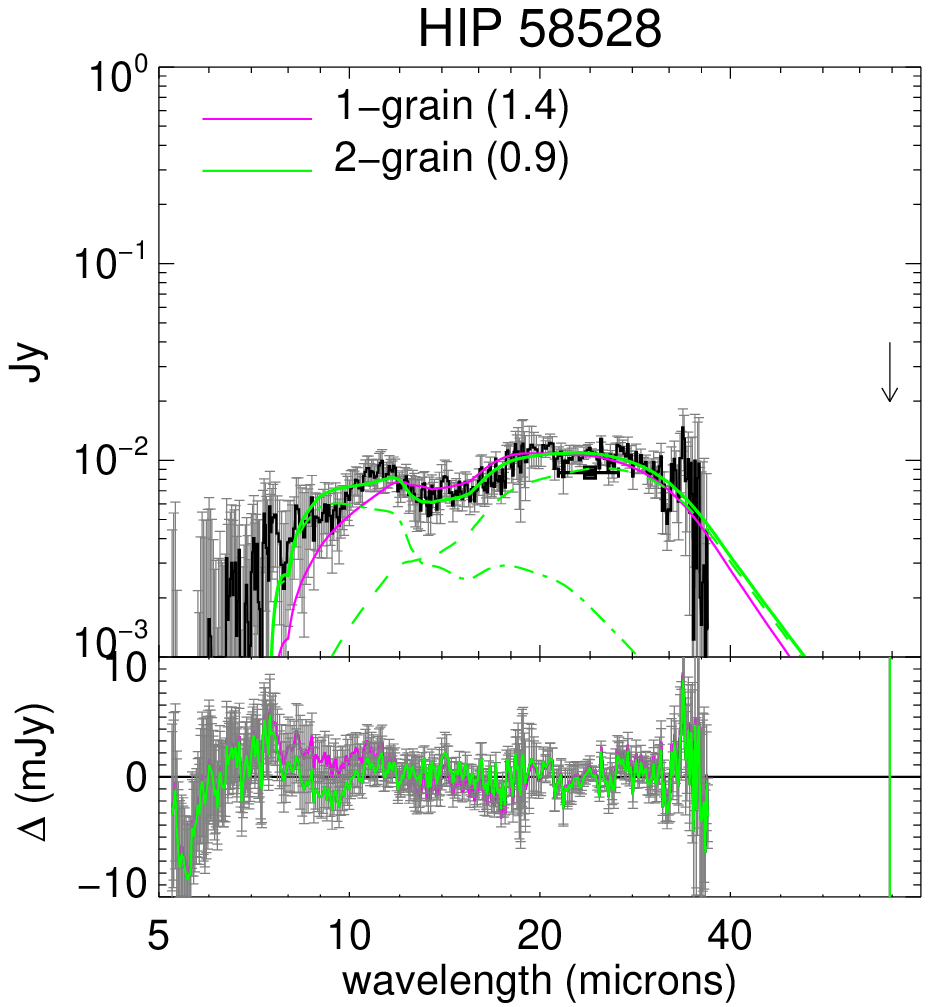} }
\\
\parbox{\stampwidth}{
\includegraphics[width=\stampwidth]{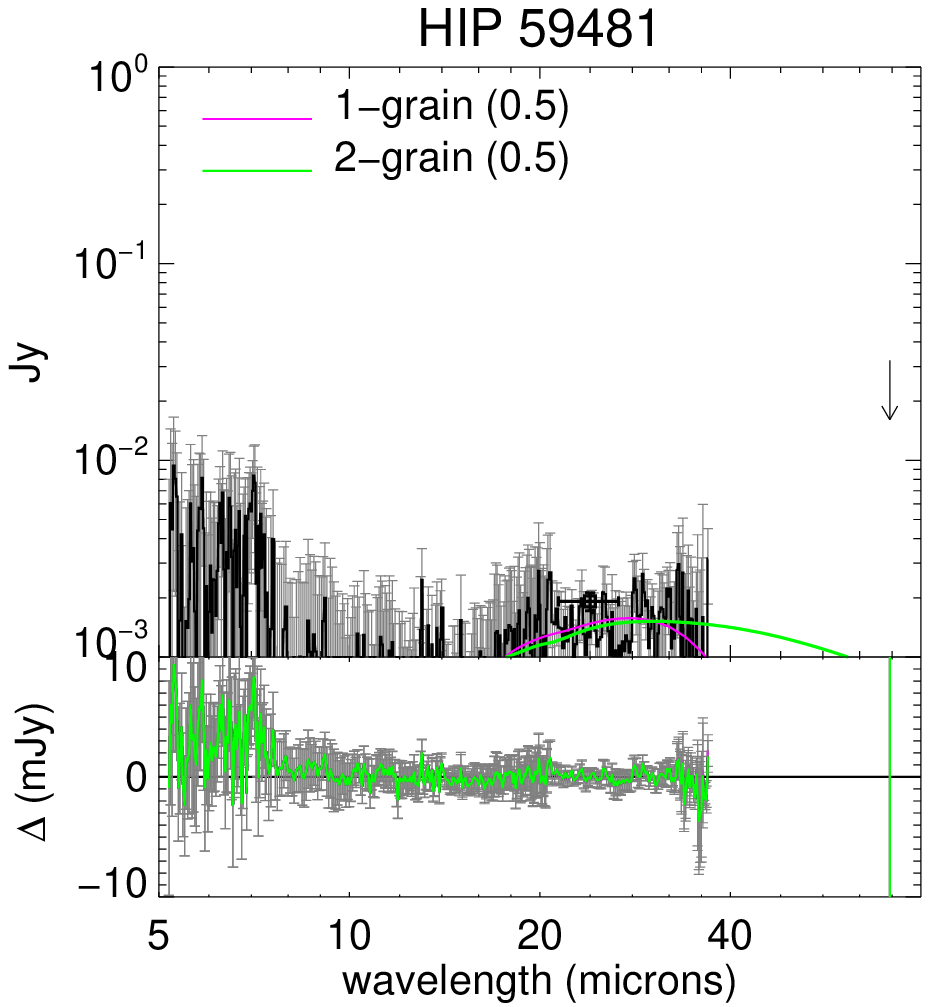} }
\parbox{\stampwidth}{
\includegraphics[width=\stampwidth]{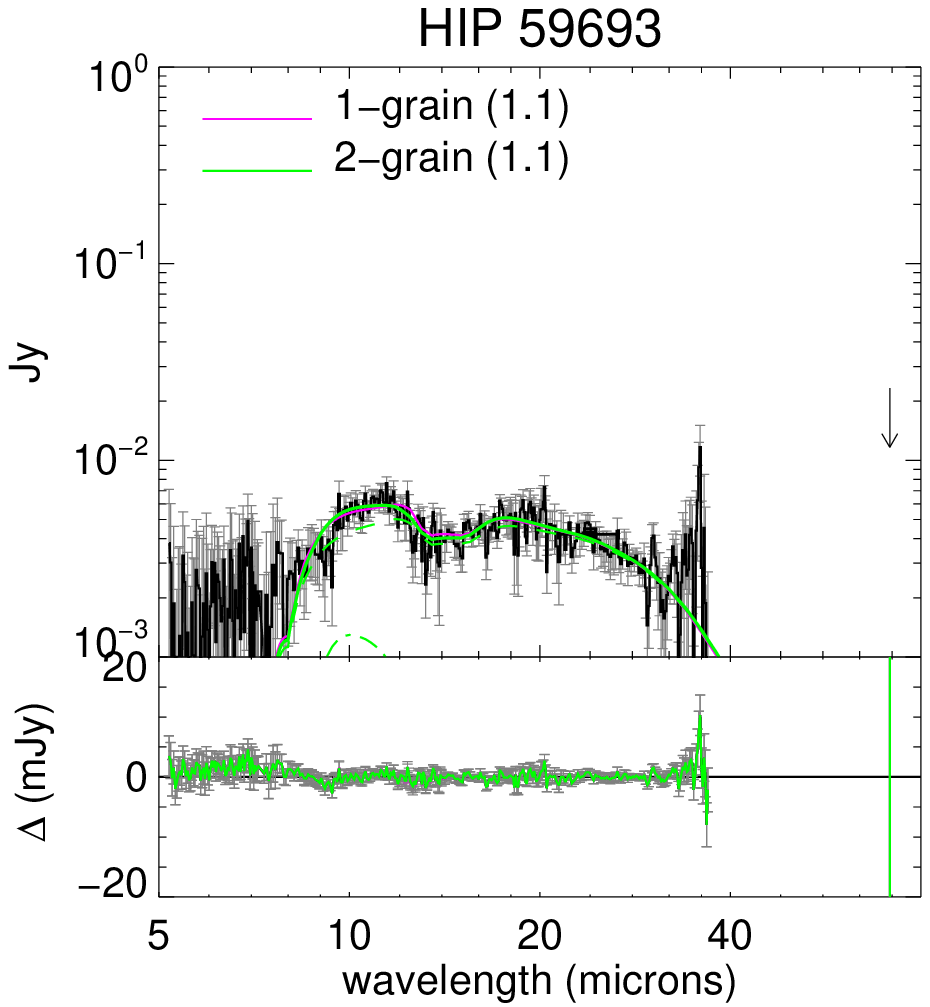} }
\parbox{\stampwidth}{
\includegraphics[width=\stampwidth]{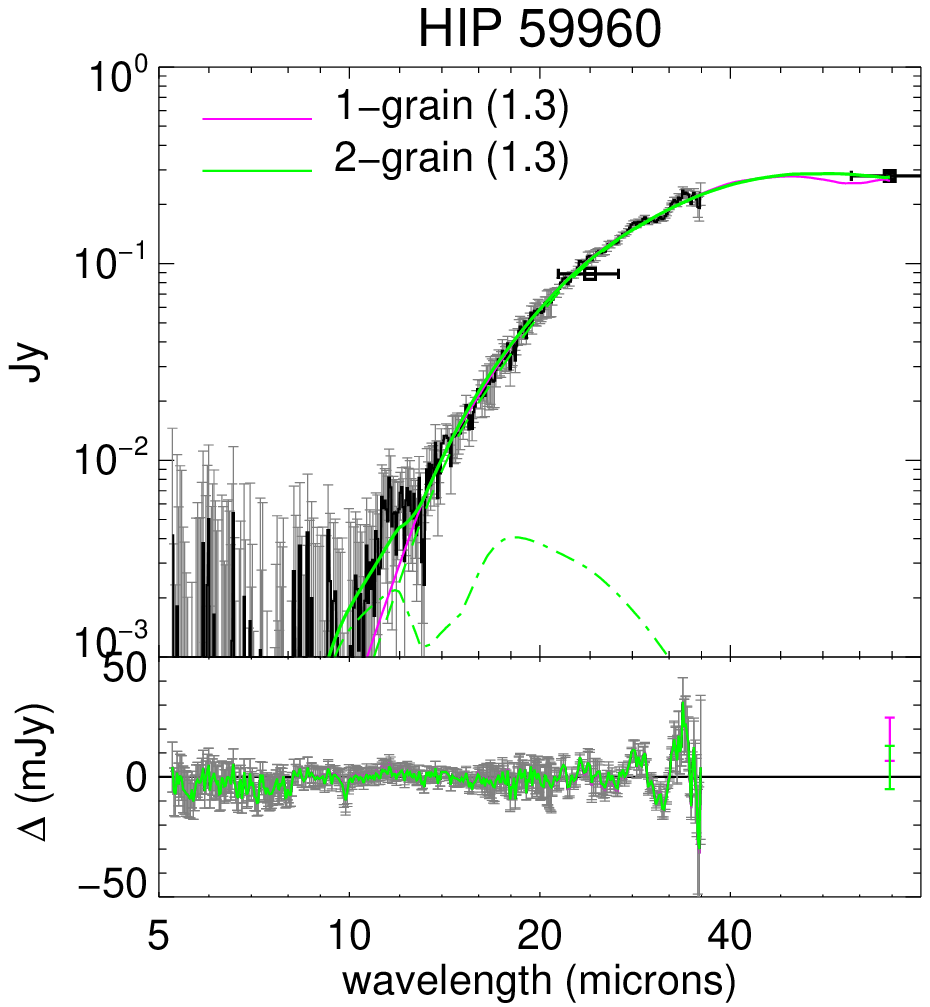} }
\parbox{\stampwidth}{
\includegraphics[width=\stampwidth]{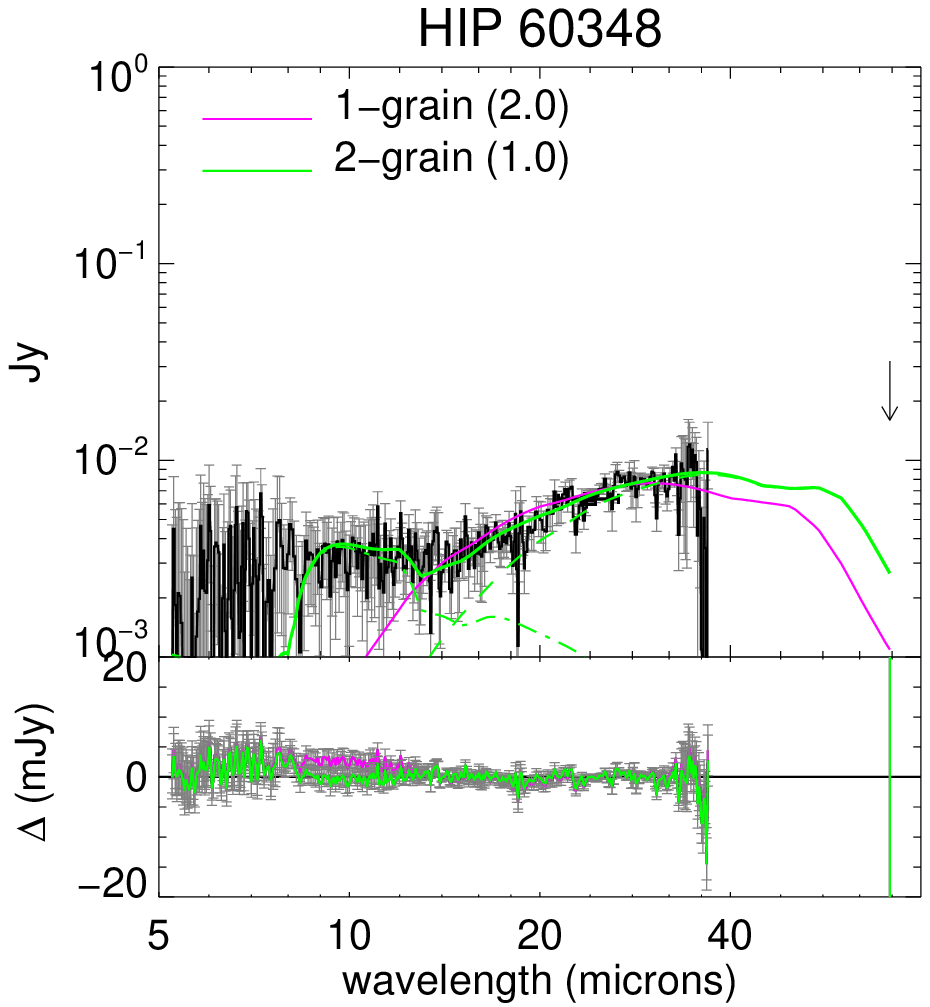} }
\\
\parbox{\stampwidth}{
\includegraphics[width=\stampwidth]{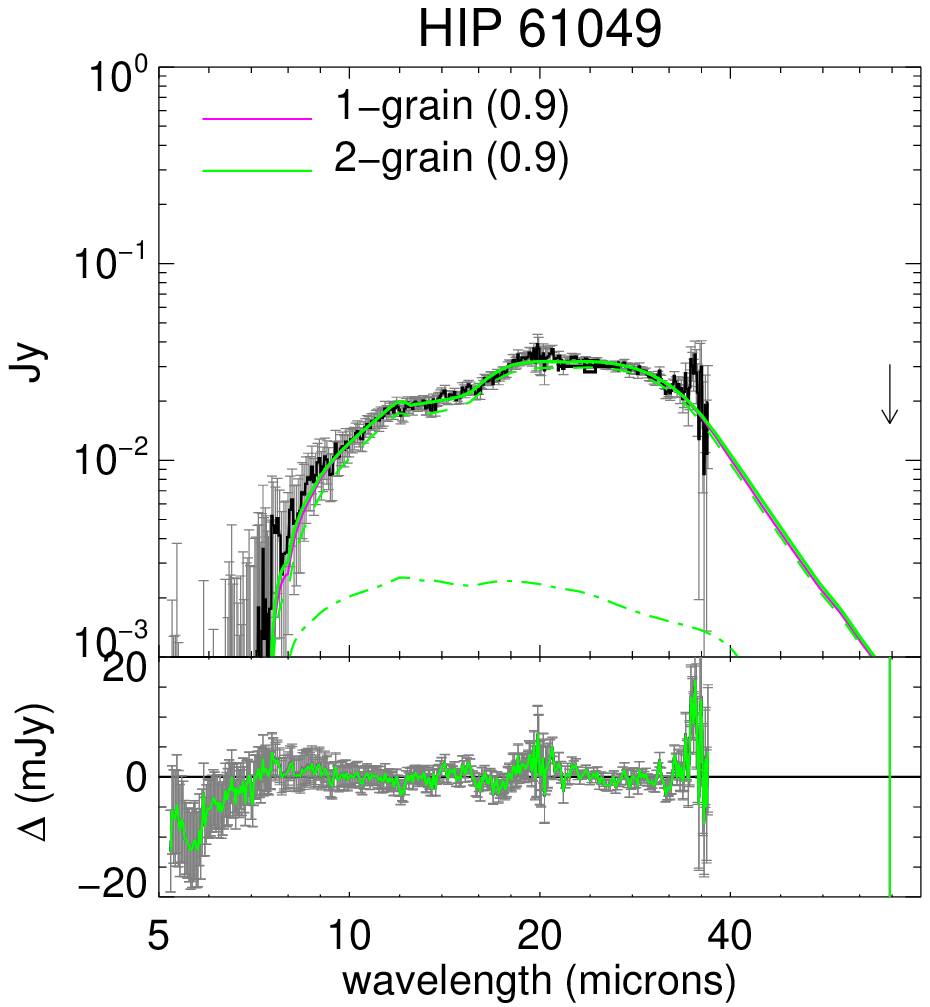} }
\parbox{\stampwidth}{
\includegraphics[width=\stampwidth]{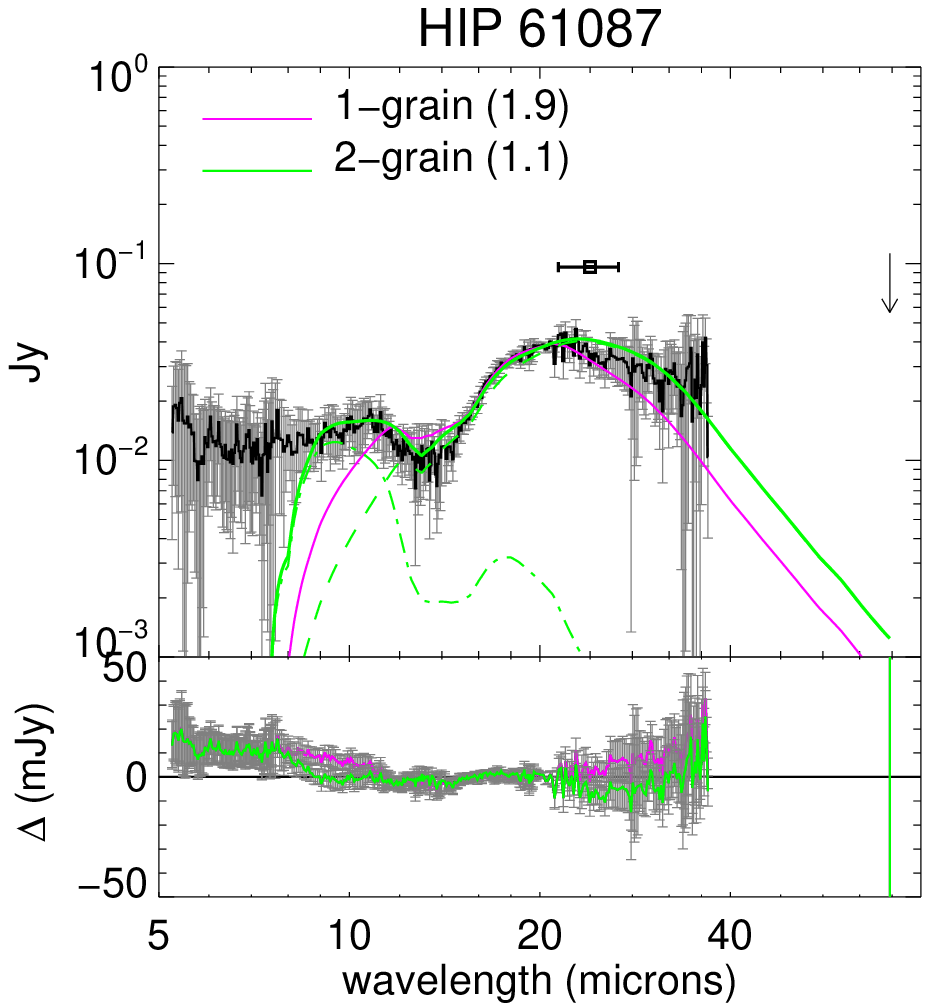} }
\parbox{\stampwidth}{
\includegraphics[width=\stampwidth]{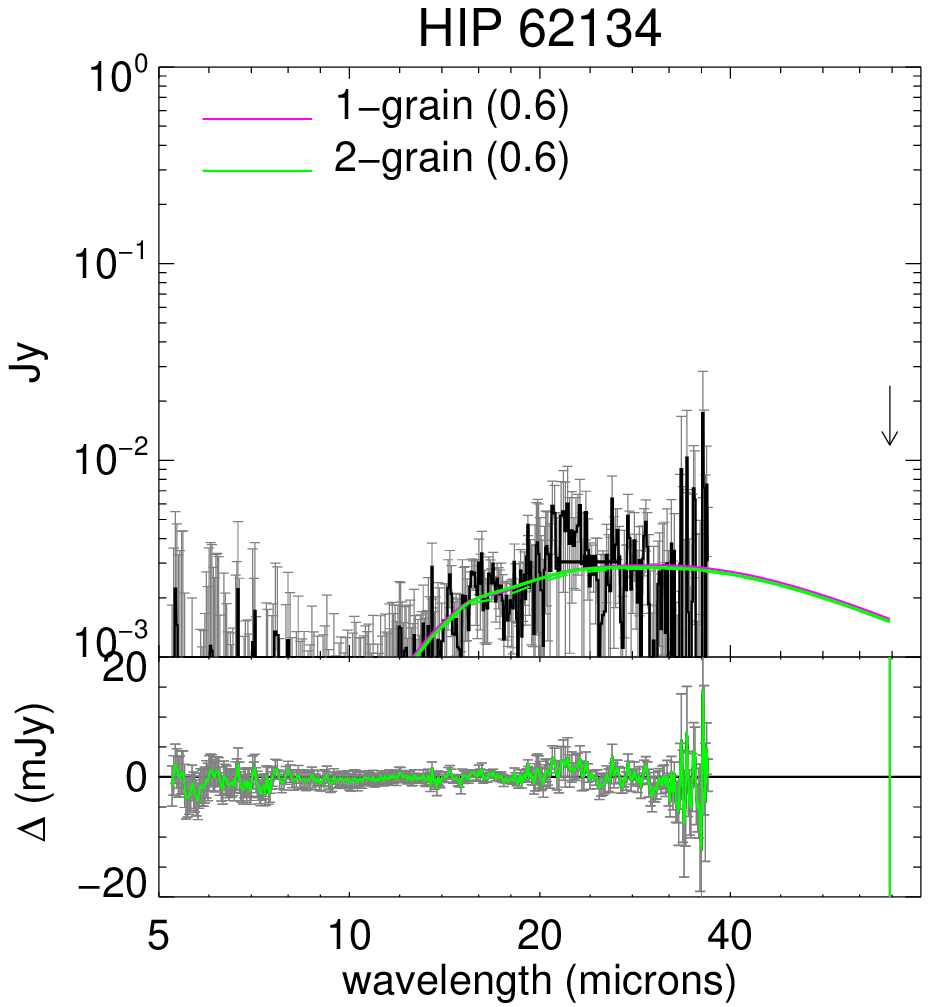} }
\parbox{\stampwidth}{
\includegraphics[width=\stampwidth]{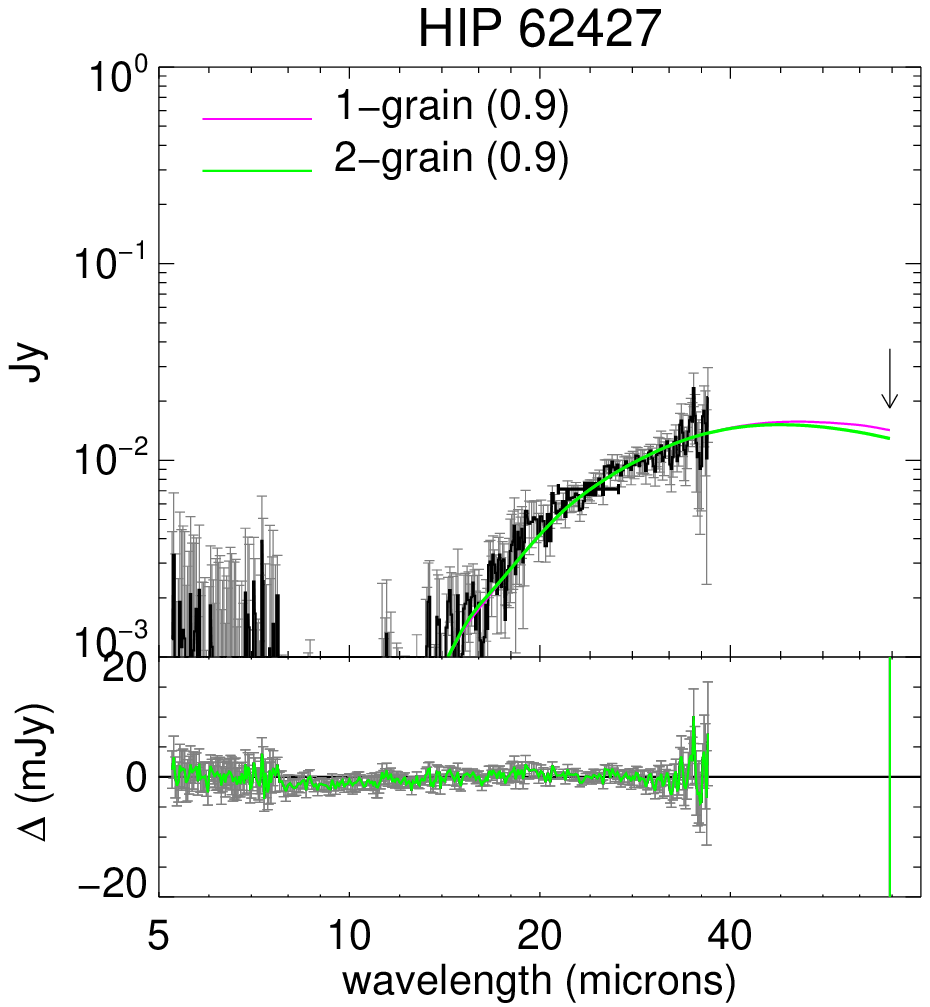} }
\\
\parbox{\stampwidth}{
\includegraphics[width=\stampwidth]{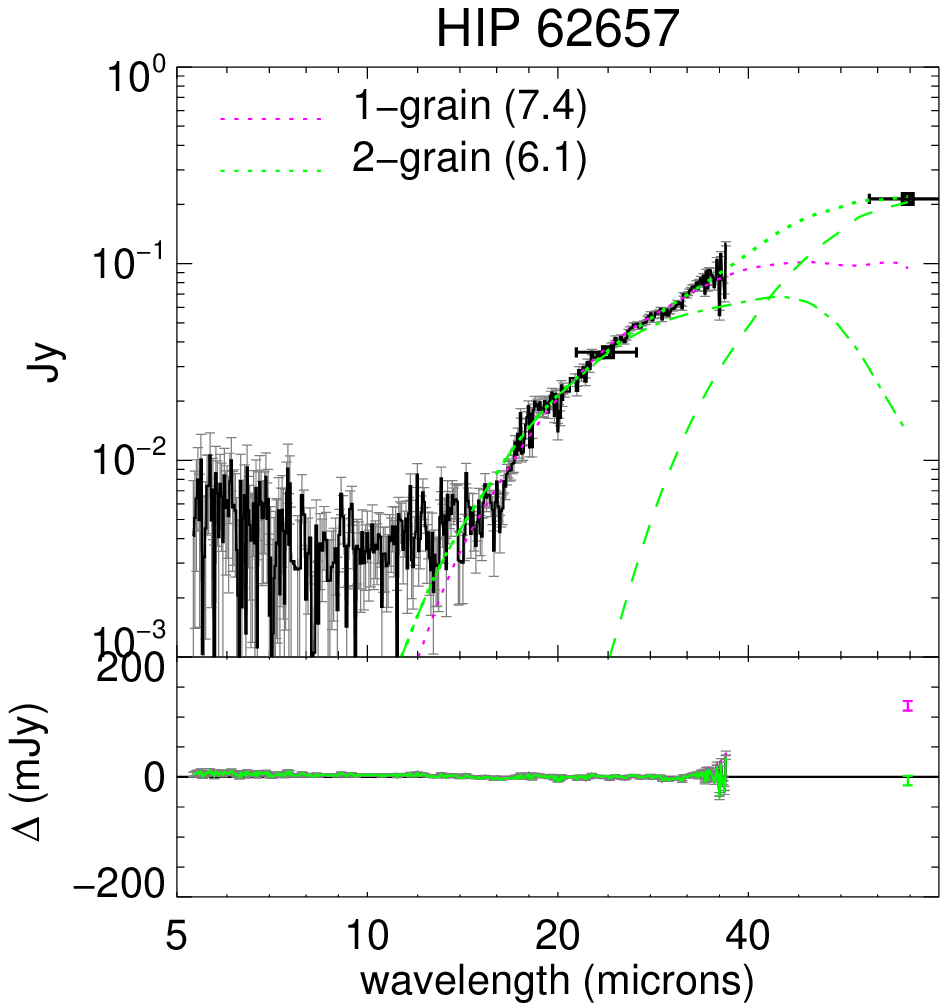} }
\parbox{\stampwidth}{
\includegraphics[width=\stampwidth]{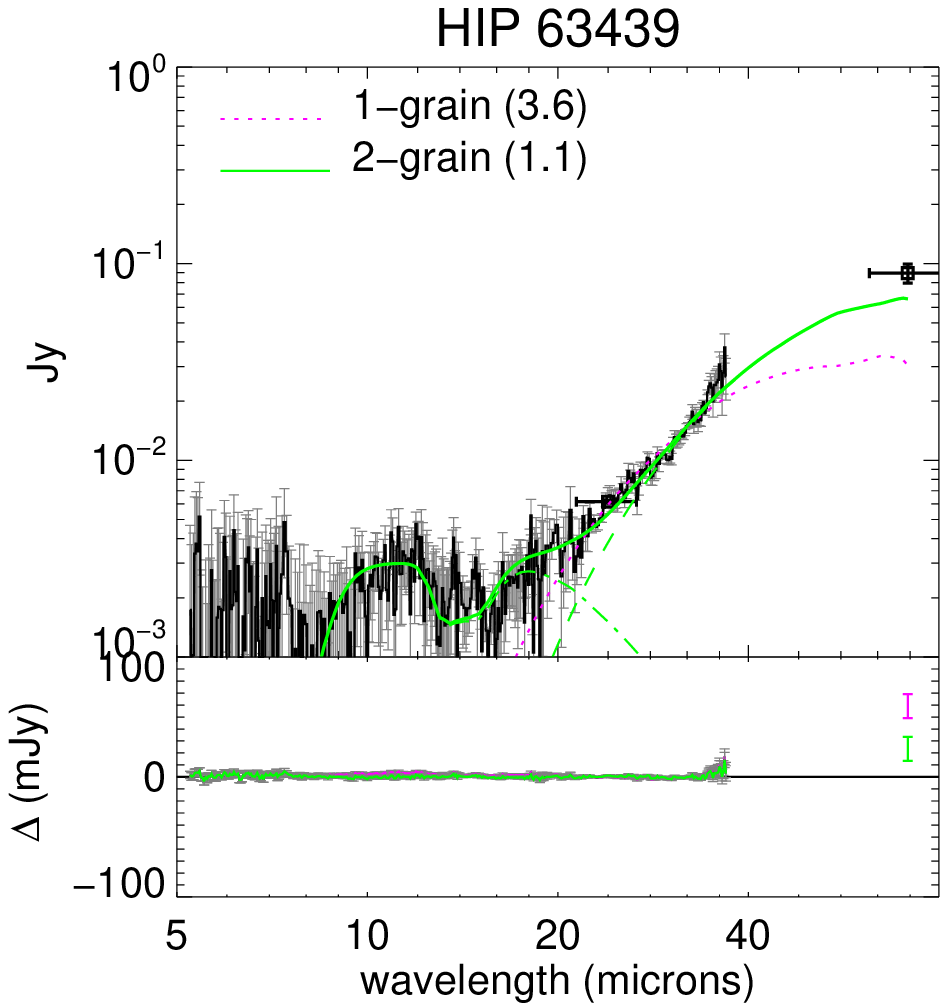} }
\parbox{\stampwidth}{
\includegraphics[width=\stampwidth]{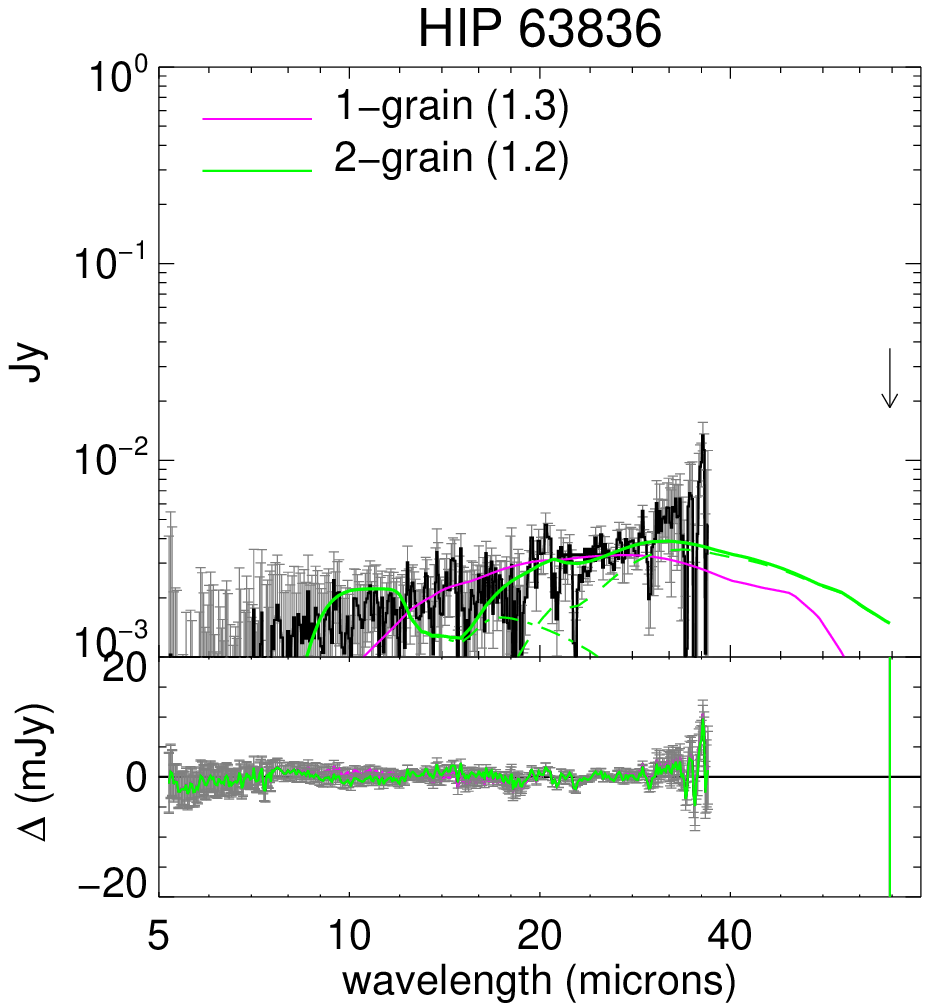} }
\parbox{\stampwidth}{
\includegraphics[width=\stampwidth]{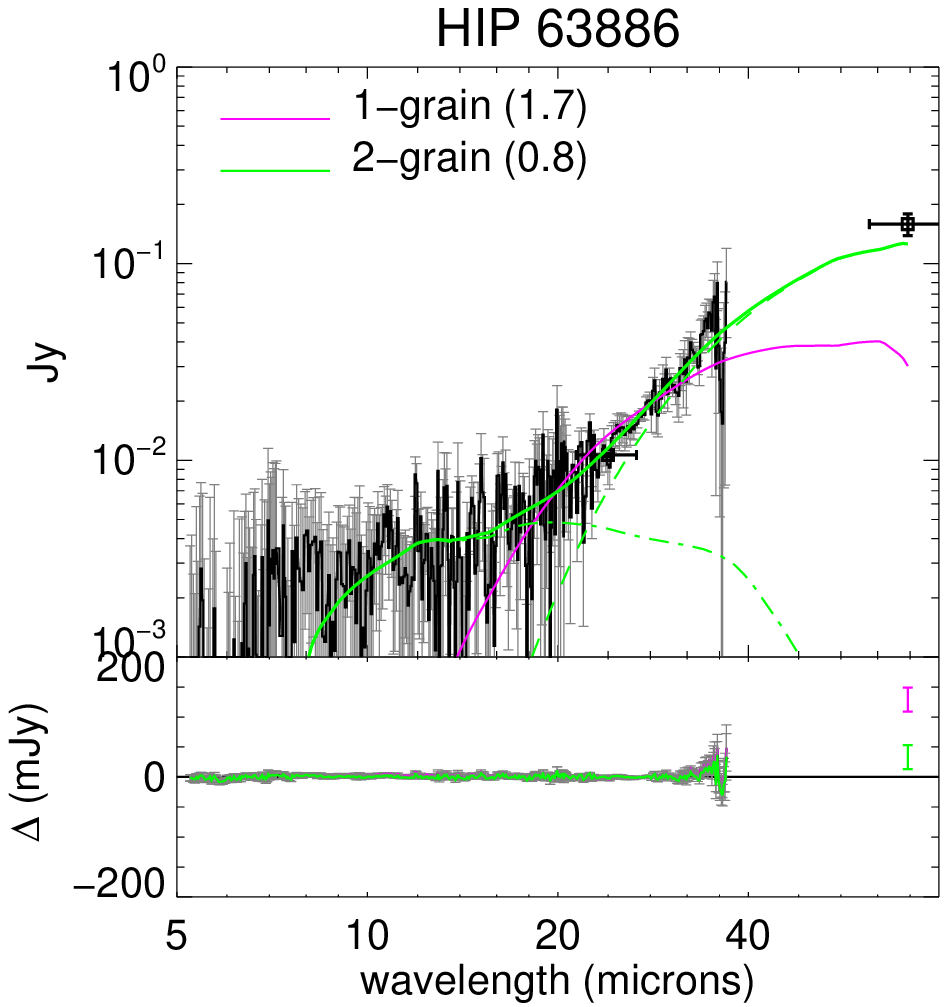} }
\\
\caption{ \label{fitfig2}
Continuation Figure \ref{fitfig0}.}
\end{figure}
\addtocounter{figure}{-1}
\stepcounter{subfig}
\begin{figure}
\parbox{\stampwidth}{
\includegraphics[width=\stampwidth]{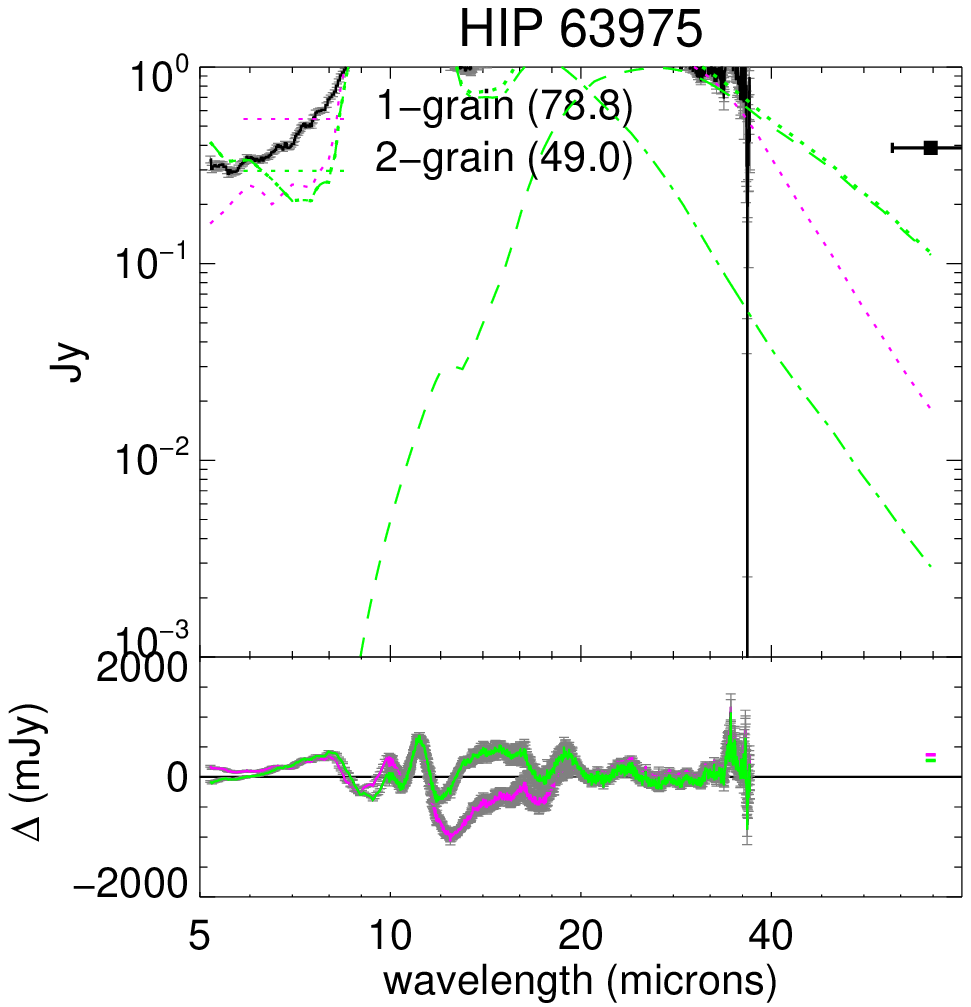} }
\parbox{\stampwidth}{
\includegraphics[width=\stampwidth]{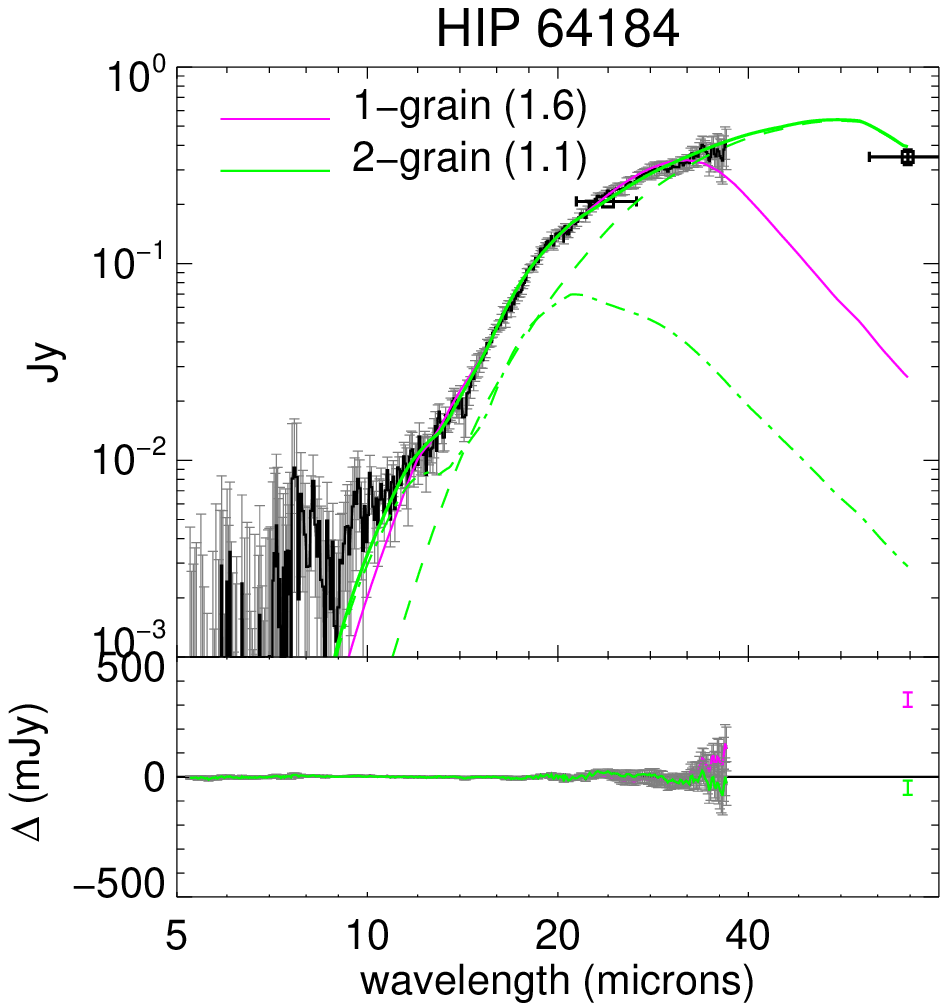} }
\parbox{\stampwidth}{
\includegraphics[width=\stampwidth]{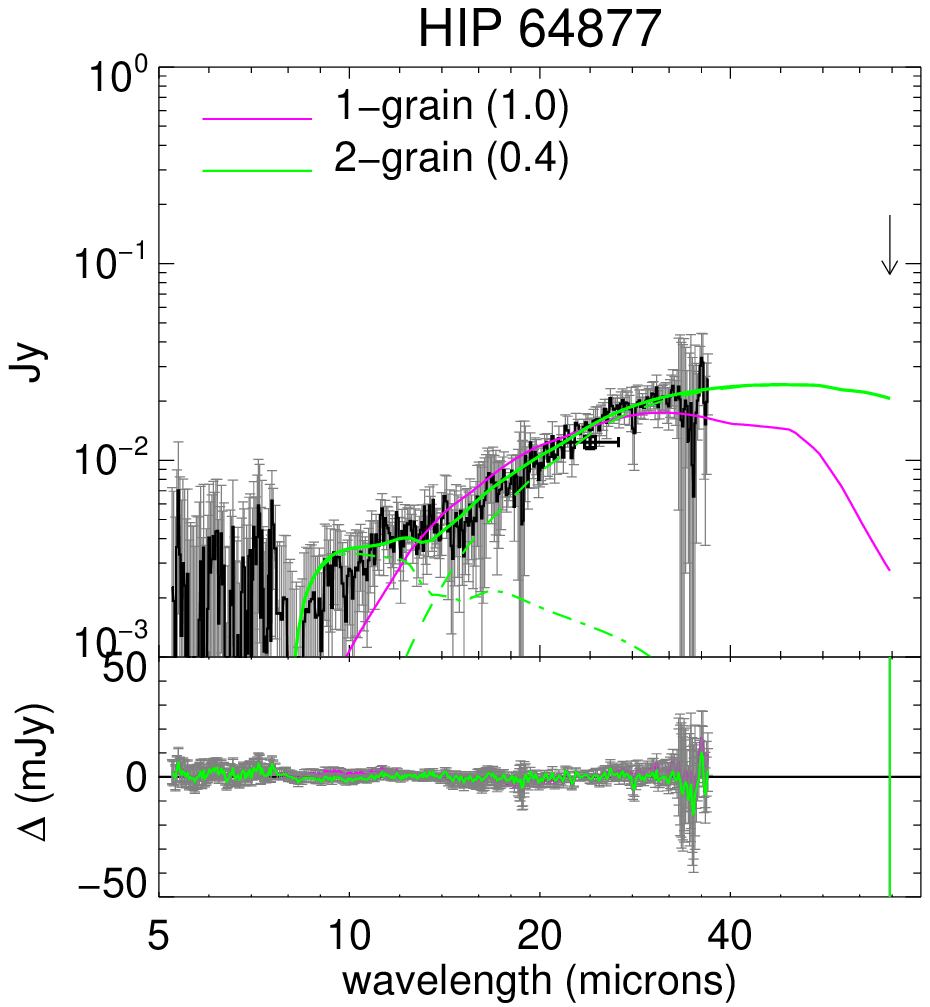} }
\parbox{\stampwidth}{
\includegraphics[width=\stampwidth]{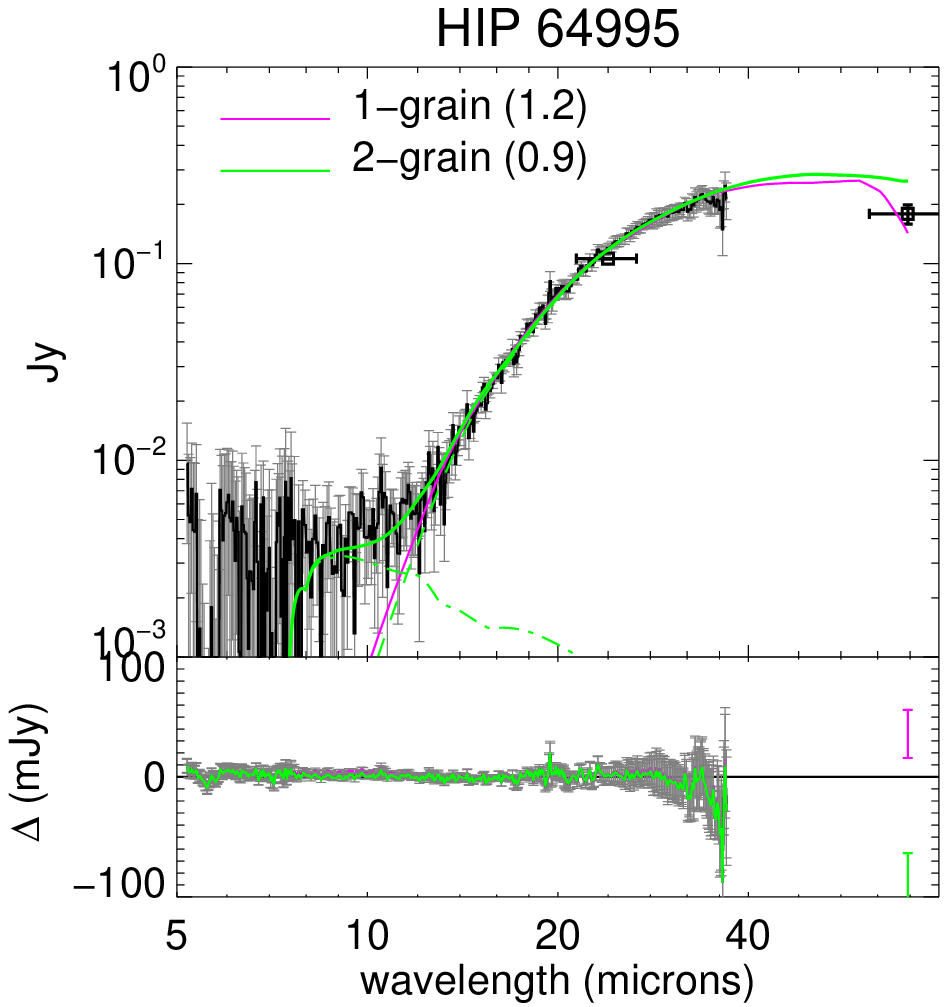} }
\\
\parbox{\stampwidth}{
\includegraphics[width=\stampwidth]{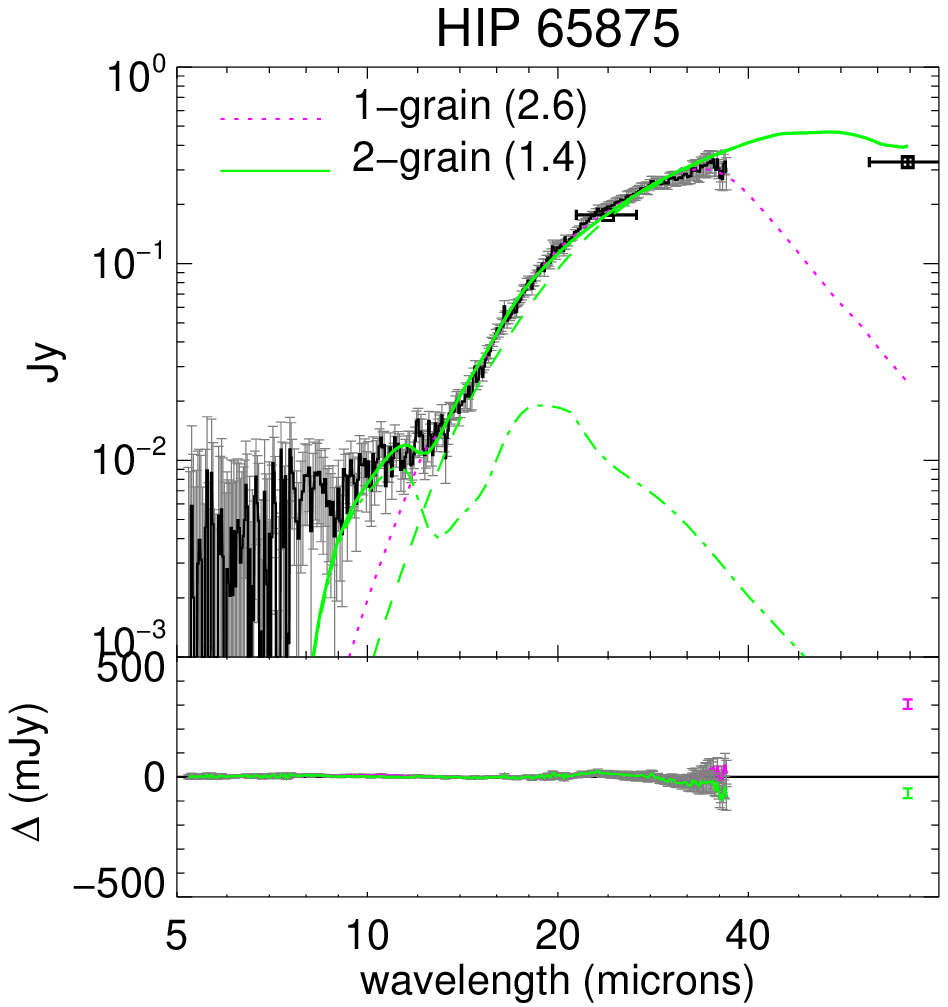} }
\parbox{\stampwidth}{
\includegraphics[width=\stampwidth]{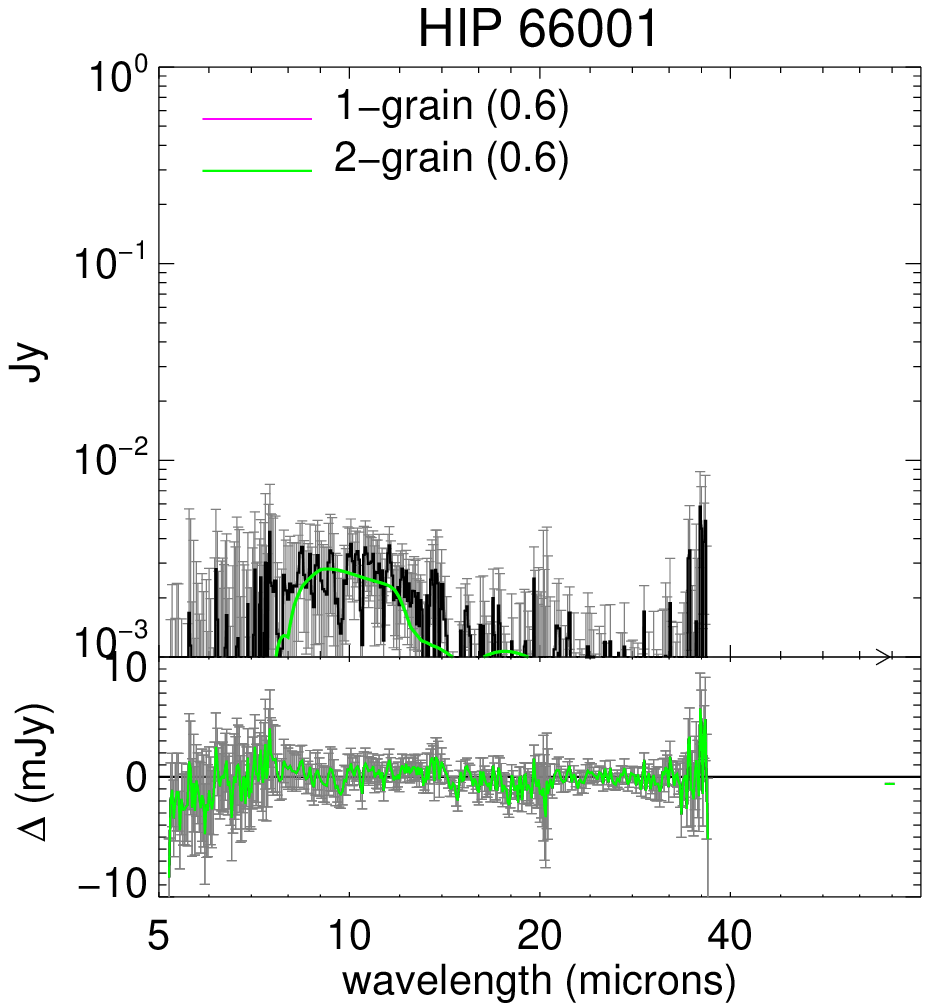} }
\parbox{\stampwidth}{
\includegraphics[width=\stampwidth]{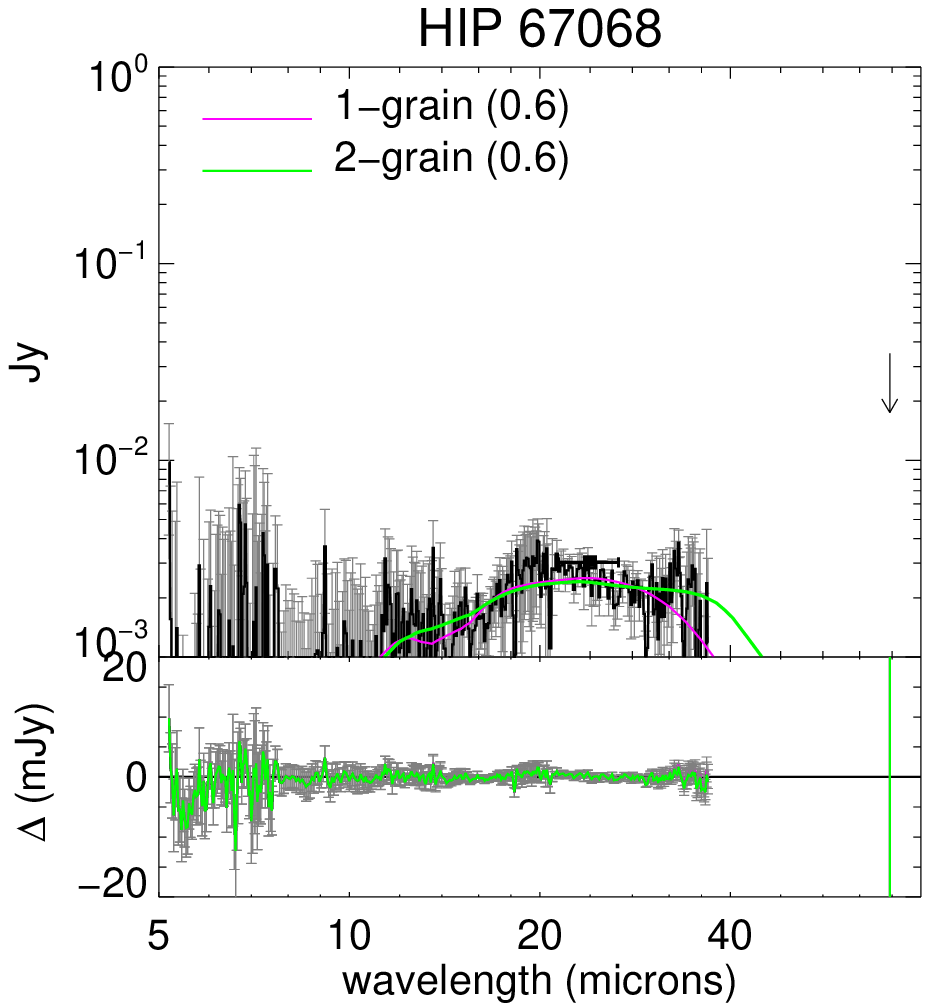} }
\parbox{\stampwidth}{
\includegraphics[width=\stampwidth]{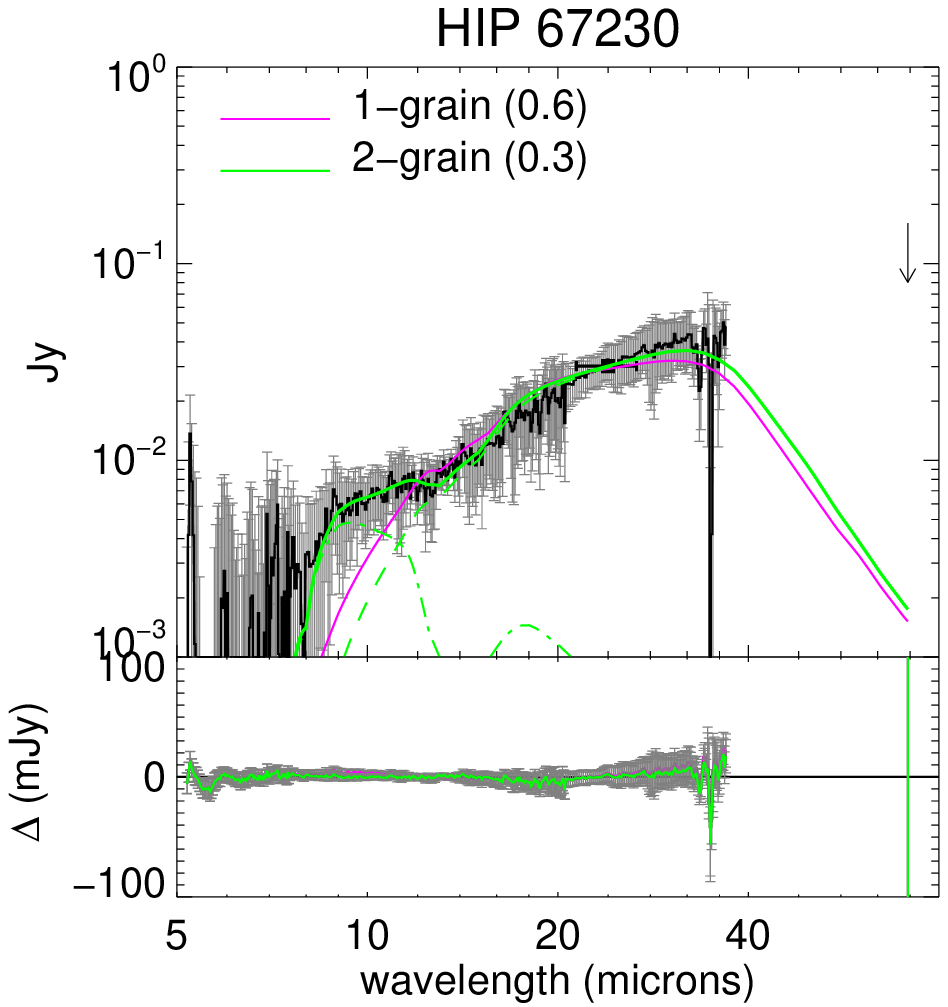} }
\\
\parbox{\stampwidth}{
\includegraphics[width=\stampwidth]{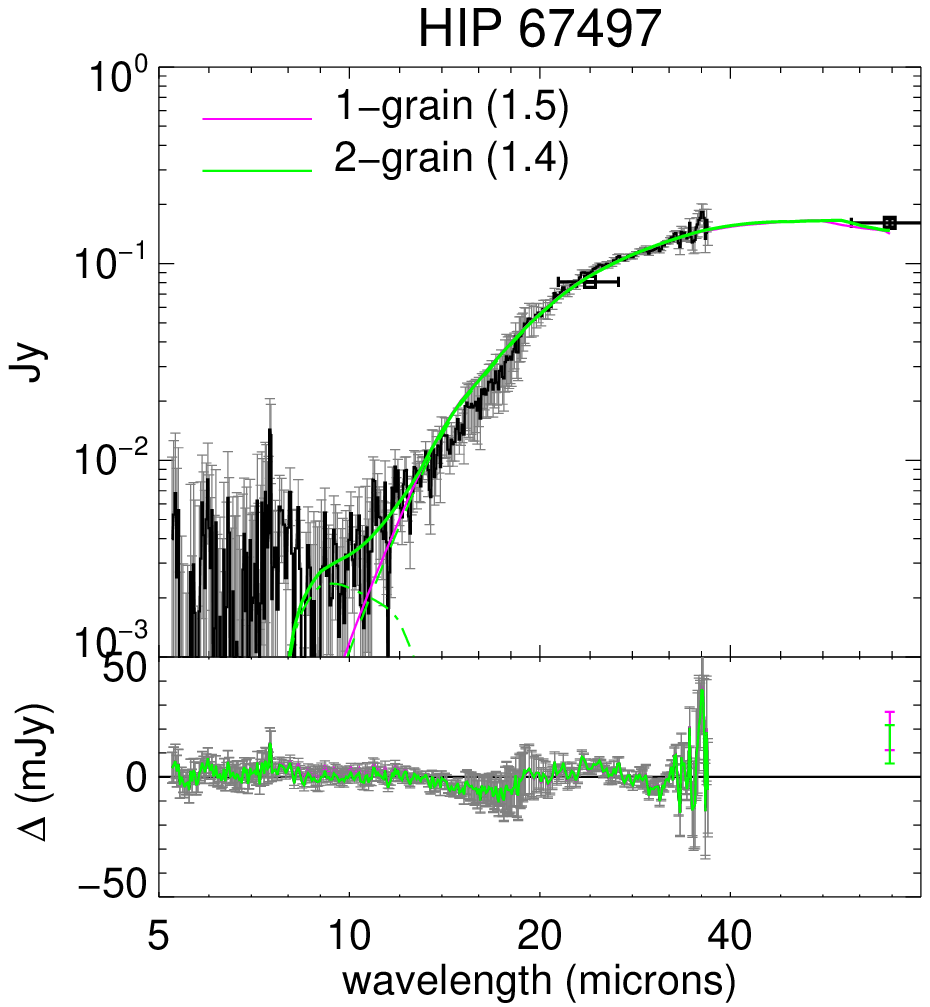} }
\parbox{\stampwidth}{
\includegraphics[width=\stampwidth]{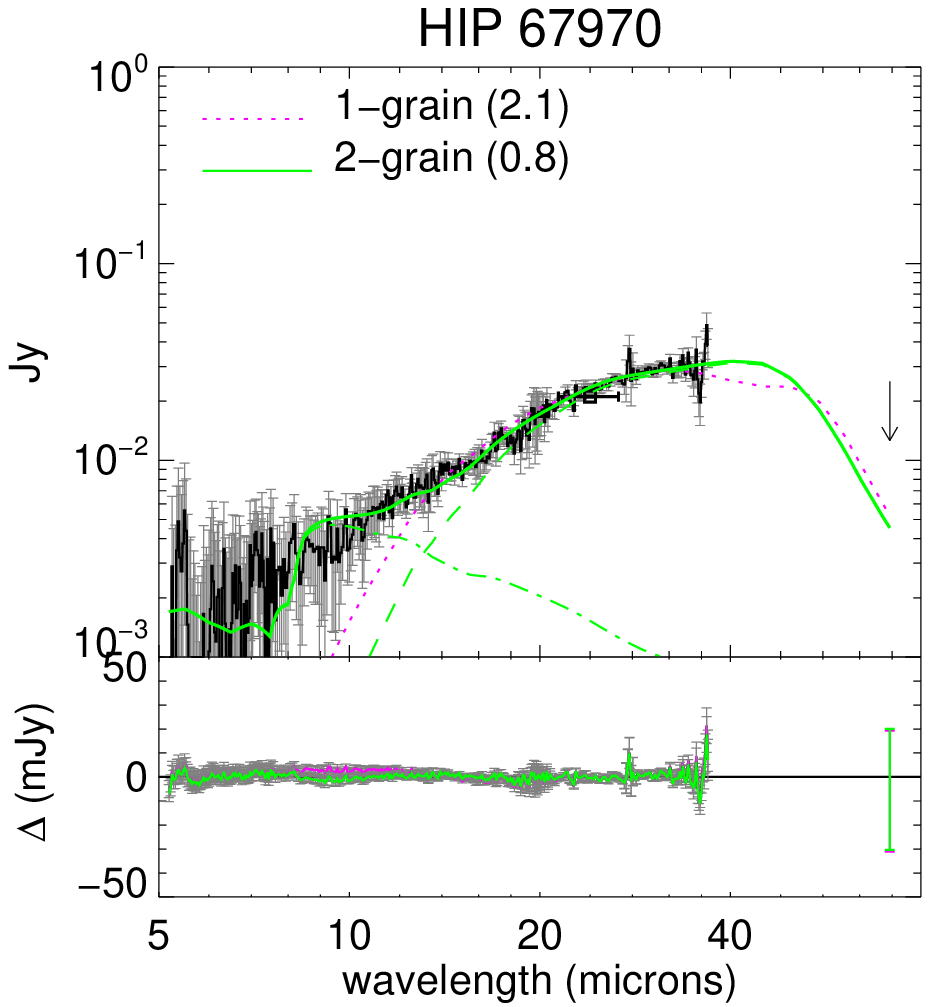} }
\parbox{\stampwidth}{
\includegraphics[width=\stampwidth]{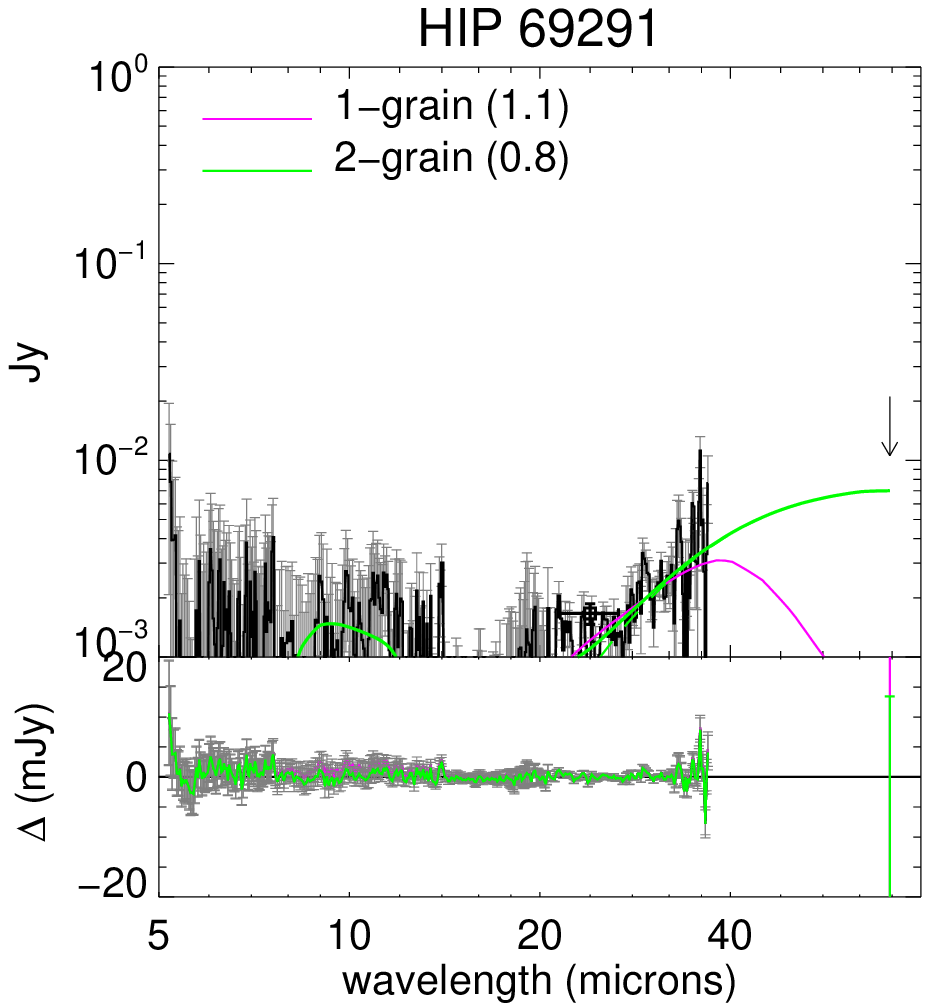} }
\parbox{\stampwidth}{
\includegraphics[width=\stampwidth]{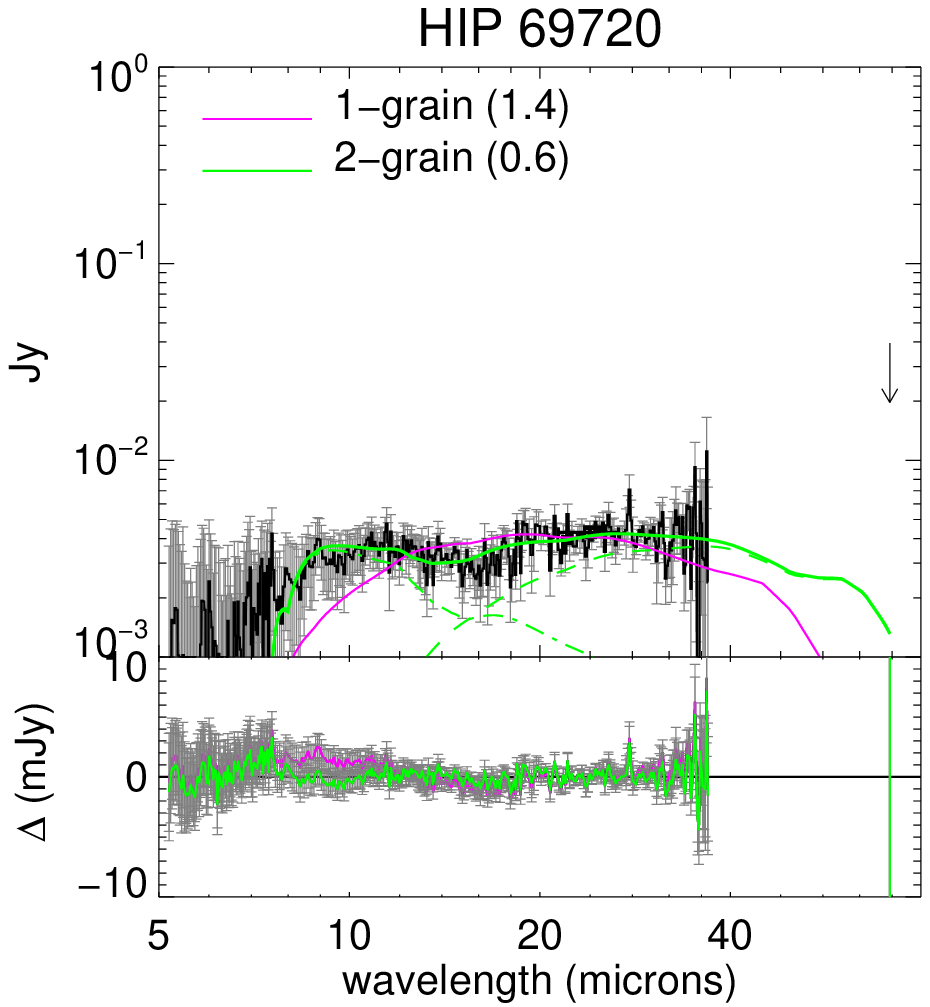} }
\\
\parbox{\stampwidth}{
\includegraphics[width=\stampwidth]{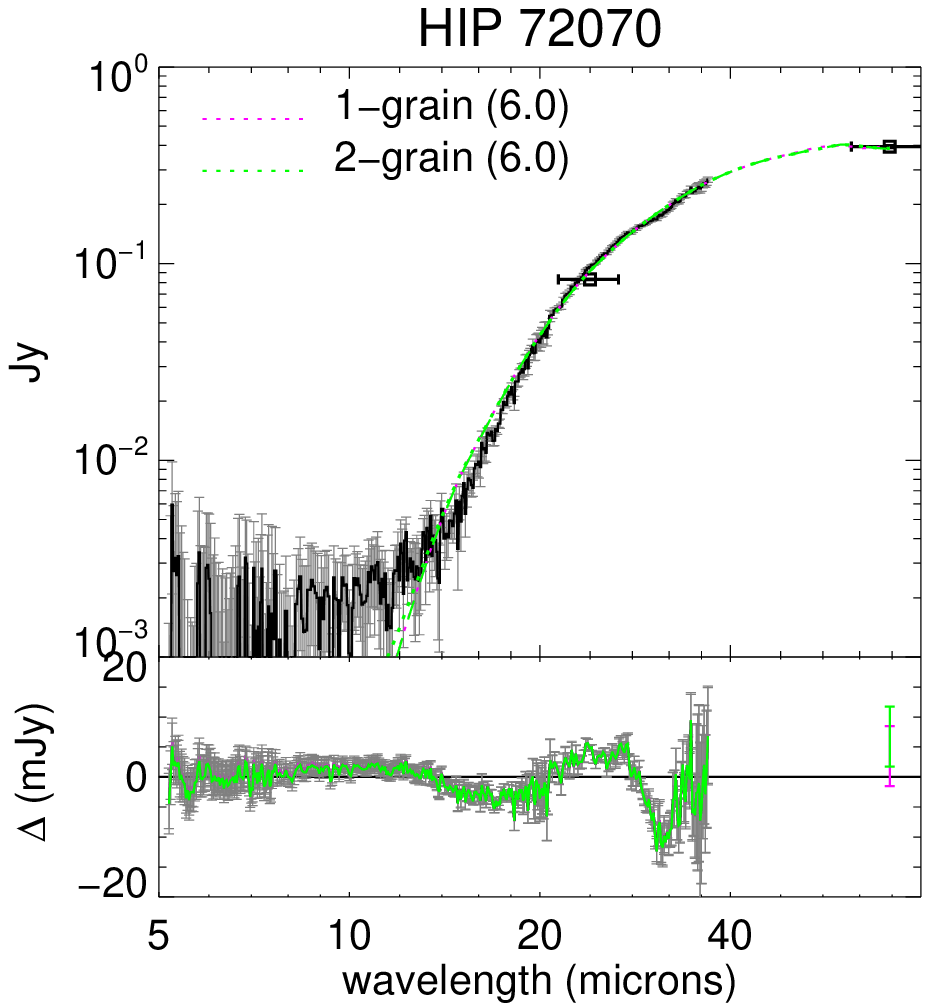} }
\parbox{\stampwidth}{
\includegraphics[width=\stampwidth]{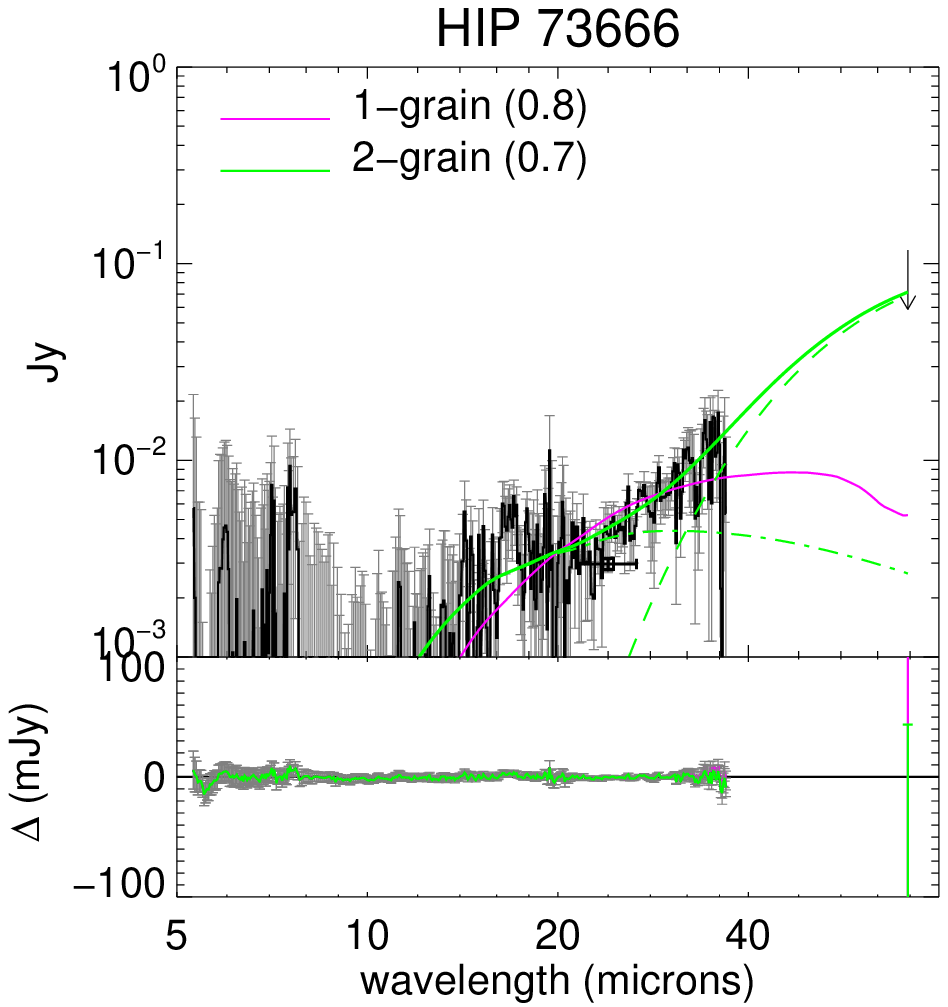} }
\parbox{\stampwidth}{
\includegraphics[width=\stampwidth]{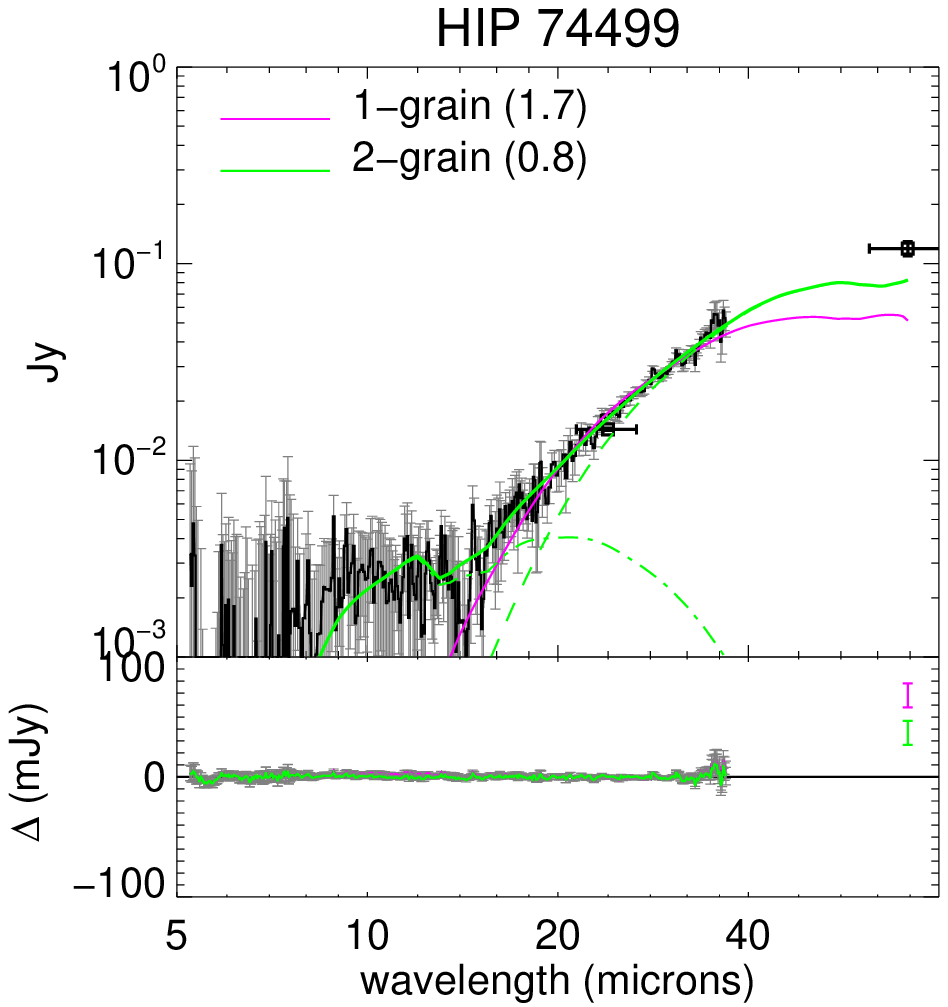} }
\parbox{\stampwidth}{
\includegraphics[width=\stampwidth]{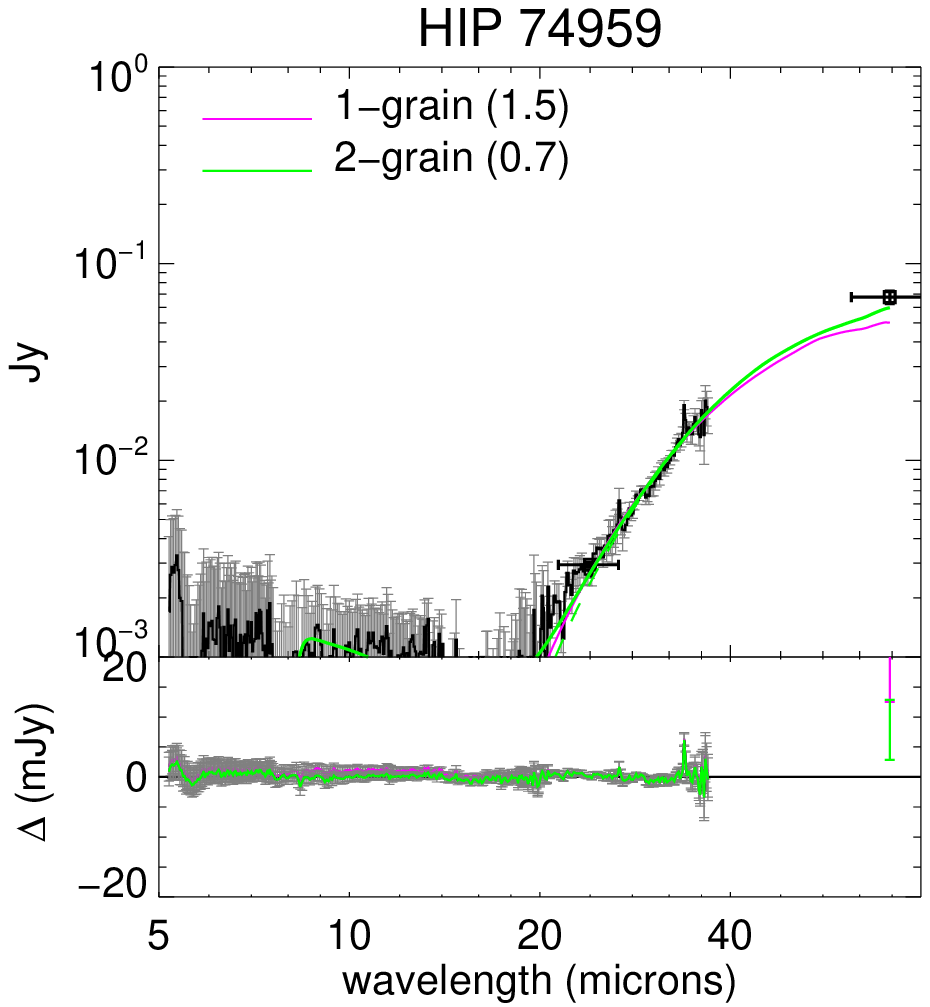} }
\\
\parbox{\stampwidth}{
\includegraphics[width=\stampwidth]{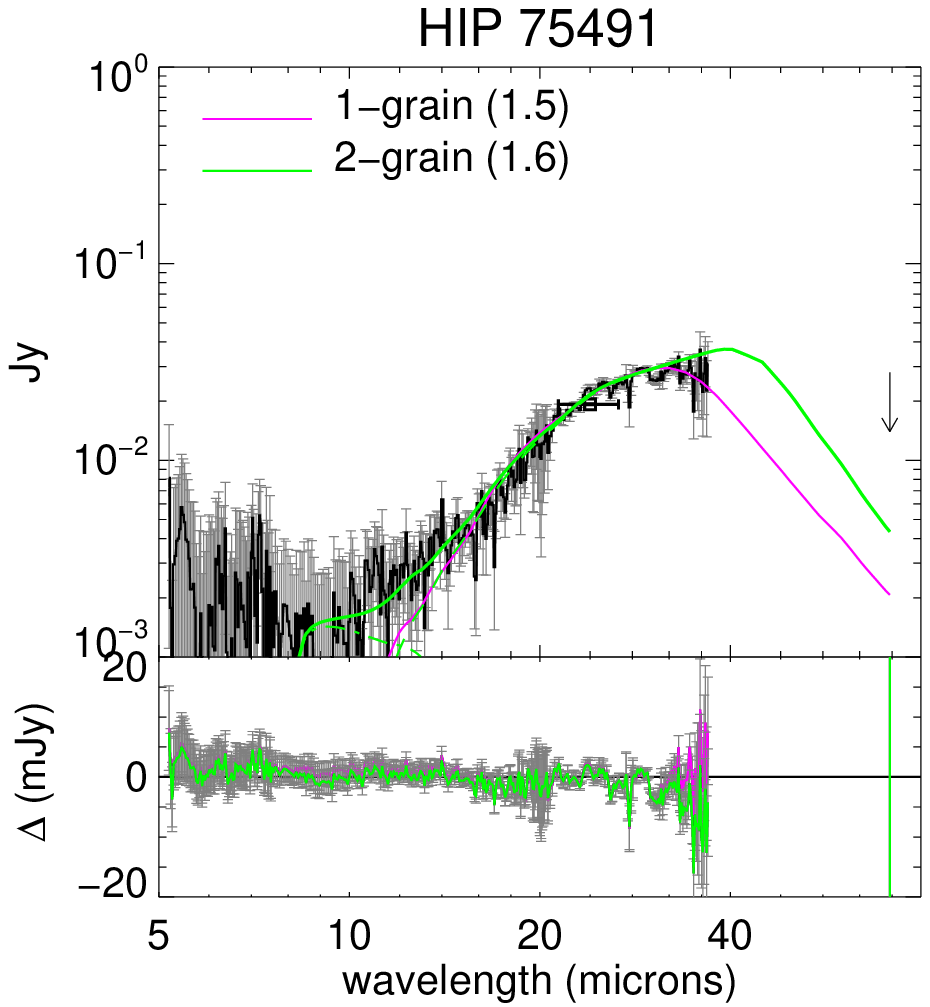} }
\parbox{\stampwidth}{
\includegraphics[width=\stampwidth]{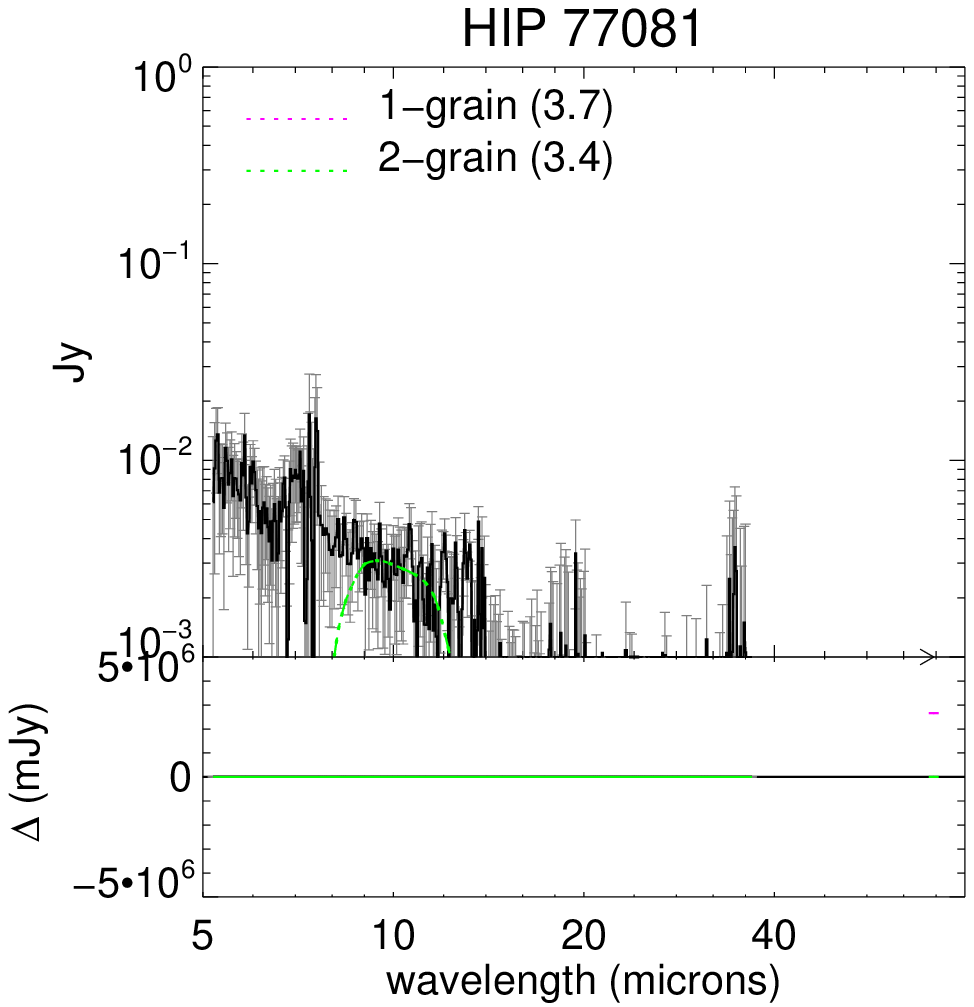} }
\parbox{\stampwidth}{
\includegraphics[width=\stampwidth]{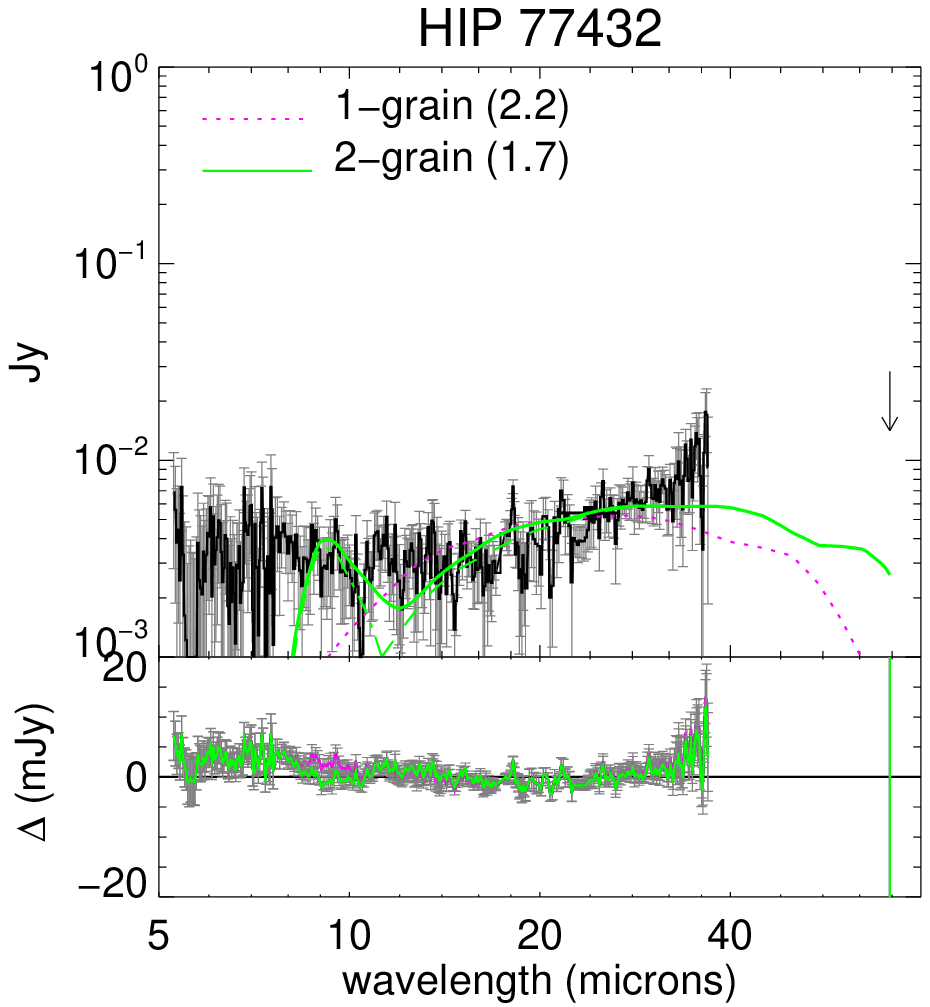} }
\parbox{\stampwidth}{
\includegraphics[width=\stampwidth]{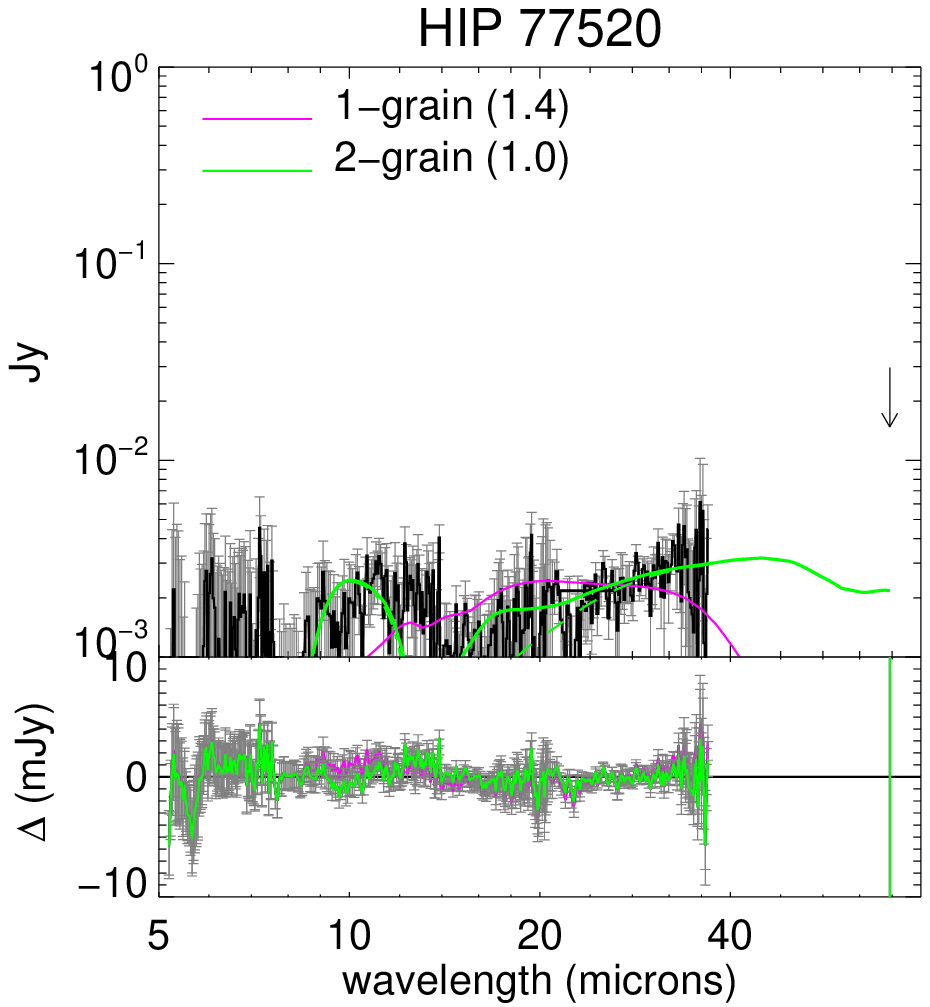} }
\\
\caption{ \label{fitfig3}
Continuation Figure \ref{fitfig0}.}
\end{figure}
\addtocounter{figure}{-1}
\stepcounter{subfig}
\begin{figure}
\parbox{\stampwidth}{
\includegraphics[width=\stampwidth]{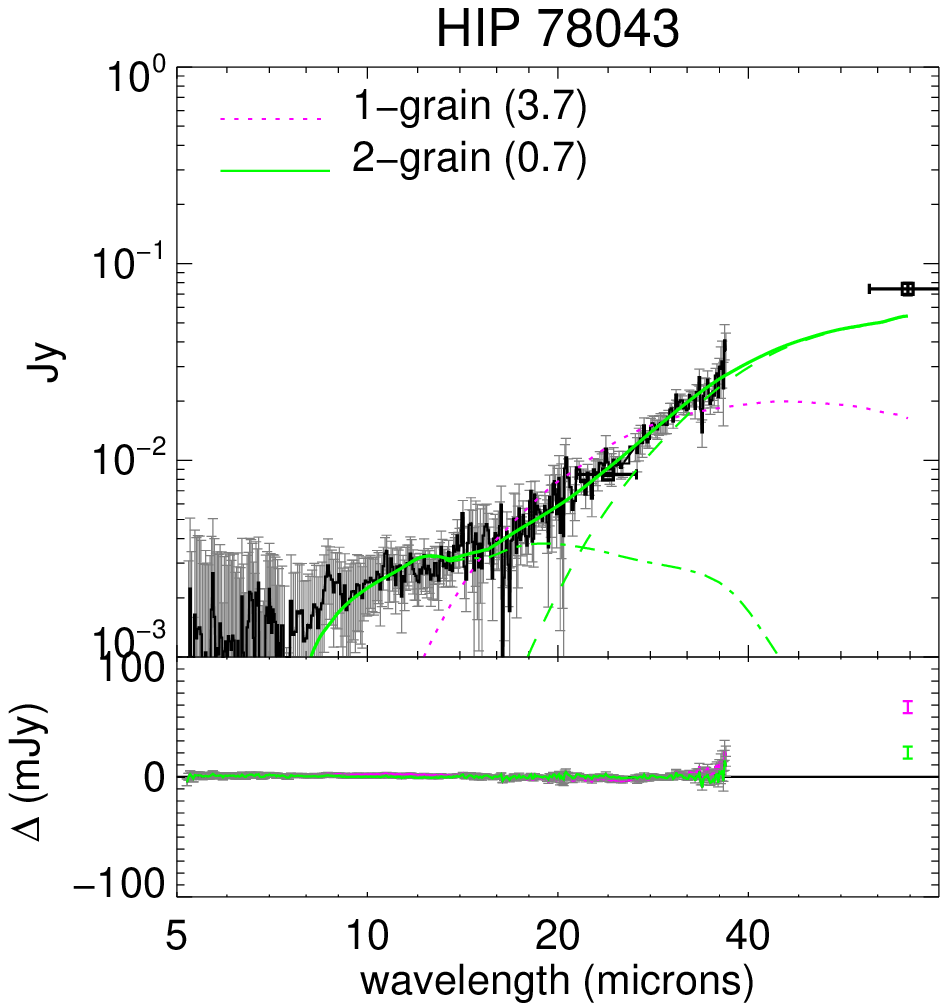} }
\parbox{\stampwidth}{
\includegraphics[width=\stampwidth]{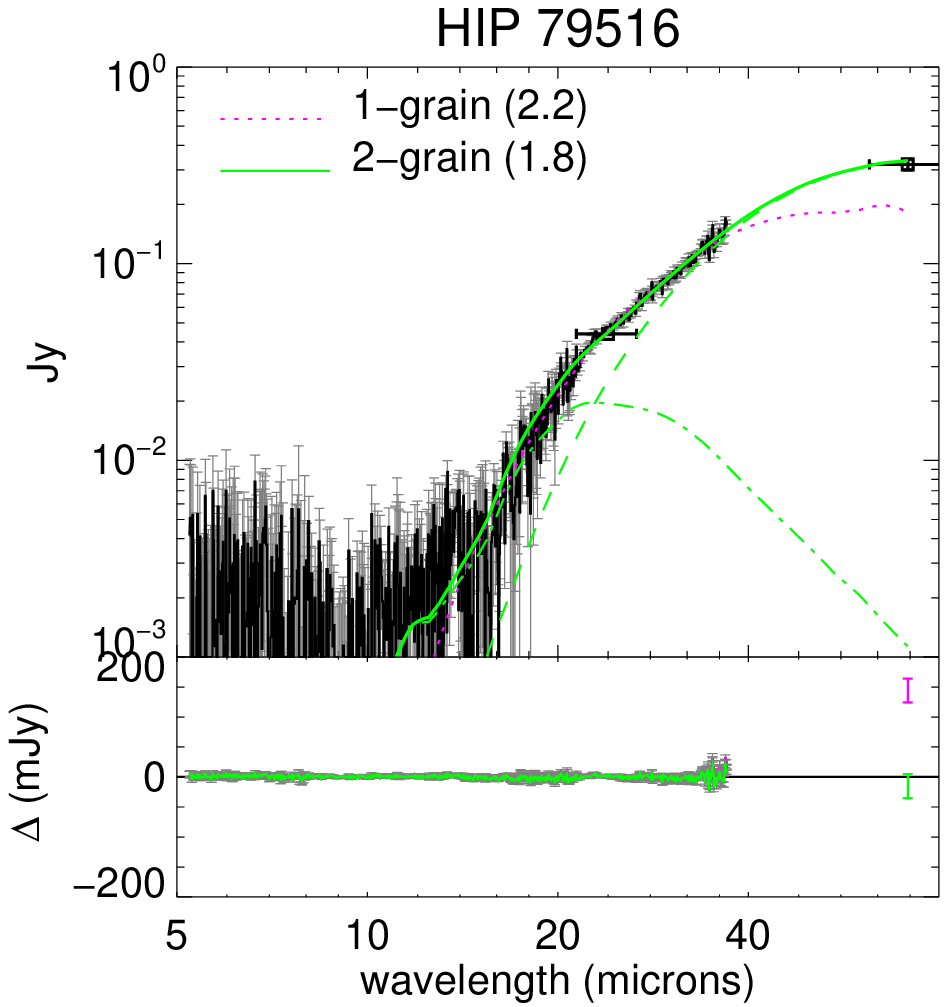} }
\parbox{\stampwidth}{
\includegraphics[width=\stampwidth]{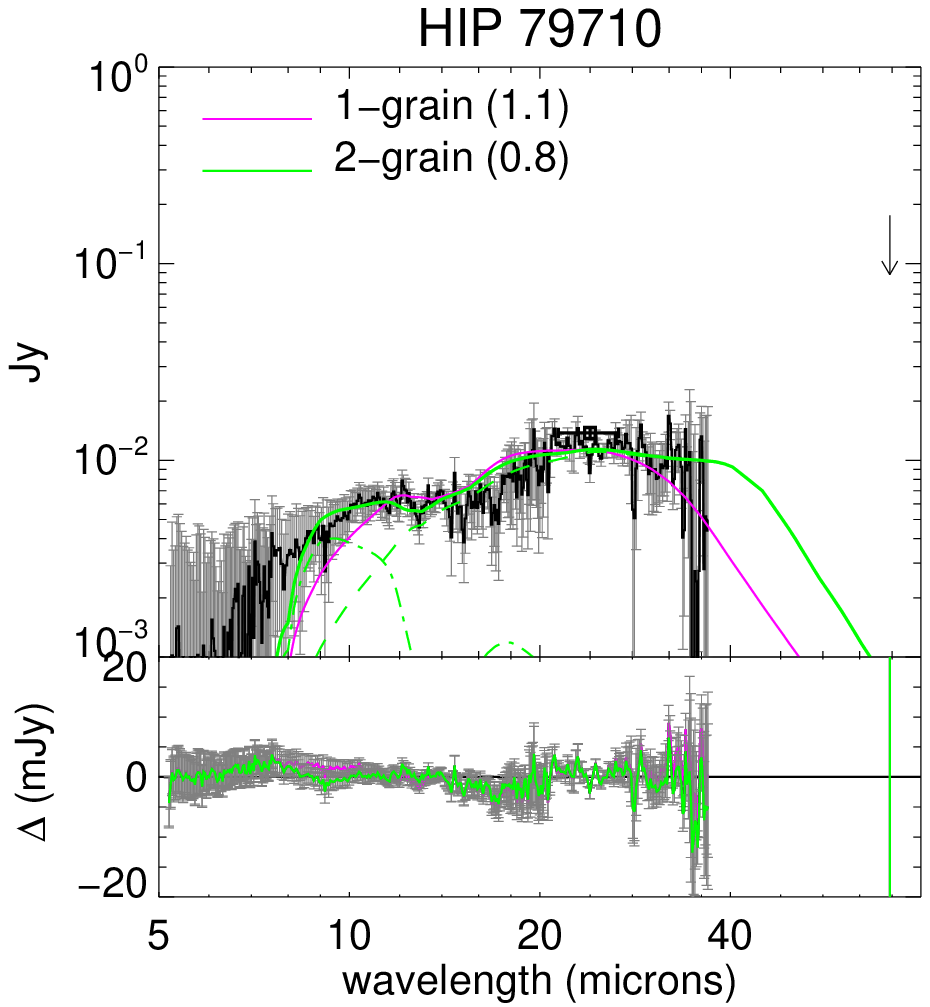} }
\parbox{\stampwidth}{
\includegraphics[width=\stampwidth]{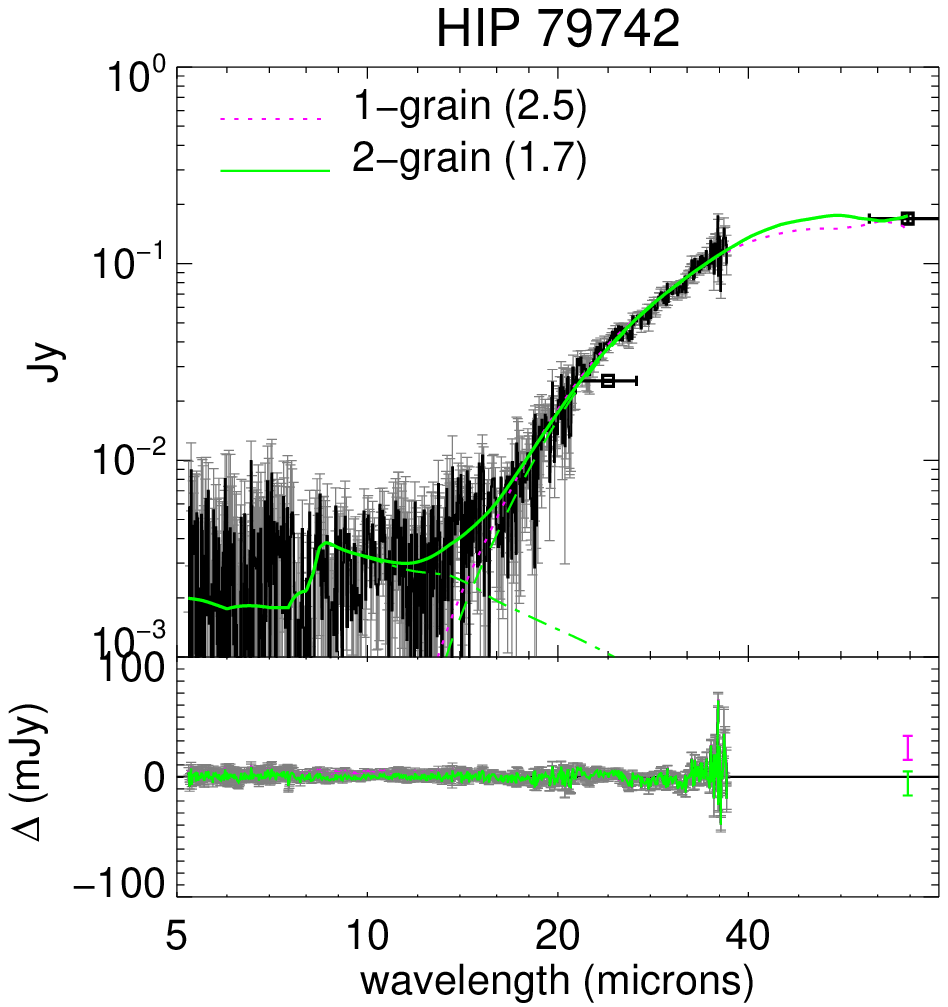} }
\\
\parbox{\stampwidth}{
\includegraphics[width=\stampwidth]{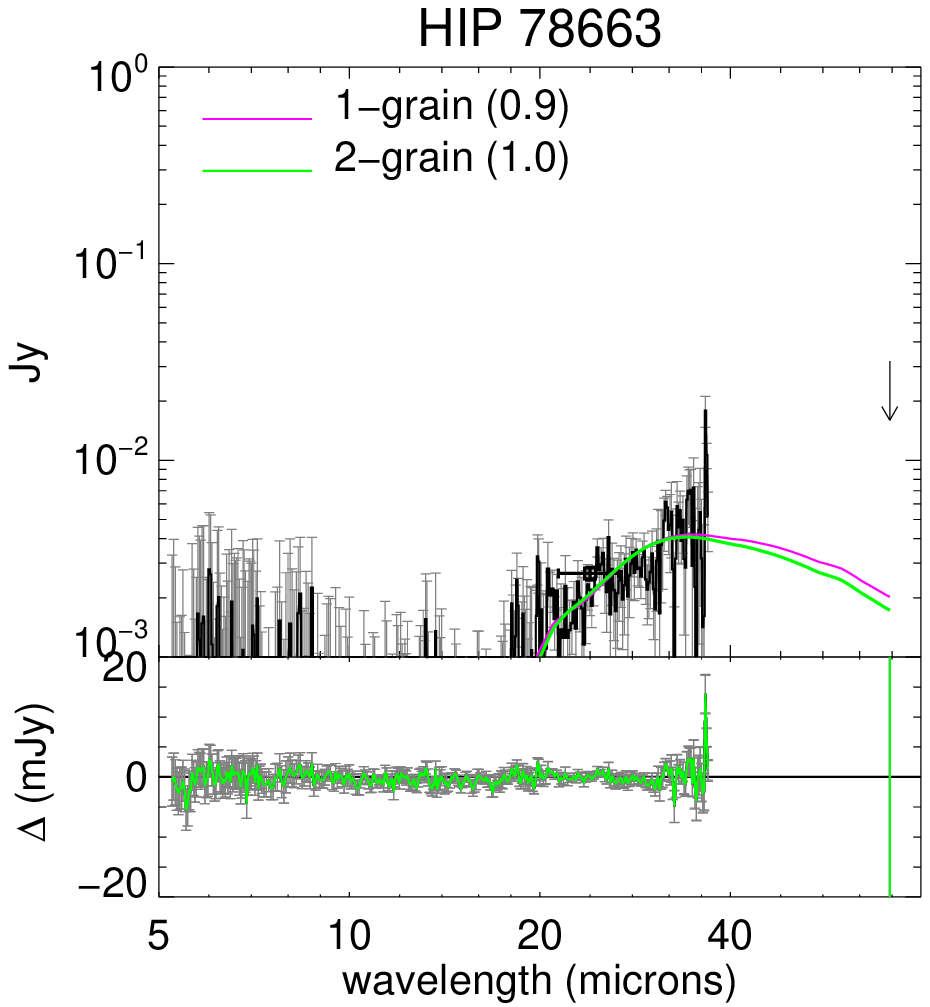} }
\parbox{\stampwidth}{
\includegraphics[width=\stampwidth]{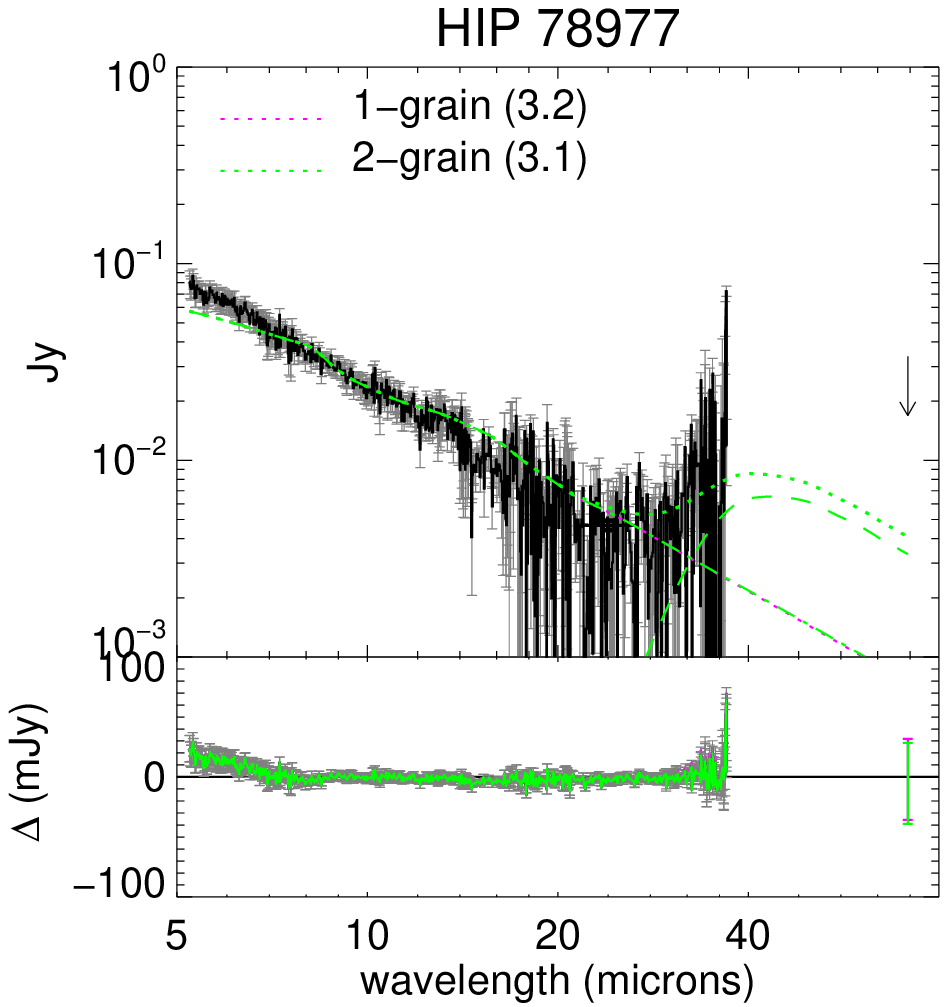} }
\parbox{\stampwidth}{
\includegraphics[width=\stampwidth]{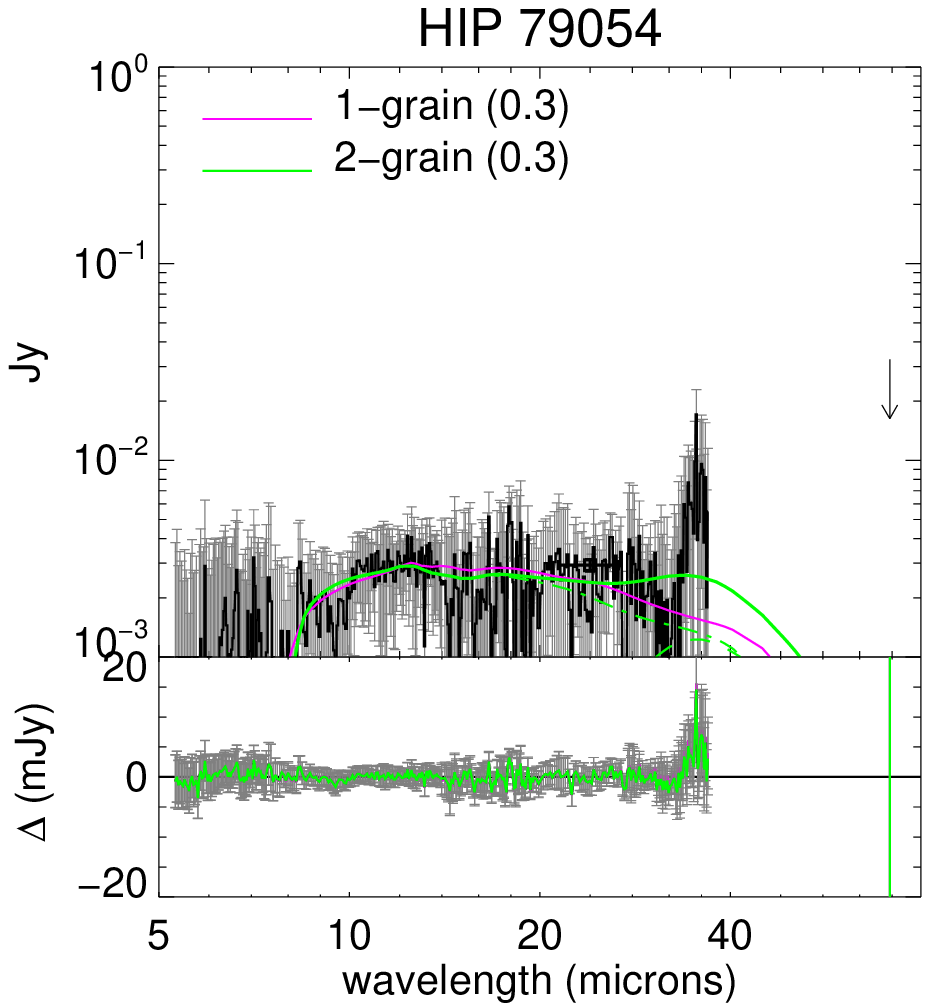} }
\parbox{\stampwidth}{
\includegraphics[width=\stampwidth]{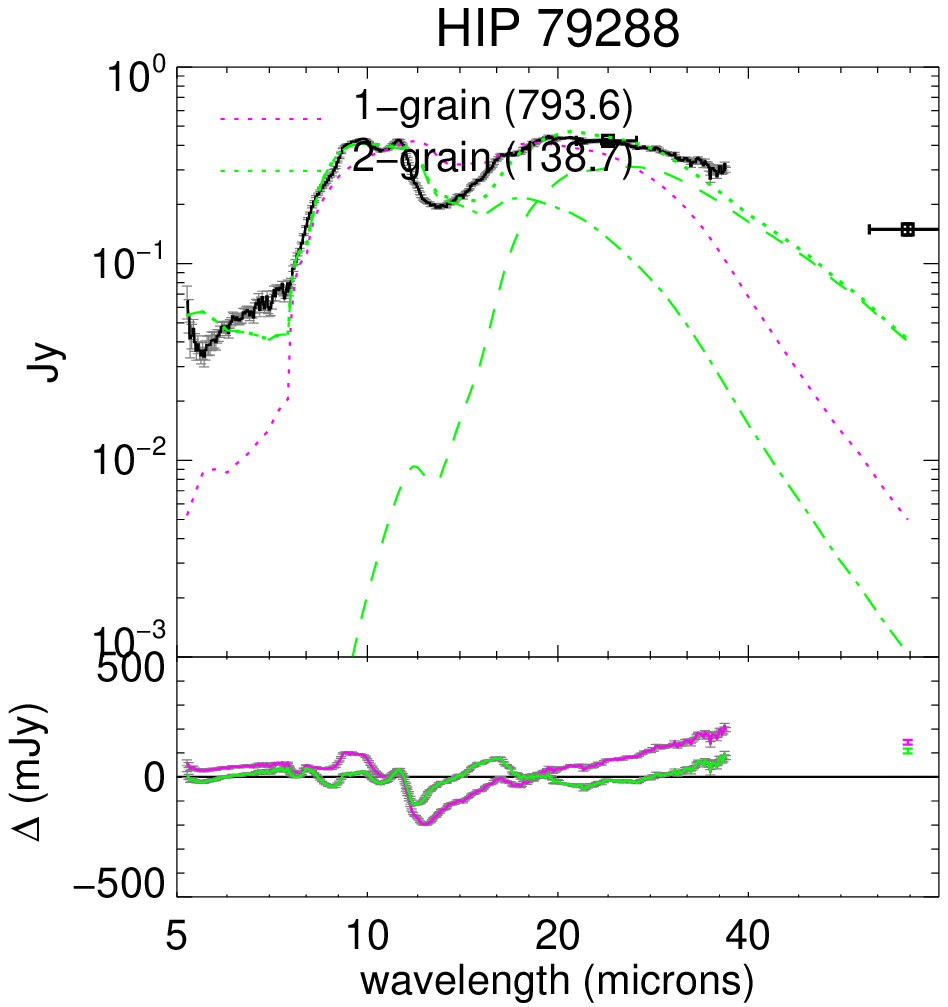} }
\\
\parbox{\stampwidth}{
\includegraphics[width=\stampwidth]{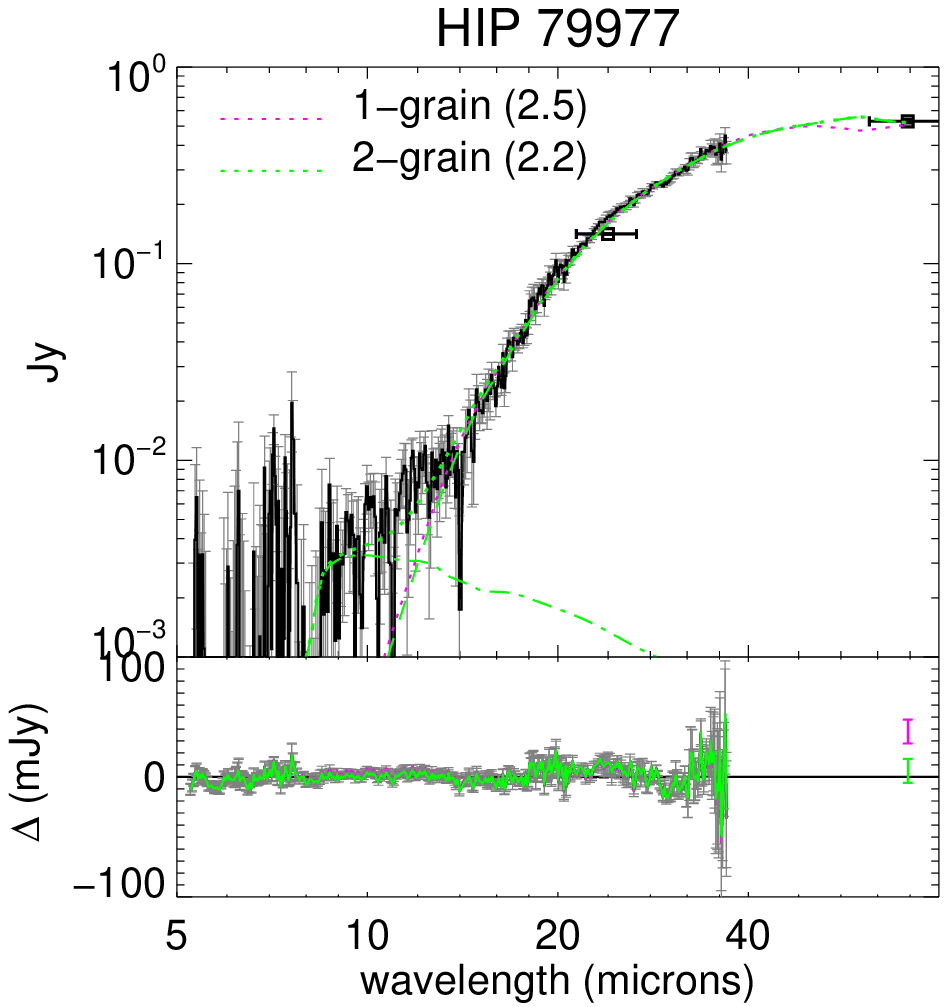} }
\parbox{\stampwidth}{
\includegraphics[width=\stampwidth]{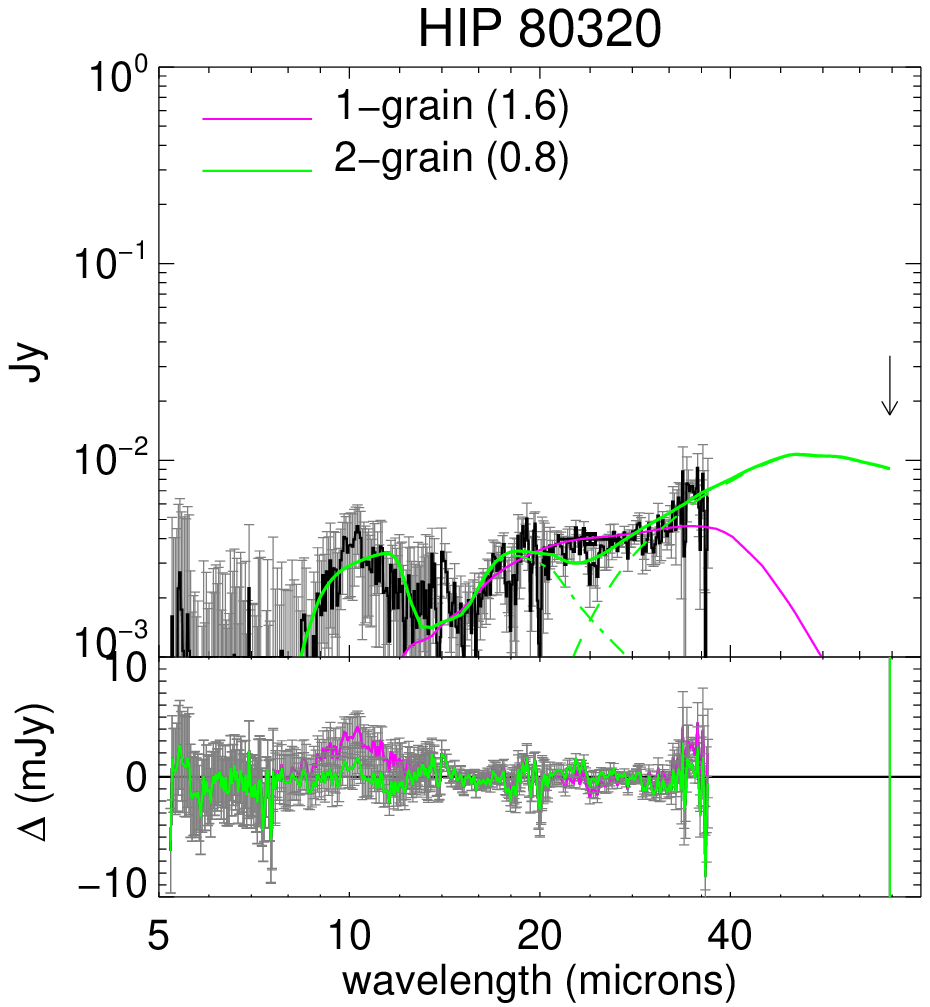} }
\parbox{\stampwidth}{
\includegraphics[width=\stampwidth]{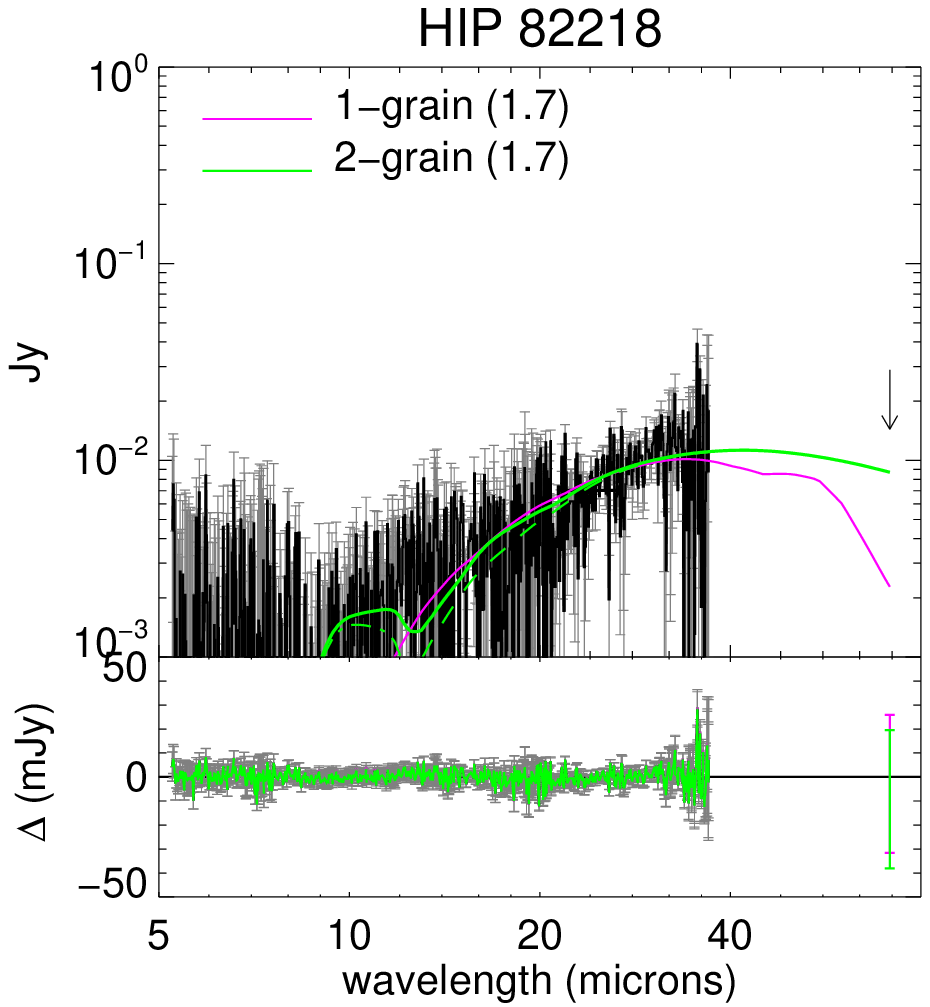} }
\parbox{\stampwidth}{
\includegraphics[width=\stampwidth]{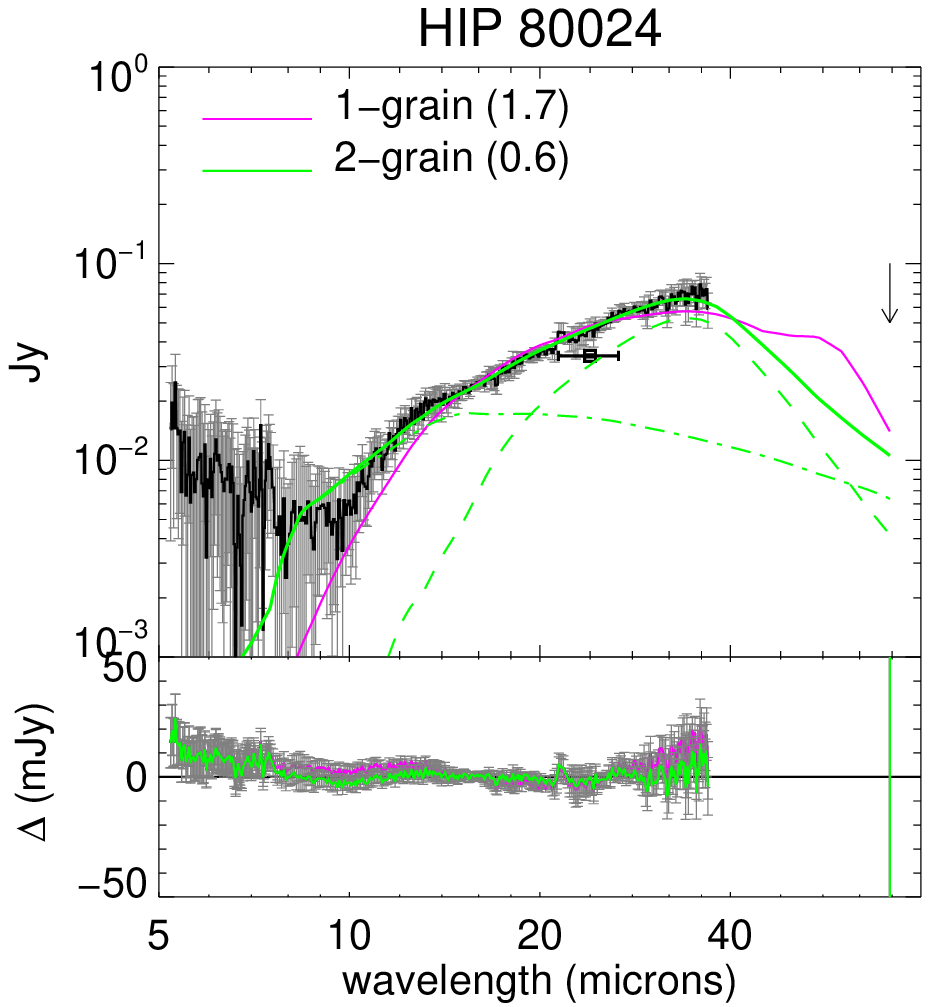} }
\\
\parbox{\stampwidth}{
\includegraphics[width=\stampwidth]{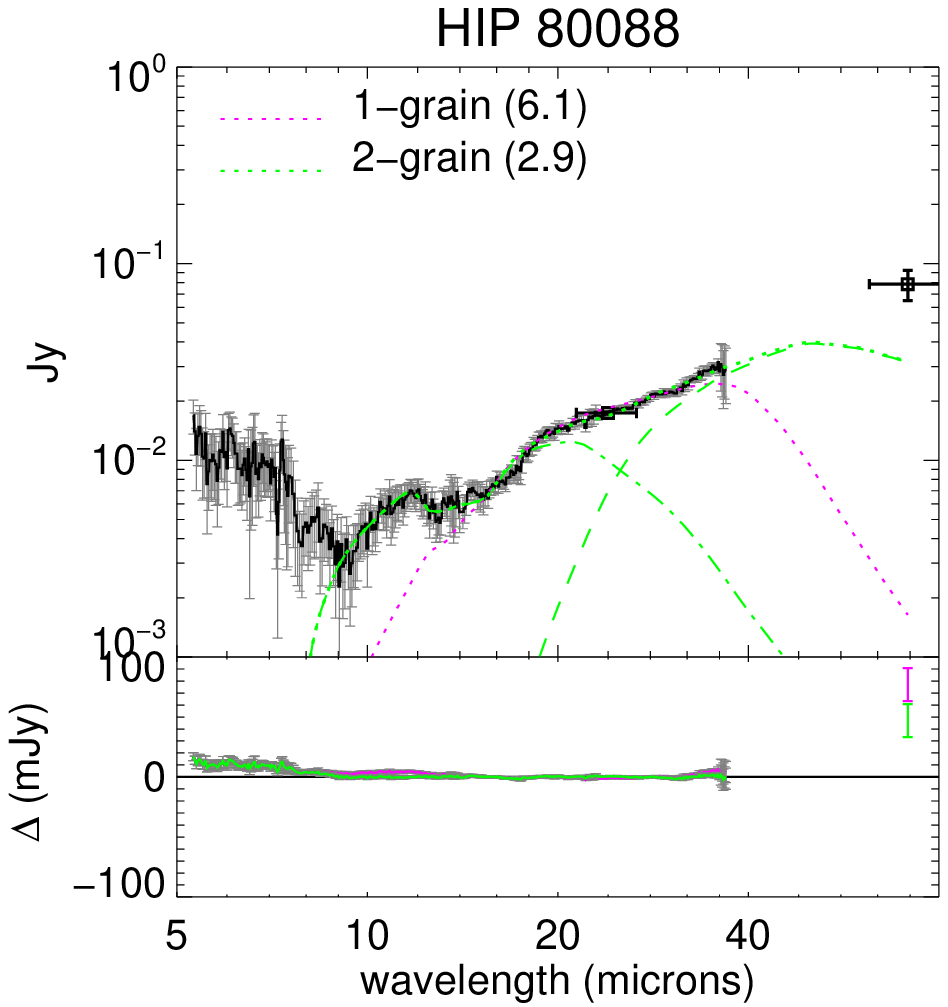} }
\parbox{\stampwidth}{
\includegraphics[width=\stampwidth]{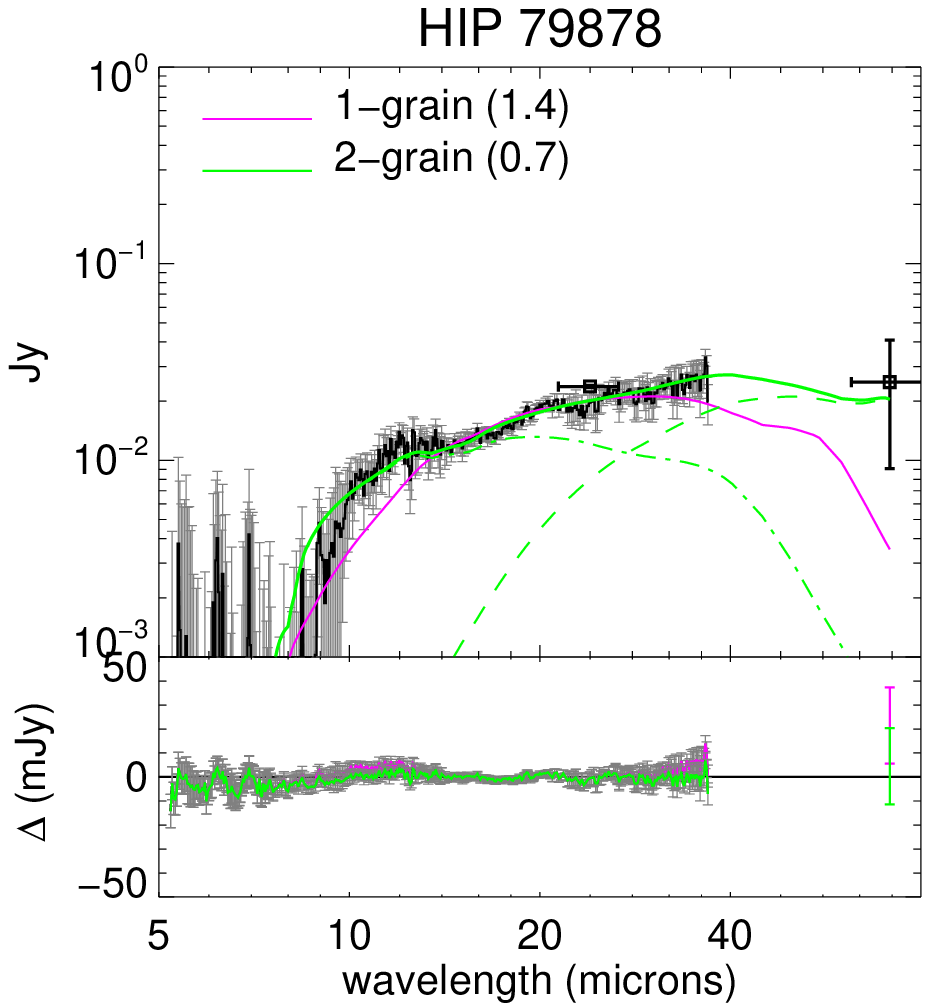} }
\parbox{\stampwidth}{
\includegraphics[width=\stampwidth]{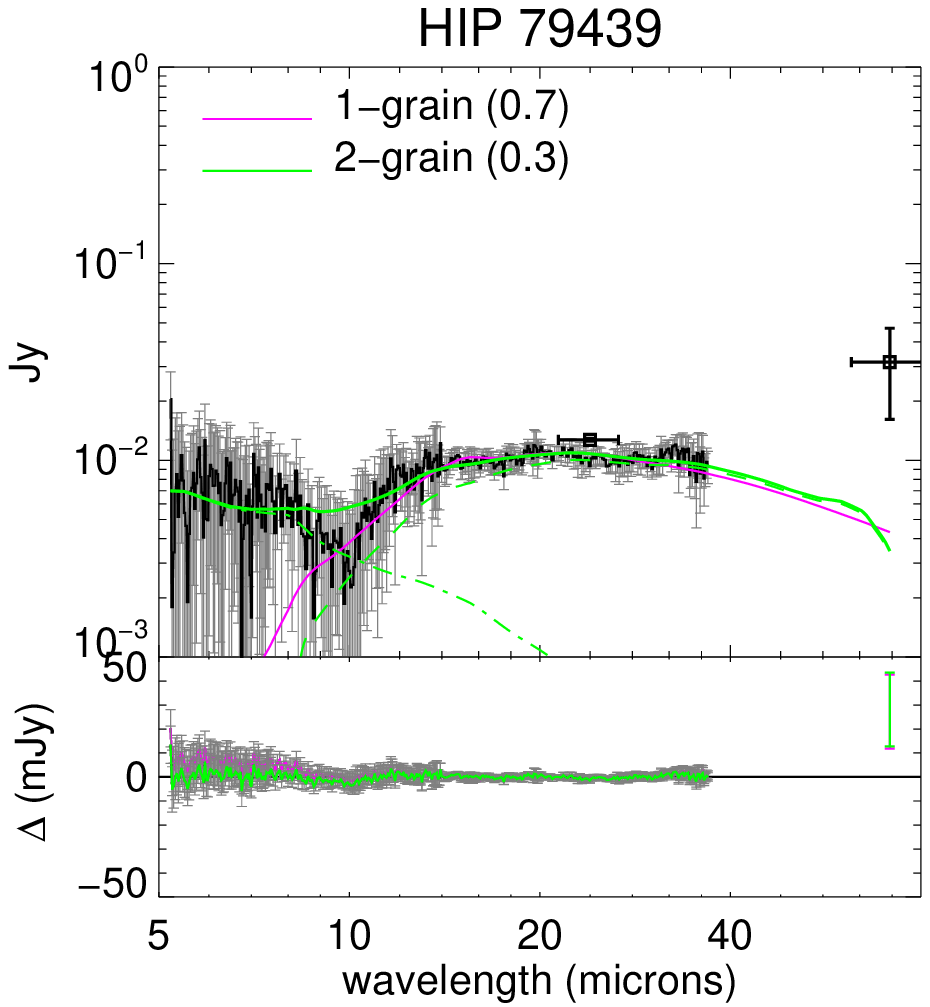} }
\parbox{\stampwidth}{
\includegraphics[width=\stampwidth]{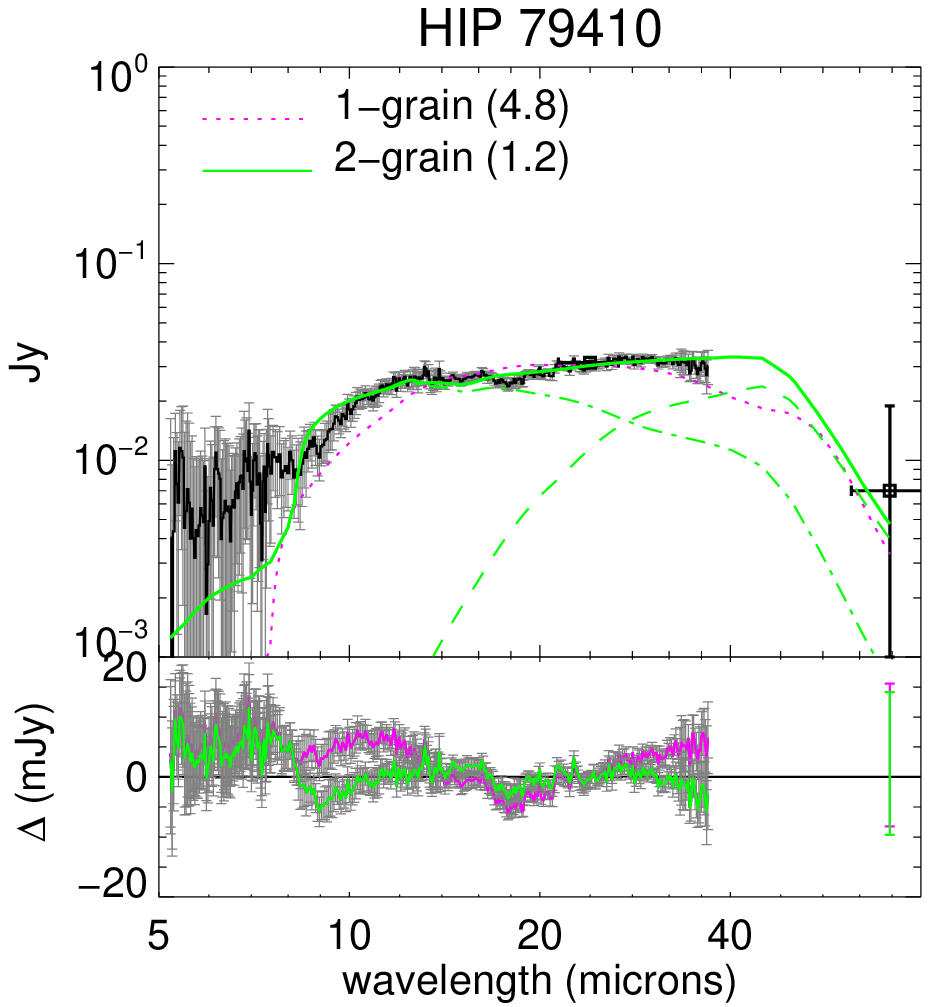} }
\\
\parbox{\stampwidth}{
\includegraphics[width=\stampwidth]{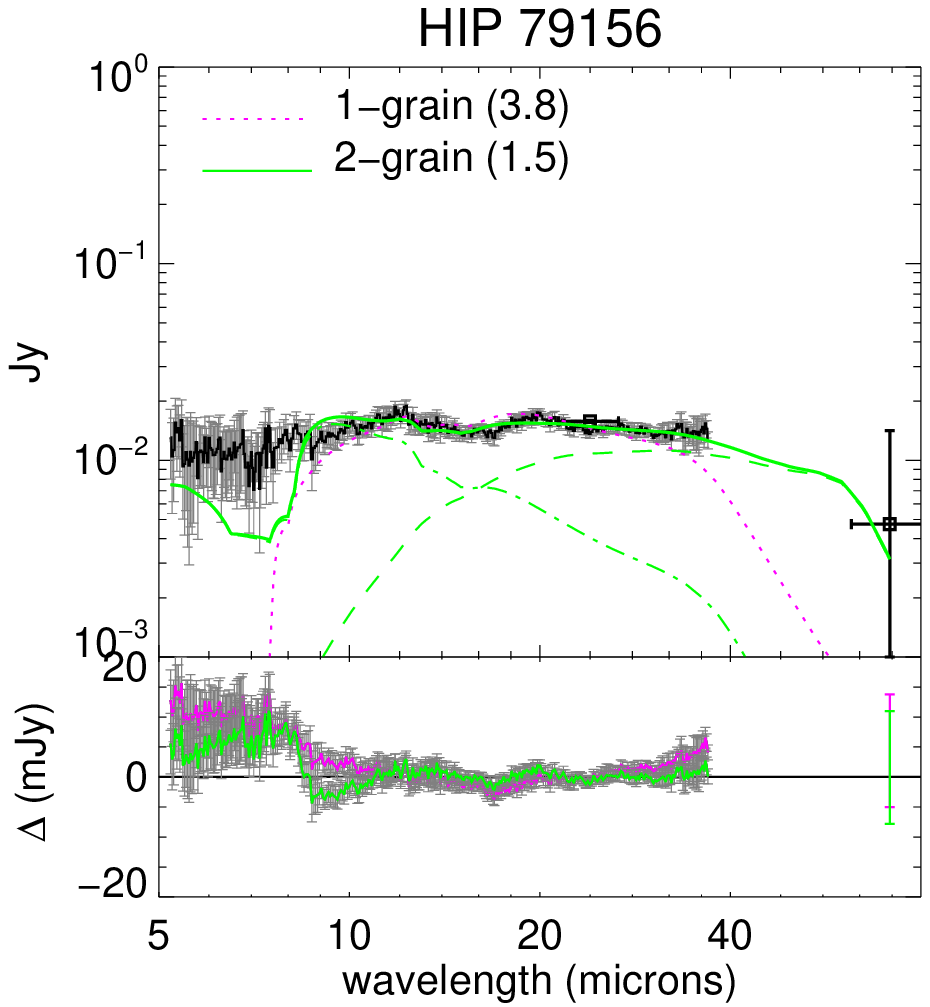} }
\parbox{\stampwidth}{
\includegraphics[width=\stampwidth]{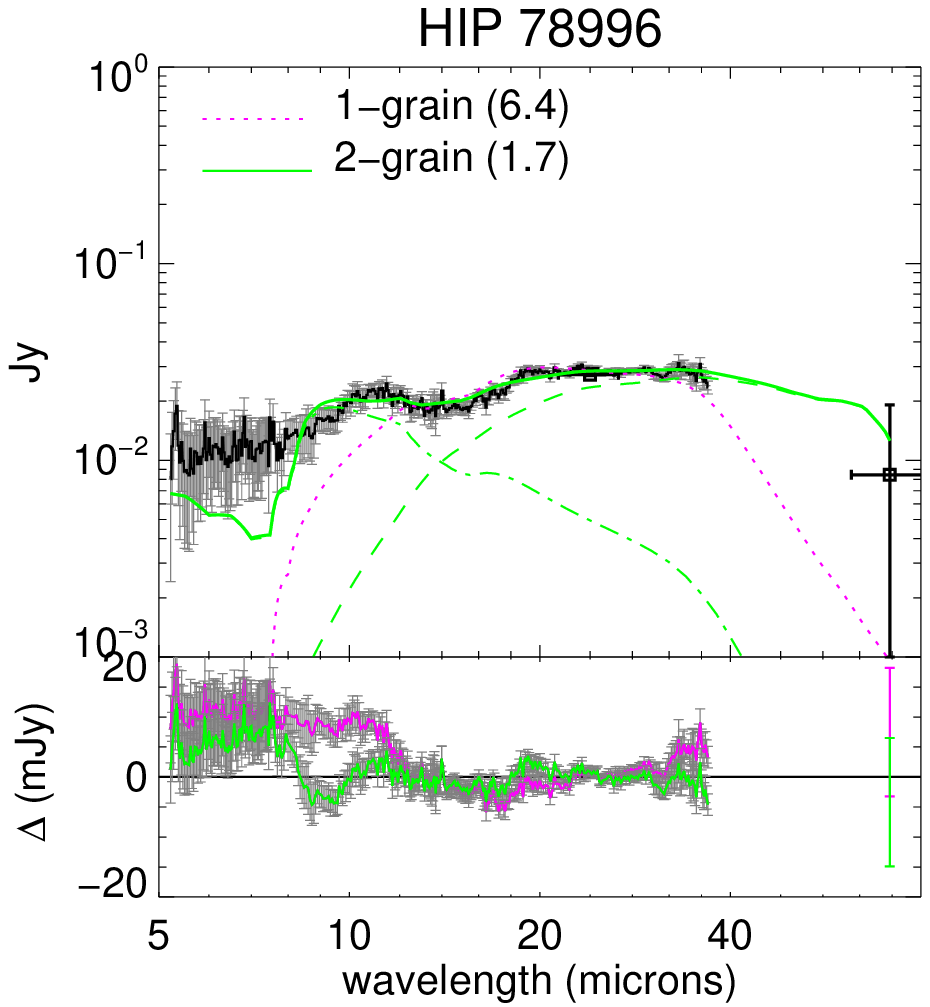} }
\parbox{\stampwidth}{
\includegraphics[width=\stampwidth]{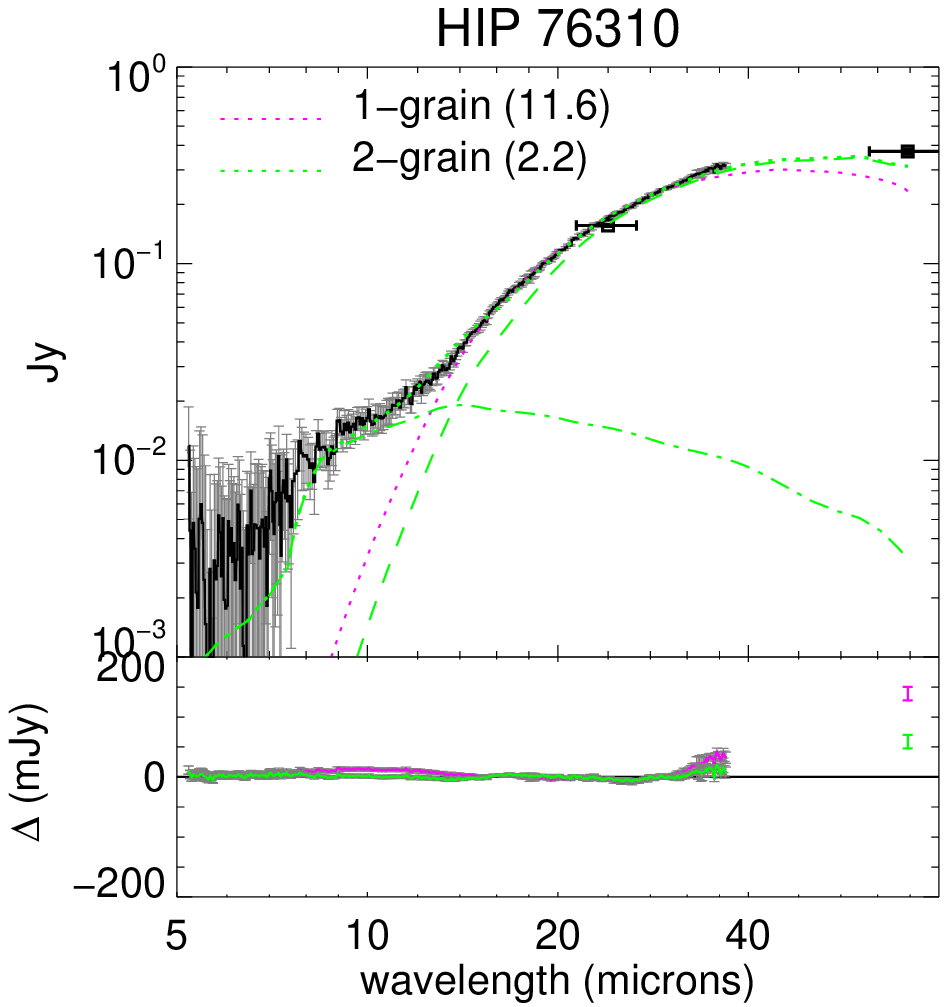} }
\parbox{\stampwidth}{
\includegraphics[width=\stampwidth]{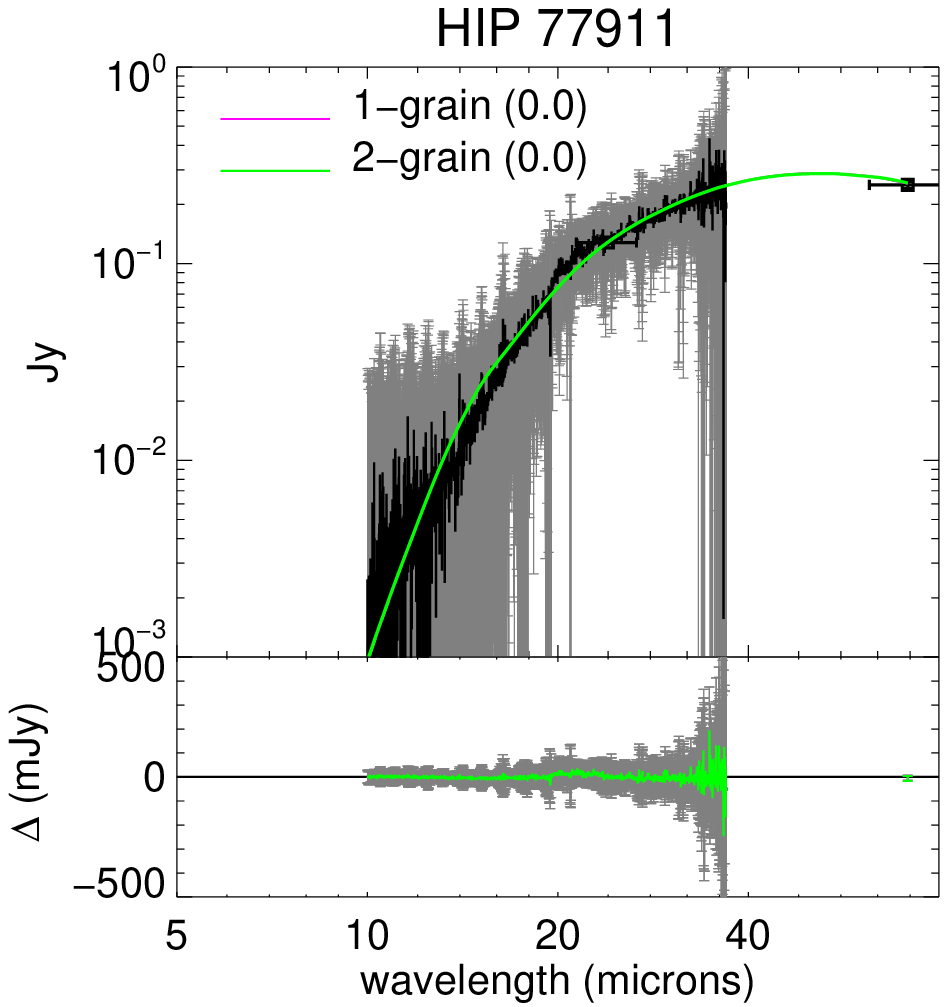} }
\\
\caption{ \label{fitfig4}
Continuation Figure \ref{fitfig0}.}
\end{figure}

\renewcommand{\thefigure}{\arabic{figure}}


If the single grain model is a reasonable fit for a given object, 
then that object is considered a single-belt debris system.  
If the single grain model is not a reasonable fit for the 
object but the two-grain model is, then we consider that object 
a two-belt debris system.  
Upon visual inspection, we found 10 objects whose formal 
$\chi_{\nu}^2$ values for a single grain model 
are less than 2, but have much better fits to the two-grain 
model, and are best explained by noisiness of the spectra.  
We re-categorize these objects as two-belt systems.  
Similarly, 5 objects have formal $\chi_{\nu}^2$ greater 
than 2 for a single grain model, but are not significantly 
better fit by a two grain model, and we re-classify these 
as single belt systems.  An additional 11 systems have 
$\chi_{\nu}^2$ for both fits larger than 2, but are mostly well-fit 
by a two grain model, and we re-categorize them as two belt 
systems.  Most of the objects in the latter two groups have 
mismatches at the short wavelength part of the spectrum, which is 
most affected by the normalization of the stellar photosphere.  
HIP 55188 (Figure \ref{fitfig0}) is one of the systems for which 
neither model produces a good fit, although it is clear that 
the two-grain model produceds a much better fit than 
the single-grain model for this particular case.  For each 
fit, the grain temperature, grain size, and amorphous silicate 
composition are allowed to be free parameters.  
(Note: HIP 55188 is well-fit if we include crystalline silicates.)
In total, we have 48 objects that are single-belt systems and 
44 that are two-belt systems.  

\begin{deluxetable}{lcr@{$\pm$}lr@{$\pm$}lcr@{$\pm$}lr@{$\pm$}l}
\tabletypesize{\footnotesize}
\tablecaption{\label{tab:fits1}Fits to single grain model}
\tablewidth{0pt}
\tablehead{
  \colhead{HIP ID} & 
  \colhead{$\chi_{\nu}^2$} &
  \multicolumn{2}{c}{$T$\sub{gr}} &
  \multicolumn{2}{c}{$a$\sub{gr}} &
  \colhead{mass} &
  \multicolumn{2}{c}{$f_o$} &
  \multicolumn{2}{c}{$r$\sub{gr}}
  \\
  \colhead{} & 
  \colhead{} &
  \multicolumn{2}{c}{(K)} &
  \multicolumn{2}{c}{($\mu$m)} &
  \colhead{($M\sub{moon}$)} &
  \multicolumn{2}{c}{} &
  \multicolumn{2}{c}{(AU)}
}
\startdata
${}^\ddagger$HIP 56673 & 0.93 & 267 & 12 & 7.59 & 0.83 &   1.75e-05 &   0.00 &  0.00 & 3.69 & 0.33\\
{*}HIP 56673 & 2.19 & 1480 & 40 & 198 & 8 &   9.38e-05 &   0.00 &  0.00 & 0.151 & 0.02\\
HIP 57524 & 1.41 & 246 & 4 & 85 & 4.3 &   2.12e-04 &   0.00 &  0.00 & 1.73 & 0.12\\
HIP 57950 & 1.14 & 203 & 2 & 8.07 & 0.25 &   4.39e-05 &   0.00 &  0.00 & 2.93 & 0.36\\
HIP 58528 & 1.39 & 274 & 1 & 4.68 & 0.08 &   1.77e-05 &   0.00 &  0.00 & 1.63 & 0.17\\
HIP 59282 & 1.50 & 274 & 2 & 6.37 & 0.12 &   2.94e-05 &   0.00 &  0.00 & 2.92 & 0.26\\
HIP 59397 & 1.27 & 216 & 1 & 5.24 & 0.05 &   1.07e-04 &   0.00 &  0.00 & 5.15 & 0.43\\
HIP 59481 & 0.47 & 176 & 4 & 4.98 & 0.42 &   1.03e-05 &   0.00 &  0.00 & 4.1 & 0.55\\
HIP 59693 & 1.11 & 438 & 4 & 4.28 & 0.1 &   3.41e-06 &   0.54 &  0.11 & 0.618 & 0.07\\
HIP 59960 & 1.29 & 108 & 0.3 & 11.2 & 0.1 &   1.20e-02 &   1.00 &  0.00 & 12.3 & 3.4\\
HIP 60348 & 1.95 & 182 & 2 & 8.87 & 0.23 &   5.64e-05 &   0.00 &  0.00 & 2.65 & 0.13\\
HIP 61049 & 0.92 & 250 & 0.5 & 4.9 & 0.03 &   5.47e-05 &   0.00 &  0.00 & 1.74 & 0.5\\
HIP 61087 & 1.85 & 208 & 1 & 3.27 & 0.07 &   7.75e-05 &   0.13 &  0.11 & 3.49 & 0.38\\
HIP 62134 & 0.60 & 175 & 6 & 1030 & 160 &   1.02e-02 &   1.00 &  0.00 & 5.3 & 1\\
HIP 62427 & 0.91 & 112 & 1 & 37.2 & 2 &   7.02e-03 &   1.00 &  0.00 & 10 & 1.4\\
HIP 63236 & 1.53 & 235 & 1 & 5.22 & 0.03 &   1.00e-04 &   0.00 &  0.00 & 5.2 & 0.61\\
HIP 63836 & 1.26 & 227 & 4 & 8.92 & 0.36 &   1.79e-05 &   0.00 &  0.00 & 1.81 & 0.3\\
HIP 64053 & 0.50 & 294 & 1 & 9.52 & 0.1 &   1.08e-04 &   0.00 &  0.00 & 6.61 & 0.87\\
HIP 64877 & 1.00 & 167 & 2 & 8.85 & 0.35 &   2.94e-04 &   0.00 &  0.00 & 4.83 & 0.48\\
HIP 64995 & 1.24 & 114 & 1 & 9.48 & 0.29 &   1.14e-02 &   0.97 &  0.28 & 10.7 & 1.4\\
{*}HIP 65089 & 2.13 & 93.1 & 0.7 & 0.102 & 0.01 &   1.70e-03 &   0.59 &  0.07 & 131 & 15\\
HIP 66447 & 0.83 & 114 & 1 & 10 & 0.3 &   2.04e-03 &   0.84 &  0.22 & 17.7 & 2\\
HIP 67068 & 0.63 & 226 & 8 & 4.48 & 0.44 &   4.62e-06 &   0.25 &  0.70 & 2.07 & 0.36\\
HIP 67497 & 1.53 & 124 & 0.4 & 15.5 & 0.2 &   9.96e-03 &   1.00 &  0.00 & 8.82 & 1\\
HIP 68781 & 1.06 & 276 & 2 & 4.9 & 0.08 &   1.36e-05 &   0.00 &  0.00 & 2.56 & 0.35\\
HIP 69291 & 1.09 & 102 & 2 & 5.89 & 0.43 &   1.71e-04 &   1.00 &  0.00 & 15.3 & 1.7\\
HIP 69720 & 1.36 & 285 & 4 & 7.92 & 0.27 &   1.63e-05 &   0.00 &  0.00 & 1.77 & 0.26\\
HIP 70455 & 0.94 & 216 & 2 & 8.93 & 0.19 &   1.95e-04 &   0.00 &  0.00 & 11.5 & 1.3\\
HIP 71453 & 0.43 & 224 & 6 & 119 & 12 &   2.18e-03 &   0.80 &  0.71 & 21.1 & 4.1\\
{*}HIP 72070 & 5.98 & 93.7 & 0.1 & 16.2 & 0.1 &   7.98e-02 &   1.00 &  0.00 & 12.2 & 2.3\\
HIP 73341 & 0.44 & 242 & 2 & 9.57 & 0.19 &   2.09e-04 &   0.00 &  0.00 & 11.9 & 1.6\\
HIP 73666 & 0.83 & 132 & 4 & 14.8 & 1.5 &   7.65e-04 &   0.00 &  0.00 & 14.5 & 1.5\\
HIP 74752 & 1.63 & 401 & 7 & 138 & 5 &   6.23e-04 &   0.00 &  0.00 & 4.1 & 0.45\\
HIP 74959 & 1.52 & 68.6 & 0.6 & 15.9 & 0.8 &   2.38e-02 &   1.00 &  0.00 & 21.8 & 2.2\\
HIP 75491 & 1.52 & 125 & 0.2 & 5.12 & 0.05 &   1.41e-03 &   0.00 &  0.00 & 13.5 & 1.3\\
HIP 75509 & 1.58 & 251 & 1 & 5.13 & 0.07 &   4.37e-05 &   0.00 &  0.00 & 3.09 & 0.5\\
HIP 76395 & 1.39 & 306 & 1 & 9.53 & 0.07 &   1.13e-04 &   0.00 &  0.00 & 5.32 & 0.73\\
{*}HIP 77432 & 2.17 & 228 & 3 & 8.75 & 0.4 &   2.19e-05 &   0.00 &  0.00 & 1.62 & 0.63\\
HIP 77520 & 1.36 & 243 & 4 & 5.63 & 0.26 &   5.68e-06 &   0.00 &  0.00 & 1.44 & 0.19\\
HIP 77523 & 1.02 & 268 & 3 & 8.96 & 0.22 &   9.43e-05 &   0.00 &  0.00 & 6.56 & 1\\
HIP 77911 & 0.04 & 112 & 1 & 29.4 & 1 &   1.02e-01 &   1.00 &  0.00 & 38.5 & 5.6\\
HIP 78663 & 0.95 & 67 & 0.9 & 2.23 & 0.41 &   6.40e-03 &   0.00 &  0.00 & 91.8 & 9.1\\
HIP 78756 & 0.68 & 204 & 1 & 5.26 & 0.13 &   5.96e-05 &   0.00 &  0.00 & 9.55 & 1\\
{*}HIP 78977 & 3.18 & 1500 & 0 & 2690 & 10 &   8.00e-04 &   0.00 &  0.00 & 0.0781 & 0.01\\
HIP 79400 & 1.24 & 238 & 2 & 9.62 & 0.2 &   7.78e-05 &   0.00 &  0.00 & 3.95 & 0.63\\
HIP 79439 & 0.69 & 236 & 0.3 & 56.2 & 1 &   8.27e-04 &   1.00 &  0.00 & 15.5 & 2.2\\
HIP 79710 & 1.08 & 245 & 1 & 4.63 & 0.09 &   3.30e-05 &   0.00 &  0.00 & 2.54 & 0.91\\
{*}HIP 79742 & 2.48 & 93.6 & 0.3 & 10.5 & 0.1 &   2.33e-02 &   1.00 &  0.00 & 14 & 2.1\\
HIP 79878 & 1.40 & 203 & 1 & 9.53 & 0.22 &   2.43e-04 &   0.00 &  0.00 & 15.9 & 3.6\\
{*}HIP 79977 & 2.54 & 102 & 0.2 & 11.1 & 0.1 &   4.44e-02 &   1.00 &  0.00 & 11.5 & 2.7\\
HIP 80024 & 1.70 & 169 & 0.1 & 10 & 0.04 &   1.81e-03 &   0.00 &  0.00 & 37.6 & 4.9\\
HIP 82218 & 1.70 & 154 & 2 & 9.54 & 0.47 &   2.74e-04 &   0.00 &  0.00 & 5.3 & 0.71\\

\enddata
\tablenotetext{*}{Fits with formal $\chi_{\nu}^2>2$ but appear 
to be fit well with a single-grain model by visual inspection.}
\end{deluxetable}

\begin{deluxetable}{lcr@{$\pm$}lr@{$\pm$}lcr@{$\pm$}lr@{$\pm$}lr@{$\pm$}lr@{$\pm$}lcr@{$\pm$}lr@{$\pm$}l}
\rotate
\tablecaption{\label{tab:fits2}Fits to two-grain model}
\tablewidth{0pt}
\tabletypesize{\footnotesize}
\tablehead{
  \colhead{HIP ID} & 
  \colhead{$\chi_{\nu}^2$} &
  \multicolumn{2}{c}{$T_1$} &
  \multicolumn{2}{c}{$a_1$} &
  \colhead{mass$_1$} &
  \multicolumn{2}{c}{$f_{o,1}$} &
  \multicolumn{2}{c}{$r_1$}  &
  \multicolumn{2}{c}{$T_2$} &
  \multicolumn{2}{c}{$a_2$} &
  \colhead{mass$_2$} &
  \multicolumn{2}{c}{$f_{o,2}$} &
  \multicolumn{2}{c}{$r_2$}  
  \\
  \colhead{} & 
  \colhead{} &
  \multicolumn{2}{c}{(K)} &
  \multicolumn{2}{c}{($\mu$m)} &
  \colhead{($M\sub{moon}$)} &
  \multicolumn{2}{c}{} &
  \multicolumn{2}{c}{(AU)} &
  \multicolumn{2}{c}{(K)} &
  \multicolumn{2}{c}{($\mu$m)} &
  \colhead{($M\sub{moon}$)} &
  \multicolumn{2}{c}{} &
  \multicolumn{2}{c}{(AU)}
}
\startdata
{*}HIP 53524 & 2.76 & 64.7 & 0.2 & 20 & 0.3 &   2.24e-01 &   1.00 &  0.00 & 42.2 & 0.2 & 413 & 23 & 3.32 & 0.56 &   2.86e-06 &   0.37 &  0.30 & 1.24 & 0.14\\
{*}HIP 55188 & 2.20 & 128 & 1 & 13.1 & 0.3 &   1.03e-02 &   1.00 &  0.00 & 13.2 & 0.7 & 535 & 30 & 3.91 & 0.62 &   4.95e-06 &   0.65 &  0.09 & 1.06 & 0.13\\
{*}HIP 58220 & 2.20 & 169 & 1 & 6.61 & 0.02 &   1.03e-04 &   1.00 &  0.00 & 4.1 & 0.39 & 1500 & 0 & 14.2 & 0.2 &   1.72e-06 &   1.00 &  0.00 & 0.068 & 0.006\\
HIP 58720 & 1.23 & 129 & 1 & 52.2 & 1.8 &   2.89e-02 &   0.32 &  0.14 & 35.5 & 4.1 & 468 & 26 & 6.87 & 1 &   1.49e-05 &   0.66 &  0.22 & 2.94 & 0.47\\
HIP 59502 & 1.20 & 104 & 0.4 & 3.62 & 0.002 &   2.67e-03 &   0.00 &  0.00 & 31.9 & 4.1 & 368 & 5 & 7.96 & 0.38 &   4.21e-05 &   0.95 &  0.14 & 1.98 & 0.26\\
{*}HIP 59898 & 2.73 & 153 & 2 & 7.46 & 0.2 &   1.48e-03 &   0.53 &  0.09 & 15.4 & 2.8 & 524 & 51 & 5.09 & 1 &   7.52e-06 &   0.03 &  0.18 & 1.78 & 0.47\\
${}^\dagger$HIP 60183 & 1.34 & 150 & 3 & 16.1 & 0.7 &   2.65e-03 &   0.35 &  0.12 & 18 & 1.5 & 314 & 56 & 11 & 9.3 &   2.21e-05 &   1.00 &  0.00 & 4.42 & 1\\
HIP 60561 & 0.78 & 134 & 3 & 25.8 & 1.6 &   2.24e-03 &   1.00 &  0.00 & 16 & 2.8 & 413 & 24 & 19.2 & 3.8 &   2.48e-05 &   0.76 &  0.36 & 1.86 & 0.38\\
HIP 61684 & 0.70 & 106 & 1 & 14.4 & 0.5 &   6.85e-03 &   1.00 &  0.00 & 15.4 & 1.1 & 267 & 2 & 6.53 & 0.02 &   1.72e-05 &   0.00 &  0.00 & 2.41 & 0.18\\
{*}HIP 61782 & 2.86 & 115 & 0.3 & 13.4 & 0.2 &   2.93e-02 &   0.73 &  0.08 & 15.2 & 1.2 & 1500 & 0 & 4.1 & 0.12 &   2.12e-07 &   0.00 &  0.00 & 0.205 & 0.02\\
{*}HIP 62657 & 6.13 & 35.9 & 0.5 & 0.539 & 0.05 &   5.58e+00 &   0.00 &  0.00 & 958 & 172 & 119 & 1 & 8.03 & 0.54 &   2.08e-03 &   0.47 &  0.22 & 7.34 & 1\\
HIP 63439 & 1.14 & 69.5 & 0.9 & 15.6 & 1.4 &   3.47e-02 &   1.00 &  0.00 & 24.3 & 2.9 & 405 & 35 & 2.69 & 0.73 &   1.50e-06 &   0.66 &  0.36 & 0.951 & 0.2\\
HIP 63839 & 1.14 & 192 & 3 & 5.67 & 0.16 &   3.20e-04 &   0.17 &  0.10 & 7.82 & 0.98 & 753 & 140 & 4.92 & 2 &   2.26e-06 &   0.47 &  0.14 & 0.817 & 0.32\\
${}^\dagger$HIP 63886 & 0.85 & 69.2 & 2 & 15.7 & 2.9 &   3.72e-02 &   1.00 &  0.00 & 28.9 & 7.9 & 292 & 21 & 6.23 & 1 &   8.30e-06 &   0.00 &  0.00 & 1.66 & 0.5\\
${}^\dagger$HIP 64184 & 1.07 & 99 & 1.3 & 15.6 & 0.9 &   3.51e-02 &   0.10 &  0.20 & 11.3 & 0.5 & 140 & 2 & 2.93 & 0.26 &   5.07e-04 &   0.15 &  0.16 & 7.38 & 0.3\\
HIP 65875 & 1.45 & 107 & 1 & 12.5 & 0.4 &   3.30e-02 &   1.00 &  0.00 & 13.2 & 1.4 & 208 & 7 & 1.86 & 0.39 &   4.43e-05 &   0.00 &  0.00 & 4.59 & 0.55\\
HIP 65965 & 1.20 & 108 & 1 & 16.5 & 0.7 &   1.39e-02 &   0.90 &  0.15 & 27.4 & 3.2 & 367 & 12 & 17.1 & 2 &   5.50e-05 &   0.66 &  0.32 & 2.76 & 0.36\\
HIP 66068 & 1.51 & 146 & 4 & 23.8 & 1.9 &   4.08e-03 &   0.65 &  0.41 & 15.9 & 4.8 & 378 & 20 & 3.91 & 0.81 &   1.34e-05 &   0.67 &  0.21 & 2.65 & 0.84\\
HIP 66566 & 0.98 & 153 & 2 & 14.6 & 0.6 &   1.73e-03 &   0.52 &  0.20 & 10.8 & 1 & 428 & 41 & 4.08 & 1 &   4.22e-06 &   0.00 &  0.00 & 1.67 & 0.35\\
${}^\dagger$HIP 67230 & 0.27 & 161 & 3 & 5.06 & 0.4 &   4.03e-04 &   1.00 &  0.00 & 7.55 & 0.84 & 1500 & 0 & 2.13 & 0.2 &   7.68e-08 &   0.00 &  0.00 & 0.213 & 0.02\\
HIP 67970 & 0.77 & 139 & 2 & 7.44 & 0.31 &   6.67e-04 &   0.65 &  0.30 & 6.4 & 0.69 & 1020 & 240 & 6.31 & 2 &   4.25e-07 &   0.82 &  0.24 & 0.188 & 0.09\\
${}^\dagger$HIP 68080 & 1.02 & 157 & 5 & 23.9 & 2.4 &   2.90e-03 &   0.49 &  0.35 & 22.4 & 3.4 & 682 & 149 & 17.8 & 9.7 &   9.61e-06 &   1.00 &  0.00 & 1.47 & 0.68\\
${}^\dagger$HIP 70149 & 0.85 & 108 & 0.5 & 7.64 & 0.01 &   7.01e-04 &   0.00 &  0.00 & 10.1 & 1.4 & 268 & 8 & 4 & 0.54 &   6.75e-06 &   0.00 &  0.00 & 1.66 & 0.26\\
HIP 70441 & 1.87 & 69.7 & 0.8 & 20.9 & 1.4 &   5.23e-02 &   1.00 &  0.00 & 46.3 & 5.8 & 268 & 5 & 3.9 & 0.26 &   1.50e-05 &   0.00 &  0.00 & 3.27 & 0.42\\
HIP 71271 & 1.13 & 84.6 & 1.4 & 21.9 & 1.9 &   3.69e-02 &   1.00 &  0.00 & 47 & 5.9 & 311 & 14 & 3.96 & 0.66 &   1.07e-05 &   0.00 &  0.00 & 3.74 & 0.56\\
{*}HIP 73145 & 2.24 & 80.4 & 0.3 & 39.1 & 0.7 &   5.60e-01 &   1.00 &  0.00 & 35.1 & 7.5 & 214 & 1 & 4.5 & 0.16 &   2.77e-04 &   0.00 &  0.00 & 4.65 & 1\\
HIP 73990 & 1.63 & 166 & 2 & 9.91 & 0.39 &   8.26e-04 &   1.00 &  0.19 & 6.62 & 0.86 & 1440 & 890 & 2.43 & 2 &   1.42e-07 &   0.46 &  0.23 & 0.173 & 0.22\\
${}^\ddagger$${}^\dagger$HIP 74499 & 0.75 & 83.7 & 0.8 & 12 & 0.7 &   7.45e-03 &   1.00 &  0.00 & 13.1 & 1.6 & 286 & 3 & 3.98 & 0.01 &   3.21e-06 &   0.00 &  0.00 & 1.14 & 0.14\\
HIP 75077 & 0.39 & 63.5 & 0.9 & 30.4 & 2.6 &   4.48e-02 &   0.00 &  0.00 & 79.4 & 14 & 291 & 14 & 4.08 & 0.63 &   5.61e-06 &   0.00 &  0.00 & 3.92 & 0.8\\
${}^\dagger$HIP 75210 & 0.81 & 151 & 4 & 139 & 10 &   2.51e-02 &   0.42 &  0.34 & 26.4 & 3.2 & 477 & 59 & 4.91 & 1 &   6.10e-06 &   1.00 &  0.00 & 2.73 & 0.74\\
{*}HIP 76310 & 2.19 & 116 & 0.4 & 17 & 0.2 &   5.46e-02 &   1.00 &  0.00 & 52.9 & 5.2 & 361 & 12 & 10.8 & 1.3 &   6.72e-05 &   0.40 &  0.27 & 6.09 & 0.71\\
${}^\dagger$HIP 77315 & 0.52 & 153 & 4 & 22.9 & 1.5 &   5.16e-03 &   0.26 &  0.28 & 18.4 & 1.8 & 440 & 48 & 7.23 & 2 &   1.63e-05 &   0.76 &  0.29 & 2.54 & 0.59\\
HIP 77317 & 0.88 & 97.3 & 1 & 54.2 & 3.9 &   6.14e-02 &   0.96 &  0.32 & 35 & 7.8 & 486 & 35 & 12.8 & 2.6 &   6.66e-06 &   0.42 &  0.51 & 1.56 & 0.41\\
HIP 78043 & 0.75 & 78.2 & 1.7 & 20.7 & 2.5 &   2.88e-02 &   1.00 &  0.00 & 23.1 & 4 & 307 & 13 & 6.05 & 0.97 &   9.96e-06 &   0.00 &  0.00 & 1.47 & 0.27\\
HIP 78641 & 1.54 & 131 & 1 & 10.5 & 0.4 &   2.47e-03 &   0.83 &  0.13 & 9.12 & 1 & 366 & 17 & 2.32 & 0.41 &   4.19e-06 &   0.06 &  0.12 & 1.5 & 0.32\\
HIP 78996 & 1.72 & 175 & 1 & 9.93 & 0.01 &   3.61e-04 &   1.00 &  0.00 & 14.1 & 2 & 1500 & 0 & 4.96 & 0.08 &   5.40e-07 &   0.71 &  0.09 & 0.306 & 0.04\\
HIP 79156 & 1.45 & 195 & 3 & 9.06 & 0.39 &   2.55e-04 &   1.00 &  0.00 & 18.9 & 2.9 & 1500 & 0 & 5.29 & 0.11 &   1.19e-06 &   0.89 &  0.09 & 0.476 & 0.07\\
HIP 79410 & 1.20 & 114 & 4 & 8.02 & 0.91 &   1.37e-03 &   0.00 &  0.00 & 57.5 & 9.9 & 387 & 9 & 6.71 & 0.51 &   3.65e-05 &   1.00 &  0.00 & 5.62 & 0.93\\
HIP 79516 & 1.77 & 72.1 & 1.5 & 24 & 2.6 &   2.51e-01 &   1.00 &  0.00 & 24.9 & 4.5 & 128 & 3 & 3.32 & 0.59 &   5.67e-04 &   0.11 &  0.15 & 9.62 & 1\\
{*}HIP 79631 & 2.33 & 76.8 & 0.8 & 35 & 0.2 &   2.90e-01 &   0.09 &  0.36 & 66 & 10 & 205 & 4 & 7.4 & 0.78 &   3.51e-04 &   0.00 &  0.00 & 8.78 & 1\\
${}^\ddagger${*}HIP 80088 & 2.86 & 51.7 & 0.4 & 1.55 & 0.11 &   2.45e-01 &   1.00 &  0.00 & 1000 & 140 & 245 & 3 & 3.4 & 0.17 &   3.10e-05 &   0.00 &  0.00 & 8.37 & 1\\
${}^\dagger$HIP 80320 & 0.84 & 50.6 & 0.9 & 0.979 & 0.16 &   8.23e-02 &   1.00 &  0.00 & 359 & 68 & 337 & 18 & 2.34 & 0.43 &   2.74e-06 &   0.23 &  0.23 & 1.21 & 0.26\\
{*}HIP 80897 & 2.90 & 108 & 1 & 93.5 & 2.2 &   1.75e-01 &   1.00 &  0.00 & 30.2 & 5.1 & 481 & 16 & 36.3 & 3.5 &   6.89e-05 &   1.00 &  0.00 & 1.54 & 0.28\\
HIP 82154 & 0.86 & 98.3 & 2.5 & 22.3 & 2.2 &   4.74e-02 &   0.00 &  0.00 & 75.7 & 19 & 257 & 8 & 6.35 & 0.88 &   1.93e-04 &   1.00 &  0.00 & 11.1 & 2.9\\

\enddata
\tablenotetext{*}{Fits with formal $\chi_{\nu}^2>2$ but appear 
to be fit well with a two-grain model by visual inspection.}
\tablenotetext{\dagger}{Formally well-fit by a single-grain 
model, but but appear to be better fit by a two-grain model by 
visual inspection.}
\tablenotetext{\ddagger}{IRS spectra well-fit by a two-grain 
model, but does not fit the 70 $\mu$m MIPS photometric value.}
\end{deluxetable}

\begin{deluxetable}{lcr@{$\pm$}lr@{$\pm$}lcr@{$\pm$}lr@{$\pm$}lr@{$\pm$}lr@{$\pm$}lcr@{$\pm$}lr@{$\pm$}l}
\rotate
\tablecaption{\label{tab:fits0}
Objects not fit well by single or two-grain models
}
\tablewidth{0pt}
\tabletypesize{\footnotesize}
\tablehead{
  \colhead{HIP ID} & 
  \colhead{$\chi_{\nu}^2$} &
  \multicolumn{2}{c}{$T_1$} &
  \multicolumn{2}{c}{$a_1$} &
  \colhead{mass$_1$} &
  \multicolumn{2}{c}{$f_{o,1}$} &
  \multicolumn{2}{c}{$r_1$}  &
  \multicolumn{2}{c}{$T_2$} &
  \multicolumn{2}{c}{$a_2$} &
  \colhead{mass$_2$} &
  \multicolumn{2}{c}{$f_{o,2}$} &
  \multicolumn{2}{c}{$r_2$}
  \\
  \colhead{} & 
  \colhead{} &
  \multicolumn{2}{c}{(K)} &
  \multicolumn{2}{c}{($\mu$m)} &
  \colhead{($M\sub{moon}$)} &
  \multicolumn{2}{c}{} &
  \multicolumn{2}{c}{(AU)} &
  \multicolumn{2}{c}{(K)} &
  \multicolumn{2}{c}{($\mu$m)} &
  \colhead{($M\sub{moon}$)} &
  \multicolumn{2}{c}{} &
  \multicolumn{2}{c}{(AU)}
}
\startdata
HIP 63975 \\
one-grain
 & 78.77 & 478 & 0.2 & 4.54 & 0.004 &   8.45e-04 &   1.00 &  0.00 & 2.51 & 0.31
 & \multicolumn{2}{c}{\nodata} & \multicolumn{2}{c}{\nodata} & \nodata & \multicolumn{2}{c}{\nodata} \\
two-grain
 & 49.00 & 101 & 0.06 & 2.62 & 0.0003 &   8.06e-02 &   0.49 &  0.01 & 74.6 & 9.1 & 1110 & 10 & 2.5 & 0.02 &   6.68e-05 &   1.00 &  0.00 & 0.65 & 0.08 \\
HIP 79288 \\
one-grain
 & 793.62 & 361 & 0.1 & 4.38 & 0.01 &   5.49e-04 &   0.00 &  0.00 & 1.35 & 0.2
 & \multicolumn{2}{c}{\nodata} & \multicolumn{2}{c}{\nodata} & \nodata & \multicolumn{2}{c}{\nodata} \\
two-grain
 & 138.68 & 100 & 0.1 & 2.04 & 0.02 &   4.71e-02 &   0.51 &  0.01 & 33.9 & 5.1 & 822 & 4 & 3.56 & 0.03 &   4.56e-05 &   0.48 &  0.00 & 0.418 & 0.06 \\

\enddata
\end{deluxetable}

\clearpage

\begin{figure}[bt]
  \centerline{\includegraphics[width=3.5in]{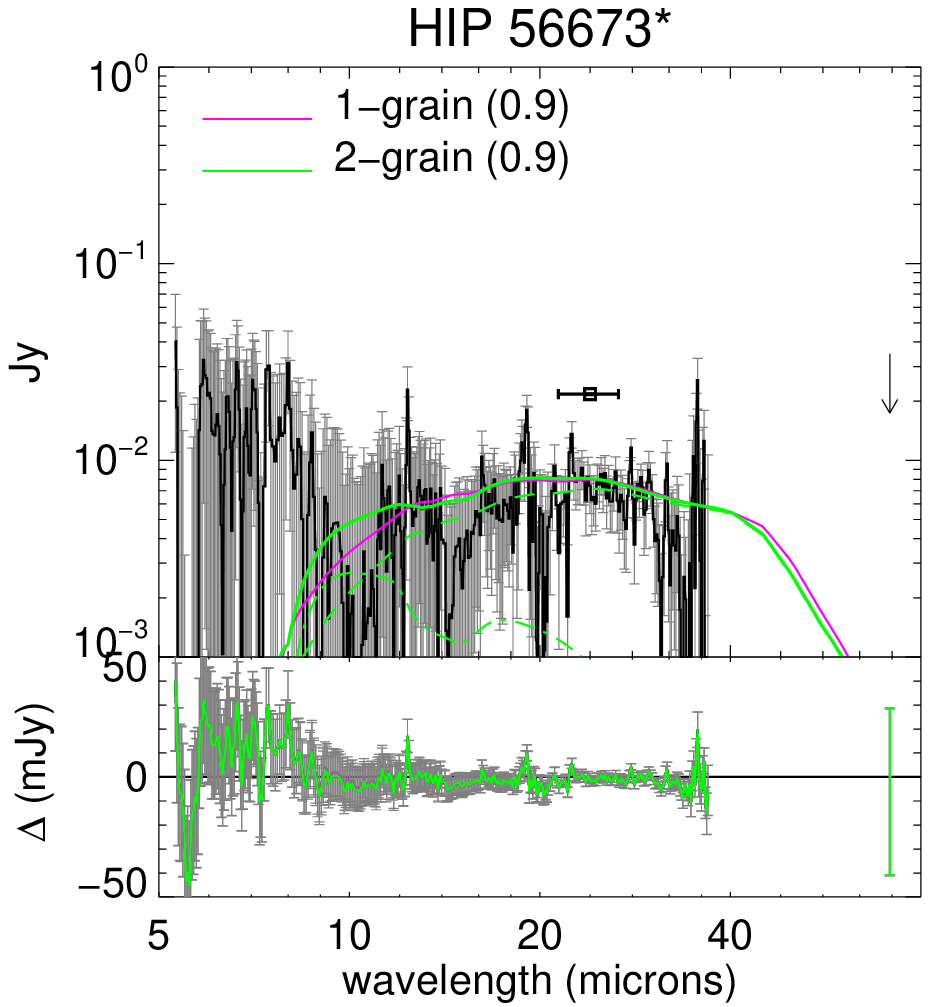}}
\caption{\label{fit_hip56673.eps}
Best fitting single grain and two-grain models to HIP 56673, with 
the photosphere fit to the 5-6 micron region of the IRS spectrum.
We find that a single component model best fits this spectrum.
}
\end{figure}

This count does not include HIP 56673 or HIP 78977.  
As discussed previously, these sources have excesses 
similar to Rayleigh-Jeans profiles, consistent with either a 
photosphere mismatch or a hot dust component.  If the photosphere is 
scaled to match the 5-6 micron region of the IRS spectrum, then 
HIP 78977 is a non-excess source, and HIP 56673 has a marginal excess 
that can be modeled well as a single belt.  In Figure \ref{fit_hip56673.eps}, 
we show the best fit to HIP 56673, after fitting the photosphere 
to the IRS spectrum instead of calibrating the photosphere 
to optical and near-IR photometry.  

Two of our targets, HIP 63975 (HD 113766) and HIP 79288 (HD 145263) 
have high amounts of excess and show evidence of silicate features 
that cannot be adequately reproduced with the simple models used in this 
paper.  The mineralogy of HIP 63975 was analyzed in detail in 
\citet{2008Lisse+}, and was found to consist of both amorphous and 
crystalline silicates, as well as Fe-rich sulfides, amorphous carbon, and 
water ice.  HIP 79288 has a similarly complex composition, and requires
silica in addition to other species to adequately model it
(Lisse, et al., in prep).

HIP 74499 and HIP 80088 both appear to require at at least two components 
to fit their IRS spectra: 
a large cold component to fit the 20-40 micron region, and a small hot 
component to fit the 10 micron region.  
However, in both these cases, the best fit to the IRS spectrum does 
not produce enough emission at 70 micron to match the MIPS photometry.  
It is likely that these two systems have an additional third cold component 
that would account for the missing 70 micron excess.
Both these sources have been classified as two-belt systems for 
further analysis.  

Recently, a debris belt has been imaged around one of our targets,
HIP 64995 (HD 115600), with a semi-major axis of $\sim48$ AU\@.
\citep{2015Currie+}.  Our modeling predicts a single belt of debris
at a temperature of 114 K, or $\sim11$ AU\@.  This belt would be
interior to the inner working angle of the coronagraphic image of 
the belt detected by \citet{2015Currie+}.  If the outer belt is an analog
of the Kuiper belt, then the belt predicted by the mid-infrared spectroscopy
would be an asteroid belt analog.
This example illustrates how mid-IR spectral analysis is complementary
to coronagraphic imaging for studying the structure of debris disks.  

\begin{figure}[p]
\begin{center}
\includegraphics[width=4in]{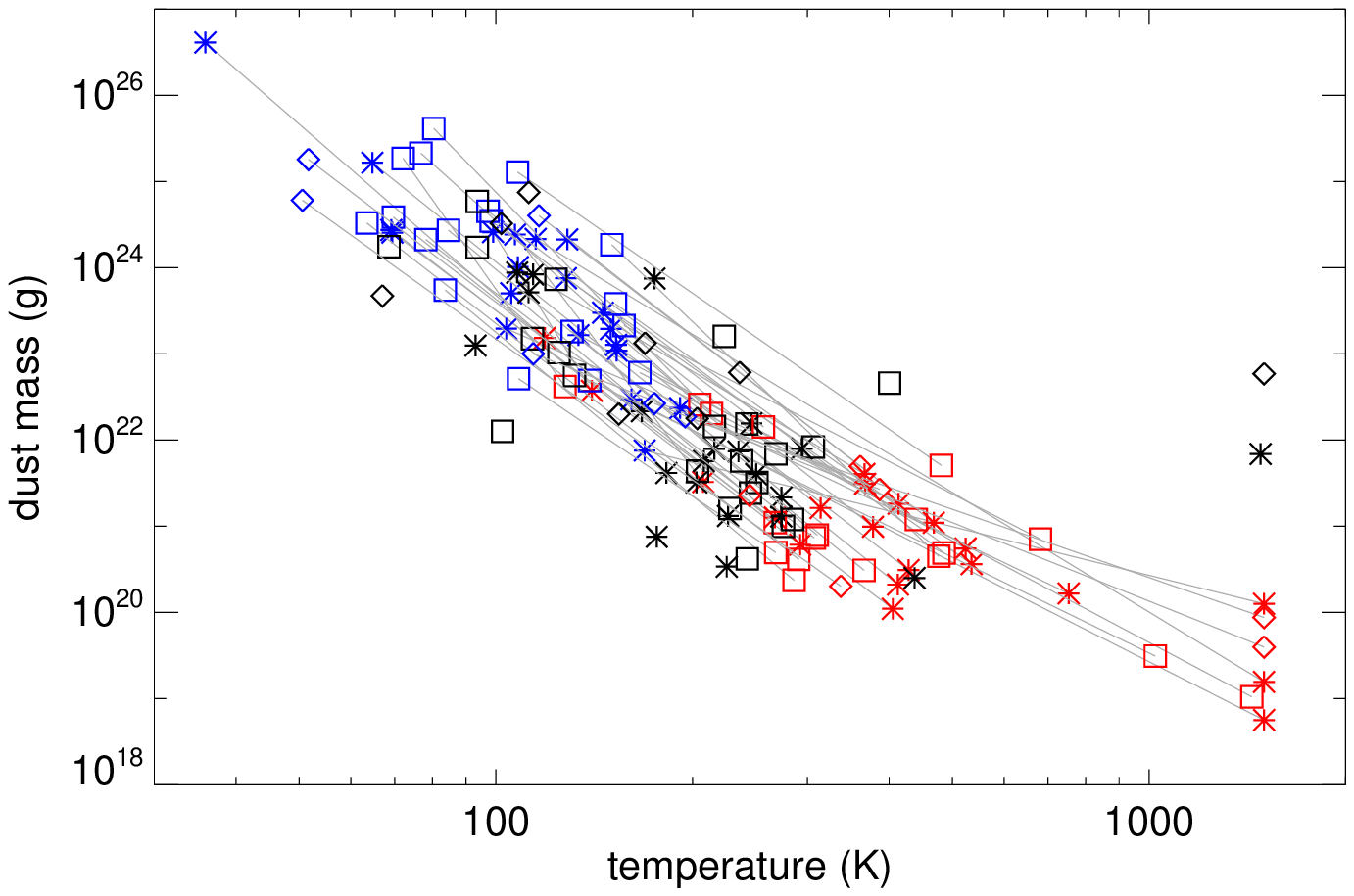}\\
\includegraphics[width=4in]{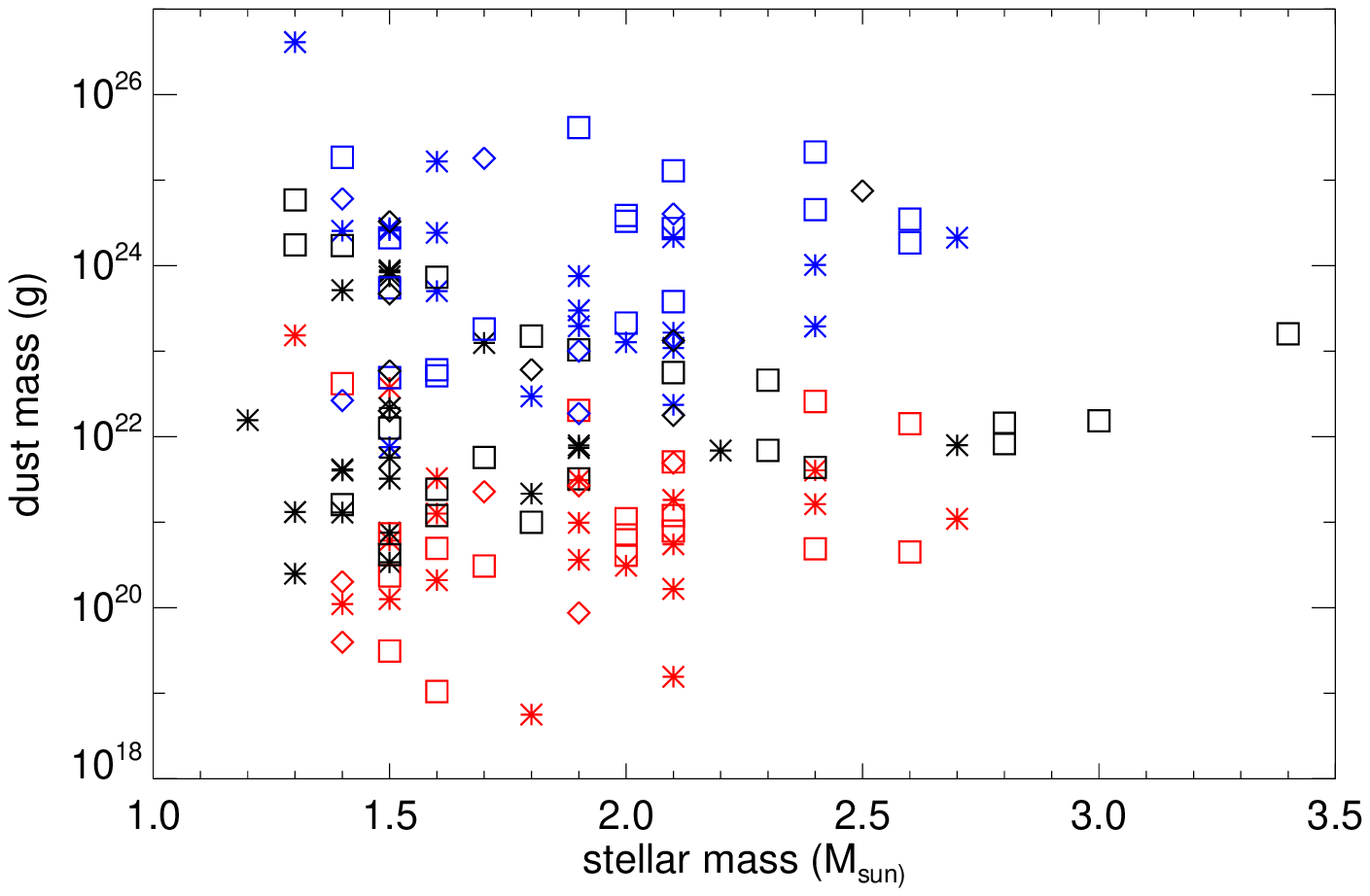}
\end{center}
\caption{\label{fig:dustmass}Dust belt masses
versus dust temperature (top) and stellar mass (bottom), 
for the single grain population disks (black)
and the two-grain population disks (red/blue).  
In the two-grain population model, the cold component is shown in 
blue while the hot component is shown in red.  
Objects belonging to LCC, UCL, and USco are labeled by 
asterisks, squares, and diamonds, respectively.
}
\end{figure}

In Figure \ref{fig:dustmass}, we show the masses of the dust belts 
in the single-belt (black) and two-belt (red/blue) debris disk systems.  
For the two-belt systems, the hot and cold belts are colored red and 
blue, respectively.
Dust mass appears to be inversely correlated to temperature. 
This is likely a selection effect, because more cold material 
must exist in order for it 
to contribute significantly to the infrared 
excess.  That is, a lower temperture blackbody 
emits less radiation total than a higher temperature one, 
holding the emitting surface area constant.  
There does not appear to be a trend in measured dust mass versus 
stellar mass.  This implies that the trend that 
$L\sub{IR}/L_*$ decreases with increasing stellar mass 
(see Figure \ref{fig:lirlstar}) is best explained by 
differences in dust temperature
(i.e.~distance from the system primary) 
rather than differences in measured dust mass.  

\begin{figure}[p]
\plottwo{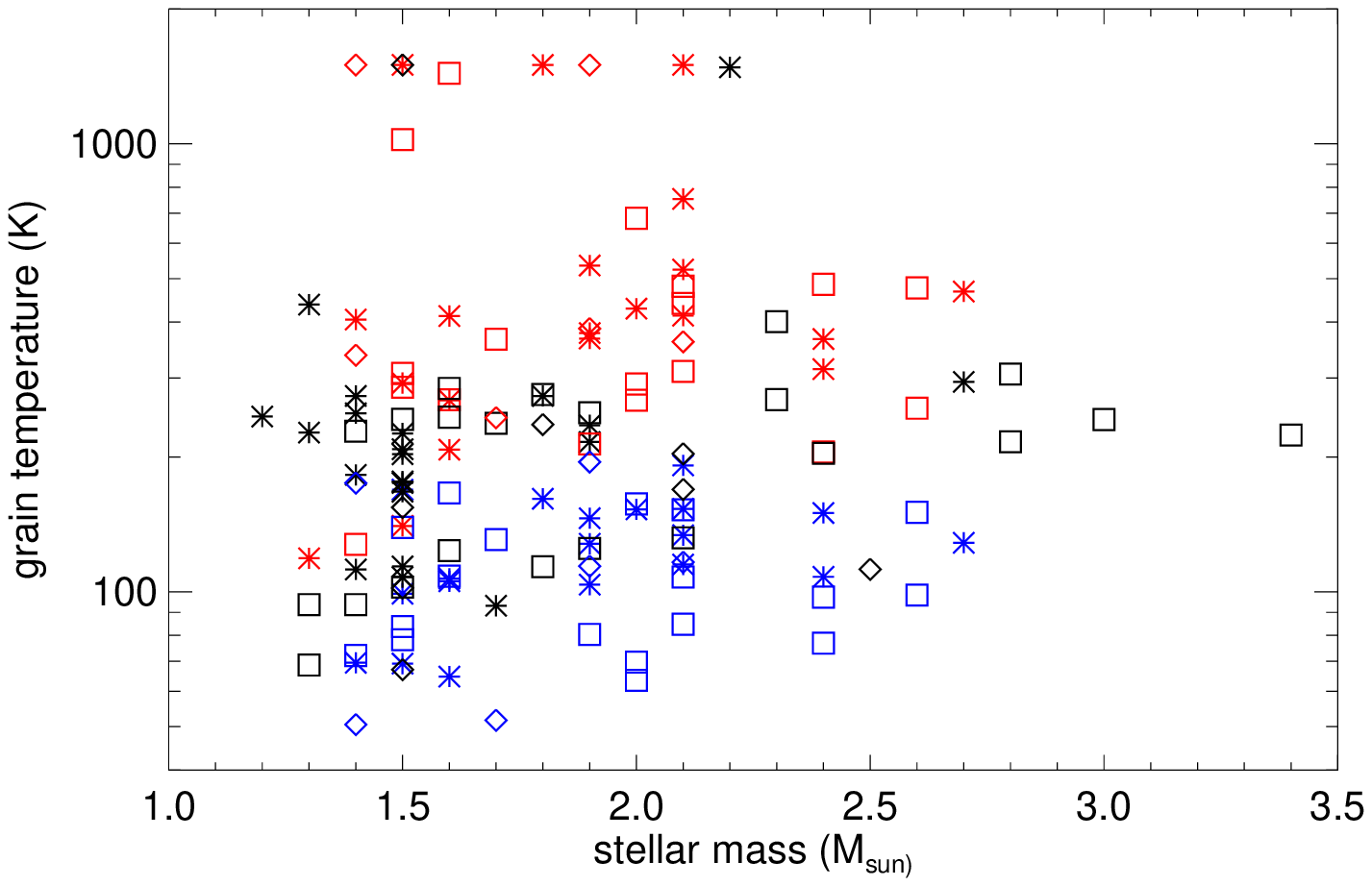}{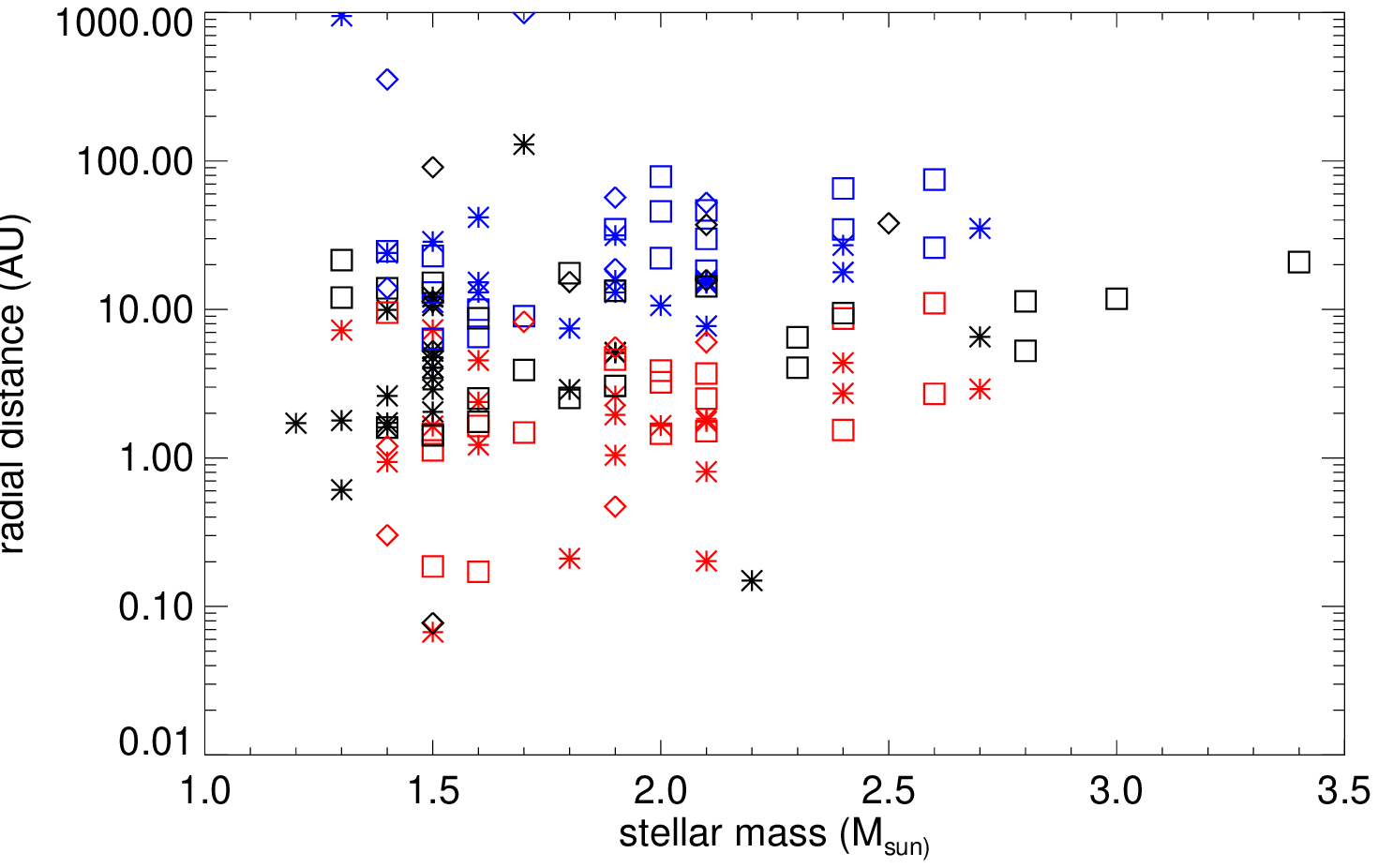}\\
\plottwo{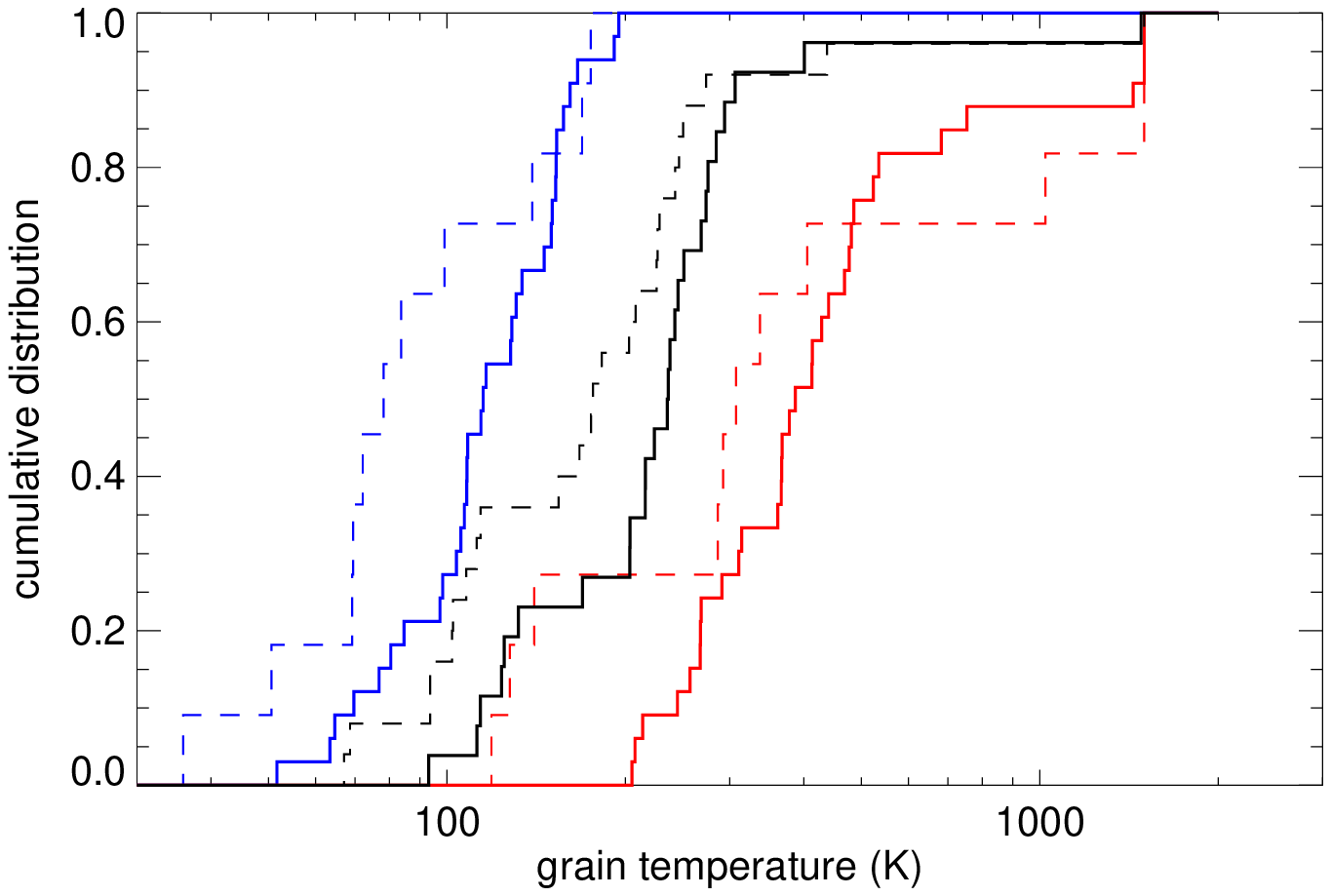}{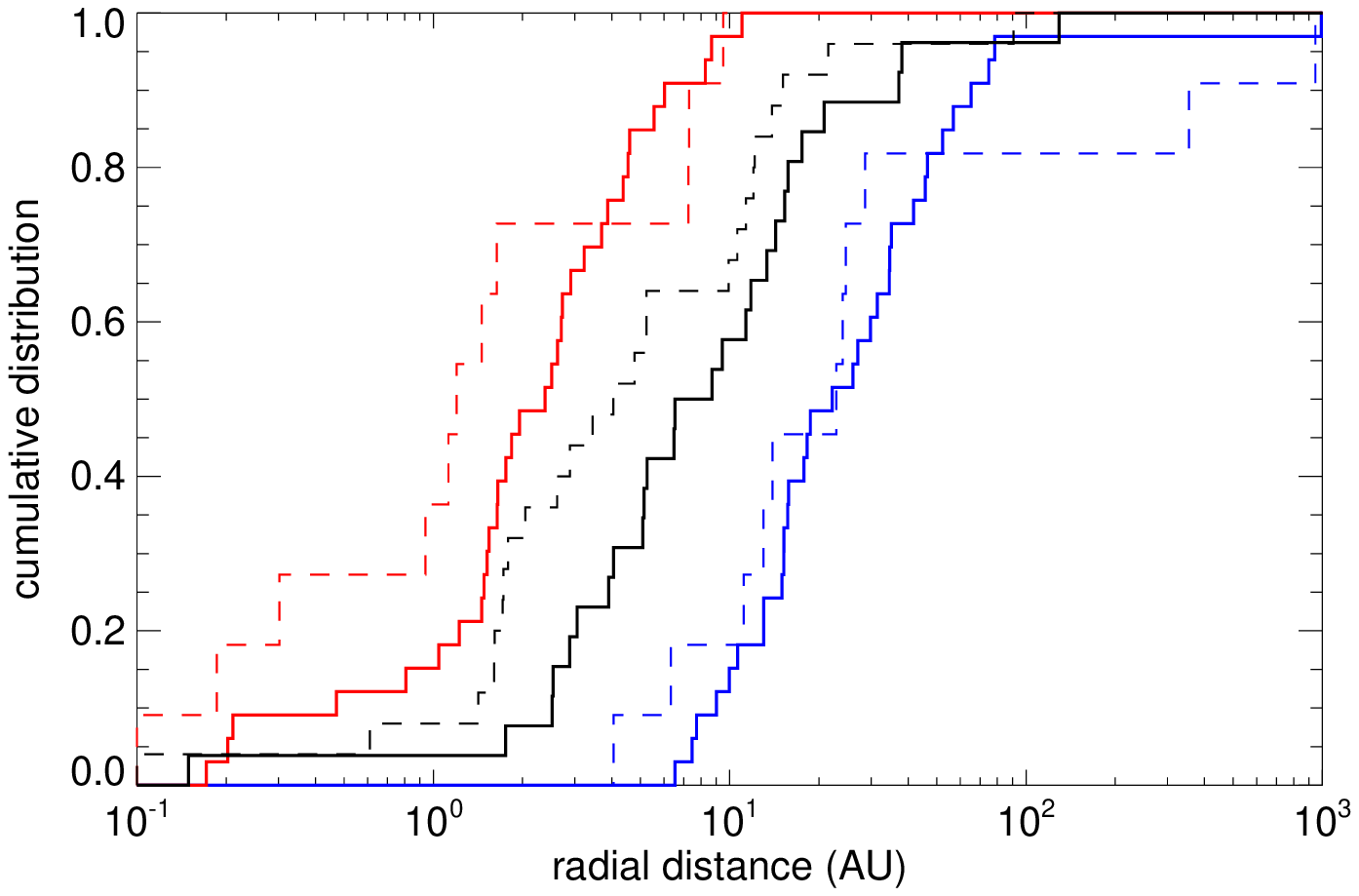}
\caption{\label{fig:graintemprad}Temperature and radial distribution of 
grain populations in single belt (black) and two-belt (blue/red)
systems.  For the two-belt systems, 
the hot component is plotted in red, the cold component in blue.
Top row: $T\sub{gr}$ (left) and $r$ (right)
versus stellar mass.  
Objects belonging to LCC, UCL, and USco are labeled by 
asterisks, squares, and diamonds, respectively.
Bottom row: Cumulative temperature distribution (left) 
and radial distribution (right) 
for the dust components of best fit disk models.  
The solid lines show stars with mass $\geq1.6$ $M_{\sun}$, 
while the dashed lines show stars with mass $\leq1.5$ $M_{\sun}$.
}
\end{figure}

In Figure \ref{fig:graintemprad}, we summarize the 
results of our models in terms of the temperatures and 
radial distance of the grain populations.  In general, the temperatures 
of the single-belt models are intermediate between 
the hot and cold components of the two-belt models.  
There appears to be more scatter in the hot components 
for lower stellar masses, which is more easily seen 
in the cumulative distributions in temperature after 
separating them out into low mass ($\leq1.5\,M_{\odot}$, dashed lines) 
and high mass ($\geq1.6\,M_{\odot}$, solid lines) stars.  

The same discrepancy between high-mass and low-mass stars 
appears in the distribution of radii of the belts, as shown 
in the right panel of Figure \ref{fig:graintemprad}.  
The belt radii are an average distance for the olivine and pyroxene 
components of the grains.  
The model spectra assume that the olivine and pyroxene 
components of the dust all have the same equilibrium 
temperatures.  The grain properties are similar enough between 
olivine and pyroxene that grains of the same temperature 
are effectively co-located as well, and the radius derived 
for pure olivine is nearly the same as that of pure pyroxene.  
Since in many cases the composition is 
either pure pyroxene or pure olivine, the difference in the 
distances of the two components is insignificant.

The grain temperatures of dust around high mass stars are 
systematically higher than around low mass stars.  
This agrees with the finding in \citet{2014Chen_etal}.
In Table \ref{tab:stats}, we show the results of the
Kolmogorov-Smirnov (KS) and Anderson-Darling (AD) statistics
for testing the similarity of the grain temperatures between
high and low mass stars.  We consider the single belt
systems, the cold component of the two-belt systems,
and the hot component of the two-belt systems separately.  
The Anderson-Darling test \citep{1954AndersonDarling}
is arguably a more sensitive statistic than the KS test because it
gives more weight to the the tails of the distributions. 
To test whether or not 
two samples $X_1, \ldots, X_n$ and $Y_1, \ldots, Y_m$
are drawn from the same populations,
we can use the two-sample Anderson-Darling statistic, given by
\begin{equation}
  A_{nm}^2 = \frac{1}{mn} \sum_{i=1}^{N-1} \frac{(M_iN-ni)^2}{i(N-i)}
\end{equation}
where $N=m+n$ and $M_i$ is the number of $X$'s less than or equal
to the $i$th smallest element in the combined sample 
\citep{1976Pettitt}.

\begin{table}[p]
  \centering
  \caption{\label{tab:stats}
    Kolmogorov-Smirnov and Anderson-Darling tests for
    grain populations}
  \begin{tabular}{lrrrr}
    \hline\hline
    population & \multicolumn{2}{c}{KS test}
    & \multicolumn{2}{c}{AD test} \\
    & $D$ & $p(D)$ & $A_{mn}^2$ & $p(A)^\dagger$ \\
    \hline
    Temperatures \\
    $\ldots$ single belt &
    0.33 &   0.097 &
    2.5 & $<0.05$ \\
    \ldots cold component & 
    0.45 &   0.045 &
    2.6 & $<0.05$ \\
    $\ldots$ hot component &
    0.30 &   0.37 &
    1.5 & $>0.15$ \\
    Distances \\
    $\ldots$ single belt &
    0.28 &     0.22 &
    2.1 &     $<0.10$ \\
    $\ldots$ cold component & 
    0.24 &     0.65 &
    0.93 &     $>0.15$ \\
    $\ldots$ hot component &
    0.39 &     0.12 &
    1.9 &     $<0.10$ \\
    \hline
  \end{tabular}
  \\$^\dagger$Based on tabulations of probabilities in
  \citet{1974Stephens}.
\end{table}

The KS test for the single belt systems 
and the cold component of the two belt systems give a
10\% and 4\% or less probability, respectively, 
of being drawn from the same population, 
while the probability of the hot components being similar 
is 37\%.  
The AD test gives comparable results: $<5\%$ probabilities
for the single belt systems and cold components,
but higher probability of similarity for the hot component ($>15\%$).
A potential explanation for this systematic shift 
is that higher mass stars heat the dust more than lower 
mass stars do, but that is not is fully borne out when we examine 
the distributions with respect to dust distance.

It appears likely that the cold grain populations for both 
the high mass and low mass stars are at similar distances:
the KS test gives a probability of 65\% that they are 
drawn from the same population.  However, both the single belt
systems and the hot component of the two-belt systems 
are systematically offset to larger distances in the 
high mass systems, despite being hotter.  
The single belt systems have a 22\% probability of being similar, 
while the probability for the hot component of the two belt systems 
being similar is 12\%.
Again, the AD test gives similar results: it is unlikely that the
single belts and hot components of low mass stars are drawn from
the same population as the high mass stars ($<10\%$).
However, it is much more likely that the cold components are similar
($>15\%$).  

The trend in $L_{IR}/L_*$ seen in Figure \ref{fig:lirlstar} also
supports the idea that lower mass stars have dust at closer radii.
This is because the infrared excess luminosity, which is attributed to
the dust, scales as $L_{IR} \propto M_d T_d^4$, where
$M_d$ and $T_d$ are the total mass and temperature of the dust, respectively.
The temperature of the dust is determined by stellar illumination, as 
$T_d^4 \propto L_*/d^2$ where $d$ is the stellocentric distance.
This gives $L_{IR}/L_* \propto M_d /R^2$.  If $M_d$ stays constant,
then the trend that lower mass stars have higher $L_{IR}/L_*$
implies that the disks in lower mass stars are more compact.  

This implies that that the low mass stars retain 
close-in dust more readily than high mass stars, suggesting 
that debris disks in high mass stars evolve faster than 
low mass stars, and that this evolution occurs inside out.  
One explanation for this is that debris disks evolve faster in 
high mass stars because the dynamical times are shorter 
\citep{2008KenyonBromley}.  Another possibility is that 
the higher mass stars (F-type and earlier) evolve onto 
the main sequence sooner.  By 15-17 Myr, the ages of LCC 
and UCL, these stars are already on the main sequence.  
The ignition of hydrogen burning in these stars could 
enable the clearing out of inner dust belts at 
$\lesssim1$ AU\@.
The retention of dust by low-mass stars could also simply be
the result of more efficient grain blow-out by more massive
stars, as indicated by their larger values of $a\sub{min}$
[Eq.~(\ref{eq:amin})].
Still another possibility is that 
the initial protoplanetary gas disk differs between high mass
and low mass stars.  



\section{Discussion}

The high degree of scatter in the distances of the dust belts 
indicates that they are not likely connected to
intrinsic properties of the primordial 
disks from which they arose.  Primordial disks
generally have continuous 
radial distributions of material rather than belts.  Material 
could potentially pile up at pressure maxima, for example, but 
the origin of the pressure maxima depends on stellar 
properties such as effective temperature and luminosity,
which should be relatively stable for a given stellar mass.
An example of this phenomenon is the T Tauri object
HL Tau, which has been seen
to have several gaps in ALMA imagery \citep{2015HLTau}.
Although it is possible that these gaps were created by planets, 
locations of the gaps also appear to be co-incident with
condensation fronts in the disk \citep{2015Zhang+}.
The locations of the fronts are based on the predicted 
temperature profile of the disk, which in turn depends on the
heating of the disk from stellar irradiation
\citep[see e.g.~][]{1987KenyonHartmann}.
  
Alternatively, dust locations could depend
on the formation of planets,
which is highly stochastic in regard to stellocentric distances.
The varying diameters of the inner holes seen in transitional disks
are often attributed to planet formation for this reason. 
Planets could also 
explain the origin of the dust belts after the dissipation 
of the primordial gas, since planets could be responsible 
for shepherding the parent bodies that produce the dust.  

Stellar companions could also affect the dust belts in our disks.  
A high fraction of stars in our sample have identified stellar companions.  
A few stars have directly imaged substellar companions that may be distant 
planets.  Since planets and binary companions are likely to play 
key roles in the sculpting of debris disks, it is important 
to put the properties of dust belts into 
context with the presence of binarity and the presence of planets.

\subsection{Binary Stars}

A number of stars in our sample have been identified to have binary companions
\citep{2013Janson_etal,2012Chen_etal,2005Kouwenhoven_etal,2007Kouwenhoven_etal}.
The projected distances of these binary companions are tabulated 
in Table \ref{tab:binaries}.  
The distances are calculated from the angular separations 
from the above references and using Hipparcos stellar distances 
from \citet{2007vanLeeuwen}.

\begin{deluxetable}{ccccc}
\tablecaption{\label{tab:binaries}
Projected Distances of Binary Companions}
\tablehead{
    \colhead{HIP} &
    \colhead{HD name} &
    \colhead{$a_1$ (AU)} &
    \colhead{$a_2$ (AU)} &
    \colhead{$a_3$ (AU)} 
}
\tablewidth{0pt}
\startdata
53524 & HD 95086 &      440.3 &   \nodata  &   \nodata \\
58220 & HD 103703 &      75.17 &   \nodata  &   \nodata \\
58528 & HD 104231 &      492.8 &   \nodata  &   \nodata \\
59502 & HD 106036 &      296.1 &      911.4 &   \nodata \\
59693 & HD 106389 &      58.49 &   \nodata  &   \nodata \\
63236 & HD 112383 &      837.2 &      1326. &   \nodata \\
63839 & HD 113457 &      427.4 &      595.4 &      627.2\\
65965 & HD 117484 &      1514. &   \nodata  &   \nodata \\
68080 & HD 121336 &      268.5 &      3902. &   \nodata \\
69291 & HD 123889 &      195.4 &   \nodata  &   \nodata \\
77315 & HD 140817 &      100.1 &      5552. &   \nodata \\
77317 & HD 140840 &      4126. &   \nodata  &   \nodata \\
77520 & HD 141254 &      223.5 &   \nodata  &   \nodata \\
78756 & HD 143939 &      1249. &   \nodata  &   \nodata \\
79400 & HD 145357 &      734.2 &   \nodata  &   \nodata \\
79631 & HD 145880 &      376.0 &      1132. &   \nodata \\
82154 & HD 151109 &      1869. &   \nodata  &   \nodata \\

\enddata
\end{deluxetable}

Stellar companions will truncate a circumstellar disk through tidal 
interactions.  
For a debris disk, where gas has little to no dynamical effect, 
the disk truncation radius can be estimated from the 
last stable orbit.  Assuming a circular orbit, 
the outermost radius of a circumstellar 
disk allowed by a binary companion can be expressed as 
\begin{equation}\label{eq:aint}
a\sub{int} = (0.464-0.380\mu)a
\end{equation}
and the inner edge of a circumbinary disk is 
\begin{equation}
a\sub{ext} = (1.60+4.12\mu)a
\end{equation}
\citep{1999HolmanWiegert}, where $\mu$ is the ratio of the
mass of the binary companion to the total masses of the two stars, 
and $a$ is the semi-major axis of the binary orbit.    

\begin{figure}[bt]
  \centerline{\includegraphics[width=4in]{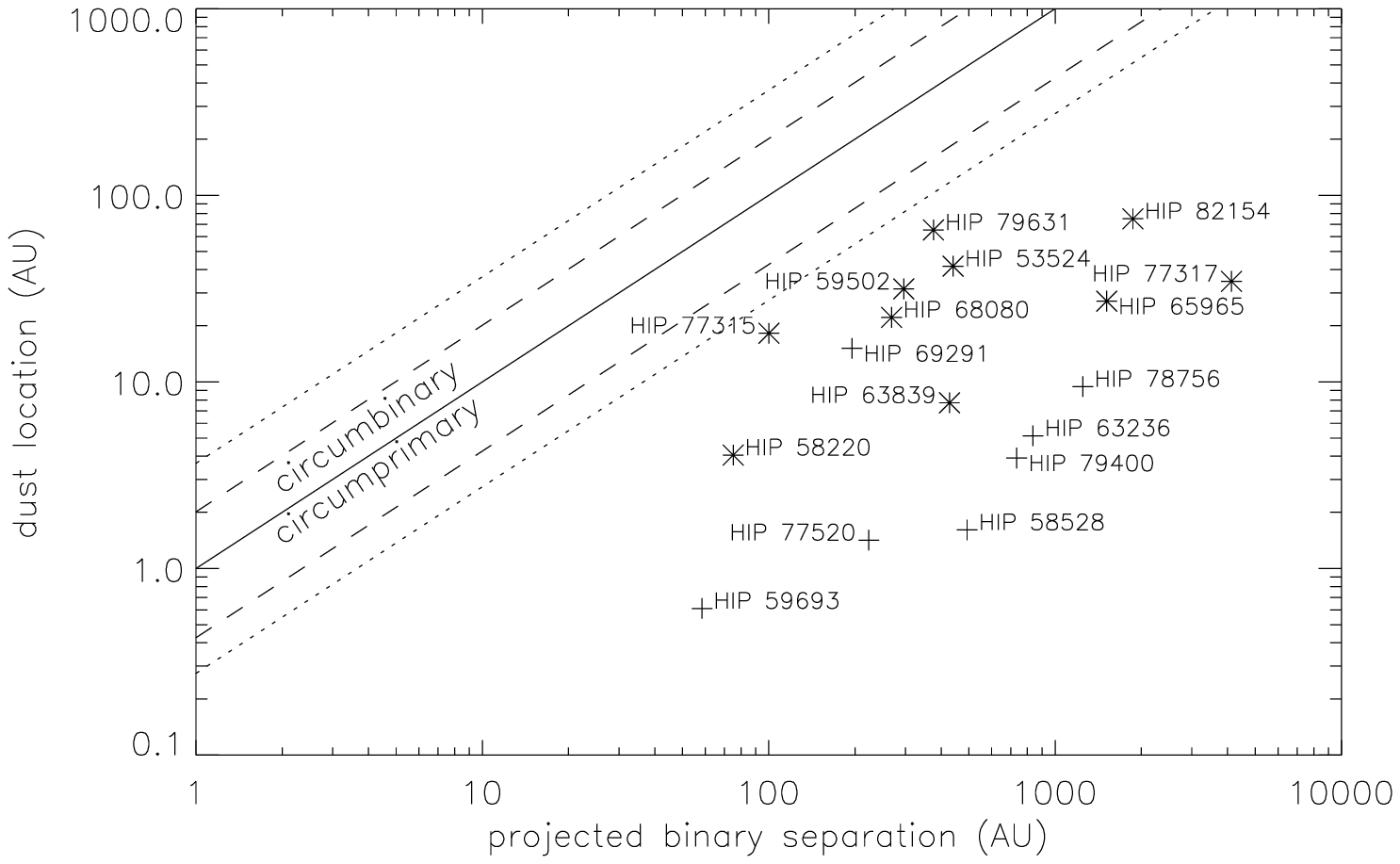}}
\caption{\label{fig:binaryvsbelt}
Outer dust belt location versus projected binary separation.
The solid line marks the 1:1 line, so circumstellar disks 
lie below the solid line while circumbinary disks lie above 
the solid line.  The dotted and dashed lines mark the disk truncation 
radii for an equal-mass binary and a $\mu=0.1$ binary, 
respectively.  
Crosses indicate single belt systems, and asterisks indicate the outer 
belt of two-belt systems.  Crosses indicate systems not well-modeled 
by either single or two-belt systems, but using the best fitting 
two-belt model regardless.  
}
\end{figure}

In Figure \ref{fig:binaryvsbelt}, we show the outer 
location of the best-fit dust belts versus the binary separation 
for those objects with binary companions. 
If the dust is best fit with a single grain population, then 
that distance is used.  For the remaining systems, the location 
of the outer belt in the two-grain fit is used.  We also indicate 
the disk truncation radius for binaries with mass ratio 0.5 
(equal mass binary) and 0.1 assuming a circular orbit.

In all cases, the dust is located interior to the binary separation, 
so the binary companion must have truncated
the disks in all these systems.  
A few dust belts appear close to the truncation radius of a 
$\mu=0.1$ binary companion, however the projected binary separation 
is affected by the inclination of the system and eccentricity of 
the orbit, so the true truncation radius might be higher.  
No circumbinary dust belts in our sample of stars has been identified. 

\subsection{Stars with planets}

Three of our sources have been identified to be planet hosts.  
HIP 53524 (HD 95086) is a binary star \citep{2012Chen_etal} 
around which a $\sim5\,M\sub{Jup}$ planet has been detected 
at a projected separation of 56 AU by direct imaging 
\citep{Rameau2013}.  
HIP 59960 (HD 106906) hosts a very distant planetary-mass companion 
at a projected separation of $\sim650$ AU, also detected by direct 
imaging \citep{Bailey2014}. 
HIP 73990 (HD 133803) has at least two planets in its system, 
at projected separations of 20 and 32 AU
\citep{2015Hinkley+}.  

HIP 53524 is best fit by a two-belt model, though 
its relatively high reduced chi-square value ($\chi_{\nu}^2=2.8$)
suggests that a few more addition parameters, such as 
finite belt widths or the addition of crystalline silicates, 
could improve the model.  It hosts a stellar companion at 
440 AU in addition to the imaged planet.  
The positions of the belts are 42 AU (65 K) and 1.2 AU (413 K), both 
interior to the projected position of the planet.  
Since the outer belt is close to the projected planet distance 
of 56 AU, it is likely that it is sculpted by the planet.  

Herschel images of HIP 53524 marginally resolve the disk, suggesting 
that the systems is surrounded by a halo that extends to 
around 800 AU \citep{2014Su_DPS}.  That analysis included 
Spitzer/MIPS data and proposed that in addition to the halo, 
the systems consists of a warm belt at 175 K and a cold belt at 55 K\@. 
The wavelength coverage of Spitzer does not 
provide much sensitivity to emission from the 
cold halo, but the 55 K belt is consistent with the model 
for HIP 53524 presented here.  Our model predicts a hotter 
inner belt, which may be a result of 
not including the third 
coldest dust distribution that is inferred from the Herschel data.  

In our analysis, HIP 59960 is well-fit by a single dust belt at 12 AU, and 
its reported planetary companion \citep{Bailey2014}
is so distant that it does not interact with 
the dust.  It is possible that additional unseen companions 
exist between the belt and the planetary companion and that 
those companions could shephered the dust.  

HIP 73990 is best fit to a two-belt model with grain temperatures
of 1440 K and 166 K, corresponding to distances of
0.17 and 6.6 AU, respectively.  Assuming that planets are able to clear
material out to the 2:1 mean motion resonance, then the inner planet
imaged at 20 AU \citep{2015Hinkley+}
should truncate the debris disk to 12.6 AU.  Both dust belts
detected from our spectroscopic modeling are interior to this distance.

Additional objects in ScoCen that have detected sub-stellar companions 
include HIP 78530, which exhibits no infrared excess; 
GSC 06214-00210, an accreting T Tauri star;
and 1RXS J160929.1-210524. 
\citep{Bailey2013}.  
These objects were not included in our study because they are 
pre-main sequence stars and their disks are protoplanetary in nature.  

Although imaging planets around debris disk systems can help 
us understand the role that planets play in sculpting debris 
disks, few such planet images exist.  On the other hand, many 
debris disks are well-studied.  We can turn the question around, 
then, and ask what can be learned about planet formation 
from debris disks.  Dynamical interactions with planets should 
sculpt and shephered debris disks.  Therefore, debris disks 
with structure, such as gaps, can imply the presence of unseen planets.

Numerical simulations have shown that a companion orbiting in a disk can create gaps via planetesimal scattering in overlapping resonances \citep[e.g.,][]{Roques1994, LecavelierdesEtangs1996}. The width of the gap is related to the mass of the companion by a power law \citep[e.g.,][]{Quillen2006, Chiang2009, Rodigas2014}, the parameters of which depend on the age of the system and the optical depth of the disk \citep{Nesvold2014}.

Following the procedure described in \citet{Nesvold2014}, we analyzed the two-belt systems in our sample to place upper limits on the mass of a possible single perturbing companion in each system. In each case, we assumed that a single body on a circular orbit equidistant in log semimajor axis between the two dust bands has cleared the gap between the bands. We used $L_{IR}/L_{\star}$ as a proxy for the face-on optical depth of each disk.

\citet{Nesvold2014} found that the largest gap size that a single body on a circular orbit can create has a full width of $\Delta r/r = 1.6$. Larger bodies tend to stir the disk; destroying it and widening it, while roughly preserving the gap edge near the location of the 2:1 mean motion resonance.  Nine disks in our sample had gaps narrower than this maximum width. Table (\ref{tab:planetmass}) summarizes inferred companion masses and semimajor axes for these nine disks. 

\begin{table}
\centering
\caption{Masses and semimajor axes of companions inferred from the disk gaps.
\label{tab:planetmass}}
\begin{tabular}{ccc}
HIP & Companion Semimajor Axis (AU) & Max Companion Mass ($M_{Jup}$) \\
\tableline
60561 & 5.4 & 28.8 \\
61684 & 6.4 & 12.2 \\
65089 & 22.9 & 33.1 \\
66068 & 5.9 & 10.4 \\
66566 & 4.2 & 15.3 \\
78641 & 3.7 & 9.8 \\
82154 & 28.7 & 42.9 \\
65875 & 7.7 & 1.1 \\
79516 & 15.3 & 0.8 
\end{tabular}
\end{table}

\begin{figure}[bt]
  \centerline{\includegraphics[width=4in]{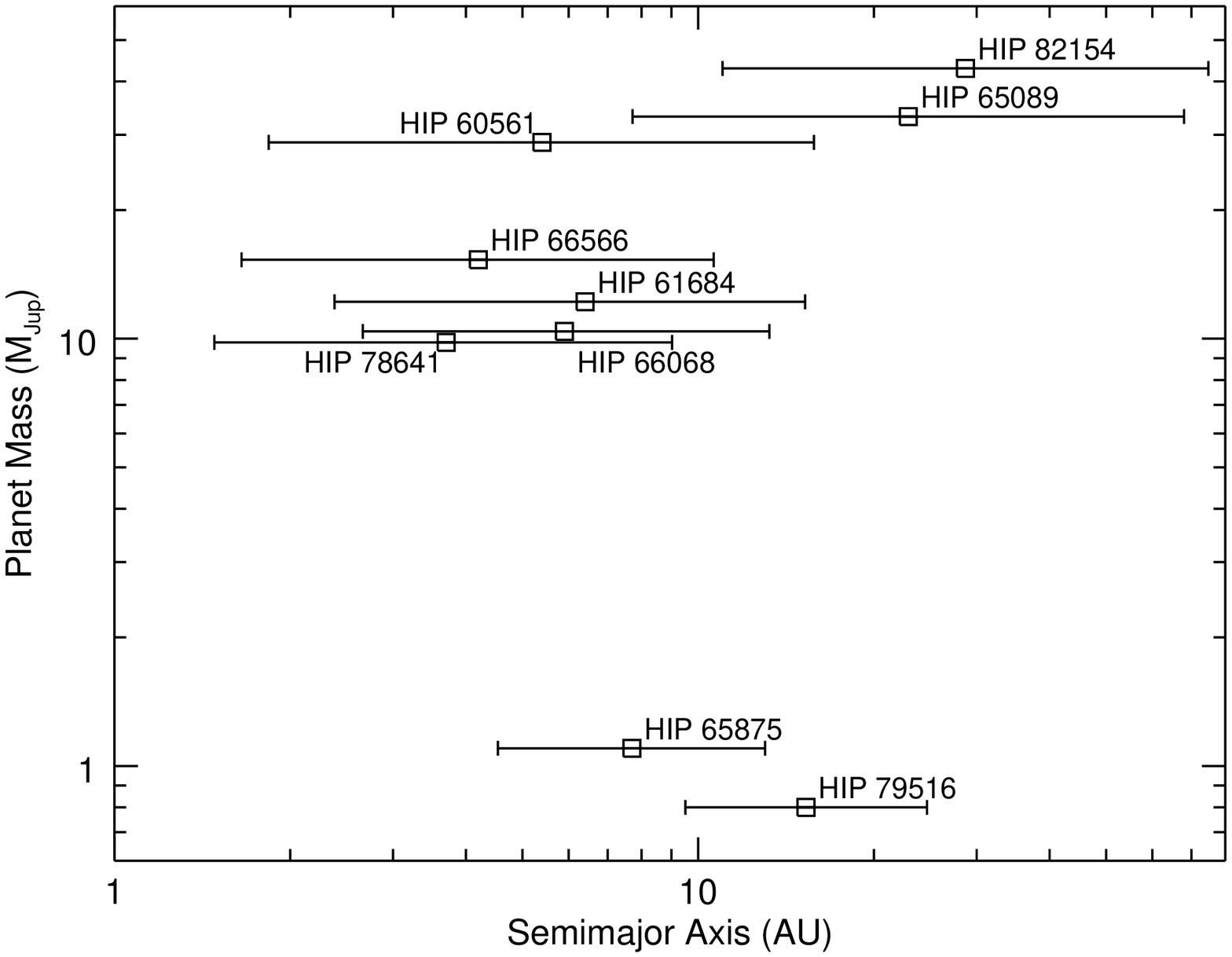}}
\caption{\label{fig:planets}
Maximum companion mass vs. estimated companion semimajor axis for the nine companions inferred from the disk gaps. The bars indicate the widths of the gaps between belts.  
}
\end{figure}

Five of these systems, HIP 61684, HIP 66068, HIP 78641, HIP 65875, and 
HIP 79516, have small enough gaps that they require a single 
perturbing body whose mass is in the range of planet masses. The other 
disks require either planets on eccentric orbits, companions in the 
brown-dwarf mass range or multiple planets. The inferred companions in 
these systems are all located within 0.2 arcseconds of their host 
stars, probably too close to detect directly with today's instruments.  
However, future observatories, such as WFIRST-AFTA, may be 
able to resolve sub-stellar companions at angular separations 
of less than 0.2\arcsec.  In addition, the next generation of large 
aperture ground-based telescopes, such as GMT, TMT, or E-ELT, 
could have the necessary resolving power and inner working angle.  
ATLAST, currently a NASA strategic mission concept study, could 
also detect these planets.  

HIP 82154 also has a binary companion at a projected distance of 1869
AU (see Table \ref{tab:planetmass}).  This companion is probably too distant to 
have any dynamical effect on the dust belts or any planet located 
between them.  

Six out of the nine inferred companions listed in Table
(\ref{tab:planetmass}) have estimated semimajor axes within $5 \pm 3$
AU\@, close to Jupiter's semi-major axis of 5.2 AU\@.  This lends credence 
to the idea that giant planets preferentially form in the 5-10 AU range.   
This is also consistent with statistics of exoplanets showing that, 
apart from hot Jupiters, the number of exoplanets increases 
toward orbital separations larger than 1 AU\@.  

\clearpage

\section{Conclusions}

Constraining our study of debris disks to those in ScoCen gives 
us the advantage of examining a cohort of debris disks of similarly 
young age.  Within this sample, 
we find evidence of mass-dependent evolution 
of the hot dust.  In particular, in systems with two belts, 
low mass stars have closer inner belts than high mass stars.  
This implies that high mass stars have less hot dust than low mass 
stars.  This could be related to the faster evolution times of 
the higher mass stars, which results in higher mass stars reaching 
the main sequence sooner than the less massive stars.  
Then the lack of hot dust could be explained by the clearing of the 
dust from the inside out as the star evolves.  

We explored how stellar and sub-stellar companions could sculpt debris 
disks.  Many of the objects in our sample have known binary companions.  
We find that the dust distances from our models predict circumstellar 
disks rather than circumbinary disks.  However, because the binaries 
are generally wide, and our data is limited to infrared 
wavelengths, our observations are not sensitive to 
any thermal emission that might come from circumbinary disks.  
The dust distances are consistent with disk truncation at 
outer radii by the binary companion.  

Two of our objects host known planets.  These planets have been 
detected by direct imaging, so they are distant planets outside 
the dust disks.  The fact that the planets that have been discovered 
in ScoCen also host debris disks suggest that debris disks and 
planet formation are correlated.  Planets can also sculpt debris 
disks, and we use this fact to consider the possibility that 
our disks might host planets.  

The locations of dust belts can put constraints on the locations 
of planets since they should clear out gaps in the disks.  
The two-belt systems found in our study could be the result of 
one or several planets carving out gaps.  If the distance 
ratios are small, then the location and mass of the potential planet 
can be narrowly constrained.  These systems are 
particularly good targets for follow-up planet searches.  

Wider gaps could be created by multiple planets or eccentric orbits.  
These are also good targets for follow-up, although 
predictions about planet propeties are less well-constrained.

\acknowledgements
This work is based on observations made with the Spitzer Space
Telescope, which is operated by JPL/Caltech under a contract with
NASA. Support for this work was provided by NASA through an award
issued by JPL/Caltech.
HJ-C acknowledges support from NASA grant NNX12AD43G.  

\bibliographystyle{apj}
\bibliography{scocen}

\end{document}